\documentclass[11pt, a4paper, fleqn]{book}
\usepackage[english,ngerman]{babel}
\usepackage[dvips]{graphicx}
\usepackage{amssymb}
\usepackage{amsmath}
\usepackage{mathrsfs} 
\usepackage[margin=0pt,font=footnotesize,labelfont=bf]{caption}
\usepackage{pdflscape}

\usepackage[table]{xcolor}

\newcommand{\ud}{\mathrm{d}}
\newcommand{\NLTE}{\mathrm{NLTE}}
\newcommand{\LTE}{\mathrm{LTE}}
\newcommand{\TE}{\mathrm{TE}}
\newcommand{\pd}[2]{\frac{\partial #1}{\partial #2}}
\renewcommand{\l}{\lambda}
\newcommand{\m}{\mu}

\renewcommand{\div}[2]{\frac{#1}{#2}}
\newcommand{\phx}{{\texttt{PHOENIX}}}

\begin{document}
\frontmatter
\begin{titlepage}
\setlength{\headheight}{15pt}
\begin{center}
\vspace{2.5cm}
\Huge{  \textbf{Massive NLTE models for X-ray novae with \phx}}\\
\vspace{5.0cm}
\huge{\textbf {Dissertation \\ zur Erlangung des Doktorgrades \\ des Fachbereichs Physik \\ der Universit{\"a}t Hamburg}}\\
\vspace{3.5cm}
\Large{\textbf{vorgelegt von \\[0.3cm] {\LARGE{Dani\"el R. van Rossum}} \\[0.3cm] aus Soest} }\\
\vspace{2.0cm}
\LARGE{\textbf{Hamburg \\ 2009}}
\end{center}
\end{titlepage}

{\phantom{test}}
\vspace{0.5\textheight}
\begin{tabular}{l l}
Dissertationsgutachter: & Prof. Dr. P. Hauschildt\\
			& Prof. Dr. S. Starrfield\\
                        & Prof. Dr. K. Werner\\
			&\\
Disputationsgutachter:  & Prof. Dr. J. Schmitt\\
                       & Prof. Dr. G. Wiedemann\\
		       & \\
Datum der Disputation: & 27. Oktober 2009\\
                       & \\
Vorsitzender des Pr{\"u}fungsausschusses: & Dr. R. Baade\\
			& \\
Vorsitzender des Promotionsausschusses: & Prof. Dr. R. Klanner\\
			& \\
Dekan des Departments Physik: & Prof. Dr. H. Graener\\
%
\end{tabular}
\newpage

{\phantom{test}}
\cleardoublepage
\section*{Abstract}
X-ray grating spectra provide much spectral detail from classical nova outbursts.
They supplied the confirmation of continued mass loss from the nova in the late super-soft source (SSS) stage of the outburst.

It is not clear a priori, what the influence of the mass loss on the spectrum is, apart from causing systematic blue shifts in the absorption lines.
In order to answer this question, and to test whether it is safe to neglect this aspect of expansion in model atmospheres for novae in the SSS stage, physically consistent models for expanding nova atmospheres have been developed in this work.

The very high temperatures of these models combinded with high expansion velocities and the accompanying large radial extension make nova in the SSS phase very interesting objects but also physically complicated objects to model.
In this work the general purpose radiative transport code \phx, designed to deal with expanding atmospheres, has been chosen for modeling X-ray novae.

\phx\ has been used for this type of objects before, but careful ana\-lysis of the old results lead to a number of new methods and improvements to the code, being the main achievement of this work.
Firstly, essential improvements to the physical treatment of NLTE have been made, including: new opacity expressions, a new rate matrix solver, a new global iteration scheme, and a new temperature correction method.
Secondly, a new hybrid hydrostatic-dynamic nova atmosphere setup has been implemented.
Thirdly, NLTE models are treated in pure NLTE, without LTE opacities.
Also, the models have been made faster to compute by at least a factor 10.

With the new framework a modest amount of models, limited by computation time, have been calculated.
These models show that systematic results are achieved from the framework for various atmospheric conditions.
They also show, that the influence of the expanding shell on the model spectrum is important and that the model spectra are sensitive to the details of the atmospheric structure.
The nova models are compared to the 10 well-exposed X-ray nova grating spectra presently available: 5$\times$ V4743 Sgr 2003, 3$\times$ RS Oph 2006, and 2$\times$ V2491 Cyg 2008.
Although the models are on a coarse grid and have not yet been tuned to the observations they do match surprisingly well.

Also, a comparison with hydrostatic models is made.
The reproduction of the data is clearly inferior to the expanding models.
But what is more important is that the interpretation of the data with hydrostatic models leads to conclusions that are opposite to those with expanding models, e.g. the former requires a sub-solar O abundance and the later a super-solar.

The models give the ability to derive accurate constraints on the physical conditions in the deep layers of novae that are visible only in the SSS phase.

\clearpage
\selectlanguage{ngerman}
\section*{Zusammenfassung}
\small
Die grating Spektren von klassischen Novae enhalten viel spektrale Information.
Sie haben bestaetigt, dass der Massenverlust weiterhin stattfindet im spaeten SSS Stadium des Nova-Ausbruches.

Man kann nicht vorhersagen, wie der Massenverlust das Spektrum beeinflusst, ausser dass Absorptionslinien blauverschoben sind.
Um diese Frage zu klaeren, und ob man den Aspekt der Expansion wirklich vernachlaessigen kann in Model Atmosphaeren fuer Novae in der SSS-Phase, wurden physikalisch konsistente Modelle fuer expandierende Nova-Atmosphaeren entwickelt.

Sehr hohe Temperaturen, hohe Expansionsgeschwindigkeiten und die damit einhergehende starke radielle Ausdehnung machen Novae in der SSS-Phase zu sehr interessanten, aber auch physikalisch schwer modellierbaren Objekten.
In dieser Arbeit wuerde der general-purpose Strahlungstransportcode \phx\ verwendet, der fuer expandierende Atmosphaeren entwickelt wurde, um X-ray Novae zu modellieren.

Fuer diese Objekte is \phx\ schon eher benutzt worden, aber die gruendliche Analyse der alten Modelle fuehrte zu einer Anzahl neuer Methoden und Verbesserungen am Code, die den Hauptbestandteil dieser Arbeit ausmachen.
Die physikalische Behandlung von NLTE wurde grundlegend verbessert mit
1) neuen Opazitaeten, einem neuen Rate Matrix Solver, einem neuen globalen Iterationsschema und einer neuen Temperatur-Korrketur-Methode;
2) der Implementation eines neuen hydrostatisch-dynamischen Nova Atmosphaerenaufbaus;
3) der Behandlung von Modellen in purem NLTE, ohne LTE Opazitaeten;
und 4) der Beschleunigung der Modellberechnung wurde zum Faktor 10.

Mit dem neuen Code wurde eine kleine Menge an Modellen berechnet, eingeschraenkt durch Rechenzeit.
Diese Modelle zeigen, dass der neue Code systematische Ergebnisse liefert fuer mannigfaltige atmosphaerische Zustaende.
Auch zeigen sie, dass die expandierende Huelle einen wichtigen Einfluss auf das Modellspektrum haben, und dass das Spektrum empfindlich ist fuer die atmosphaerische Struktur.
Die Modelle werden verglichen mit den zehn vorhandenen gut belichteten X-ray Nova Beobachtungen: 5$\times$ V4743 Sgr 2003, 3$\times$ RS Oph 2006, and 2$\times$ V2491 Cyg 2008.
Obwohl das Modellgitter grob ist und noch nicht getuned wurde auf die Beobachtungen, ist die Uebereinstimmung ueberraschend gut.

Auch wurden hydrostatische Modelle verglichen.
Diese reproduzieren die Daten deutlich schlechter als die expandierenden.
Was aber noch wichtiger ist, ist dass die Interpretation der Daten mit hydrostatischen Modellen zu Schlussfolgerungen fuehrt, die denen mit expandierenden Modellen widersprechen.
Zum Beispiel findet man, dass diese eine sub-solare O Haeufigkeit erfordern und jene eine super-solare.

Die Modelle ermoeglichen es, praezise Einschraenkungen abzuleiten ueber die physikalischen Bedingungen in den tiefen Schichten einer Nova, die sich nur zeigen in der SSS-Phase.
\normalsize
\selectlanguage{english}

\clearpage
\tableofcontents

\mainmatter

\clearpage
\chapter{Introduction} \label{sec:Introduction}
The phenomenon of novae has long been known, the name ``nova stella'' being given by Tycho Brahe (1572).
The basic principles responsible for the observed behavior of novae have firstly been clearly formulated in \cite{Grotrian37}.
The light originates from a continuously expanding atmosphere where the optical properties are determined by the density and velocity fields.
Classical reviews are \cite{Payne57}, \cite{McLaughlin57} and \cite{Gallagher78}.
Numerical models based on the theory of a thermonuclear runaway that cause a nova explosion, developed by \cite{Starrfield87}, were found to be in excellent agreement with the optical and UV observations available at that time.
A recent collection of work on the topic of classical novae is \cite{Bode08}.
The following short summary is based on these publications and the references therein.

\section{Novae: the classical theory} \label{sec:NovaTheory}
The commonly accepted model for a nova is a close binary system with one member being a white dwarf and the other a larger, cooler star that fills its Roche lobe.
Because it fills its lobe, any tendency for it to grow in size, because of evolutionary processes or for the lobe to shrink because of angular momentum losses, will cause a flow of gas through the inner Lagrangian point into the lobe of the white dwarf.
The size of the white dwarf is small compared to the size of its lobe and the high angular momentum of the transferred material causes it to spiral into an accretion disk surrounding the white dwarf.
Some viscous process, as yet unknown, acts to transfer mass inward and angular momentum outward through the disk so that a fraction of the material lost by the secondary ultimately ends up on the white dwarf.
Over a long period of time, the accreted layer grows in thickness until the bottom reaches a temperature that is high enough to initiate thermonuclear fusion of hydrogen.
Given the proper conditions a thermonuclear runaway (TNR) will occur at the bottom of the accreted hydrogen-rich envelope.

Once the energy released by the thermonuclear reactions has reached the surface of the white dwarf its effective temperature rapidly increases.
With a luminosity of the order of the Eddington luminosity and the radius of a white dwarf, temperatures at peak will exceed $10^6$ K, depending on the mass of the white dwarf \cite{Starrfield96}.
The nova becomes a very strong X-ray emitter with a soft spectral energy distribution of a hot stellar atmosphere.
After ejection of the nova shell, which is expanding and cooling adiabatically, the temperature of the expanding effective photosphere rapidly drops.
When hydrogen starts to recombine the X-ray flux decreases rapidly as the expanding envelope becomes opaque to X-rays within a few hours.
At this stage, called the `fireball phase', the maximum in visual luminosity is reached.

After the rise to visual maximum the shell continues to expand so that the density and opacity decrease and the effective photosphere moves inward.
In the initial outburst only a part of the accreted material is ejected. 
The part of material that doesn't reach velocities higher than the escape velocity slows down, and gradually turns back into hydrostatic equilibrium on top of the white dwarf surface, providing enough fuel for hydrogen burning during the so-called `phase of constant bolometric luminosity'.
With the constant luminosity and the effective photosphere contracting the effective temperature rises to a level where a second phase of X-ray emission begins, the so called `supersoft source phase'.

At some point in time the hydrogen burning turns off.
The atmosphere of the white dwarf is then expected to cool at constant radius, and as a consequence the X-ray luminosity would drop.
However, this `turn-off phase' is, as yet, the least studied and least understood.
A possible explanation could be that due to mass losses in a stellar wind the burning shell becomes too thin to sustain the hydrogen burning.

\begin{figure}
 \centerline{\includegraphics[height=.97\textwidth,angle=90]{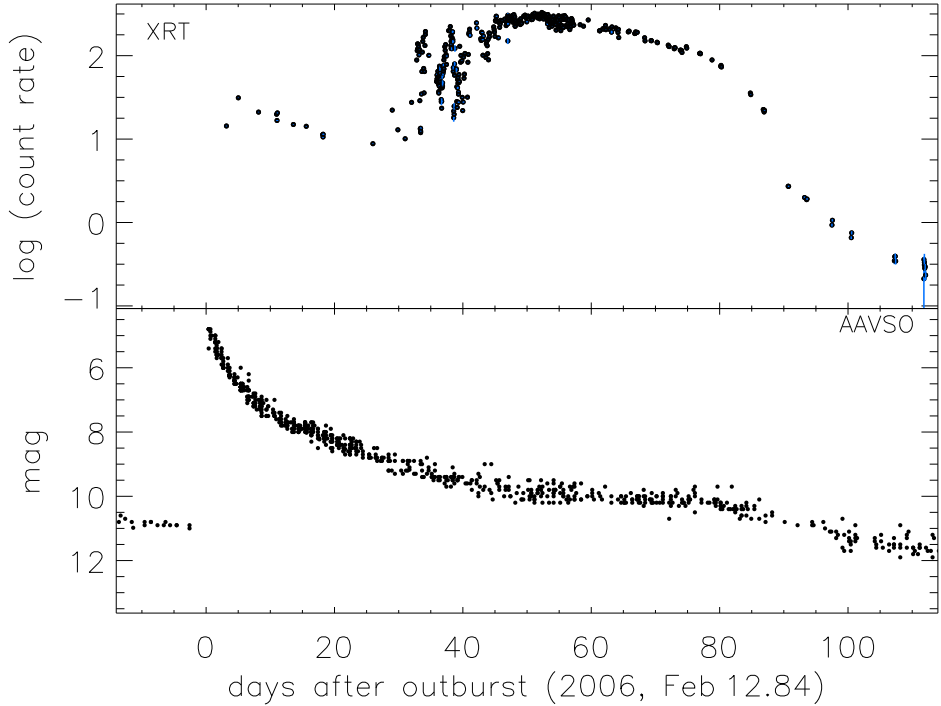}} 
 \caption{ \label{fig:RSOphLightcurve}
 Light curves of the classical nova RS Oph 2006 for X-ray (Swift XRT) and optical (AAVSO amateurs) wavelength bands.
 Unfortunately, the simultaneous UV observations made with Swift are overexposed.
 About 32 days after visual maximum the X-ray count rate increases strongly, and is initially very variable.
 Around day 45 it stabilizes soon after which it starts to decline gradually. (from \cite{JanUwePrivat})
 }
\end{figure}

\begin{figure}
 \centerline{\includegraphics[height=.97\textwidth,angle=90]{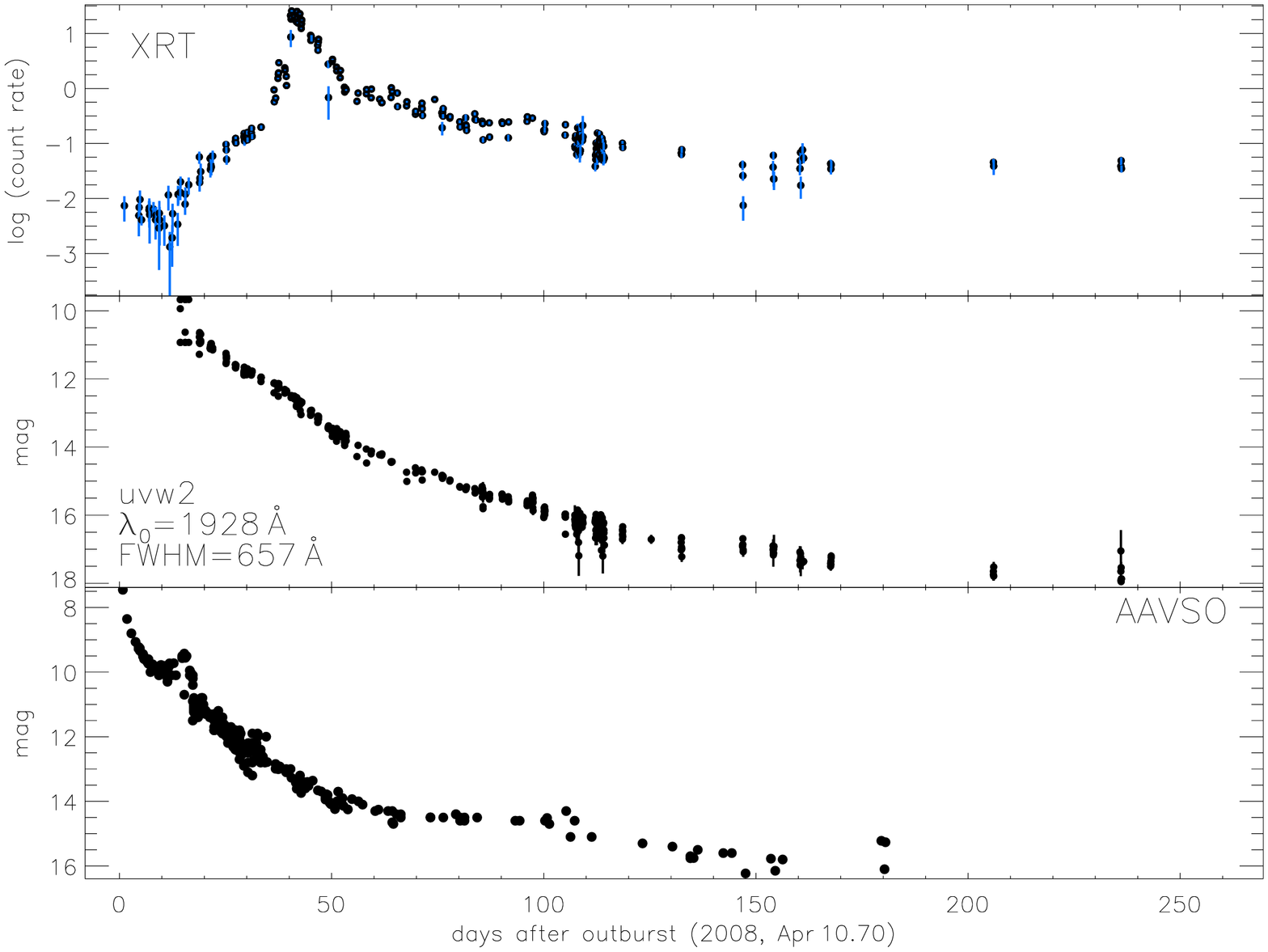}} 
 \caption{ \label{fig:V2491Lightcurve}
 The simultaneous lightcurves of V2491 Cyg 2008 have a spectacular time resolution and accuracy for three wavelength bands.
 The supersoft X-ray phase lasts from day $\sim$36 to day $\sim$53.
 Note the steep decline in the X-ray lightcurve after its maximum and the difference to RS Oph 2006 in figure \ref{fig:RSOphLightcurve}.
 (from \cite{JanUwePrivat})
 }
\end{figure}

Two examples of lightcurves measured in different wavelength bands for recent novae are shown in figures \ref{fig:RSOphLightcurve} and \ref{fig:V2491Lightcurve}.

\section{The advent of `high-res' X-ray spectra}
Before the advent of the `high-resolution' X-ray spectra from the gratings on board XMM-Newton and Chandra theory and observations were in excellent agreement.
The lightcurves could be reproduced with evolution models based on a TNR process \cite{Starrfield87} and the optical and UV spectra were reproduced with radiative transport models \cite{Hauschildt95}.
A classification of objects could be made based on the time scales of typical developments in the lightcurves of novae (the speed classes).
Differences in chemical compositions for the speed classes could be determined from the spectra and these classes could be related to distinct initial conditions for the TNR.

With the very unexpected results from the grating spectra this picture changed.
Many of the details revealed in these X-ray spectra cannot yet be explained by the established theory.
So far, each new X-ray observation of a nova revealed new facets and no nova has shown the same behavior as preceding ones.
Therefore, no classifications could yet be made, which gravely complicates the development of theoretical descriptions.

\section{This work} \label{sec:ThisWork}
One important facet in the development of the theory is detailed spectral analysis.
The comparison of synthetic spectra with observational data can yield a lot of insight in the structure of the atmosphere that produces the radiation.
\phx\ is a general-purpose, state-of-the-art stellar and planetary atmosphere code \cite{Hauschildt92}, \cite{Hauschildt99}, \cite{Hauschildt03}, \cite{Hauschildt04}.
It can calculate atmospheres and spectra of stars all across the HR-diagram including main sequence stars, giants, white dwarfs, stars with winds, TTauri stars, Novae, Supernovae, brown dwarfs and extrasolar giant planets.
The universal applicability makes the code well tested for a large diversity of atmospherical conditions.
\phx\ is one of the most advanced atmosphere codes that is capable of calculating expanding, optically thick atmospheres.

\phx\ has been applied before to X-ray novae for the object V4743 Sgr \cite{Petz05}.
In the process of finding suitable models for other nova observations (RS Oph 2005 and V2491 Cyg 2008) a number of new methods and improvements to the code have been developed.
With these new developments a framework is obtained from which it is possible to compute physically realistic, massive NLTE models for geometrically extended and expanding nova atmospheres.
This has been the major achievement of this work.
In the little time left after completion and verification of the framework, a small number of models have been computed.
Although just a few first results were obtained from the new models, the agreement with the high-res X-ray observations is good.
Also, large improvements in the spectral fits to observations of V4743 Sgr are obtained with respect to results from the old code.

After an overview of the theory of radiation transport and ``stellar'' atmosphere modeling (chapter \ref{sec:RadiationTransport}), the new construction procedure of the nova atmospheres used in this work is described (chapter \ref{sec:NovaStructure}).
Then a detailed description of the new methods developed in this work is given (chapter \ref{sec:GoodNLTE}) and the first results obtained with the new models are shown (chapter \ref{sec:Results}).


\clearpage
\chapter{Atmosphere modeling} \label{sec:RadiationTransport}
\section{Radiation transport basics}
The definitions given here comprise only some of the most important radiation quantities, namely those which are needed in the following sections.
It is mainly a summary of \cite{Rybicki79}.
Another useful description of the processes that form and affect the light a stellar atmosphere emits is given in \cite{Rutten95}.

In the following, all quantities subscripted with $\l$ denote wavelength dependence.

\paragraph{Intensity}
Light can be described by rays that follow geodesic paths.
The specific intensity $I_\l$ of a ray, at wavelength $\l$, is defined as
\begin{equation} \label{eq:Intensity}
 \ud E_\l \equiv I_\l(\vec{r},\vec{l},t)(\vec{l},\vec{n}) \,\cos \theta \, \ud A \, \ud t \, \ud \l \, \ud \Omega
\end{equation}
with d$E_\l$ the amount of energy transported through the area $\ud A$ at the location $\vec{r}$, with $\vec{n}$ the normal to d$A$, between times $t$ and $t+\ud t$,
in the frequency band between $\l$ and $\l + \ud \l$, over the solid angle $\ud \Omega = \sin \theta \,\ud \theta \ud \phi$ around the direction $\vec{l}$ with spherical coordinates $\theta$ and $\phi$.

\paragraph{Emissivity}
The contribution of the local emissivity $\eta_\l$ (defined per cm$^3$) to the intensity of a ray is
\begin{equation} \label{eq:Emission}
 \ud I_\l(s) \equiv \eta_\l(s) \,\ud s
\end{equation}
where $s$ is the path length along the ray.
There are different sources of emissivity, as explained in section \ref{sec:Opacity}.

\paragraph{Extinction coefficient}
The extinction coefficient $\chi_\l$ specifies the energy fraction taken from a ray per unit path length.
It can be interpreted as a geometrical cross-section per unit volume.
\begin{equation} \label{eq:Extinction}
 \ud I_\l \equiv - \chi_\l I_\l \,\ud s
\end{equation}
There are different processes that cause radiative extinction, as explained in section \ref{sec:Opacity}.

\paragraph{The radiative transfer equation}
The rate of change of the intensity of a ray is given by the combined effects of emission (equation \eqref{eq:Emission}) and extinction (equation \eqref{eq:Extinction})
\begin{equation} \label{eq:TransportEquation}
 \frac{\ud I_\l}{\ud s} = \eta_\l - \chi_\l I_\l
\end{equation}
This basic equation expresses that the intensity along a ray changes if photons are added or taken from it.
Generally this monochromatic differential equation is complicated by the fact that the opacities ($\chi_\l$ and $\eta_\l$) depend in a direct and/or indirect way on $I$ for all wavelengths.
This is described in section \ref{sec:Opacity} and \ref{sec:SSRTE}.

\section{Common radiation definitions}
In the process of solving the radiative transfer equation \eqref{eq:TransportEquation} the following common definitions are important.

\paragraph{Source function}
The source function is the ratio of the emissivity (equation \eqref{eq:Emission}) and extinction coefficient (equation \eqref{eq:Extinction})
\begin{equation}
 S_\l \equiv \eta_\l / \chi_\l
\end{equation}

\paragraph{Mean intensity}
The mean intensity $J_\l$ is the angle integral of $I_\l$
\begin{equation}
 J_\l \equiv \frac{1}{4 \pi} \int I_\l \,\ud \Omega = \frac{1}{2} \int^1_{-1} I_\l \,\ud \m
\end{equation}
In the second step 1D spherical coordinates are used (see equation \eqref{eq:Intensity}) and $\mu\equiv\cos\theta$.
$J$ is called the zeroth angular moment of the intensity.

\paragraph{Flux}
The flux $H_\l$ is the energy flow outwards, in radial direction $\hat{r}$
\begin{equation}
 H_\l \equiv \frac{1}{4 \pi} \int I_\l (\vec{l} \cdot \hat{r}) \,\ud \Omega = \frac{1}{2} \int^1_{-1} \m I_\l \,\ud \m
\end{equation}
where $\vec{l}$ is the unit vector pointing from the coordinate center into direction $(1,\phi,\theta)$.
$H$ is called the first angular moment of the intensity.

\paragraph{K integral}
In analogy to the other moments, the second angular moment of intensity $K_\l$ is defined as
\begin{equation}
 K_\l \equiv \frac{1}{2} \int^1_{-1} \m^2 I_\l \,\ud \m
\end{equation}
$K$ is proportional to the radiation pressure.

\paragraph{Optical length and optical depth}
The optical length is defined as
\begin{equation}
 \ud \tau_\l \equiv \chi_\l \,\ud s
\end{equation}
This definition is valid for any line of sight.
A special case is the radial optical depth
\begin{equation} \label{eq:OpticalDepth}
 \tau_\l(r) = \int\limits_r^\infty \chi_\l(r) \,\ud r
\end{equation}
which measures the optical depth along the radial line of sight from $\tau_\l=0$ at the observer's eye located at $r=\infty$.

\section{Thermodynamic equilibrium}
\subsection{Radiation}
In thermodynamic equilibrium (TE) the radiation field in an adiabatic enclosure is isotropic and its wavelength dependence is given by the \emph{Planck function} $B$ \cite{Unsoeld38}, that is parameterized by the temperature $T$
\begin{equation} \label{eq:Planck}
 I_{\l}^\TE = B_{\l}(T) \equiv \frac{2 h c^2}{\l^5} \frac{1}{e^{h c/\l k T}-1}
\end{equation}

\subsection{Matter}
Just like the radiation, the energy distribution of matter depends on the temperature $T$ only.
Each atomic species (chemical element) has its own energy-level distribution described in the following.

\subsubsection{Excitation}
The electronic excitation of atoms is described by the Boltzmann equation
\begin{equation} \label{eq:Boltzmann}
 \left(\frac{n_{js}}{n_{is}}\right)_\TE = \frac{g_{js}}{g_{is}}\, \exp\left[-\frac{\chi_{js} - \chi_{is}}{kT}\right] = \frac{g_{js}}{g_{is}}\, \exp\left[-\frac{hc/\l_{ij} }{kT}\right]
\end{equation}
where $n_{is}$ is the number density (${\rm cm}^{-3}$) of the atom under consideration in ionization stage $s$ and in excitation level $i$, $g$ is the statistical weight of the level, $\chi$ the excitation energy measured from the ground state, and $hc/\l_{ij}$ the energy of the photon that corresponds to the energy difference between the levels.

The total TE number density $N_s^\TE$ of the atom in ionization stage $s$ is then the sum over all excitation levels
\begin{equation} \label{eq:TEIonizationBalance}
 N_s^\TE = \sum_i n_{is}^\TE = \frac{n_{0s}^\TE}{g_{0s}} \sum_i g_{is} \,e^{-\chi_{is}/kT} = \frac{n_{0s}^\TE}{g_{0s}} \, Q_s^\TE(T)
\end{equation}
where $0s$ denotes the ground state of ion $s$.
The TE partition function $Q_s^\TE(T)$ is defined as
\begin{equation}
 Q_s^\TE \equiv \sum_i g_{is} \,\exp(-\chi_{is}/kT)
\end{equation}

\subsubsection{Ionization} \label{sec:TEIonization}
The ionization balance is described by the Saha equation
\begin{equation} \label{eq:Saha}
 \left(\frac{N_{k+1}}{N_k}\right)_\TE = \frac{2}{N_{\rm e}} \frac{Q_{k+1}}{Q_k} \left(\frac{2\pi m_{\rm e} kT}{h^2}\right)^{3/2} e^{-\chi_k/kT}
\end{equation}
where $N_{\rm e}$ is the electron number density and $m_{\rm e}$ the electron mass.
The factor of 2 is the statistical weight of the electron.
If chemical element $E$ has $n$ ionization stages, then equation \eqref{eq:Saha} gives $n-1$ equations.
The non-linear system of equations is closed by the constraint
\begin{equation} \label{eq:SahaConstraint}
 N_E = \sum_{s=1}^n N_s
\end{equation}
Solving this system is straight forward.
One starts with a guessed value for $N_1$, uses the Saha equation for the next higher $N_s$ recursively.
The summation on the right-hand side of equation \eqref{eq:SahaConstraint} yields the guess-related value, say $a$.
Finally, all $N_s$ are scaled by the factor $N_E/a$.

The ionization balance is updated for all chemical species, which yields a new value for the electron density.
And since the ionization balance depends on the electron density, the system of equations \eqref{eq:Saha} and \eqref{eq:SahaConstraint} is solved iteratively for each species, until the electron density is converged to a prescribed precision.

\subsubsection{Saha-Boltzmann distribution}
The combination of equations \eqref{eq:Boltzmann} and \eqref{eq:Saha} gives the TE population density of a particular level $i$ of ionization stage $s$
\begin{equation} \label{eq:SahaBoltzmann}
 n_{is}^\TE = n_{cs}^\TE \frac{g_{is}}{g_{cs}} \frac{N_{\rm e}}{2} \left(\frac{h^2}{2\pi m_{\rm e} kT}\right)^{3/2} \exp\left[-\frac{\chi_{is} - \chi_{cs}}{kT}\right]
\end{equation}
where $cs$ denotes the ``continuum'' state to which the ionization takes place.
In this work it is assumed that all ionization and recombination processes involve the ground state $g$ of the next higher ionization stage $n_{cs} = n_{g,s+1}$.
Therefore, all $n_{is}$ (of the element under consideration) can be computed with equation \eqref{eq:SahaBoltzmann}, starting from the uppermost continuum state, when the ionization balance (equation \eqref{eq:SahaConstraint}) is solved.
The uppermost continuum state is the bare nucleus.

\subsection{Interaction of radiation with matter} \label{sec:DetailedBalance}
\subsubsection{Detailed radiative balance}
Since in TE the radiation field inside a closed system is given by the Planck law, equation \eqref{eq:Planck}, it must be reproduced by interactions of the radiation field with matter.
This condition is called \emph{detailed radiative balance}.
It states that in TE the number of photons which are locally absorbed in the momentum range $[\vec{p},\vec{p}+\ud\vec{p}]$ must be equal to the number of photons emitted in that range.

\subsubsection{Detailed rate balance}
Another balance can be derived for TE, based on the Saha-Boltzmann distribution, the \emph{detailed rate balance}.
It states that in TE the number of atoms leaving the excitation state $i$ must be equal to the number that enter state $i$.

\subsubsection{Kirchhoff-Planck relation} \label{sec:KirchhoffPlanck}
From detailed radiative balance it follows that
\begin{equation} \label{eq:DetailedBalanceEta}
 \eta^\TE_\l = \chi^\TE_\l I_\l
\end{equation}
A classical interpretation of the interaction between radiation and matter is that the interaction processes can be divided into pure scattering and true absorption + emission processes.
Here `true' means that in the process of absorption the photon is destroyed and its energy is added to the thermal energy of the gas.
In contrast, in pure scattering processes the photon interacts with the scatterer and propagates in a new direction.
This interaction can be elastic or inelastic.
In the latter case a change in the scatterer's internal excitation state is produced, but the assumption for scattering is that no energy of the photon is converted to kinetic energy of the particles in the gas.

In this classical picture the macroscopic emissivity $\eta$ and extinction coefficient $\chi$ can be written as
\begin{align}
 \eta_\l &= \eta_\l^{\rm th} + \sigma_\l I_\l\\
 \chi_\l &= \kappa_\l + \sigma_\l
\end{align}
where 'th' means thermal, $\kappa$ is the true absorption and $\sigma$ the pure scattering coefficient.

Inserting these $\eta$ and $\chi$ in equation \eqref{eq:DetailedBalanceEta} the pure scattering terms cancel and one obtains
\begin{equation} \label{eq:KirchhoffPlanck}
 \eta^{\rm th}_\l \stackrel{\rm TE}{=} \kappa_\l B_\l
\end{equation}
This is called the \emph{Kirchhoff-Planck} relation.

\section{Local thermodynamic equilibrium} \label{sec:LTE}
In stellar atmospheres there is a net flux outwards $H(\tau) \neq 0$ so that the assumptions of TE are not met.
Local thermodynamic equilibrium (LTE) is a less stringent requirement than TE, because the radiation field is allowed to depart from the isotropic TE field (equation \eqref{eq:Planck}), as long as it does not change the state of matter.
This may occur for example if locally collisional processes play a dominant role over photon processes.

Unlike in the TE case, the radiation field is not known a priori.
Generally
\begin{equation}
 I_\l \neq B_\l(T)
\end{equation}
but $I_\l$ is described by the radiation transport equation \eqref{eq:TransportEquation}.

For this reason detailed radiative balance is \emph{not} valid in LTE.
The Kirchhoff-Planck relation \eqref{eq:KirchhoffPlanck} does not hold exactly.
However, it is a reasonable approximation (with $B_\l$ replaced by $I_\l$) if the gradient of the radiation field is small over the mean free path of a photon, which is true in LTE.

In LTE the local state of matter is the same as for TE, given by equation \eqref{eq:SahaBoltzmann}, but now with a local temperature $T$
\begin{equation}
\begin{split}
 n_{is}^\LTE &= n_{is}^\TE \\
  &= n_{cs}^\LTE \frac{g_{is}}{g_{cs}} \frac{N_{\rm e}}{2} \left(\frac{h^2}{2\pi m_{\rm e} kT}\right)^{3/2} \exp\left[-\frac{\chi_{is} - \chi_{cs}}{kT}\right] \label{eq:SahaBoltzmannLTE}
\end{split}
\end{equation}

In LTE detailed rate balance still holds.

\section{Radiative rates and opacities} \label{sec:Opacity}
There are different sources of opacity, with physically different underlying processes.
The radiative processes accounted for in this work are:
\begin{itemize}
 \item Atomic Bound-Bound (bb) processes (excitation, deexcitation)
 \item Atomic Bound-Free (bf) processes (ionization, recombination)
 \item Atomic Free-Free (ff) processes (Bremsstrahlung)
 \item Thomson scattering
\end{itemize}
The emissivity (equation \eqref{eq:Emission}) and the extinction coefficient (equation \eqref{eq:Extinction}) can therefore be written as
\begin{align}
 \label{eq:DetailedEmissivity}
 \eta_\l &= \eta_\l^{\rm bb} + \eta_\l^{\rm bf} + \eta_\l^{\rm ff} + \sigma_\l J_\l \\
 \label{eq:DetailedExtinction}
 \chi_\l &= \chi_\l^{\rm bb} + \chi_\l^{\rm bf} + \chi_\l^{\rm ff} + \sigma_\l
\end{align}
where $\sigma$ is the scattering coefficient.
In this work only Thomson scattering is considered.
This is an elastic scattering process.
Compton and inverse-Compton scattering are inelastic processes that occur for high-energy photons and high-energy electrons respectively.
A self consistent treatment would introduce an inter-wavelength coupling of the radiation field.
This gravely complicates the requirements for the radiative transport method and are therefore not treated in this work.
Rayleigh scattering can be assumed to be negligible for hot stellar atmospheres, see \cite{BoehmVitense89}.

Equations \eqref{eq:DetailedEmissivity} and \eqref{eq:DetailedExtinction} are the macroscopic quantities describing the interaction of the radiation field with matter.

\subsection{Thomson scattering}
In general Thomson scattering depends on the angle between the incident and the outgoing photon, but usually, as in this work, it is assumed to be isotropic.
This simplification allows to write the contribution of scattering to the local emissivity $\eta_\l$ in equation \eqref{eq:DetailedEmissivity} proportional to the mean intensity $J_\l$.

The amount of radiation scattered out of the ray is $\sigma_\l I_\l$, and a fraction of the local mean intensity $\sigma_\l J_\l$ is scattered into the ray.
For small values\footnote{
If $\l = 1 {\textrm{\AA}}$ then $\gamma$ has the numerical value $\gamma = 0.024$.
}
of $\gamma \equiv (h c/\l)/(m_0 c^2) \ll 1$ the Thomson scattering coefficient is \cite{Pomraning73}
\begin{equation}
 \sigma_\l = n_e \frac{8 \pi e^4}{3 m_0^2 c^4} \left( 1 - 2\gamma + \frac{26}{5} \gamma^2 + \dots \right)
\end{equation}
with $n_e$ being the electron density, $e$ the electron charge and $m_0$ the electron mass.

\subsection{Bound-Bound} \label{sec:BBFrequency}
There are three forms of radiative bound-bound (bb) processes: absorption, induced emission and spontaneous emission.
The radiative rates $R$ at which these processes occur (per second and) per particle are expressed in the Einstein coefficients $B_{ij}$ $B_{ji}$ and $A_{ji}$, where $i$ is the lower and $j$ the upper level of the transition \cite{Mihalas84}.

\subsubsection{Radiative rates}
The \emph{absorption rate} per particle is proportional to the local radiation field
\begin{equation} \label{eq:FrequencyBBRate}
 R_{ij} 
  = B_{ij} \int_0^\infty \phi_\nu J_\nu \,\ud \nu
\end{equation}
with the absorption profile $\phi$.
The profile function is area normalized
\begin{equation}
 \int_0^\infty \phi_\nu \,\ud \nu = 1
\end{equation}

Note that $B_{ij}$ is defined as the rate per units of \emph{intensity in the frequency representation}\footnote{
In the frequency representation the analogon of equation \eqref{eq:Intensity} is 
\begin{math}
 \ud E_\nu \equiv I_\nu(\vec{r},\vec{l},t)(\vec{l},\vec{n}) \,\cos \theta \, \ud A \, \ud t \, \ud \nu \, \ud \Omega
\end{math}
}.
The reason for adopting this definition, and the comparison with a definition based on the wavelength representation of the intensity, is discussed in section \ref{sec:WavelengthVsFrequencyRepresentation}.

Conversion of equation \eqref{eq:FrequencyBBRate} from the frequency representation to the wavelength representation is achieved by using the equivalence of the integrated radiation fields in both representations
\begin{equation}
 \int_0^\infty J_\l \,\ud \l = \int_0^\infty J_\nu \,\ud \nu
  = \int_\infty^0 J_\nu \frac{\ud \nu}{\ud \l} \,\ud \l
  = \int_0^\infty \frac{c}{\l^2} J_\nu \,\ud \l
\end{equation}
Since the right and the left hand side both are integrals over the same variable $\l$ their integrands must be equivalent.
The same holds for the integrated profile functions in both representations
\begin{equation}
 \int_0^\infty \phi_\l \,\ud \l = \int_0^\infty \phi_\nu \,\ud \nu
  = \int_0^\infty \phi_\nu \frac{c}{\l^2} \,\ud \l
\end{equation}
The representation conversion rules are thus found to be
\begin{align}
 J_\l &= \frac{c}{\l^2} J_\nu \label{eq:JConversion} \\
 \phi_\l &= \frac{c}{\l^2} \phi_\nu \label{eq:PsiConversion}
\end{align}
where $\nu = c/\lambda$ is substituted in the expressions for the frequency representation.
Note that the profile function is still normalized after the change of the integration variable.

Using equations \eqref{eq:JConversion} and \eqref{eq:PsiConversion} the absorption rate equation \eqref{eq:FrequencyBBRate} can be written in wavelength representation as
\begin{equation} \label{eq:BBUpRates}
 R_{ij}
  = B_{ij} \int_0^\infty \frac{\l^2}{c} \phi_\l \frac{\l^2}{c} J_\l \frac{\ud \nu}{\ud \l} \,\ud \l
  = B_{ij} \int_0^\infty \phi_\l \frac{\l^2}{c} J_\l \,\ud \l
\end{equation}

The \emph{emission rate} per particle consists of a spontaneous emission term and an induced emission term.
The latter is proportional to the local radiation field.
\begin{equation}
 R_{ji} = R_{ji}^{\rm spon} + R_{ji}^{\rm ind}
  = A_{ji} + B_{ji} \int_0^\infty \psi_\l \frac{\l^2}{c} J_\l \,\ud \l
\end{equation}
Here again $B_{ji}$ is defined as the rate per units of intensity in the frequency representation, and the factor $\l^2/c$ comes from the conversion to the wavelength representation.

These rates $R$ are total rates per particle, where total means integrated over all angles and all wavelengths.
\begin{align}
 R_{ij} &= \int \mathscr{R}_{ij,\l}(\l,\Omega) \,\ud\Omega \ud\l \\
 R_{ji} &= \int \mathscr{R}_{ji,\l}(\l,\Omega) \,\ud\Omega \ud\l
\end{align}
The angle and wavelength dependent radiative rates $\mathscr{R}_{ij,\l}$ and $\mathscr{R}_{ji,\l}$ are given by
\begin{align}
 \mathscr{R}_{ij,\l} &= \frac{1}{4\pi} B_{ij} \phi_\l \frac{\l^2}{c} I_\l \label{eq:SpecificUpRate} \\
 \mathscr{R}_{ji,\l} &= \mathscr{R}_{ji,\l}^{\rm spon} + \mathscr{R}_{ji,\l}^{\rm ind}
  = \frac{1}{4\pi} \left( A_{ji} \varphi_\l + B_{ji} \psi_\l \frac{\l^2}{c} I_\l \right)
\end{align}
It can be shown from quantum mechanical considerations, that the profiles for induced emission and spontaneous emission must be the same \cite{Dirac47}.
\begin{equation}
 \varphi_\l = \psi_\l
\end{equation}

\subsubsection{Einstein relations}
In TE the radiation field must be Planckian.
This forces the rates to satisfy detailed radiative balance (see section \ref{sec:DetailedBalance}).
The rate at which photons are removed from the field must be equal to the rate at which photons are added.
In general, at frequency $\nu$ all bound-bound processes, i.e. the transitions between all levels $i$ and $j$, interact with the radiation field according to their profile functions $\phi_{ij,\nu}$ and $\psi_{ij,\nu}$ (in the following simply written as $\phi_\nu$ and $\psi_\nu$).
Bound-free and other processes can be neglected for the moment, as justified below.
The detailed radiative balance condition thus becomes
\begin{equation} \label{eq:DetailedBalance}
 \sum_{ij} n_i B_{ij} \phi_\nu J_\nu =
  \sum_{ij} n_j A_{ji} \psi_\nu   +   \sum_{ij} n_j B_{ji} \psi_\nu J_\nu
\end{equation}
Solving for $J$ and inserting the TE radiation field $J_\nu = B_\nu$ yields
\begin{equation} \label{eq:EinsteinRelationDeriv1}
 B_\nu = \frac{\sum n_j A_{ji} \psi_\nu}{\sum n_j B_{ji} \psi_\nu}
  \frac{1}{ \frac{\sum n_i B_{ij} \phi_\nu}{\sum n_j B_{ji} \psi_\nu} -1 }
\end{equation}
If one assumes that the profile functions are narrow, then in the line centers $\nu_{ij}$ the summations over all transitions reduce to only one line.
It can be shown \cite{Mihalas78}, that in TE for each transition the profile functions can be assumed to be $\psi_{ij,\nu} = \phi_{ij,\nu}$.
Thus it follows
\begin{equation} \label{eq:EinsteinRelationDeriv}
 B_{\nu_{ij}} = \frac{A_{ji}}{B_{ji}} \frac{1}{ \frac{n_i B_{ij}}{n_j B_{ji}} -1 }
  = \frac{A_{ji}}{B_{ji}} \frac{1}{ \frac{g_i B_{ij}}{g_j B_{ji}} e^{h\nu_{ij}/kT} -1 }
\end{equation}
where in the last step the Boltzmann level distribution, equation \eqref{eq:Boltzmann}, is used.
The equivalence to the Planck function \eqref{eq:Planck} yields the \emph{Einstein relations}
\begin{align} \label{eq:EinsteinRel1}
 A_{ji} &= \frac{2h\nu_0^3}{c^2} B_{ji}
  = \frac{2hc}{\l_0^3} B_{ji} \\
 B_{ji} &= g_i/g_j B_{ij} \label{eq:EinsteinRel2}
\end{align}
These relations reflect, that the Einstein coefficients are properties of the atoms only and are independent of the radiation field.

Formally, other interactions of the radiation field, like bound-free and free-free transitions, must be considered in the detailed radiative balance equation \eqref{eq:DetailedBalance}.
But in general, since the derivation holds for any gas at any temperature, one can find a situation, where the other processes are negligible at the exact center of a sharp bound-bound transition line.
Furthermore, if the Einstein relations (between atomic properties, and their independence of the radiation field) are valid for strong bound-bound transitions, then it must be valid for weaker transitions too, since their interaction with the radiation field is not different.

Using the Einstein relations (equations \eqref{eq:EinsteinRel1} and \eqref{eq:EinsteinRel2}) the downward rates can be written as functions of $B_{ij}$, the same Einstein Coefficient as the upwards rates
\begin{align} \label{eq:BBDownRates}
 R_{ji} &= B_{ij} \frac{g_i}{g_j} \left( \frac{2hc}{\l_0^3} + \int \psi_\l \frac{\l^2}{c} J_\l \,\ud\l \right) \\
 \mathscr{R}_{ji,\l} &= \frac{1}{4\pi} B_{ij} \frac{g_i}{g_j} \psi_\l \left( \frac{2hc}{\l_0^3} + \frac{\l^2}{c} I_\l \right) \label{eq:SpecificDownRate}
\end{align}

\subsubsection{Opacities}
The bound-bound opacities of equations \eqref{eq:DetailedEmissivity} and \eqref{eq:DetailedExtinction} are in fact macroscopic rates at which energy is removed or added to the ray.
They can be expressed as the microscopic photon rates per particle $\mathscr{R}$, multiplied by the particle density $n$ and the energy per photon.
However, the macroscopic picture in equation \eqref{eq:TransportEquation}, is that the extinction depends on the intensity of the ray $I$ and the emission does not depend on $I$.
In order to match the two descriptions, the induced emission rate is treated as negative absorption instead of positive emission.
\begin{align}
 \eta_{ij,\l}^{\rm bb} &= \frac{hc}{\l_0} n_j \mathscr{R}_{ji,\l}^{\rm spon} \\
 \chi_{ij,\l}^{\rm bb} \,I_\l &= \frac{hc}{\l_0} \left( n_i \mathscr{R}_{ij,\l} - n_j \mathscr{R}_{ji,\l}^{\rm ind} \right)
\end{align}
Inserting the expressions for the rates, equations \eqref{eq:SpecificUpRate} and \eqref{eq:SpecificDownRate}, the emissivity and the extinction coefficient become
\begin{align}
 \eta_{ij,\l}^{\rm bb} &= \frac{hc}{4\pi \l_0} B_{ij} n_j \frac{g_i}{g_j} \frac{2hc}{\l_0^3} \psi_\l \\
 \chi_{ij,\l}^{\rm bb} &= \frac{hc}{4\pi \l_0} B_{ij} \frac{\l^2}{c} \left( n_i \phi_\l - n_j \frac{g_i}{g_j} \psi_\l \right)
\end{align}
and the bound-bound opacities of equations \eqref{eq:DetailedEmissivity} and \eqref{eq:DetailedExtinction} are given by the sum over all possible transitions $ij$.

Using these expressions for the line opacities, the single line source functions $S_{ij,\l}^{\rm bb}$ can be written as
\begin{equation}
\begin{split}
 S_{ij,\l}^{\rm bb} \equiv \frac{\eta_{ij,\l}^{\rm bb}}{\chi_{ij,\l}^{\rm bb}}
  &= \frac{2hc^2}{\l_0^3 \l^2} \left( \frac{n_i g_j \phi_\l}{n_j g_i \psi_\l} -1 \right)^{-1} \\
  &= \frac{2hc^2}{\l_0^3 \l^2} \left( \frac{b_i}{b_j} e^{\frac{hc}{\l_0 kT}} -1 \right)^{-1}
\end{split}
\end{equation}
where in the last step complete redistribution $\psi_\l = \phi_\l$ was assumed.
When LTE holds, then in the line center the single line source function adopts the Planck value $S_{ij,\l_0}^{\rm bb} = B_{\l_0}$.
In the wings, the contribution to the (total) source function $S_\l$ is very small, so that generally, in the continuum the source function is dominated by the bound-free terms.

\subsubsection{Line profiles}
The energy of a bound-bound transition is not sharp.
It is expressed in the rate for such a transition with a profile function with a specific width.
The profile function is normalized, meaning that the rate originates from a range of wavelengths and that the integral over the range yields the total rate.
The precise form of the profile functions has a very important influence on the run of the total opacity over wavelength (equations \eqref{eq:DetailedEmissivity} and \eqref{eq:DetailedExtinction}), and thus on the radiation field.

\paragraph{Redistribution}
In the classical interpretation of the interaction between radiation and matter, described in section \ref{sec:KirchhoffPlanck}, the bound-bound transitions are divided into pure scattering and true absorption + emission processes.
In the true absorption + emission processes there is no correlation between the incoming and outgoing photon.

In the pure scattering processes both the direction and the frequency of a photon may change.
These changes are described by a \emph{redistribution function} $R$ \cite{Mihalas70}
\begin{equation} \label{eq:RedistributionFunction}
 \oint  \oint  \int_0^\infty \int_0^\infty  R(\l',\l,\vec{n}',\vec{n}) \,\ud \l' \,\ud \l\frac{\ud \Omega'}{4 \pi}\frac{\ud \Omega }{4 \pi} = 1
\end{equation}
where $R(\l',\l,\vec{n}',\vec{n}) \,\ud \l' \,\ud \l\frac{\ud \Omega'}{4 \pi}\frac{\ud \Omega }{4 \pi}$ gives the probability that a photon will be scattered - from direction $\vec{n}'$ in solid angle $\ud \Omega'$ and wavelength range $(\l',\l'+\ud\l')$ - into solid $\vec{n}$ in solid angle $\ud \Omega$ and wavelength range $(\l,\l+\ud\l)$.
The redistribution function $R$ contains within it both a normalized absorption profile $\phi_{\l'}$ and a normalized emission profile $\psi_\l$
The absorption profile $\phi_{\l'}$ is obtained by integration of $R$ over the outgoing wavelengths and solid angle, and the emission profile $\psi_\l$ by integration of $R$ over the ingoing wavelengths and solid angle
\begin{align}
 \phi_{\l'} (\vec n') &= \oint \int R(\l',\l,\vec n',\vec n) \,\ud \l \frac{\ud \Omega }{4 \pi} \label{eq:GeneralPhi} \\
 \psi_\l (\vec n) &= \oint \int R(\l',\l,\vec n',\vec n) \,\ud \l' \frac{\ud \Omega'}{4 \pi} \label{eq:GeneralPsi}
\end{align}

Equation \eqref{eq:RedistributionFunction} gives the full angle-frequency dependence of the emission profile.
It is difficult to treat the radiative transfer problem in the degree of generality implied here, but some useful simplifications of the problem can be made.

For example, one can assume a \emph{angle independent} redistribution function $R(\l',\l)$.
From equations \eqref{eq:GeneralPhi} and \eqref{eq:GeneralPsi} and the normalization by equation \eqref{eq:RedistributionFunction} it follows that
\begin{equation}
 R(\l',\l) = \phi_{\l'} \psi_\l
\end{equation}
In the special case that $R(\l',\l) = R(\l,\l')$ this reduces to
\begin{equation}
 \phi_\l = \psi_\l
\end{equation}
which is called \emph{complete redistribution}.
A good approximation to this case occurs, for example, when the time of interaction is long enough for collisions to occur.
These redistribute the exited electron to the degenerate sub states of the upper level and thus completely remove the angle and frequency correlation to the initial absorption process.

A sophistication of complete redistribution is \emph{partial redistribution}, e.g. \cite{Mihalas78}.
Such methods have the disadvantage that they introduce another direct inter-wavelength coupling of the radiation field, which tremendously complicates the radiation transport problem (section \ref{sec:SSRTE}).

In the microscopic description of this section \ref{sec:BBFrequency} the clear distinction between true absorption + emission and pure scattering processes cannot be made.
Every interaction, either absorption or emission, is treated individually based on its profile function.
So the profile functions statistically account for all redistribution effects.
Generally, such profile functions should be derived from quantum mechanics.

In this work the simplifying assumption of complete redistribution is made and the profile functions $\phi$ and $\psi$ are assumed to be given by a Voigt-profile.

\paragraph{Voigt profile}
The profile functions used in this work are defined in the frequency representation, which is consistent with the definitions of the Einstein coefficients (see equation \eqref{eq:FrequencyBBRate}).
A comparison with other definitions is given in the appendix (section \ref{sec:BBWavelength}).
The Voigt profile is given by
\begin{align}
 \psi(a,u) = \psi(a,u) &= \frac{H(a,u)}{\Delta \nu_D \sqrt{\pi}} \label{eq:FrequencyVoigt} \\
 \nu_D &\equiv \frac{\nu_0}{c} \sqrt{2kT/m + \xi^2} = \frac{\sqrt{2kT/m + \xi}}{\l_0} \\
 H(a,u) &\equiv \frac{a}{\pi} \int^\infty_{-\infty} \frac{e^{-y^2}}{a^2 + (u-y)^2} \,\ud y
\end{align}
with the dimensionless frequency offset $u$ and the damping parameter $a$ determined by
\begin{align}
 u &= \frac{\nu-\nu_0}{\Delta \nu_D} = \frac{c/\l - c/\l_0}{\Delta \nu_D} \\
 a &= \frac{\Delta \nu_d}{\Delta \nu_D} = \frac{\Gamma}{4 \pi \Delta \nu_D} \label{eq:VoigtDampingParameter}
\end{align}
$\Delta \nu_D$ being the Doppler width, $\xi$ the microturbulence parameter, and the dispersion width $\Delta \nu_d$ determined by the parameter $\Gamma$ is given by
\begin{equation}
 \Delta \nu_d = \Gamma/(4\pi)
\end{equation}
$H$ is called the Voigt function.
The Voigt-profile results from the convolution of a Gauss profile with width $\Delta \nu_D$
\begin{equation}
 \psi(\Delta \nu_D,u) = \frac{1}{\sqrt{\pi} \Delta \nu_D} \, e^{-u^2}
\end{equation}
and a Lorentz profile with width $\Delta \nu_d$
\begin{equation} \label{eq:LorentzProfile}
 \psi(\Delta \nu_d,u) = \frac{1}{\pi} \frac{\Delta \nu_d}{\Delta \nu^2 + \Delta \nu_d^2}
\end{equation}
The Gaussian part in the Voigt profile describes line broadening by thermal motions of the atoms, which causes Doppler shifts in the wavelength of absorbed and emitted photons.
The Lorentzian part is the exact profile shape resulting from radiative damping, which is caused by the limited lifetime of excited quantum states.
Currently the other broadening processes, the Van der Waals broadening and the quadratic Stark effect, are all described by Lorentz profiles.
This simplification allows to use the fact that the convolution of two Lorentz profiles again yields a Lorentz profile with a width equal to the sum of the original widths.

Collision broadening is a very wide field of work, quoting Rob Rutten: "A physicist may easily spend a whole career on collisional line broadening" \cite{Rutten95}.
Extensive work on line broadening by electron collision processes was done by \cite{Griem74}.

In the inner regions of the atmospheres modeled in this work, where the electron densities are high and the Stark effect is a very effective broadening process, the current description yields $\Gamma$-factors that are too small.
In the outer regions the radiative damping factor gets more important, so that the approximations for the collisional damping are less important there.
And for the shapes of individual lines the outer region of the atmosphere is more important than the inner region.
However, the opacity in the inner regions do influence the overall structure of the atmosphere, so that the indirect influence of the too small $\Gamma$-factors might yet be important.

\subsection{Bound-Free}
\subsubsection{Radiative rates}
\begin{align}
 R_{ic} &= ... \label{eq:BFUpRates} \\
 R_{ci} &= ... \label{eq:BFDownRates}
\end{align}
Not yet...

\subsubsection{Opacities}
In analogy to the bound-bound opacities the bound-free opacities follow from the rates.
\begin{align}
 \eta_{ic} &= ... \label{eq:BFChi} \\
 \chi_{ci} &= ... \label{eq:FBEta}
\end{align}
Not yet...

\subsection{Free-Free}
Free-free transitions are transitions between unbound states.
The rates of these transitions do not influence the population numbers of the bound states.
Thus they do not influence the statistical equilibrium (see section \ref{sec:StatisticalEquilibrium}) and the rates do not need to be computed.

The free-free opacities are calculated from Gaunt factors from \cite{Sutherland98} and \cite{Itoh00} with routines from the publicly available CHIANTI 4 database (see section \ref{sec:AtomicDatabases}).
These opacities are assumed to occur at TE rates, so that the emissivities follow from the Kirchhoff-Planck relation \eqref{eq:KirchhoffPlanck}.

\section{Collisional rates}
The collisional rates $C_{ij}$ are independent of the radiation field.
They are determined from theoretical collision strengths that are input data to \phx.
The data are part of the atomic databases described in section \ref{sec:AtomicDatabases}.
These contain data for electronic collisions and protonic collisions.

The collisional rates occur at thermodynamic equilibrium values as long as the gas of colliders has a Maxwellian velocity distribution \cite{Mihalas78}.
This is a reasonable assumption for the atmospheric conditions treated in this work.
Therefore, the contribution of the collisional rates to the total rates $P_{ij}$, equation \eqref{eq:RateMatrixElements}, always drives the population numbers towards the TE (Saha-Boltzmann) distribution.

\section{Atomic data} \label{sec:AtomicDatabases}
The calculation of the X-ray emission and the spectral features of a hot stellar plasma requires atomic transition rates and energy levels of highly ionized chemical elements.
In \phx\ multiple high energy databases are available
\begin{itemize}
\item CHIANTI 3, see \cite{Dere97}
\item CHIANTI 4, see \cite{Dere01}
\item CHIANTI 5, see \cite{Landi06}
\item APED 1.3.1, see \cite{Smith01}
\end{itemize}
In this work primarily the CHIANTI 5 database has been used, since it is the most modern and most complete of the four.
For many ions the APED database contains only CHIANTI data.
A systematic comparison of the different databases with \phx\ was made by \cite{PetzPhd}.

\section{Non-LTE}
The assumption of LTE is more realistic than TE, but is not generally valid in stellar atmospheres.
When the assumption of LTE is dropped, it is called Non-LTE (NLTE).
In NLTE the state of matter is no longer given by the Saha-Boltzmann distribution and is no longer fixed by the local temperature.

\subsection{Statistical equilibrium} \label{sec:StatisticalEquilibrium}
In this work \emph{statistical equilibrium} is assumed.
This means that the population numbers and the radiation field adjust so quickly, that the upwards and downwards rates are in balance.
This allows to write down the \emph{rate equations} for a snapshot of the atmosphere
\begin{equation} \label{eq:RateEquation}
 0 = \frac{\ud}{\ud t} n_i = \sum_{j \neq i} n_j P_{ji} - n_i \sum_{j \neq i} P_{ij}
\end{equation}
which describe the net change of occupation number $n_i$ of level $i$ depending on the transitions (bound-bound and bound-free) from and to all possible levels $j$.
$P_{ij}$ denotes the rate of the transition from $i$ to $j$ per particle per second.
The rates $P$ contain the radiative $R$ and the collisional rates $C$
\begin{equation} \label{eq:RateMatrixElements}
 P_{ij} = R_{ij} + C_{ij}
\end{equation}
The radiative rates are given by the sum over all bound-bound (equations \eqref{eq:BBUpRates} and \eqref{eq:BBDownRates}) and bound-free rates (equations \eqref{eq:BFUpRates} and \eqref{eq:BFDownRates}).

\subsection{Rate matrix equation} \label{sec:RateMatrixEquation}
One rate equation \eqref{eq:RateEquation} exists for each state $i$.
Its variables are all $n_j$ with $j \ne i$.
This is a linear system with a dimension equal to the number of levels $N$ in the atom.
However, the system is not yet of maximum rank because one equation factors out to an identity.
The system is closed by replacing a redundant equation with the constraint of number conservation
\begin{equation} \label{eq:RMEConstraint}
 N_E = \sum_{i=1}^N n_i
\end{equation}
Here the summation is performed over all the $N$ levels of the atom under consideration in all ionization stages.
All levels of all stages are numbered continuously, starting from the ground state of the neutral atom.
In this notation the last level $N$ is the bare nucleus.
The matrix equation can be written in the general form
\begin{equation} \label{eq:RateMatrixEquation}
 \mathbf{P} \vec{n} = \vec{b}
\end{equation}
with $\mathbf{P}$ being the rate matrix, $\vec{n}$ the vector of occupation numbers (number densities) to solve for, and $\vec{b}$ the vector that has only one non-zero ($N_E$) element from the constraint equation.
An explicit rate matrix is shown in section \ref{sec:RMESolver}, figure \ref{fig:ExplicitRateMatrix}, for the example of a simplified atom with four ionization stages, the last stage being the bare nucleus.

This simple treatment changes the ionization balances of all elements independently.
This results in a change of the electron density.
Because all rates depend on the electron density, the ionization balance of all elements are coupled.

A rigorous treatment requires the electron density as an additional variable.
However, the rate equation is then no longer linear and the rank of the system becomes so large, that it is not feasible anymore for numerical computation.

Furthermore, the radiative rates depend on the radiation field, which in turn depends on the occupation numbers.
The radiation field dependence dominates the dependence on the electron density.
Therefore, it is preferable to \emph{not} treat electron density changes directly.
In this work the consistency between the occupation numbers, the radiation field and electron density is obtained iteratively.

\subsection{Departure coefficients}
The departure coefficient $b$ of a atomic level is defined as the ratio of the NLTE (or true) occupation number and the LTE occupation number
\begin{equation} \label{eq:Ni2Bi}
 b_i \equiv \frac{n_i}{n_i^\LTE}
\end{equation}
for any level $i$ in any ionization stage.
In literature this is called the Zwaan definition \cite{Zwaan72}.
In the original Menzel definition \cite{Menzel37}, the departure coefficients\footnote{here denoted with a $*$ for notational convenience} $b_i^*$ are not based upon the LTE occupation numbers, but on occupation numbers $n_i^*$.
These are computed per ionization stage with a LTE Saha-Boltzmann distribution based on the true NLTE continuum, i.e. equation \eqref{eq:SahaBoltzmannLTE} with $n_{cs}^\LTE$ replaced with $n_{c}$.
The subscript $s$ for the ionization stage in equation \eqref{eq:SahaBoltzmannLTE} is omitted here.
For the uppermost continuum, the bare nucleus, $n_i^*$ is defined as the true NLTE occupation number.
\begin{align}
 n_{i}^*
  &\equiv n_c \frac{g_i}{g_c} \frac{N_{\rm e}}{2} \left(\frac{h^2}{2\pi m_{\rm e} kT}\right)^{3/2} \exp\left[-\frac{\chi_i - \chi_c}{kT}\right] \label{eq:NiStar} \\
 n_C^* &\equiv n_C \label{eq:NiStarC}
\end{align}
with $c$ being the continuum state to which the actual level $i$ ionizes, and $C$ the uppermost continuum.
The $b_i^*$ are thus given by
\begin{equation} \label{eq:BiStar}
 b_i^* \equiv \frac{n_i}{n_i^*}
  = b_i \frac{n_c^\LTE}{n_c} 
  = b_i / b_c
\end{equation}
except for the last continuum, in which case the departure coefficient is converted as
\begin{equation}
 b_C^* = 1 = b_C \frac{n_C^\LTE}{n_C}
\end{equation}

\subsection{Ionization balance} \label{sec:NLTEIonization}
When the rate matrix equation \eqref{eq:RateMatrixEquation} is solved the ionization balance is determined.
The ionization balance is important, as it determines the electron donation of each species.
In TE and LTE changes to the ionization balance exactly follow from changes to the (local) temperature via the Saha-Boltzmann law.
In NLTE the influence of the temperature to the ionization balance is indirect, via the radiation field, and the rates.
It cannot be determined directly.
When changing the temperature, an assumption must be made about the change to the ionization balance, and therefore about the $n_i$ for all levels of all ionization stages of all elements.

For this assumption two extreme approximations can be made.
In the first approximation the $n_i$ are completely independent of the local temperature $T$.
The $n_i$ are kept fixed, the LTE energy level distribution is computed for the new temperature, see page \ref{sec:TEIonization}, and from that the $b_i$ follow.
With the fixed ionization balance also the electron density becomes independent of the temperature.

In the second approximation the $n_i$ adapt as much to the temperature as they would in LTE.
The $b_i$ are kept fixed and the $n_i$ follow from the new LTE distribution for the new temperature.
In this approximation it is convenient to directly compute the electron density from the $b_i$, for which a NLTE partition function can be introduced, in analogy to the TE partition function (equation \eqref{eq:TEIonizationBalance}).
The total number density per ionization stage $s$ is
\begin{equation} \label{eq:NLTEIonization}
 N_s = \sum_i n_{is}
  = \sum_i b_{is} n_{is}^\LTE
  = \frac{n_{0s}^\LTE}{g_{0s}} \, Q_s^\NLTE(T,b_i)
\end{equation}
where $i$ is the excitation state of ionization stage $s$ and $0s$ denotes its ground state.
The NLTE partition function $Q_s^\NLTE(T)$ is defined as
\begin{equation}
 Q_s^\NLTE(T,b_i) \equiv \sum_i b_{is} g_{is} \,\exp(-\chi_{is}/kT)
\end{equation}
From equation \eqref{eq:NLTEIonization} the number density of the ground state of ionization stage $k$ can be written as
\begin{equation} 
 n_{0k} = \frac{N_{k}}{Q_{k}^\NLTE} g_{0k} b_{0k} e^{-\chi_{0k}/kT}
\end{equation}
Using equation \eqref{eq:NLTEIonization} for $s=k+1$ and the Saha-Boltzmann distribution (equation \eqref{eq:SahaBoltzmann}) it follows
\begin{equation} \label{eq:NLTESaha}
 \left(\frac{N_{k+1}}{N_k}\right)_\NLTE = \frac{Q_{k+1}^\NLTE}{Q_k^\NLTE} \frac{2}{N_{\rm e}} \left(\frac{2\pi m_{\rm e} kT}{h^2}\right)^{3/2} e^{-\chi_k/kT}
\end{equation}
This equation is similar to the Saha equation \eqref{eq:Saha} for TE.
The ionization balance is solved analog to TE, which is described in section \ref{sec:TEIonization}.

The impact of these approximations is discussed in section \ref{sec:NiScaling}

\section[Spherically symmetric radiative transfer eqn.]{The spherically symmetric 1D radiative transfer equation} \label{sec:SSRTE}
The time-independent, Lagrangian radiative transfer equation \eqref{eq:TransportEquation} in the case of a spherically symmetric atmosphere (\mbox{SSRTE}) can be written as \cite{Mihalas84} \cite{Hauschildt99}
\begin{equation} \label{eq:SSRTE}
      \alpha_r \pd{I_\l}{r}
    + \alpha_\m \pd{I_\l}{\m}
    + \alpha_\l \pd{}{\l} \left( \l I_\l \right)
    + 4 \alpha_\l I_\l
    = \eta_\l - \chi_\l I_\l
\end{equation}
in which
\begin{align*}
   \alpha_r  &= \gamma (\m+\beta) \\
   \alpha_\m &= \gamma (1-\m^2)
              \left[\div{1+\beta\m}{r}-\gamma^2(\m+\beta)\pd{\beta}{r} \right]
                         \\
   \alpha_\l &= \gamma \left[\div{\beta(1-\m^2)}{r}
              +\gamma^2\m(\m+\beta)\pd{\beta}{r} \right]
\end{align*}
with velocity $v$, $\beta = v/c$ and $\gamma^2 = 1/(1-\beta^2)$.

On the right hand side of equation \eqref{eq:SSRTE}, $\eta_\l = \eta_\l(r)$ and $\chi_\l = \chi_\l(r)$ are given by equations \eqref{eq:DetailedEmissivity} and \eqref{eq:DetailedExtinction}.

The basic problem of radiative transfer is that the evaluation of a particular $I_\l (\tau_\l,\mu)$ requires $J_\l$, and therefore $I_\l$ in all directions, and therefore the source function $S$ at many locations and many wavelengths, and therefore $J$ on these locations and wavelengths.
With equation \eqref{eq:DetailedEmissivity} the SSRTE becomes an integro-differential-equation.
In this work a monotonic velocity field is assumed.
Then the SSRTE becomes a boundary value problem in spatial coordinates and an initial value problem in wavelength.
The SSRT equation \eqref{eq:SSRTE} is solved with the operator splitting method as described in \cite{Hauschildt92} and \cite{Hauschildt04}.

\subsection{The Operator splitting method}
The mean intensity $J_\l$ is obtained from the source function $S$ by a formal solution of the SSRTE, which is symbolically written using the $\Lambda$-operator $\Lambda_\l$ as
\begin{equation} \label{lambda_iteration}
 J_{\lambda} = \Lambda_{\lambda} S_{\lambda}
\end{equation}
In the following the subscript $\l$ is dropped to ease the notation.
The usual Lambda iteration method
\begin{align}  \label{J_new_original}
 J_{\rm new} &= \Lambda S_{\rm old} \\
 \label{S_new}
 S_{\rm new} &= (1 - \epsilon) J_{\rm new} + \epsilon B
\end{align}
fails in the case of large optical depths and small $\epsilon$ (thermal coupling parameter, B is the Planck function).
The range of the Lambda operator is only in the order of $\Delta \tau \sim 1$.
At large optical depths, the mean intensity calculated with the Lambda iteration must be $J_{\lambda} = B_{\lambda} + O(e^{-\tau_\lambda})$ and the convergence is very poor.
A physical description of this effect can be found in \cite{Mihalas80}.

Therefore, the operator splitting method with an approximate lambda operator $\Lambda^*$ (ALO) is used, which is defined by
\begin{equation} \label{lambda_star}
  \Lambda = \Lambda^* + (\Lambda - \Lambda^*)
\end{equation}
and equation \eqref{lambda_iteration} is rewritten as
\begin{equation} \label{J_new}
  J_{\rm new} = \Lambda^* S_{\rm new} + (\Lambda - \Lambda^*) S_{\rm old}
\end{equation}
Using equation \eqref{S_new} it follows
\begin{equation} \label{eq:ALISolution}
  \left[1 - \Lambda^* (1 - \epsilon)\right] J_{\rm new} = J_{\rm fs} - \Lambda^* (1 - \epsilon) J_{\rm old}
\end{equation}
where $J_{\rm fs} = \Lambda S_{\rm old}$ is the formal solution.
With equations \eqref{lambda_star} - \eqref{eq:ALISolution} new values of the mean intensity $J_{\rm new}$ can be obtained and with equation \eqref{S_new} the new source function can be calculated.

\subsection{The Approximate Lambda Operator} \label{sec:ALI}
The formal solution is performed along characteristic rays \cite{Olson87}.
Along the characteristic rays the SSRT equation \eqref{eq:SSRTE} has the form \cite{Mihalas80}
\begin{equation}
\pd{I}{s} + \alpha_\l \pd{\l I}{\l} = \eta - (\chi + 4\alpha_\l) I
\end{equation}
where $s$ is the path length along a ray.
Before the SSRTE can be solved the characteristic rays have to be calculated.
The source function is interpolated piecewise linearly or parabolically along each ray.
For the specific intensity $I(\tau_i)$ the following expressions are obtained along a ray $k$
\begin{align}
 I^k(\tau_i^k) & =      I^k(\tau_{i-1}^k) \exp(\tau_{i-1}^k - \tau_i^k) + \int\limits_{\tau_{i-1}^k}^{\tau_i^k} \hat{S}(\tau) \exp(\tau - \tau_i^k) \ud \tau  \label{eq:IFormalSolution} \\
 I_i^k         & \equiv I_{i-1}^k \exp(-\Delta\tau_{i-1}^k) + \Delta{}I_i^k
\end{align}
$\tau_i^k$ is the optical depth along the ray $k$ with $i$ as the running index of the optical depth points.
$\hat{S}$ is a generalized source function which contains terms of the wavelength derivative.
It is defined in \cite{Hauschildt92}.
With $\tau_1 = 0$ and $\tau_{i-1}^k \le \tau_i^k$ it follows for the calculation of $\tau^k$:
\begin{equation}
 \Delta\tau_{i-1}^k = (\hat{\chi}_{i-1} + \hat{\chi}_i) |s_{i-1}^k - s_i^k| / 2
\end{equation}
$\hat\chi_i = \chi_i + 4\alpha_{\l,i}$ is the effective extinction coefficient at point $i$ and $|s_{i-1}^k - s_i^k|$ is the path length between point $i$ and $i-1$ along the ray $k$.

For $\Delta{}I_i^k$ the following expression applies
\begin{equation} \label{eq:DILinearCombination}
 \Delta{}I_i^k = \alpha_i^k \hat{S}_{i-1} + \beta_i^k \hat{S}_i + \gamma_i^k \hat{S}_{i+1}
\end{equation}
$\alpha_i^k$, $\beta_i^k$, and $\gamma_i^k$ are interpolation coefficients.
The expressions for the coefficients can be found in equations (23) to (25) in \cite{Hauschildt92} and in \cite{Hauschildt04}.
If $\hat{S}$ is interpolated linearly then $\gamma_i^k$ is zero.
For parabolic interpolation all three coefficients are non-zero.

The choice of the approximate $\Lambda$-operator is very important to achieve rapid convergence.
The ALO used in \phx, a generalization of the tri-diagonal operator of \cite{Olson87}, is given in \cite{Hauschildt94}.
It is in some sense exact, as its elements are exactly the elements of the $\Lambda$ operator.
The approximation that is used is that only band diagonal elements of the full $\Lambda$ matrix are used in the construction of $\Lambda^*$.
The expressions for the elements of $\Lambda$ are given in equations (27) to (32) of \cite{Hauschildt92}.

The fastest convergence of equation \eqref{eq:ALISolution} is achieved with $\Lambda^* = \Lambda$, but to construct the full $\Lambda$ is very time consuming.
In \cite{Hauschildt94} the operator splitting method with band-diagonal ALO operators is described and analyzed.
In general, a ALO with a larger bandwidth will lead to a higher convergence rate, and therefore less iterations are needed to reach a given accuracy.
But the ALO construction takes more time as more coefficients need to be computed.
The optimal value for the bandwidth of $\Lambda^*$ depends on several conditions, described in \cite{Hauschildt94}, like the computer architecture used and the type of model atmosphere that is computed.
However, the strongest dependence is the quality of the initial guess to the solution.
Therefore, the optimal ALO bandwidth is to be determined experimentally.

\section{Radiative equilibrium} \label{sec:RadiativeEquilibrium}
If an atmosphere is in \emph{radiative equilibrium} (RE) then there are no net sources or sinks of radiation in the atmosphere and the radiative flux through each shell of the atmosphere is constant.
Usually a target luminosity $L$ is specified or an effective temperature $T_{\rm eff}$ and a reference radius $R_*$.
The RE condition in the Euler frame is
\begin{equation}
\begin{split}
  4\pi r^2 H^{\rm obs}(r)
  &\equiv 4\pi r^2 \int H^{\rm obs}_\l(r) \,\ud\l \\
  &= 4 \pi r^2 H_0^{\rm obs}(r)
  \equiv 4\pi R_*^2 \int B_\l(T_{\rm eff}) \,\ud\l \\
  &= L
  = 4\pi R_*^2 \sigma_{\rm SB} T_{\rm eff}^4 \label{eq:StefanBoltzmannLaw}
\end{split}
\end{equation}
where $H^{\rm obs}$ is the Eulerian flux, $H_0^{\rm obs}$ is the target flux, related to the Luminosity $L$.
In last step the \emph{Stefan-Boltzmann law} has been used, and
\begin{equation}
 \sigma_{\rm SB} = \frac{2\pi^5 k^4}{15 h^3 c^2}
\end{equation}
is the Stefan-Boltzmann constant.

The RE condition in the Euler frame may also be written in differential form
\begin{equation} \label{eq:EulerRE}
 \frac{\ud (r^2 H^{\rm obs})}{\ud r} = 0
\end{equation}
In the Lagrange frame this becomes \cite{Hauschildt03} \cite{Mihalas84}
\begin{equation} \label{eq:LagrangeRE}
 \frac{\partial (r^2 H)}{\partial r}
  + \beta \frac{\partial (r^2 J)}{\partial r}
  + \frac{\beta}{r} r^2 (J - K)
  + \gamma^2 \frac{\partial \beta}{\partial r} r^2 (J + K + 2\beta H) = 0
\end{equation}
Using the first angular moment of the radiation transport equation \eqref{eq:TransportEquation} one obtains from both \eqref{eq:EulerRE} and \eqref{eq:LagrangeRE} an expression that is valid in both the Euler and the Lagrange frame
\begin{equation} \label{eq:RadiativeEquilibrium}
 \int (\eta_\l - \chi_\l J_\l) \,\ud\l = 0
\end{equation}
In the Eulerian case $\eta$, $\chi$ and $J$ are replaced by the Eulerian $\eta^{\rm obs}$, $\chi^{\rm obs}$ and $J^{\rm obs}$.

\subsection{Model convergence criteria} \label{sec:ModelCompletion}
The first step in the modeling process is to make an initial guess for the atmosphere structure $\{T,n_i\}$ as function of the radius $r$.
RE is not imposed at this stage.
Subsequently, the atmosphere is corrected for deviations from RE iteratively using a temperature correction method (see section \ref{sec:TemperatureCorrection}).

Inserting $\eta_\l$ and $\chi_\l$ from equations \eqref{eq:DetailedEmissivity} and \eqref{eq:DetailedExtinction} into equation \eqref{eq:RadiativeEquilibrium} the scattering term $\sigma$ cancels out
\begin{equation}
 \int ( (\eta_\l - \sigma_\l J_\l) - (\chi_\l - \sigma_\l) J_\l ) \,\ud\l = 0
\end{equation}
so that the relative deviation from RE can be written as
\begin{equation} \label{eq:REErrorDif}
  \varepsilon_{\rm RE,dif} \equiv \frac {\int (\eta_\l - \chi_\l J_\l) \,\ud\l } {\int (\chi_\l - \sigma_\l) J_\l \,\ud\l }
\end{equation}
This is the differential form of the relative RE error.
The definition is made in the Lagrange frame, because that is the frame the opacities are computed in.

The integral form follows from equation \eqref{eq:StefanBoltzmannLaw}
\begin{equation} \label{eq:REErrorInt}
 \frac { H^{\rm obs} - H_0^{\rm obs}} {H_0^{\rm obs}} = \varepsilon_{\rm RE,int}
\end{equation}
Note that this definition is made in the Euler frame.
Defining it in the Lagrange frame is almost equivalent.
But the target flux is formally specified in the Euler frame and the Lagrangian $H_0$ is derived from it.
Therefore, the $H_0^{\rm obs}$ is the more fundamental quantity to check RE against.

If the deviations from RE get below some threshold, typically $\varepsilon_{\rm RE} < 1\%$, then the model is called converged.
However, this condition is not very strict, since errors in single layers, as well as errors in the innermost or outermost regions (optically thick and thin respectively) are not as important as broad range errors in the middle region, that has the most important influence on the resulting spectrum.

\section{The temperature correction method} \label{sec:TemperatureCorrection}
One way to obtain radiative equilibrium is to use equation \eqref{eq:REErrorDif} in a simple linear Newton-Raphson procedure.
In the first step the temperature is changed arbitrarily.
Subsequent temperature corrections can then be computed according to the induced changes in $\varepsilon_{\rm RE}$.
However, this scheme is not very robust.
It takes a lot of steps to converge to the required precision or it does not converge at all.

A different approach is the Uns\"old-Lucy (UL) temperature correction scheme \cite{Unsoeld55} \cite{Lucy64}.
In this method a direct correction $\Delta T$ to the temperature $T$ is computed.
This method is found to be much more stable than the Newton-Raphson scheme \cite{Hauschildt03}.

In this section all unprimed quantities are Eulerian and the primed quantities are Lagrangian.

\subsection{The Uns\"old-Lucy method} \label{sec:ULTC}
The original procedure was developed for plane-parallel atmospheres.
The derivation of the spherically symmetric generalization begins with the time independent, static, spherically symmetric transfer equation with an approximate LTE source function \cite{Hauschildt03}.
From equation \eqref{eq:SSRTE} with $\beta = 0$ follows
\begin{equation} \label{eq:ULBasic}
 \mu \frac{\partial}{\partial r} I_\l  +  \frac{1-\mu^2}{r} \frac{\partial}{\partial \mu} I_\l
  = \eta_\l - \chi_\l I_\l
\end{equation}
The zeroth and first angular moments are
\begin{align}
 \frac{\partial r^2 H_\l}{\partial r} &= r^2 \chi_\l (S_\l - J_\l) \\
 -\chi_\l H_\l &= \frac{\partial K_\l}{\partial r}  +  \frac{3K_\l - J_\l}{r}
\end{align}
Integration over all wavelengths and assuming the approximate expression for the source function (see section \ref{sec:LTE})
\begin{equation} \label{eq:ULTC_LTE_Source}
 S_\l = (\kappa_\l B_\l + \sigma_\l J_\l)/(\kappa_\l + \sigma_\l)
\end{equation}
yields
\begin{align}
 \frac{\partial r^2 H}{\partial r} &= r^2 (\kappa_B B - \kappa_J J) \label{eq:ULTC2a} \\
 -\chi_H H &= \frac{\partial K}{\partial r}  +  \frac{3K - J}{r} \label{eq:ULTC2b}
\end{align}
Here $B$, $J$ and $H$ are the integrals in wavelength of $B_\l$, $J_\l$ and $H_\l$ and
\begin{align}
\kappa_B &\equiv \frac{1}{B} \int \kappa_\l B_\l \,\ud\l \label{eq:OpacityIntegral1} \\
\kappa_J &\equiv \frac{1}{J} \int \kappa_\l J_\l \,\ud\l \label{eq:OpacityIntegral2} \\
\chi_H &\equiv \frac{1}{H} \int \chi_\l H_\l \,\ud\l \label{eq:OpacityIntegral3}
\end{align}
are the mean opacities.
Using the \emph{Eddington factor}
\begin{equation}
 f = K/J
\end{equation}
and multiplying equation \eqref{eq:ULTC2b} with the \emph{sphericity factor}, introduced by \cite{Auer71}
\begin{displaymath}
 q = \frac{r_c^2}{r^2} \exp \int_{r_c}^r \frac{3f-1}{fr'} \ud r'
\end{displaymath}
and writing $\mathscr{B} = r^2 B$, $\mathscr{J} = r^2 J$ and $\mathscr{H} = r^2 H$ one obtains from equations \eqref{eq:ULTC2a} and \eqref{eq:ULTC2b}
\begin{align}
 \frac{\partial \mathscr{H}}{\partial r} &= \kappa_B \mathscr{B} - \kappa_J \mathscr{J} \label{eq:ULTC3a} \\
 \frac{\partial \, qf\mathscr{J}}{\partial r} &= - q \chi_H \mathscr{H} \label{eq:ULTC3b}
\end{align}
It is assumed that the opacity means $\kappa_B$, $\kappa_J$ and $\chi_H$, as well as the Eddington factor $f$ are independent of the temperature.
Accordingly they do not change from iteration to iteration.
Equations \eqref{eq:ULTC3a} and \eqref{eq:ULTC3b} for the current iteration can then be subtracted from those for the next iteration:
\begin{align}
 \frac{\partial \Delta \mathscr{H}}{\partial r} &= \kappa_B \Delta \mathscr{B} - \kappa_J \Delta \mathscr{J} \label{eq:ULTC4a} \\
 \frac{\partial \, qf\, \Delta \mathscr{J}}{\partial r} &= - q \chi_H \Delta \mathscr{H} \label{eq:ULTC4b}
\end{align}
with $\Delta \mathscr{H} = \mathscr{H}_{\rm new} - \mathscr{H}_{\rm old}$ and $\mathscr{H}_{\rm new}$ being the target flux and $\mathscr{H}_{\rm old}$ the flux in the current iteration.
Equation \eqref{eq:ULTC4a} is solved for $\Delta \mathscr{B}$ and equation \eqref{eq:ULTC4b} is integrated from the outermost layer $R$ to $r$
\begin{align}
 \Delta \mathscr{B} &= \frac{1}{\kappa_B} \left( \frac{\partial \mathscr{H_{\rm new}}}{\partial r} - \frac{\partial \mathscr{H_{\rm old}}}{\partial r} \right) + \frac{\kappa_J}{\kappa_B} \Delta \mathscr{J} \label{eq:ULTC5a} \\
 \begin{split}
 \Delta \mathscr{J}(r) &=
  - \frac{1}{q(r) f(r)} \Bigg{(} q(R) f(R) \,\Delta \mathscr{J}(R) \\
  & \qquad + \int_{R}^r q(r') \chi_H(r') \Delta \mathscr{H}(r') \,\ud r' \Bigg{)} \label{eq:ULTC5b}
 \end{split}
\end{align}
The \emph{second Eddington factor}
\begin{equation}
 g = J/H
\end{equation}
can be used to express the $\Delta \mathscr{J}(R)$ term in equation \eqref{eq:ULTC5b} in $\mathscr{H}$.
This factor is assumed to be independent of the temperature for the outermost model layer at radius $R$.
Inserting equation \eqref{eq:ULTC2a} (for the $\mathscr{H_{\rm old}}$ term) and \eqref{eq:ULTC5b} in \eqref{eq:ULTC5a} and dividing by $r^2$ yields
\begin{align}
 \Delta B(r) \equiv& \Delta B_1(r) + \Delta B_2(r)
  + \frac{1}{r^2 \kappa_B} \frac{\partial \mathscr{H}_0}{\partial r} \label{eq:ULTC6} \\
 \Delta B_1(r) =& \frac{\kappa_J}{\kappa_B} J(r) - B(r) \label{eq:ULTC6a} \\
 \begin{split}
  \Delta B_2(r) =& - \frac{\kappa_J}{\kappa_B} \frac{1}{r^2 q(r) f(r)} \,\cdot \\
  & \bigg{(} g(R) q(R) f(R) \left[ \mathscr{H}_0(R) - \mathscr{H}(R) \right] \\
  & + \int_R^r q(r') \chi_H(r') \left[ \mathscr{H}_0(r') - \mathscr{H}(r') \right] \,\ud r' \bigg{)} \label{eq:ULTC6b}
 \end{split}
\end{align}
where the target flux is written as $\mathscr{H}_{\rm new} = \mathscr{H}_0$.

A static atmosphere was assumed so that the radiative equilibrium condition implies a constant flux $\mathscr{H}_0(r) = \mathscr{H}_0$ or equivalently $(\partial \mathscr{H}_0)/(\partial r) = 0$.

All quantities at the right hand sides of equations \eqref{eq:ULTC6a} and \eqref{eq:ULTC6b} are available upon completion of each temperature correction iteration, so that $\Delta B$ can be determined.
The temperature correction $\Delta T$ follows from $\Delta B$ using the Stefan-Boltzmann law (equation \eqref{eq:StefanBoltzmannLaw})
\begin{equation} \label{eq:TempCorr}
 \Delta T(r) = \frac{\ud T}{\ud B} \, \Delta B
             = \frac{T}{4} \frac{\Delta B}{B}
\end{equation}

In the inner regions $J_\l \approx B_\l$ so that the $\Delta B_1$ term becomes small.
In a radially extended atmosphere, the $1/r^2$ factor makes in the $\Delta B_2$ term small in the outer regions.


\subsection{The non-static generalization} \label{sec:ULNonStatic}
The non-static generalization of the UL temperature correction method must be derived from the complete SSRTE \eqref{eq:SSRTE}.

However, if the expansion velocities are small ($\beta \ll 1$) then the result for the static atmosphere (equations \eqref{eq:ULTC6a} and \eqref{eq:ULTC6b}) can be used as a reasonable approximation when the target flux for the Lagrange frame $H'_0$ is used.
For an atmosphere in radiative equilibrium the flux in the Lagrangian frame is not constant, see equation \eqref{eq:LagrangeRE}.
The value of the Eddington flux $H_0$, as measured by an external observer, is a given model constraint.
The following transformation formulae relate the Lagrangian (primed quantities) and Eulerian systems for the wavelength integrated moments of the radiation field \cite{Mihalas84}
\begin{align}
 J &= \gamma^2 \left( J' + 2\beta H' + \beta^2 K' \right) \\
 H &= \gamma^2 \left( (1 + \beta^2)H' + \beta(J'+K') \right) \label{eq:HLagrange2EulerH} \\
 K &= \gamma^2 \left( K' + 2\beta H' + \beta^2 J' \right)
\end{align}
The Eulerian target flux $H_0$ can be converted to the Lagrangian, using the inverse of equation \eqref{eq:HLagrange2EulerH}.
\begin{equation} \label{eq:TargetFluxLagrange}
 H' = \gamma^2 H \left( (1 + \beta^2) - \beta(j + k) \right)
\end{equation}
where $j \equiv J/H$ and $k \equiv K/H$.
If $j$ and $k$ are independent of the temperature then $j_{\rm new} = j_{\rm{old}}$ and $k_{\rm new} = k_{\rm{old}}$.
For the next iteration (denoted with \emph{new}) equation \eqref{eq:TargetFluxLagrange} can be written as
\begin{equation} \label{eq:HTargetLagrange}
 H'_0 = \gamma^2 H_0 \left( (1 + \beta^2) - \beta \frac{J + K}{H} \right)
\end{equation}
Here the target fluxes $H_0$ and $H'_0$ are inserted $H_{\rm new} = H_0$ and $H'_{\rm new} = H'_0$.
$J$, $H$ and $K$ are the values obtained in the current iteration (denoted with \emph{old}).

This equation yields the target Lagrange flux $H'(r)$ for all radial points $r$ in the atmosphere.
The temperature correction for moving atmospheres is then given by equation \eqref{eq:TempCorr} using the Lagrangian variables $J'$, $H'$ and $K'$.

\clearpage
\chapter{Nova atmosphere construction} \label{sec:NovaStructure}
In the theoretical description of nova outbursts, described in section \ref{sec:NovaTheory}, multiple phases can be discerned.
Whereas the agreement between theory and observations in the optical and UV was very good, the X-ray observations do not fit into the picture.
This requires further investigation in order to improve our physical understanding of the phenomenon ``nova''.
It was described in that same section that X-rays gradually start to emerge from the atmosphere after the ``fireball'' phase when the optical lightcurve has already been steeply declining for a while.
The phase identified with strong X-ray luminosity is called the ``supersoft source (SSS) phase''.
It is this phase that is modeled in this work.

According to theory \cite{Starrfield89} the density profile in earlier stages can be described by a power law
\begin{equation}
 \rho(r) \propto r^{-N}
\end{equation}
In early phases the parameter $N$ is of the order of 15.
In later phases the value of $N$ becomes much smaller, around 3.
Successful modeling of early optical and UV lightcurves were obtained with \phx\ based on density structures of this type \cite{Hauschildt95}.

In the SSS phase the total optical thickness of the envelope has decreased further, so that the radiation emerges from deeper regions in the extended nova atmosphere.
It was described in section \ref{sec:NovaTheory} that the part of material that is accelerated in the outburst below the escape velocity falls back onto the white dwarf atmosphere and reestablishes hydrostatic equilibrium.
In the SSS phase a stellar wind-type mass loss occurs, driven by radiation pressure on atomic lines \cite{Starrfield89}.
This is observationally confirmed by finding systematic blueshifts in absorption line centers \cite{Ness03} and \cite{Ness07}.

Line driven stellar winds are theoretically described by a beta law, see \cite{Castor75}, \cite{Kudritzki89}, and \cite{Lamers97}.
\begin{equation} \label{eq:BetaLaw}
 v(r) = v_\ast + (v_\infty - v_\ast) \left(1 - \frac{r_\ast}{r} \right)^\beta
\end{equation}
with $v_\ast$ being the velocity at the bottom of the wind $r_\ast$, and $v_\infty$ the terminal velocity.
The value for $\beta$ obtained for hot giant winds from theory and observations is $\beta=0.8$.
For cool stellar winds the values determined from the observations are typically larger \cite{Baade92}, \cite{Baade96}.

It cannot be known a priori how optically thick the wind is that escapes from the hydrostatic white dwarf atmosphere.
Therefore, in this work a hybrid-type atmospheric structure is assumed, composed of a hydrostatic base and an expanding envelope.
This is based on an idea of \cite{Aufdenberg00}.
In the models the optical thickness of the wind can be varied freely and the hydrostatic core provides proper boundary conditions at the base of the wind.

In the rest of this chapter the subsequent steps involved in the atmosphere construction procedure for the hybrid-type nova atmospheres are described.

\section{Atmosphere construction procedure}
\subsection{Envelope: Density}
Following \cite{Hauschildt95} the envelope is considered expanding but stationary, and spherically symmetric.
Therefore, it is assumed that all time dependent terms both in the hydrodynamics and in the radiative transfer equation can be neglected and all quantities depend only on the radial coordinate (except the specific intensity $I_\l$ of the radiation field which, in addition, depends on the angle to the radial direction and $\l$).
These assumptions are justified since the hydrodynamic timescales are much larger than the timescales for photon thermalization and for the establishment of the statistical equilibrium (see section \ref{sec:StatisticalEquilibrium}).
The assumption of spherical symmetry is only an approximation since novae are contained in a binary system which is clearly not spherically symmetric.
Whereas the 1-dimensional treatment presented in this work is already very computationally expensive with the currently available computer power, the requirements for a 3-dimensional generalization are hardly met by the largest supercomputers presently available.
Furthermore, although 3D radiation transport is doable with \phx\ \cite{Hauschildt08}, a 3D `stellar' structure is needed as input, and at present no self-consistent 3D hydrodynamical nova evolution models are known in literature that could provide such a structure.
Until such input structures become available the most realistic 1D models can provide the best possible constraints that are important for the development of nova evolution theory, which in turn is the basis for detailed 3D hydrodynamical models.

Since the wind is assumed to be time-independent and stationary, the rate of mass loss is constant with time and therefore with radius
\begin{equation} \label{eq:Continuity}
 \dot{M}(r) = 4\pi r^2 \rho(r) v(r) = \dot{M}
\end{equation}
This is called the \emph{continuity equation} and represents mass conservation in the envelope.
With the velocity field being prescribed by equation \eqref{eq:BetaLaw} the density structure of the envelope is fixed
\begin{equation} \label{eq:RhoStructure}
 \rho(r) = \frac{\dot{M}}{4 \pi r^2 v(r)}
\end{equation}

\subsection{Envelope: Radial sampling}
The radial sampling of the atmosphere should provide an optimally smooth grid for the source function to be interpolated over, see section \ref{sec:ALI}.
Large steps in optical depth from one layer to the next when integrating the specific intensity along characteristic rays leads to inaccurate solutions for the radiation field, see section \ref{sec:SourceInterpolation}.
Improper radial sampling can in principle be remedied by increasing the number of layers the atmosphere is divided in, but that is computationally expensive.
So in order to optimize the computation time \emph{and} accuracy of the radiation transport the radial sampling is important to consider.
A number of radial sampling prescriptions based on power laws or the cosh function can be found in literature (e.g. \cite{Steffen97}).
But the sampling rates obtained from these prescriptions are far from ideal.

The radial extension of the envelope can be very large (typically a factor 100-1000) and the opacity scales roughly with the density.
Therefore, the radial grid is constructed thus that the density is divided into logarithmically equally spaced parts.
\begin{align} \label{eq:DeltaRho}
 \rho(l) = \rho_\ast (\Delta \rho)^{l_\ast-l} \\
 \Delta \rho = \left(\frac{\rho(1)}{\rho_\ast}\right)^{\frac{1}{l_\ast -1}}
\end{align}
where $l=1,2,\dots,l_\ast$ is the layer number increasing inwards, and $l_\ast$ the bottom layer of the wind (and simultaneously the top layer of the hydrostatic core).
The determination of $\Delta \rho$ is possible because the density structure is fixed by the parameters of the wind, equation \eqref{eq:RhoStructure}.
Now the radial grid follows from the continuity equation \eqref{eq:Continuity}
\begin{equation}
 0 \stackrel{!}{=} (r(l))^2 - \frac{\dot{M}}{4\pi \rho(l) v(l)}
\end{equation}
which is straight forward to solve numerically, layerwise, starting from the inner boundary, for example with a Newton-Raphson scheme.

\subsection{Envelope: Optical depth scale}
Once the density and radial grids for the expanding envelope are fixed, the optical depth grid\footnote
{This $\tau$-scale plays no direct role for the expanding envelope, but it provides the boundary condition for the construction of the hydrostatic core of the model (see section \ref{sec:HydrostaticCore}).
}
$\tau_{\l_{\rm ref}}$ for that range $(l=1,\dots,l_\ast)$ can be integrated inwards using
\begin{equation} \label{eq:dTau}
 \ud \tau_{\l_{\rm ref}} = -(\chi_{\l_{\rm ref}} - x\sigma_{\l_{\rm ref}}) \,\ud r
\end{equation}
where $\chi$ is defined by equation \eqref{eq:DetailedExtinction} \emph{except} that the cores of atomic lines are excluded.
It is important to include the opacities of far line wings in the construction of the $\tau_{\l_{\rm ref}}$ scale, since these make up for a significant part of the total opacities in some parts of the atmosphere.
This has a major impact on the $\tau_{\l_{\rm ref}}$ scale.
$x$ is a free parameter $0\le x \le1$ that is used to turn off the scattering contribution to the extinction in the optical depth reference scale, which mainly influences equation \eqref{eq:dTau} of the outermost layers.
The values of $x$ and $\l_{\rm ref}$ depend on the physical conditions of the atmosphere.
For the models computed in this work the typical values are $x=1$ and $\l_{\rm ref} = 40$\AA{}.

\subsection{Core: Optical depth scale} \label{sec:HydrostaticCore}
For the hydrostatic layers $(l=l_\ast+1,\dots,N)$ the structure is computed on a fixed logarithmically spaced optical depth grid
\begin{equation} \label{eq:HydrostaticTauGrid}
 \log (\tau(l)) = \log (\tau(l_\ast)) + (l - l_\ast)
   \left[ \frac{\log (\tau(N)) - \log (\tau(l_\ast))} {N - l_\ast} \right]
\end{equation}
starting from the value of the dynamic region's bottom layer $\tau(l_\ast)$.
Here $\tau = \tau_{\l_{\rm ref}}$ to ease the notation.
In the following for all occurrences of $\tau$ referring to the optical depth grid $\tau=\tau_{\l_{\rm ref}}$ holds.
The lower boundary is defined for an optical depth $\tau(N) = \tau_\mathrm{max}$ with a typical value of $\tau_\mathrm{max}=1000$ to ensure it is set well below the thermalization depth\footnote
{The thermalization depth is defined as the optical depth at which the radiation field "thermalizes" $J_\lambda = B_\lambda$ as seen from outside. It is not easy to pin down the exact location at which this is satisfied, therefore mostly a thermalization proximity is meant.
}.

\subsection{Both: Temperature structure}
The temperature structure $T(l)$ is provided as input to the model.
Deviations from \emph{that} temperature structure that fulfills the condition of radiative equilibrium are iteratively corrected for, see sections \eqref{sec:ULTC}, \eqref{sec:GlobalIterationScheme} and \eqref{sec:ULNLTE}.

The effective temperature specifies the Eulerian target flux $H_0$ according to equation \eqref{eq:StefanBoltzmannLaw} (see also equations \eqref{eq:HTargetLagrange} and \eqref{eq:ULTC6b}) via the Stefan-Boltzmann law
\begin{equation} \label{eq:EffectiveRadius}
  H_0^{\rm obs}(r) = \frac{R_*^2}{r^2} \sigma_{\rm SB} T_{\rm eff}^4
\end{equation}
In this definition the 'effective radius' (the radius specified together with the effective temperature in the Stefan-Boltzmann law) is not related to the optical depth scale but fixed to the outer radius of the white dwarf, being the base of the wind.
This choice has a large advantage in the convergence properties and the stability of the models.
The optical depth scale changes during the iterative process, so that also a target flux that depends on it would fluctuate.

Care must be taken with the physical interpretation of the ``effective temperature'', since in fact it only directly specifies the model's luminosity (and not any real temperature).

\subsection{Both: Structure completion}
Now the complete atmospheric structure can be computed for each layer.

For the expanding envelope the gas pressure $P_g(l)$ is calculated from the ideal gas law using the given density and temperature structures
\begin{equation} \label{eq:IdealGasLaw}
 P_g = \frac{\rho k T}{\mu m_H}
\end{equation}
with $\mu \equiv \bar{m}/m_H$ being the mean atomic weight per free particle (including free electrons) in atomic units, and $P_g = \sum_i P_i = \frac{\rho k T}{m_i}$ being the gas pressure is the sum of all partial pressures, where $i$ specifies all types of particles including free electrons.

For the hydrostatic core this is done by numerically integrating the hydrostatic equation
\begin{equation} \label{eq:Hydrostatic}
 \frac{\ud P_g}{\ud \tau_{\rm std}} = - \frac{g}{\kappa_{\rm std}}
\end{equation}
downwards, starting from the transition layer $l_\ast$ at which the pressure, density and thus the mass extinction coefficient are already known.
Finally, the radius for every layer is calculated from equation \eqref{eq:dTau}.

\section{Structure check}
Thus far no restrictions were made to ensure a smooth transition from the expanding to the hydrostatic region.
The hydrostatics is parameterized by the gravity log($g$).
The wind is parameterized by mass loss rate $\dot{M}$, the terminal velocity $v_\infty$, the wind parameter $\beta$ and the base velocity $v_\ast$.

\begin{figure}
 \centerline{ \includegraphics[height=\textwidth,angle=90]{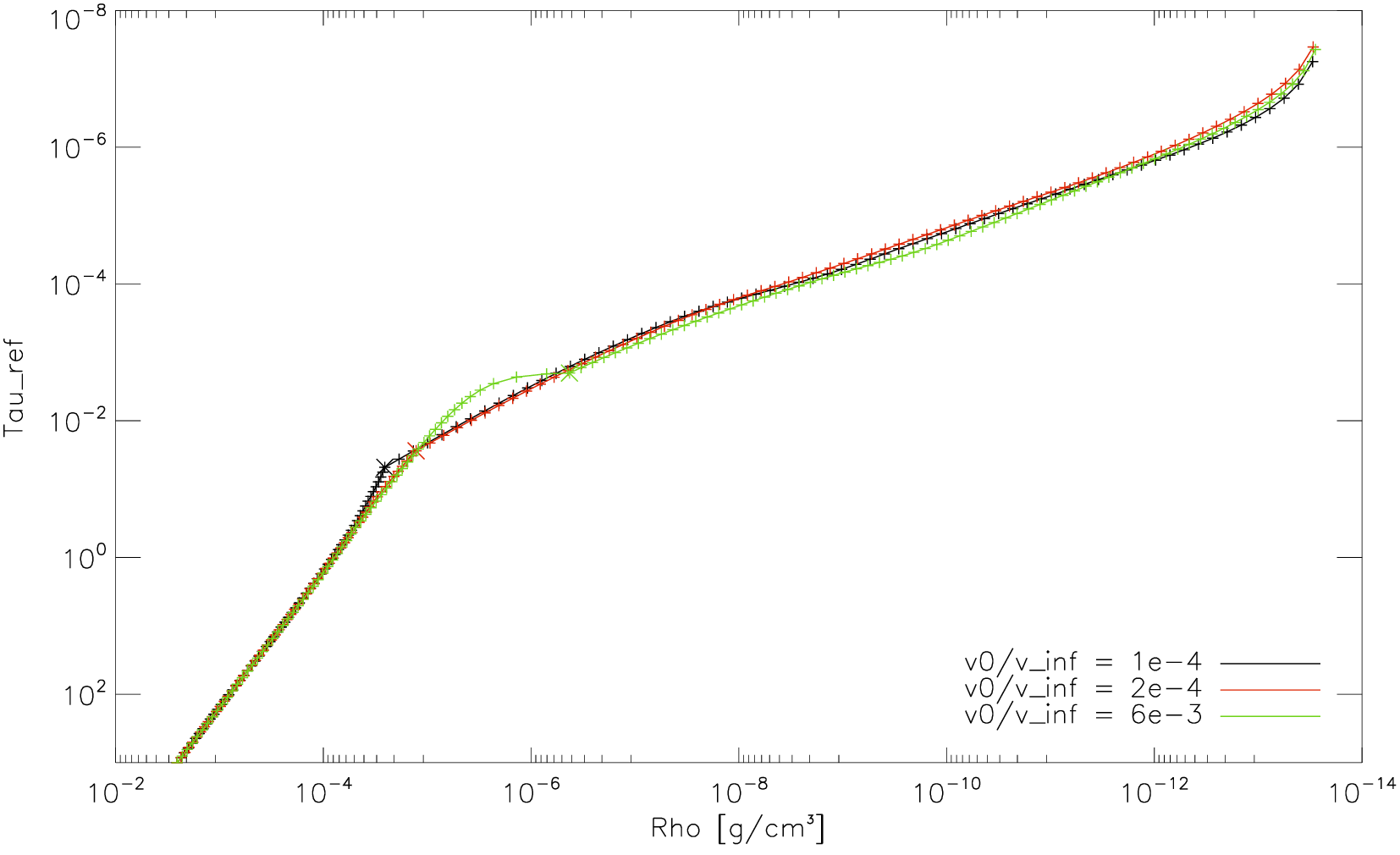}}
 \caption{ \label{fig:TransitionLayer1}
 The optical depth scale, here plotted against the density, shows the transition from core to envelope around $\rho=10^{-5}$ g/cm$^4$ as a kink in the curve.
 The bottom wind layer is marked with a cross sign, the other layers with a plus.
 \newline
 The base velocity of the expanding envelope $v_\ast$ has an important impact on the transition.
 The proper value (red curve) yields a curve that is straight at both sides of the intersection point (transition layer).
 If the wind starts too far `in' or `out' (black and green) then the boundary condition for the hydrostatics is inappropriate, and the hydrostatics are distorted.
 }
\end{figure}

\begin{figure}
 \centerline{ \includegraphics[height=\textwidth,angle=90]{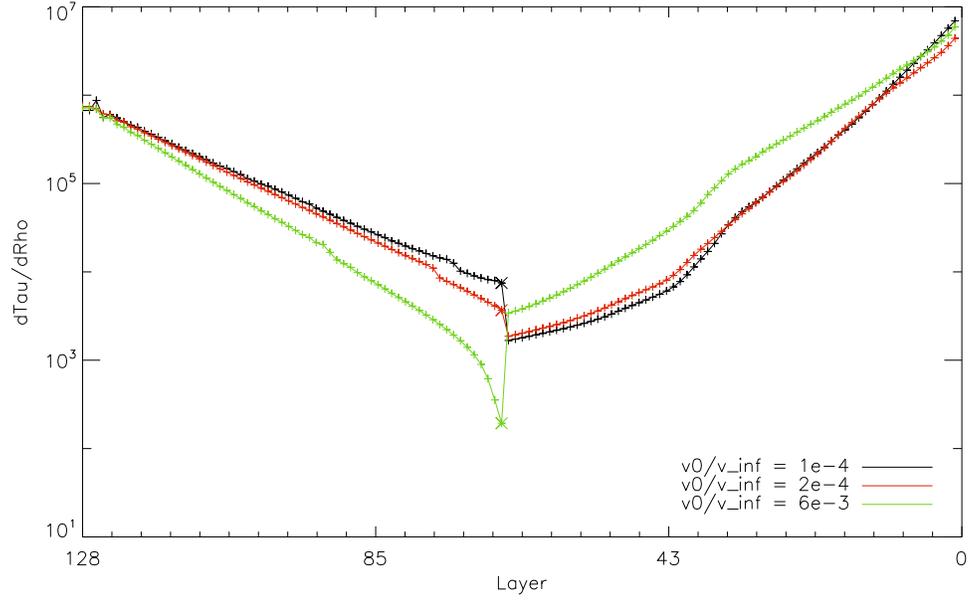}}
 \caption{ \label{fig:TransitionLayer2}
 Here $\ud \tau_{\rm ref} / \ud \rho$, being the slope of the graphs in figure \ref{fig:TransitionLayer1}, is plotted logarithmically against the layer number.
 The discontinuity at layer 67 is inherent to the construction with an ad-hoc boundary between hydrostatics and hydrodynamical expansion.
 The bottom wind layer provides the boundary condition for the hydrostatic core.
 The parameter $v_\ast$ must be tuned in order to provide the proper conditions at the boundary that meet the requirements of hydrostatics.
 In this example $v_\ast/v_\infty = 2 \cdot 10^{-4}$ (red curve) is the proper value.
 With too low or too high values (black and green) the hydrostatics (on the left side of the transition) are distorted.
 Close to the transition, the lines get curved away from the straight, undistorted run for hydrostatics.
 \newline
 In the undistorted case the slope of the hydrostatic curve only depends on the radial sampling rate (number of layers per optical length) and log($g$).
 In this example the log($g$) is equal for the three models and the different slopes are caused by different sampling rates due to a shift `out'-, or `in'-wards of the transition layer corresponding to the value of $v_\ast$.
 }
\end{figure}

\begin{figure}
 \centerline{ \includegraphics[height=\textwidth,angle=90]{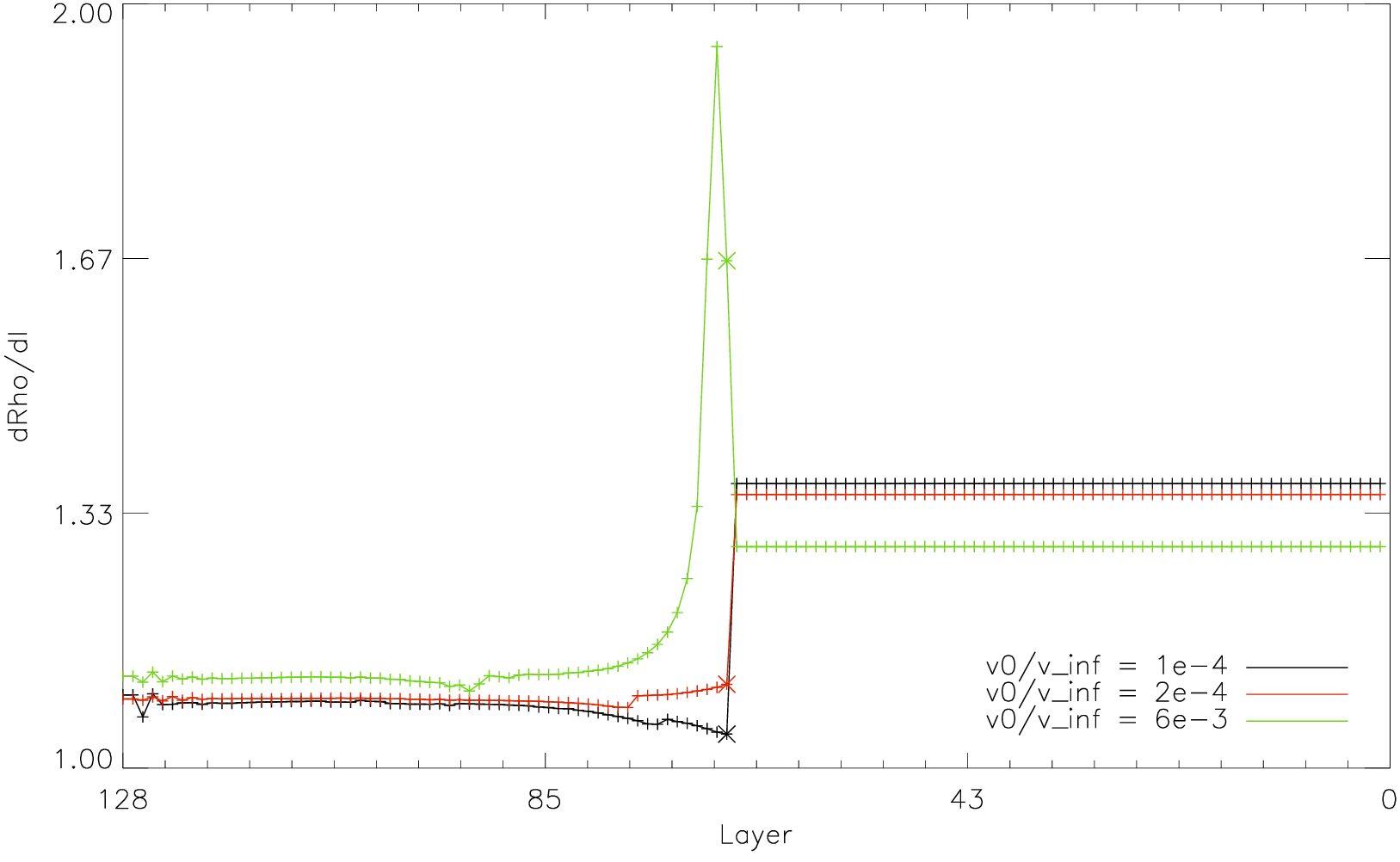}}
 \caption{ \label{fig:TransitionLayer3}
 This plot shows the sampling rate of the density by the radial grid $\ud \rho / \ud l$.
 The sampling of the expanding envelope ($l \le 67$) is fixed by construction (equation \eqref{eq:DeltaRho}).
 The sampling of the core ($l>67$) follows from the hydrostatics on a logarithmic $\tau_{\rm ref}$ scale with the transition layer as boundary condition.
 \newline
 With the proper boundary condition (red curve), the density sampling rate for the hydrostatic range is approximately constant.
 For too small values of $v_\ast/v_\infty$ the density sampling $\ud \rho / \ud l$ can become smaller than 1, in which case the density structure is no longer monotonically increasing.
 For too large values, the changes in density from layer to layer become large which is disadvantageous for the computation of the radiation field (especially, the interpolation of the source function over large optical lengths can introduce large inaccuracies, see sections \ref{sec:ALI} and \ref{sec:SourceInterpolation}).
 }
\end{figure}

It is right \emph{this} last parameter $v_\ast$ that is responsible for a gradual transition in the boundary layer $l_\ast$.
The effect of this parameter on the transition is shown in figures \ref{fig:TransitionLayer1} to \ref{fig:TransitionLayer3}.
Apparently, a bad value for $v_0$ distorts the structure of the core.
Whereas in a hydrostatic atmosphere the density (layer-)gradient (figure \ref{fig:TransitionLayer2}) is constant, this is not the case if the boundary conditions supplied by the wind, i.e. the opacity and density of the transition layer, are not appropriate.
This distortion can have significant influence on the structure and thus also on the resulting spectrum.
Another effect of a unsuitable value for $v_0$ is that the atmosphere cannot be accurately brought into radiative equilibrium.

As a last step in the atmosphere construction procedure, this transition is checked, and corrected if too far off.
This is done by checking if the density sampling rate (see figure \ref{fig:TransitionLayer3}) is approximately constant for the hydrostatic region.
If $\ud \rho /\ud l$ at the outermost hydrostatic layer is off from the mean over the 10 inner layers (excluding the innermost, with optical depth of $\tau > 10^2$, as these are not reliable) by more than 20\%, then the value of $v_\ast$ is corrected and the atmosphere construction scheme is repeated.

\clearpage
\chapter{Methods for good NLTE models} \label{sec:GoodNLTE}

\section{Global iteration scheme} \label{sec:GlobalIterationScheme}
It is a complex task to attain a fully converged NLTE atmosphere model.
The procedure used in \phx\ is described in \cite{Hauschildt99}.
Basically, it is a four-steps iterative process.
It starts from an initial guess for the atmospheric structure, something like $[T,\rho,p_{\rm gas},n_i](r)$.
\begin{enumerate}
\item[1] In the first step the \emph{radiation field} is computed using the SSRTE \eqref{eq:SSRTE}.

\item[2] Then, from the radiation field the deviation from \emph{radiative equilibrium} (equation \eqref{eq:REErrorDif}) can be checked and, if necessary, the \emph{temperature correction} (equation \eqref{eq:StefanBoltzmannLaw}) can be computed.

\item[2b] For NLTE models the radiative rates are calculated from the radiation field, and the rate matrix equation is solved.
This updates the \emph{occupation numbers} $n_i$.
From these $n_i$ the $b_i$ are determined.

\item[3] The \emph{new temperature} $T_{\rm new}(r)$ is applied.

\item[3b] For NLTE models the $n_i$ for this temperature follow from $n_i^\LTE(T_{\rm new})$ and the $b_i$ (computed in step 2b), using equation \eqref{eq:Ni2Bi}.

\item[4] The \emph{structure} is adapted to the new temperature.
\end{enumerate}
With the updated structure, after step 4, a new cycle starts.
This is continued, until in step 2 the deviation from RE is smaller than a prescribed threshold.

In this work, the conventional iteration scheme used in \phx\ was found to have a poor convergence performance for NLTE models.
Often the solution diverges or very many cycles are needed to attain the prescribed convergence criteria.
This is sometimes referred to as apparent convergence as opposed to true convergence \cite{Petz05}.

Relaxing the assumption of LTE to Non-LTE gravely complicates the coherence in atmospheric models.
In LTE the atomic level populations of the gas are fully determined by one single quantity, the temperature $T$ (section \ref{sec:LTE}).
In such models the temperature structure is one of the most important properties of the atmosphere, it characterizes the spectrum.

In NLTE the level populations are no longer directly determined by the temperature, but by the rates.
In the outer parts of the atmosphere, where the densities are low, the radiative rates play a dominant role compared to the collisional rates.
The bound-bound radiative rates are functions of the mean radiation field $J$ only (equations \eqref{eq:BBUpRates} and \eqref{eq:BBDownRates}).
The bound-free radiative rates primary depend on $J$, but also on the electron density $n_e$ and the temperature $T$ (equations \eqref{eq:BFUpRates} and \eqref{eq:BFDownRates}).

In order to improve the convergence performance for NLTE models the iteration procedure must be altered.
It is important that the radiation field is fully consistent with the population numbers, before the temperature corrections are derived and the structure is updated.
In the following this is called \emph{radiation-matter consistency}.
In LTE models, this consistency is immediately given.
In NLTE models, for this consistency an inner iteration loop is needed, comprising of step 1 and step 2(b) only.
Or, to put it another way, in every iteration the structure must not be updated (step 3 and 4) before this consistency is given.
In the following, the iterations in which the structure is not updated are called \emph{idle cycles}.
Whether this condition of (approximate) consistency between the radiation field and the level populations is given can be checked by the changes in the $J$ and the $n_i$ from iteration to iteration.

An implicit condition for this consistency is that the ionization balances of all NLTE species do not change significantly from iteration to iteration.
Here, ions with small ion number densities are less significant than those with large.
In order to check for this condition the following definition of a weighted average of ionization change is useful
\begin{equation}
\begin{split}
 \overline{\Delta N}_E &= \frac{\sum N_s \, {\rm max}(N_s/N_{s,{\rm old}},N_{s,{\rm old}}/N_s)}{\sum N_s} \\
  &= \frac{\sum{\rm max}(N_s^2/N_{s,{\rm old}},N_{s,{\rm old}})}{\sum N_s} \label{eq:MeanDeltaN}
\end{split}
\end{equation}
where the sums extend over all ionization stages $s$ of chemical element $E$, $N_s$ is given by equation \eqref{eq:NLTEIonization}, and the ${\rm max}()$ function returns the maximum value of its arguments.

\begin{figure}
 \centerline{ \includegraphics[height=\textwidth,angle=90]{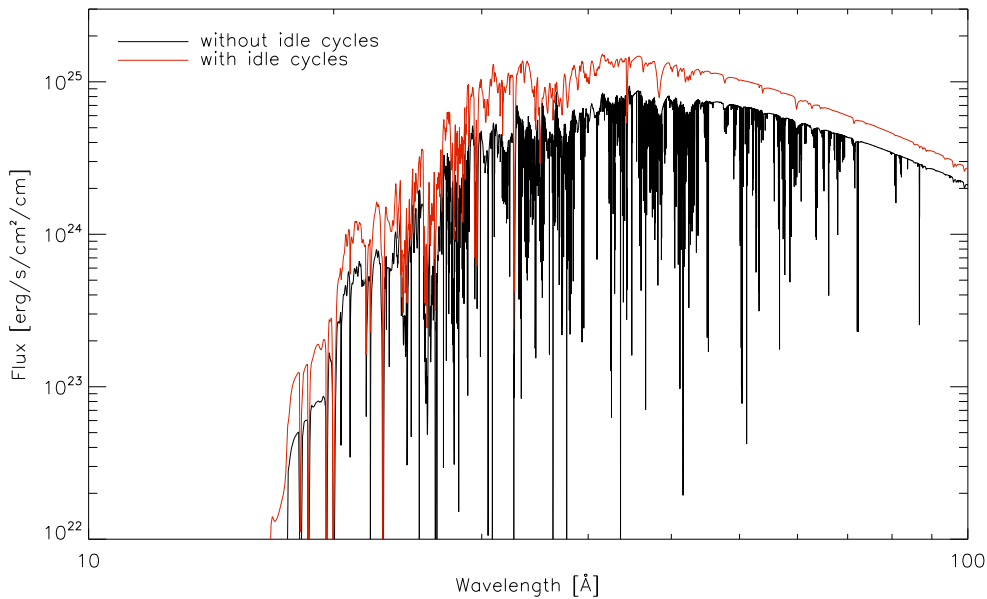}}
 \caption{In NLTE models the radiation field and the population numbers must be consistent before the atmospheric structure is updated.
  This consistency is achieved using idle cycles, in which the radiation field, the radiative rates and the population numbers are updated, but the structure is held fixed.
  Without idle cycles NLTE models hardly converge to RE or often even diverge.\newline
  This plot shows two spectra, for which the same number of updates to the structure were performed.
  The models are identical except for using idle cycles or not.
  The model with idle cycles (red) is in RE, whereas the model without idle cycles (black) diverges (becomes worse after more iterations).
 } \label{fig:ULIdle1}
\end{figure}

In figure \ref{fig:ULIdle1} an example is given of a model spectrum resulting from poor radiation-matter consistency.
The model performed the same number of updates to the structure as the reference model, performed with idle cycles.
The model without idle cycles diverges and the resulting spectrum becomes worse in every step.
In the reference model the number of idle cycles per full cycle is one.

Typically, one or two idle cycles are needed, depending on the changes of the radiation field.
If the changes are small, the radiation-matter consistency is quickly attained.

\section{Pure NLTE} \label{sec:PureNLTE}
NLTE models are commonly treated as a gradual refinement to LTE models.
That means that an LTE model is used as the initial guess, on which more and more species are subsequently treated in NLTE.

This approach may be good for models that are close to LTE in all regions of the atmosphere.
But if the deviations from LTE are large, then LTE level populations are far off from the actual populations.
Consequently LTE opacities are far off from the actual opacities.
The opacities dictate the radiation field and the atmospheric structure.
It is a matter of experimentation to find out if, for the model under consideration, the LTE opacity is more realistic than a null-opacity (no opacity at all) for the specific species.
This depends on the actual deviation from LTE.
\begin{figure}
 \centerline{ \includegraphics[height=1.0\textwidth,angle=90]{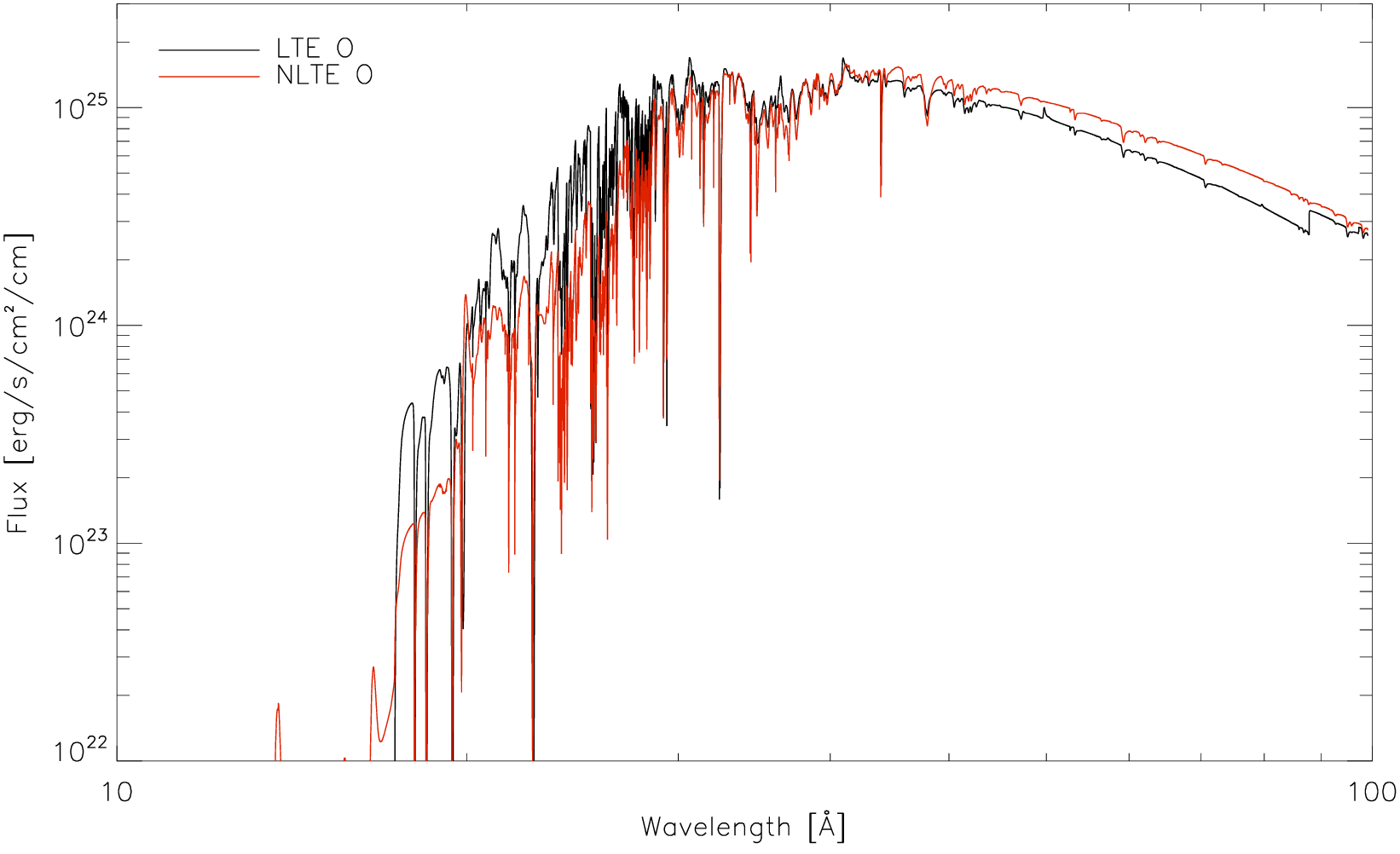}}
 \centerline{ \includegraphics[height=1.0\textwidth,angle=90]{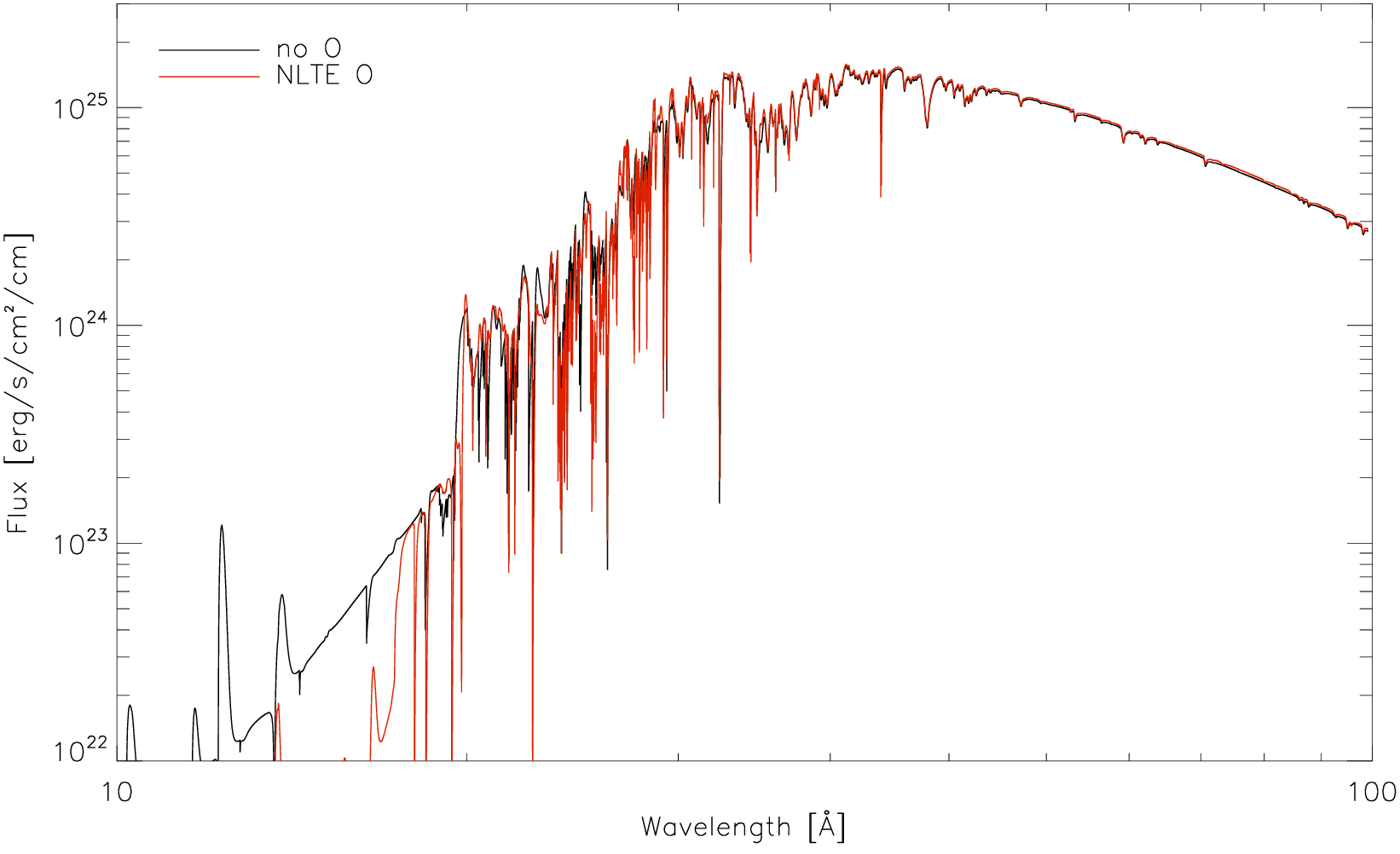}}
 \caption{Two test models (black) are compared to one reference model (red).
  In The reference model the following elements are treated in NLTE: H, He, CNO, Ne, Mg, Al, Si, S, Ar and Ca.
  It has no LTE opacities.
  In the test model in the upper graph oxygen is treated in LTE.
  In the lower graph oxygen is not considered at all, so no oxygen opacities are included.\newline
  Clearly, the oxygen LTE model departs more from the reference model than the model with no O at all.
  Note that still the far majority of the opacities is in NLTE.\newline
  In general, adding LTE opacities to NLTE models of the kind presented in this work deteriorates the results.
 } \label{fig:LTEvsNull.7}
\end{figure}
\begin{figure}
 \centerline{ \includegraphics[height=1.0\textwidth,angle=90]{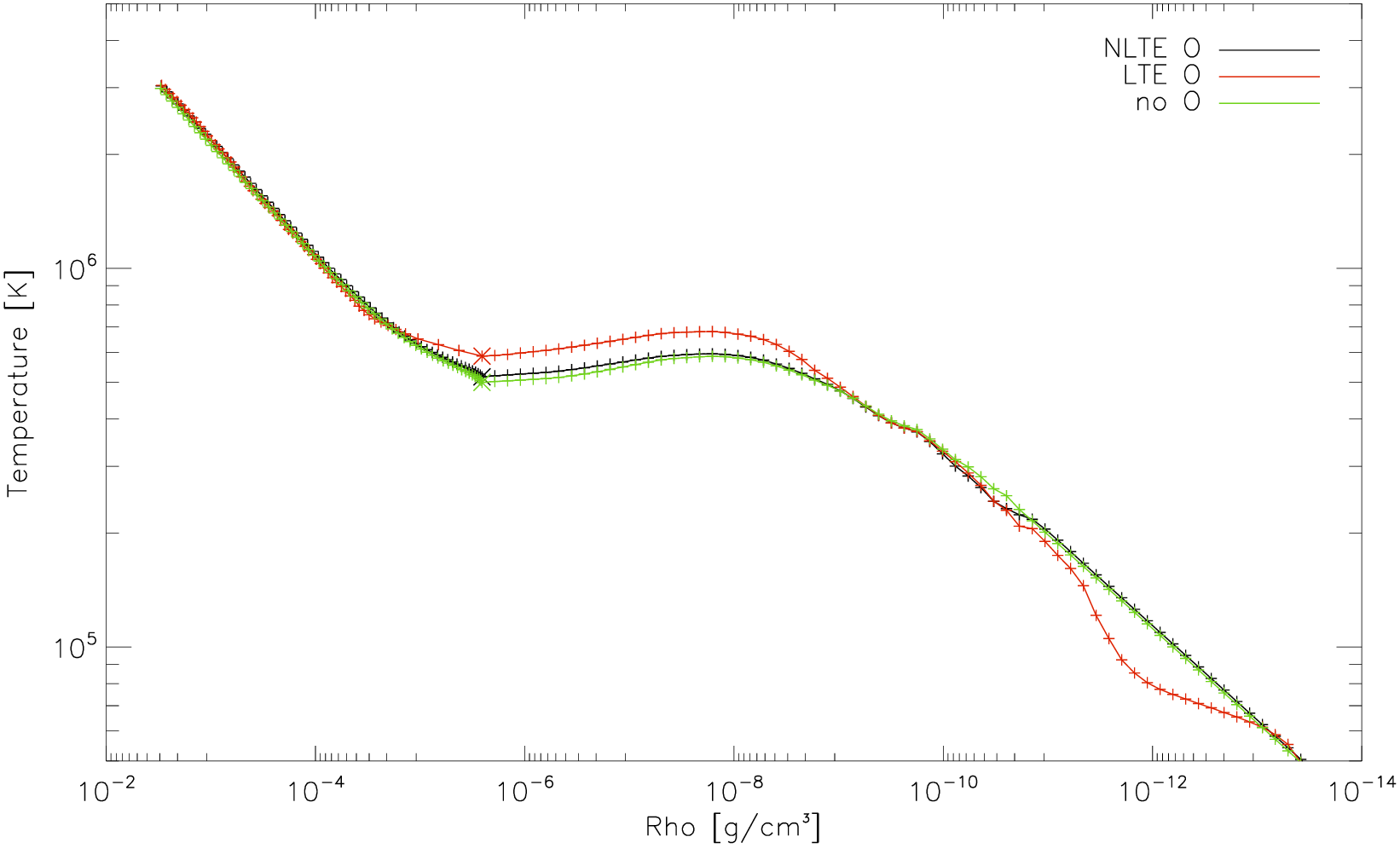}}
 \centerline{ \includegraphics[height=1.0\textwidth,angle=90]{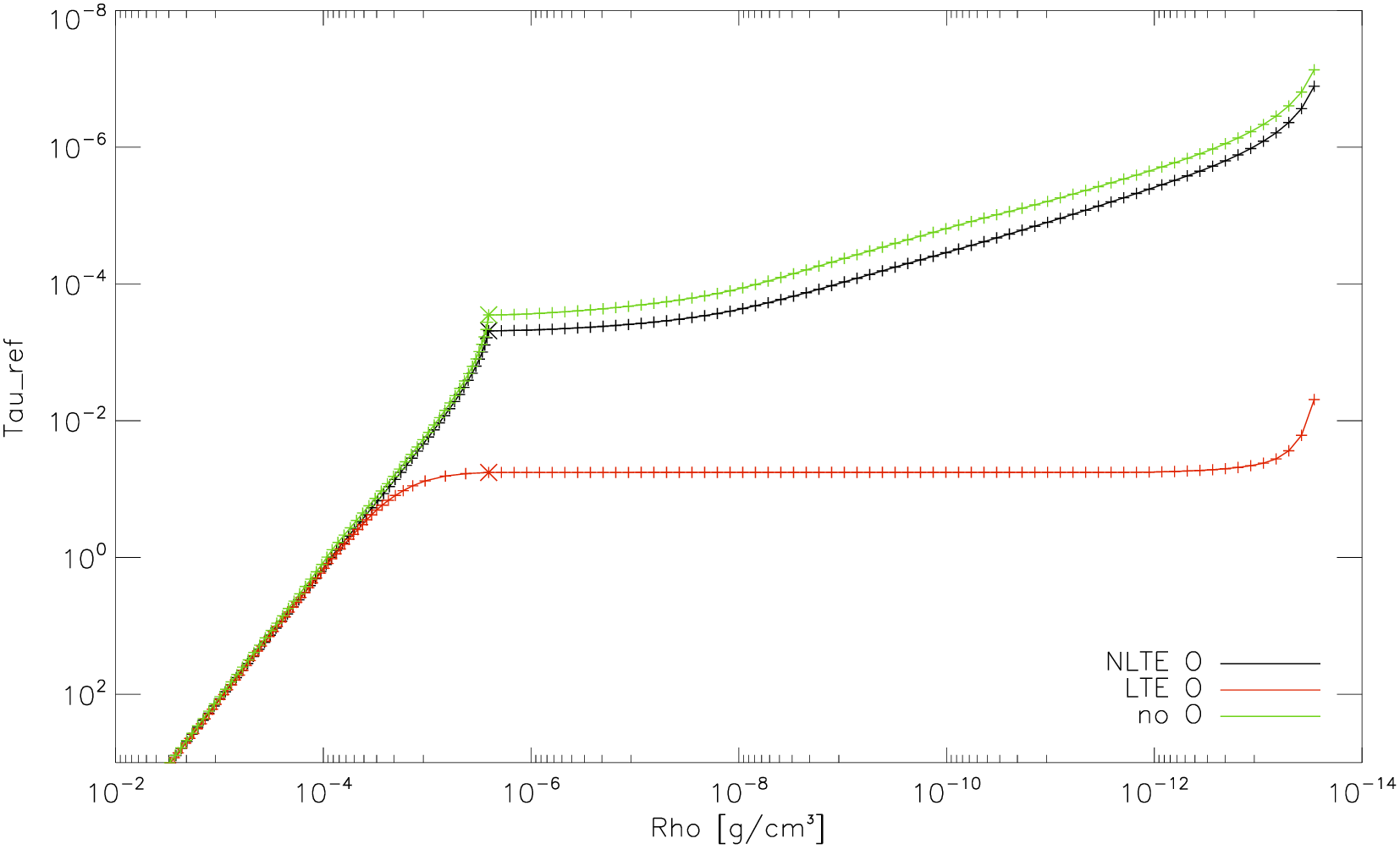}}
 \caption{Two test models (red and green) are compared to one reference model (black).
  The models are described in the previous figure caption.
  The upper graph shows the temperature structure against the density.
  The lower graph shows the optical depth at a reference wavelength (40\AA) $\tau_\mathrm{ref}$ against the density.
  This $\tau_\mathrm{ref}$ is used to construct the atmosphere.\newline
  Clearly, the oxygen LTE model departs more from the reference model than the model with no O at all.
  Note that still the far majority of the opacities is in NLTE.\newline
  In general, adding LTE opacities to NLTE models of the kind presented in this work deteriorates the results.
 } \label{fig:LTEvsNull.20}
\end{figure}

For the models performed in this work, it is found that adding LTE opacities is a worse approximation than leaving those specific opacities out altogether.
This effect is shown by an example of a typical model in figures \ref{fig:LTEvsNull.7} and \ref{fig:LTEvsNull.20}.
A comparison between pure NLTE and pure LTE models is made in section \ref{sec:LTEvsNLTE}.

The comparison is made between a full stage 2 model, which treats all elements up to Calcium in NLTE (see Table \ref{tab:NLTEStages}), and a model in which oxygen is not treated at all or treated in LTE.
The two test models started from the reference model, and passed the same number of iterations.
In all models RE is obtained.
Clearly, the model without oxygen opacities is more similar to the full model than the one with LTE oxygen opacities.
Note that in the test model with oxygen in LTE still the far majority of the opacities was treated in NLTE.

Generally, adding LTE opacities to the type of NLTE models considered in this work detoriates the resulting structure and the model spectrum.
Therefore, only pure NLTE models are computed, unless stated otherwise.

\section{Modeling steps} \label{sec:ModelingSteps}
\subsection{Multi-stage approach}
The number of iterations needed to let a model converge to RE depends significantly on how close the initial guess is to the final model structure.
The time needed for an iteration depends on which chemical species are treated in NLTE, especially the number of atomic lines (step 1) and levels (step 2b).
In order to minimize the total computation time the modeling process is divided into several stages.
In every stage the number of species is increased, see Table \ref{tab:NLTEStages}, and thus the model comes closer to the final result.
This is a typical approach, see \cite{Petz05}.
\begin{table}
\centerline{
\begin{tabular}{c|l|r|r}
stage & atomic species & \# lines & \# levels \\
\hline
1 & H He C N O                        &   2800 &  600 \\
2 & H He C N O Ne Mg Al Si S Ar Ca    &  46000 & 4000 \\
3 & H He C N O Ne Mg Al Si S Ar Ca Fe & 160000 & 7600
\end{tabular}}
\caption{Every model is computed in three subsequent stages with an increasing number of chemical elements treated in NLTE.
 The table shows the typical number of atomic levels and lines considered in each stage.} \label{tab:NLTEStages}
\end{table}

\begin{figure}
 \centerline{ \includegraphics[height=1.0\textwidth,angle=90]{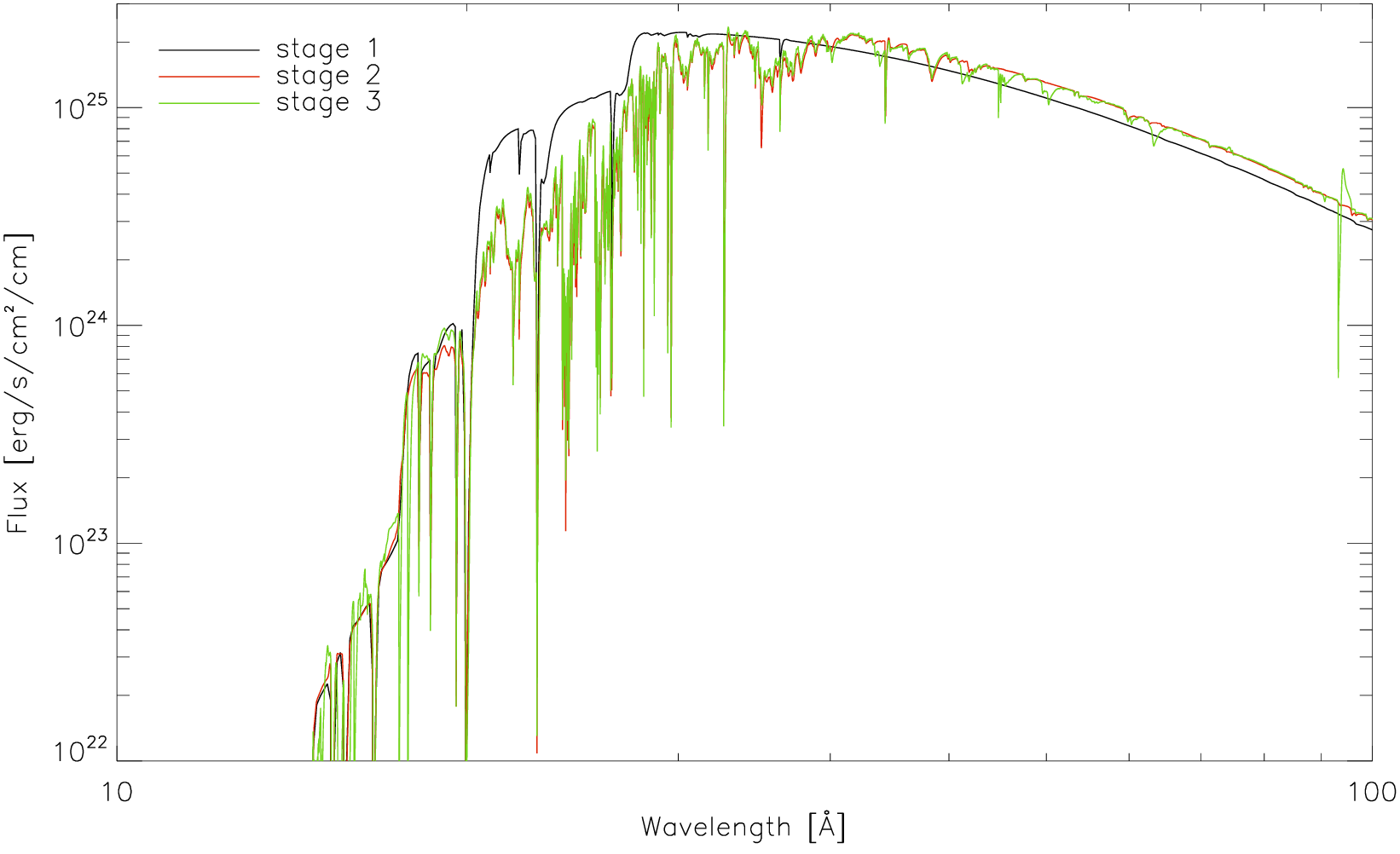}}
 \centerline{ \includegraphics[height=1.0\textwidth,angle=90]{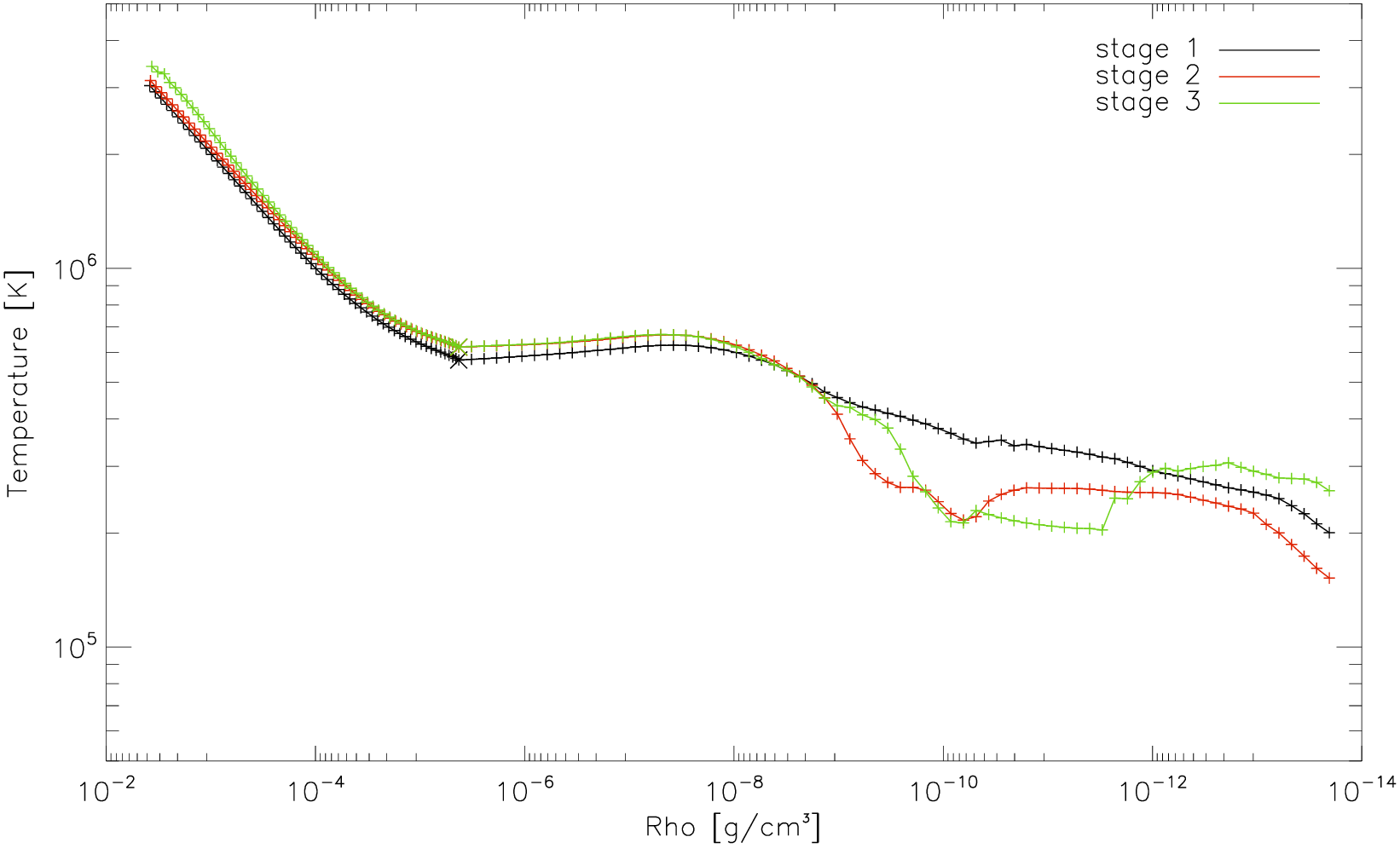}}
 \caption{The modeling process is done in three subsequent steps (black, red and green), in which the number of atomic species is increased, and with that the computation time.
  In every step the model comes closer to the final model (green).
  This procedure provides a computation time saving way to obtain a converged complex NLTE model.
  The upper graph shows the spectra, the lower graph shows the temperature structures for the three subsequent stages.
} \label{fig:ModelSteps}
\end{figure}

The first stage quickly delivers a rather good initial guess for the second stage.
And the second delivers relatively fast a good initial guess for the last stage, being the final model, which is very computationally expensive.
An example of the resulting spectra and temperature structures for the tree stages are shown in figure \ref{fig:ModelSteps}.
Note that in every step the model comes closer to the final result.

\subsection{Adding new species in NLTE}
If new species are added in NLTE, then the occupation numbers $n_i = n_i^\LTE$ for those species are far off.
That means that initially the opacities, therefore the radiation field, and therefore the radiative rates are far off from their right values.
The $n_i$ of the NLTE species of the previous stage are already close to the converged solution.
Keeping those $n_i$ fixed for a few idle cycles when adding new species in NLTE prevents them from being detoriated by the yet unrealistic new opacities.
Furthermore, this improves the radiation field which helps the new species to converge to the final occupation numbers.

When the $n_i$ of the new species have converged to an acceptable accuracy, i.e. when their changes from one idle iteration to the next become small, then the $n_i$ of the old species need to adapt to the altered opacities.
The same trick is applied again, keeping the new $n_i$ fixed and let the old converge.
After that, normal model iteration can proceed, as described in section \ref{sec:GlobalIterationScheme}.

\subsection{Countering micro-inconsistencies} \label{sec:MicroInconsistencies}
In NLTE models there is a huge number of variables, for each radial point in the atmosphere, e.g. temperature, electron density, and occupation numbers of all atomic levels.
Using the iterative modeling process, see section \ref{sec:GlobalIterationScheme}, their coupling is treated in an indirect way, via the radiation field.

Even with the greatest caution, it can occur that during the iterative refinement procedure some occupation numbers enter a wrongly or weakly coupled region.
In such a case, once a (couple of) $n_i$ is in such a region it will not be able to approach the right value soon.
In this work this is called \emph{micro-inconsistency}, as generally their influence is small on wavelength integrated quantities, and thus on the atmospheric structure.
Especially, if the initial structure is far from the final solution there is a long way for such micro-inconsistencies to develop.

A rigorous way to deal with micro-inconsistencies is to sequentially reset parts of the variables to default values and let them iteratively converge again, keeping all others fixed.
In principle, this can be done with any part of the variables, like the outer part of the temperature structure.
But the temperature is directly coupled to the gas pressure and density via the ideal gas law.
Therefore, it is easier restrict to a \emph{$n_i$-reset method}, resetting the $n_i$ of one or a few species at a time.
As default values the LTE values can be used.
A $n_i$-reset is in fact a set of idle-cycles in which not only the structure is kept fixed, but also the $n_i$ of all but a small, subsequently changing set of the atomic species.
The number of idle cycles needed per set of species to let the $n_i$ to approach the right values depends on the model atom and on the deviation from LTE.

\begin{figure}
 \centerline{ \includegraphics[height=1.0\textwidth,angle=90]{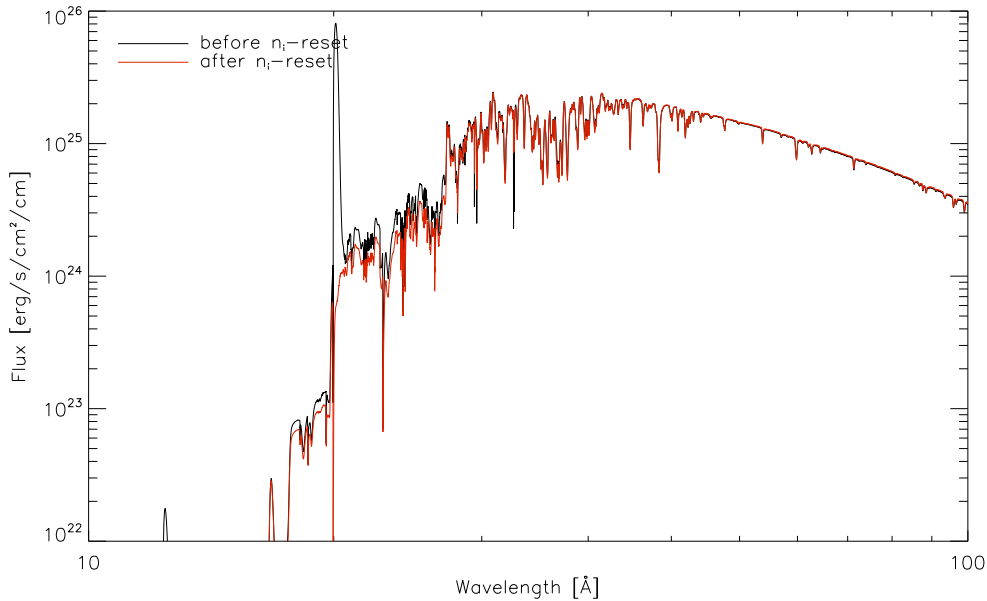}}
 \caption{In the process of iterative refinement of an initial structure to the final solution for the model micro-inconsistencies can occur.
  That means that some of the large number of $n_i$ in the model end up with a wrong value and diverge or slowly converge.
 Most often, micro-inconsistencies are not very noticeable, but can yet have a significant influence on the spectra. \newline
  In the example shown here, the existence is noticeable: a strong emission line appears due to a wrong $n_i$ (black curve).
  After a $n_i$-reset (see section \ref{sec:MicroInconsistencies}) the problem is solved (red curve).
 } \label{fig:MicroInconsitencies}
\end{figure}

Micro inconsistencies sometimes show up as strong emission lines, as in the example shown in figure \ref{fig:MicroInconsitencies}.
In such cases, their existence is evident.
However, mostly they are inconspicuous but can still have a large overall effect on the spectrum.
In any case, their existence can be checked for and eliminated by the $n_i$-reset method.
Therefore, in each of the three stages, a seemingly converged model is checked with the $n_i$-reset method.
After the reset, the model is iterated further until RE is reached.
This process is repeated until the changes to the structure, and thus to the spectrum, between two $n_i$-resets are small.



\section{Transferring temperature corrections to the $n_i$} \label{sec:NiScaling}
In the previous sections of this chapter \ref{sec:GlobalIterationScheme}, \ref{sec:PureNLTE} and \ref{sec:ModelingSteps} it has been described, that bringing a model into statistical, radiative and hydrostatic(or dynamic) equilibrium is a numerically complex task, due to the large number of indirectly coupled variables.
In order to attain radiative equilibrium the temperature structure is subsequently improved, and after each correction, the $n_i$ must find their balance again.

As described in section \ref{sec:NLTEIonization} the change in the local NLTE ionization balance due to a change of the local temperature can not generally be predicted.
In the limit of small departures from LTE, the NLTE ionization balance changes according to the Saha-equation.
In the opposite limit of large deviations from LTE, no coupling to the local temperature is expected.

\begin{figure}
 \centerline{ \includegraphics[width=\textwidth]{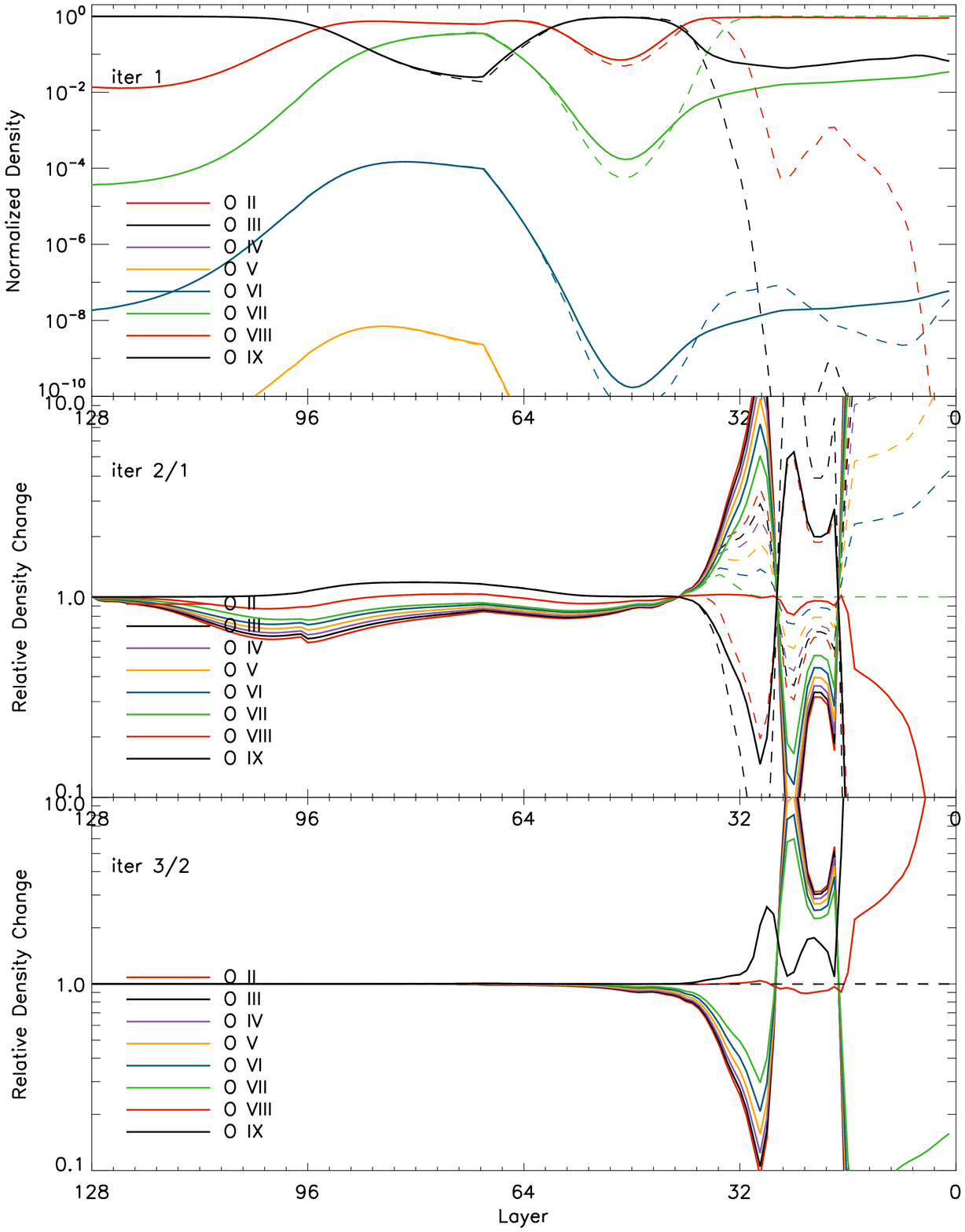}}
 \caption{
  Ordinarily, the population numbers are scaled in a LTE-like manner with the temperature, keeping the $b_i$ or the $b_i^*$ fixed.
  That often leads to strong overcorrections in regions where the $n_i$ are weakly coupled to the local temperature. \newline
  The upper graph shows the ionization balance of oxygen, with the deepest atmosphere layers at the left.
  The solid curves show the NLTE and the dashed curves the LTE balance.
  In the middle graph, the relative changes from iteration 1 to iteration 2 are plotted on a logarithmic scale.
  The changes in the LTE balance show that the temperature changed.
  The changes in the NLTE balance follow the LTE changes, but are not identical since not the $b_i$ but the $b_i^*$ were kept fixed.
  Obviously, overcorrections from iteration 2 in the outer 40 layers are largely canceled again in the third iteration (bottom).
 } \label{fig:NiOvercorrections}
\end{figure}

Ordinarily, in \phx, when the temperature structure is corrected, the $b_i^*$ are kept fixed, and the new $n_i$ follow from equations \eqref{eq:NiStar} to \eqref{eq:BiStar} after solving the NLTE ionization balance, equation \eqref{eq:NLTESaha}, for the new temperature.
This is a good approximation for the inner regions of the atmosphere.
But analysis shows that in the outer regions this often leads to overcorrections, from which the $n_i$ partly recover in following idle iterations, and finally approximate the balance before the overcorrection was made.
An example of this effect is given in figure \ref{fig:NiOvercorrections}.
These overcorrections needlessly destabilize the $n_i$ and therefore increase the number of idle iterations needed to stabilize it again.
Also, this increases the problem of micro-inconsistencies.

\begin{figure}
 \centerline{ \includegraphics[height=1.0\textwidth,angle=90]{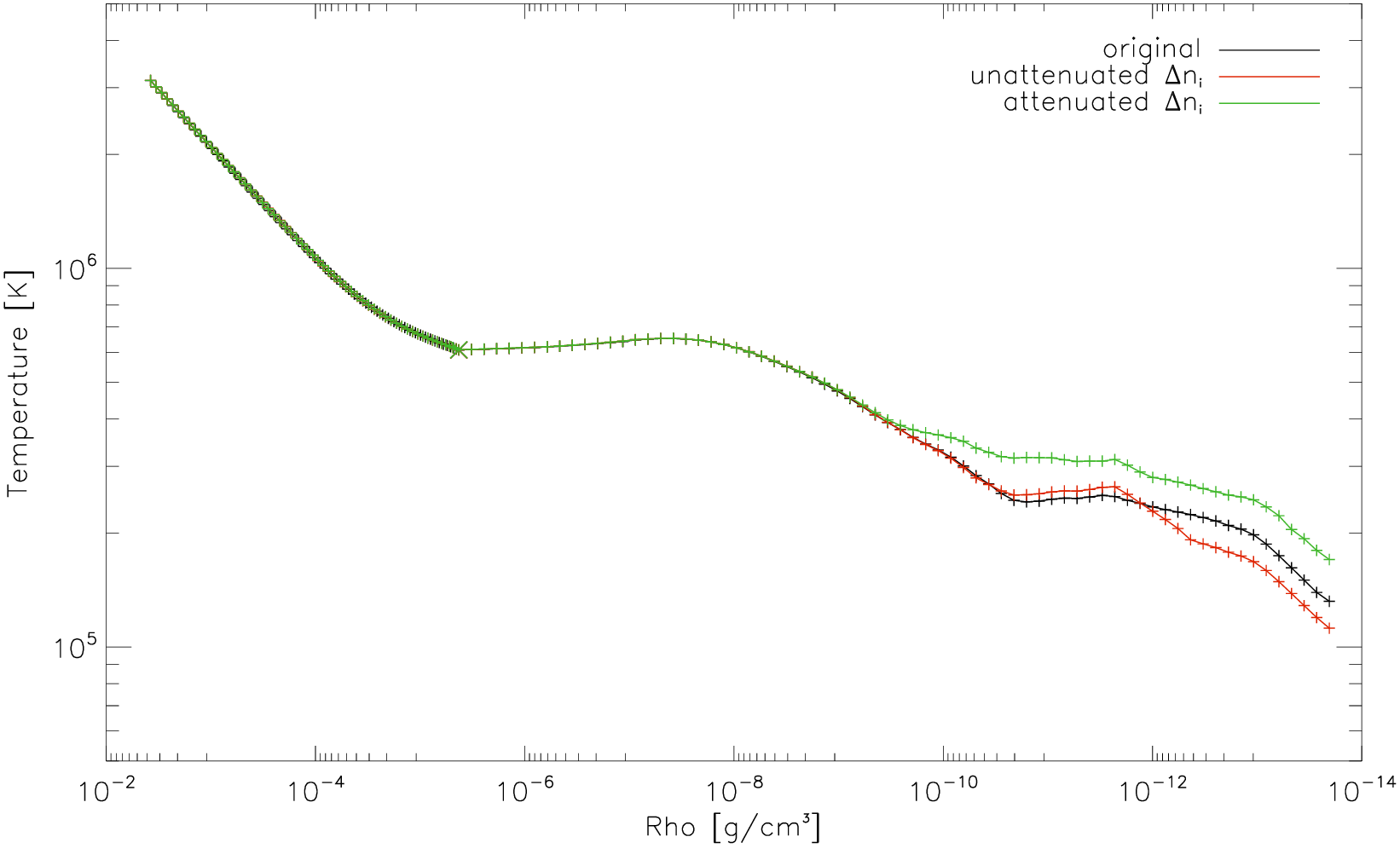}}
 \caption{
  If the NLTE occupation numbers in the outer regions of the atmosphere are assumed to depend on the local temperature like the LTE numbers, then a number of problems are introduced.
  One of the problems is that the temperature correction method becomes inaccurate, which leads to wrong temperature structures.\newline
  This plot shows the temperature structures of two models in comparison with one original structure (black) that was used as starting point for the models.
  The model with the ordinary full LTE correction to the $n_i$ is shown in red.
  In the other model (green) the corrections were attenuated by a factor of 10 ($x_E = 0.1$) in regions where the $b_i$ were large, as described in the text.
 } \label{fig:NiCorr.20}
\end{figure}

Also, the $n_i$ overcorrection affects the temperature correction method.
The method is based on the assumption, that the ratios of the integrated opacities are insensitive to the temperature, equations \eqref{eq:ULTC3a} to \eqref{eq:ULTC4b}.
In the outer regions, where the densities are low, the contributions of strong lines become more significant in the opacity integrals.
If the $n_i$ of these lines are destabilized by temperature corrections, then the opacity integrals are found to become sensitive to the temperature, and the method becomes inaccurate.
In figure \ref{fig:NiCorr.20} the resulting temperature structures are shown for a model in which the $n_i$ are scaled with temperature corrections the ordinary (unattenuated) way, compared to a more suitable attenuated way, which is described below.

In order to improve the way the temperature corrections are transferred to the $n_i$ the "small departure from LTE" approximation must be alleviated for layers in which the departures are large.
For that purpose an approximate measure is needed for the departure from NLTE for each chemical element $E$ that is treated in NLTE.
One obvious way to define this departure is to average the $b_i$ of each element, weighting each state $i$ by its number density.
\begin{equation}
 \bar{b}_E = \frac{\sum b_i n_i}{\sum n_i}
  = \frac{\sum b_i^* b_c n_i}{\sum n_i}
\end{equation}
But since the $b_i$ can be very large or very small numbers they dominate the sum, by the means that unimportant states can dominate the average.
Therefore, it is better to average the logarithms of the $b_i$
\begin{equation}
 \bar{b}_E = \frac{\sum \log(b_i) n_i}{\sum n_i}
  = \frac{\sum \log(b_i^* b_c) n_i}{\sum n_i}
\end{equation}
This measure $\bar{b}_E$ is then used to attenuate the full, LTE like, corrections to the $n_i$.
The $n_i$ can then be computed using a linear combination of the two limiting cases
\begin{equation}
 n_i = x_E \left[b_i n_i^\LTE(T_{\rm new})\right] + (1-x_E) \left[b_i n_i^\LTE(T_{\rm old})\right]
\end{equation}
where $x_E$ is the attenuation factor for chemical element $E$.
There are unlimited possibilities to relate $\bar{b}_E$ to $x_E$, with the restriction that in the limiting case of $\bar{b}_E \approx 0$ (note the logarithm in the definition) the attenuation must be off: $x_E = 0$.

\begin{figure}
 \centerline{ \includegraphics[width=\textwidth]{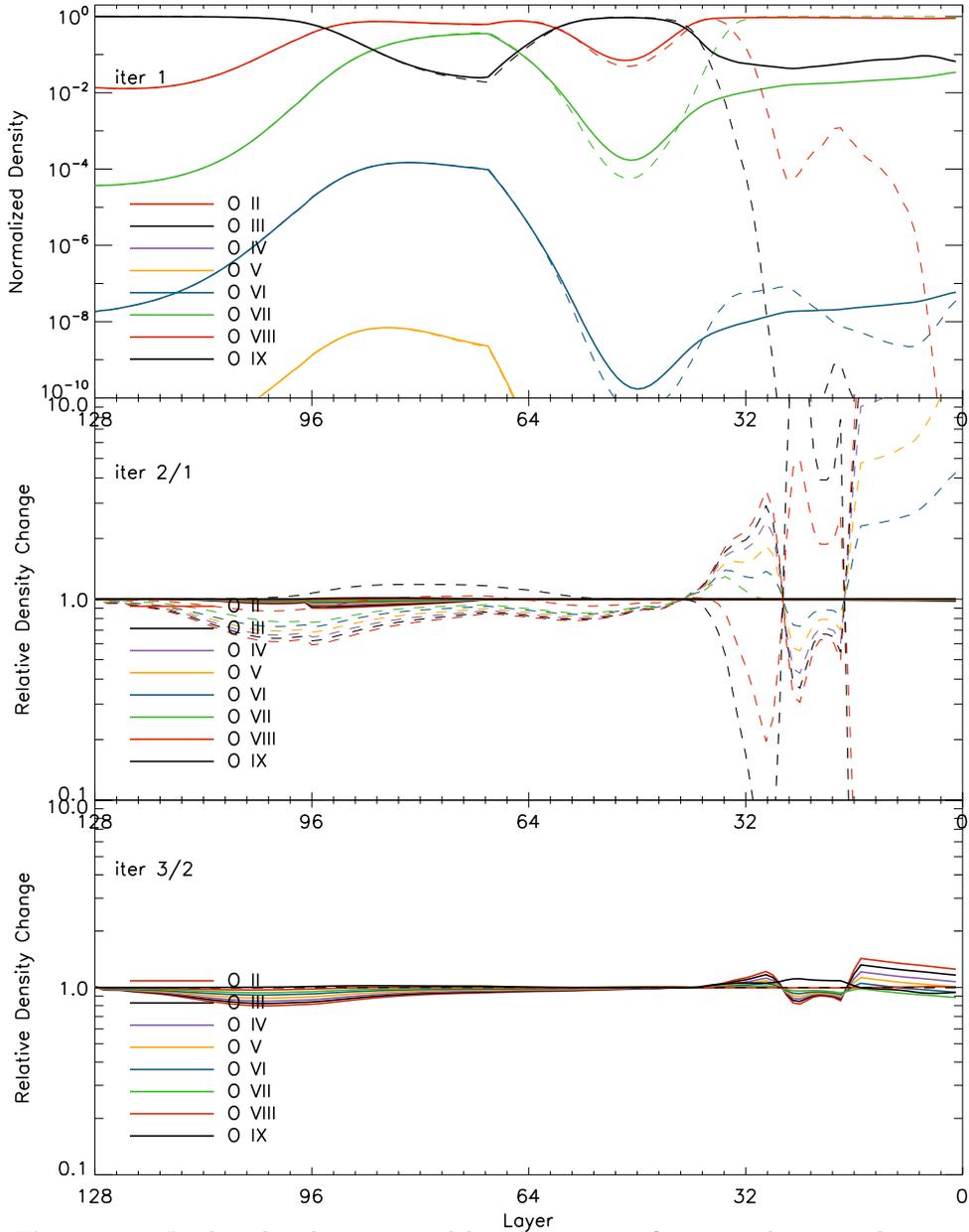}}
 \caption{
  In this plot the same model is shown as in figure \ref{fig:NiOvercorrections}, but now the non-thermal limit approximation was used for the whole atmosphere.
  From the first to the second iteration the temperature changed, but no direct changes are induced in the NLTE ionization balance.
  In the third iteration the changes to the balance in the outer regions (layer numbers smaller than 40) are still small.
  The strong overcorrections in these layers, as seen in figure \ref{fig:NiOvercorrections}, are avoided.
  This not only saves idle cycles, but is also very important for attaining a consistent set of population numbers.
 } \label{fig:NiOvercorrectionsDamped}
\end{figure}

\begin{figure}
 \centerline{ \includegraphics[width=.5\textwidth]{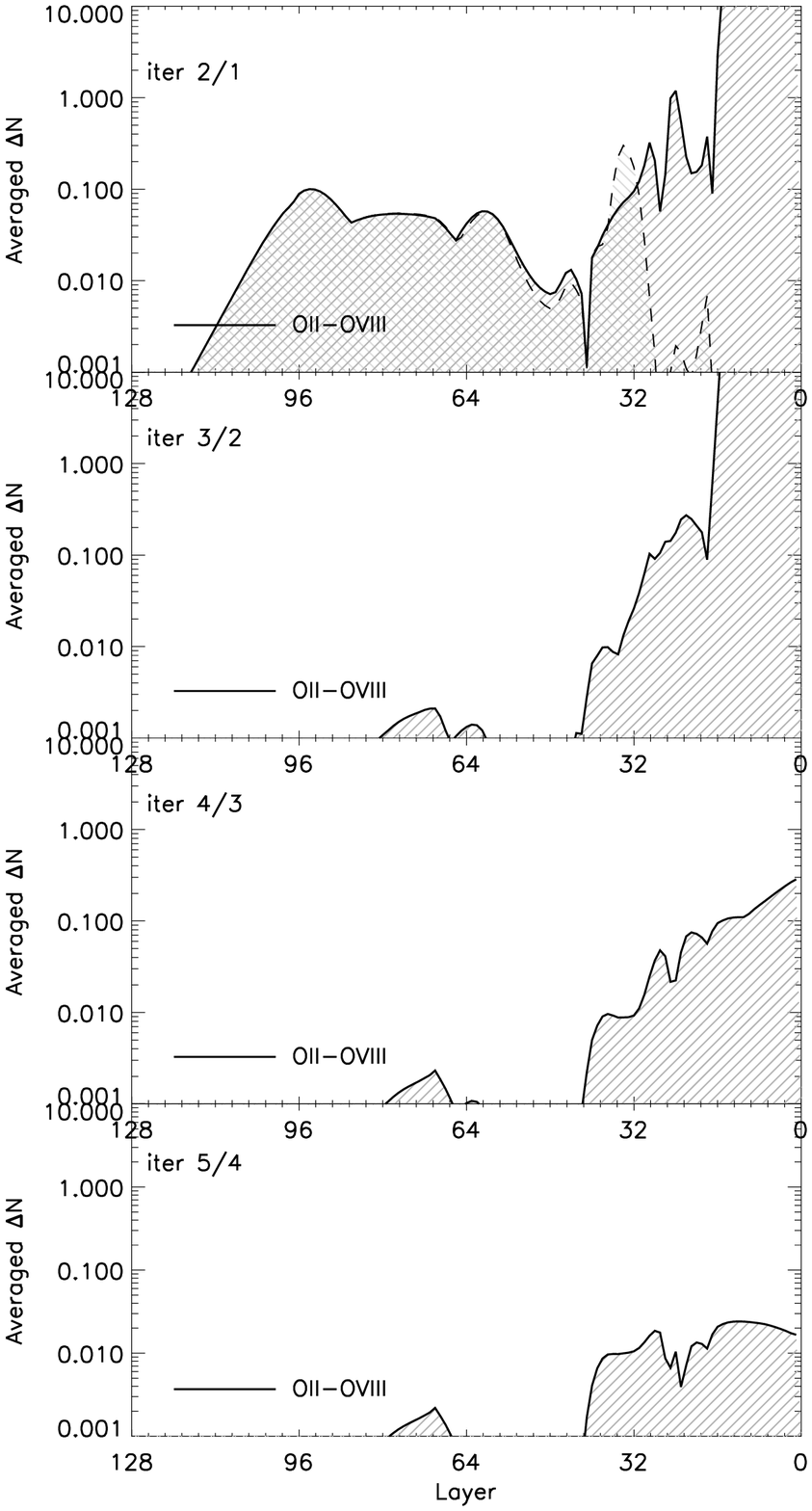}
              \includegraphics[width=.5\textwidth]{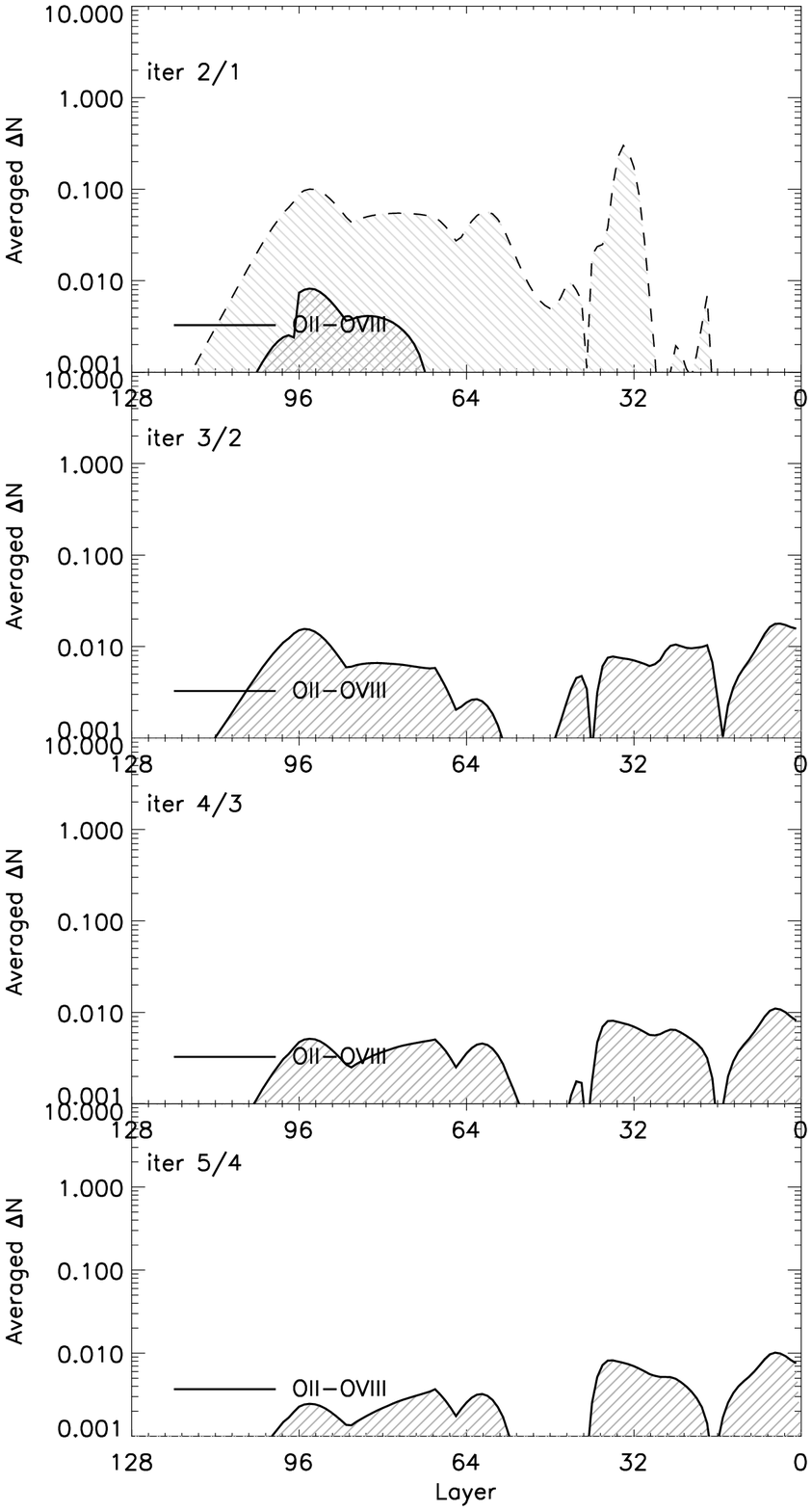}}
 \caption{
  The weighted averaged density corrections $\overline{\Delta N}_E$, as defined by equation \eqref{eq:MeanDeltaN}, are compared for oxygen in five subsequent iterations between the fully thermal $n_i$ coupling approximation (left) and the non thermal approximation (right).
  In the first iteration the temperature was corrected, and in the fully thermal approximation the $n_i$ were corrected accordingly.
  In the non thermal approximation the $n_i$ are not corrected for the temperature change.
  The changes visible in the top right plot are due to gas density changes in the atmosphere setup. \newline
  In the fully thermal approximation even in the third iteration after the temperature correction the $\overline{\Delta N}_E$ for oxygen is still 30\% in the outermost layers.
  In the non thermal approximation the $n_i$ of the outermost layers are corrected only by 10\% in total.
  In the lower layers the fully thermal approximation is superior, but the corrections are small and not problematic in either case.
 } \label{fig:NiOvercorrMeanDamped}
\end{figure}

While experimenting with relations between $\bar{b}_E$ and $x_E$ it was found that using the very simple $x_E = 0$ globally, i.e. the limit of no thermal coupling of the population numbers for all regions of the atmosphere, gives good global convergence.
In those regions of the atmosphere where the departures from LTE are small the $n_i$ are found to quickly converge to their new balance for the new temperature.
In contrast, in the regions with large departures the $n_i$ converge slowly.
Using the approximation that is well suited for the complex regions still leaves enough time for the simpler LTE like regions to converge, even if the $n_i$-adaption approximation is bad.
The performance of this approximation is shown in figure \ref{fig:NiOvercorrectionsDamped}, which is the same model as in figure \ref{fig:NiOvercorrections}.
In figure \ref{fig:NiOvercorrMeanDamped} the weighted averaged density corrections $\overline{\Delta N}_E$, as defined by equation \eqref{eq:MeanDeltaN}, are compared between the fully thermal coupling approximation and the no thermal coupling approximation.
In the inner regions the thermal approximation is a bit better, but in the outer regions the non thermal approximation is far superior.

\section{Interpolation of the Source function} \label{sec:SourceInterpolation}
In order to compute the specific intensity along a characteristic ray, see section \ref{sec:ALI}, the source function is integrated, equation \eqref{eq:IFormalSolution}.
The source function is only computed on the intersection points of the ray with the concentric shells, here often called layers, of the model.
For the integral an approximation of the source function between the intersection points is required.
If $\hat{S}$ is interpolated linearly or parabolically, then the integral can be written as a linear combination of $\hat{S}$ on either two or three intersection points, equation \eqref{eq:DILinearCombination}.

It is not possible to generally favor the linear or the parabolic interpolation based on physical arguments.
In \phx\ a threshold optical depth value can be set.
Below the threshold, for small optical depths along a characteristic ray, the linear interpolation is used, and above the threshold the parabolic.
This allows to compare results with the linear and parabolic interpolation methods.
The reason for the threshold switch method is that the parabolic interpolation occasionally gives very unrealistic values, so that the formal solution for the characteristic ray becomes inaccurate.
This leads to inaccurate mean intensities which not only degrade the convergence properties of the operator splitting method, but also lead to an inaccurate solution of the whole radiation field.
These unrealistic interpolation values usually occur in small optical depths along rays.
Setting the threshold high enough enables to avoid these problems.

Comparative tests performed in this work showed that the linear and parabolic $\hat{S}$-interpolation method can lead to very different solutions of the radiation field.
In the search for a more sophisticated method, that should have been able to determine the preferred interpolation method for each individual situation, a different approach was found.

\begin{figure}
 \centerline{ \includegraphics[height=1.0\textwidth,angle=90]{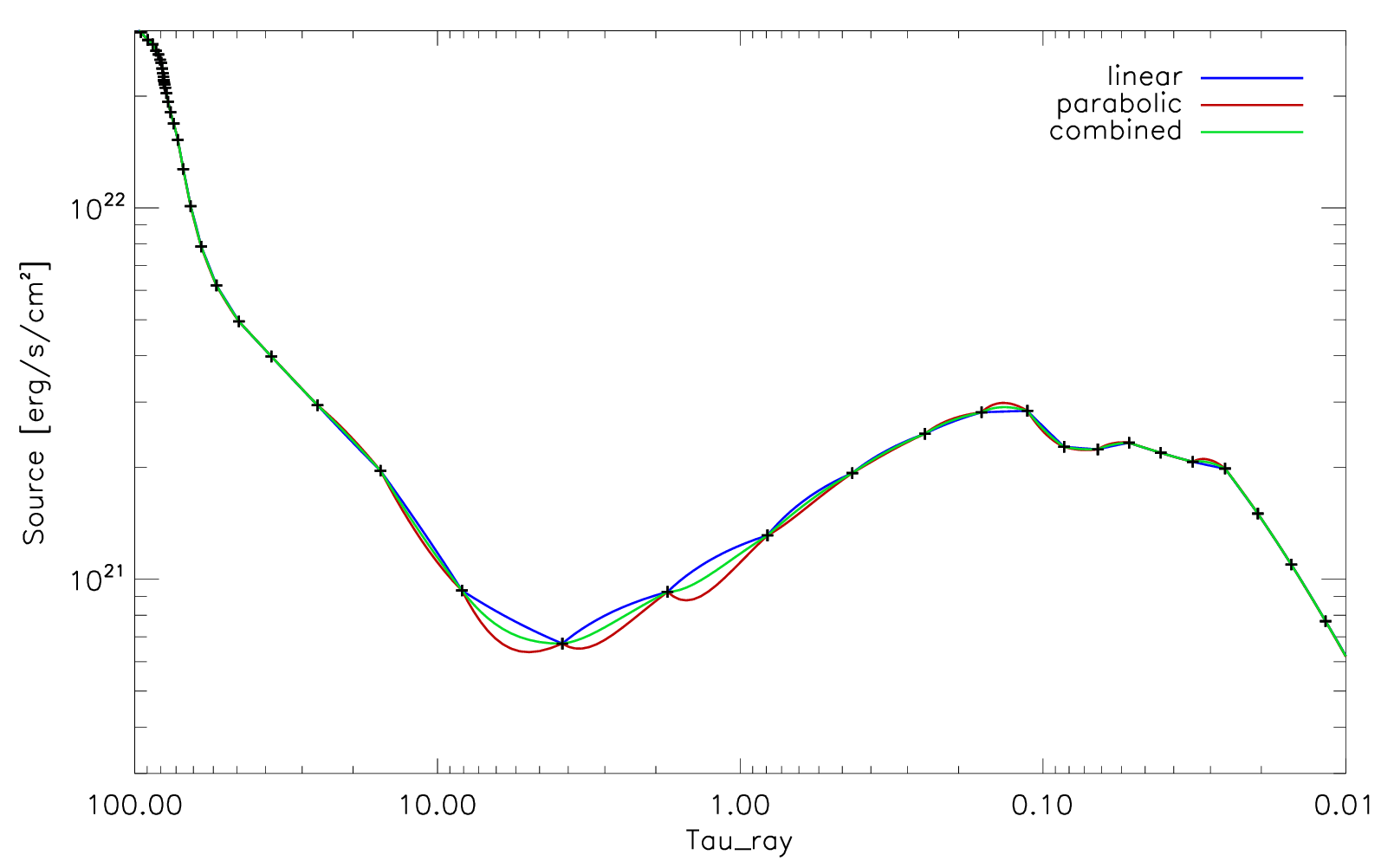}}
 \caption{The source function along a characteristic ray is only computed on intersection points of the ray with the concentric shells of the model.
 When the source function is integrated along the ray, in order to compute the formal solution of the radiative transport equation, the run of the source function between the points is needed.
 \newline
 The plot shows an example of the source function against the optical depth along the ray on a double logarithmic scale.
 The intersection points are marked with black crosses.
 If the linear interpolation (blue) overestimates the source function then the parabolic interpolation (red) underestimates it, and vice versa.
 A linear combination of the linear and parabolic case, here to equal parts (equation \eqref{eq:SourceCombinedInterpolation}), combines these tendencies and gives a much better approximation (green).
 } \label{fig:RaySource}
\end{figure}

Plotting the source function at the intersection points against the optical depth along a ray, shows a contrarily tendency of both interpolation methods: if the linear method overestimates $\hat{S}$ between two points then the parabolic method underestimates it, or opposite.
An example of these tendencies is given in figure \ref{fig:RaySource}.
The plot shows the intersection points as black crosses, the linear and parabolic interpolations and a combined interpolation.
The combined case shows
\begin{equation} \label{eq:SourceCombinedInterpolation}
 \hat{S}(\tau) = (\hat{S}_{\rm lin}(\tau) + \hat{S}_{\rm par}(\tau)) / 2
\end{equation}

Since the integral in equation \eqref{eq:IFormalSolution} can be written as linear combination of $\hat{S}$ on a few intersection points, it can also be written as a linear combination of the solutions with both interpolation methods.
\begin{equation}
 \Delta I_i^k = \frac{
   (a \alpha_{i\,{\rm lin}}^k + b \alpha_{i\,{\rm par}}^k) \hat{S}_{i-1}
   + (a \beta_{i\,{\rm lin}}^k + b \beta_{i\,{\rm par}}^k) \hat{S}_i
   + b \gamma_{i\,{\rm par}}^k \hat{S}_{i+1}}
  {a + b}
\end{equation}
This method has the advantage, that the over- and underestimating tendencies of the linear and parabolic method are canceled for good parts.
In most of all situations the combined interpolation yields a better approximation of $\Delta I_i^k$ than one of the methods alone does.
Furthermore, this method gives an unambiguous solution for the source interpolation problem.

An even better alternative would be to compute the $\alpha$, $\beta$ and $\gamma$ coefficients for a fixed combined linear and parabolic interpolation of the source function, as in equation \eqref{eq:SourceCombinedInterpolation}.
The advantage is that the improved interpolation, as shown in figure \ref{fig:RaySource}, directly enters the analytic integral, instead of taking the mean of two inaccurate integrals.
This has not yet been accomplished in the scope of this work, but is planned for future work.

\section{Solving the rate matrix equation} \label{sec:RMESolver}
\subsection{Computational weight} \label{sec:RMESolverWeight}
As described in section \ref{sec:RateMatrixEquation} the rate matrix equation is a linear system of rank $N$, where $N$ is the number of levels of all ionization stages of an atomic species.
The number of levels per atom varies considerably between the species.
In fact, the number is only limited by the atomic data that is used as input.
The procedure for solving a linear system is straight forward, but for large systems this becomes computational expensive.

The dynamic range in the solution of the system is very large $n_i \in \langle 0,N_E \rangle$.
In praxis the dynamic range in the rate matrix is large too, and consequently the occupation numbers are sensitively coupled.
For example, when determining $n_i$, the transitions into level $i$ from a weakly coupled, highly populated level $j$ might be as important as the transitions from a strongly coupled, lowly populated level $k$.

Examination of the solution of the rate matrix equations shows that the precision of {\tt FORTRAN} 64-bit double precision variables is not high enough to obtain accurate results.
For example, the insufficient precision causes the solution of the matrix equation \eqref{eq:RateMatrixEquation} to contain negative $n_i$.
Therefore, in \phx\ the rate matrix equation is solved using quad-double precision (QD).
This is supported by an extension to {\tt FORTRAN-90}, \cite{Hida00}.
A QD number is an unevaluated sum of four double precision numbers, capable of representing at least 212 bits of significand, or 64 digits.

The higher precision causes large computational overhead.
The memory required to store the rate matrix is a factor of 4 larger, which also degrades the CPUs cache performance (the cache hits/misses balance).
Also, significantly more floating point operations are needed per arithmetic operation.
Added up, solving the rate equations is one of the most time consuming parts in the atmosphere calculation when large, realistic model atoms (with many energy levels) are used.

\subsection{Explicit rate matrix equation}
In order to discuss methods for solving the rate matrix equation \eqref{eq:RateMatrixEquation} (RME) an explicit example of a rate matrix is shown in figure \ref{fig:ExplicitRateMatrix}.
The matrix corresponds to a simplified atom with four ionization stages.
\begin{figure}
\input{explicitRateMatrix.tex}
\caption{This is the explicit rate matrix for a simplified model atom, consisting of four ionization stages $j$.
 The neutral and doubly ionized stages ($j=1$ and $j=3$), apart from the ground states, have one excited state.
The singly ionized stage has three excited states.
The last stage is the bare nucleus.
 Within each ionization stage $j$ the levels are sequentially numbered after their excitation energy $i$, starting from $1$ for the ground level.
 The rates $P$ are defined by equation \eqref{eq:RateMatrixEquation}.
 \newline
 The black lines show the origin of the rates from the different ionization stages.
 In the squares the lines form around the diagonal, the rates originate from transitions between states of one specific ionization stage only.
 Rates between ionization stages only occur for the \emph{key levels} that are ground state and continuum simultaneously.
 The key levels in this example are $1_2$ and $1_3$ (omitting the bare nucleus that has no further excitation levels).
 \newline
 Note that the sum over each columns matrix elements is zero by construction.
} \label{fig:ExplicitRateMatrix}
\end{figure}

The matrix elements $P$ are given by equation \eqref{eq:RateMatrixElements}.
These matrix elements connect states of different ionization stages.
A colorized schematic version of the explicit RME is shown in figure \ref{fig:ColorRME}.
This is the raw system that is not yet closed by the constraint equation (see section \ref{sec:RateMatrixEquation}).
\begin{figure}
\input{colorRateMatrix.tex}
\caption{
 The explicit rate matrix of figure \ref{fig:ExplicitRateMatrix} shows matrix elements that couple levels of different ionization stages.
 In this schematic version of the RME each ionization stage $j$ is coded with its own color: red, green, blue and red for $j=1,2,3,4$ respectively.
 The matrix elements show the colors of the ionization stages of the levels that they couple.
 Also, each $n_i$ has the color of the ionization stage it belongs to.
 \newline
 All diagonal elements have colors of multiple stages, yellow, turquoise, purple and gray, from an overlay of red-green, green-blue, blue-red and red-green-blue respectively.
 Also, the rows and columns corresponding to the key levels show mixed colors.
 The most mixing elements are the diagonal elements that correspond to the key levels.
 All other non-zero elements show only a single color.
 \newline
 Note that this is the raw system that still contains one redundant equation (see section \ref{sec:RateMatrixEquation}).
 } \label{fig:ColorRME}
\end{figure}
Each ionization stage has its own color, and the matrix elements feature the colors of the ionization stages of the levels they are coupling.

\subsection{The old system}
In order to make solving the RMEs (one for each atomic species) computationally treatable in \phx\ a simplifying approximation is made.
The rate matrix is almost block diagonal and can be made block diagonal with the following approximation.
The ground states, which are at the same time the continuum state of the next lower ionization stage, are split up and treated in two parts: in one group as the ground state, and in another group as the continuum state.
The block diagonal form of the rate matrix is
\begin{equation}
 \mathbf{P} =
 \begin{pmatrix}
  \mathbf{P}_1 & 0 & \cdots &0 \\
  0 & \mathbf{P}_2 & \ddots &\vdots \\
  \vdots & \ddots & \ddots & 0 \\
  0 & \cdots & 0 & \mathbf{P}_j \\
 \end{pmatrix}
\end{equation}
with $j$ the last ionization stage of the atom.
The off diagonal blocks are zero matrices.

The RMEs fall apart into independent blocks, one for each ion of the atomic species under consideration.
This system is shown schematically in figure \ref{fig:ColorRMEOld} using the same color coding as in figure \ref{fig:ColorRME}.
\begin{figure}
\input{colorRateMatrixOld.tex}
\caption{
 The rate matrix is almost block diagonal, see figure \ref{fig:ExplicitRateMatrix}, and with a simplifying assumption it can be made block diagonal.
 The system falls apart into a small systems, one for each ionization stage.
 Those are easier to solve, deducing memory consumption and computation time.
 \newline
 A great disadvantage of this method is that the simplification is not generally valid.
 The larger the deviations from LTE are, the more inaccurate the solution becomes.
 \newline
 This equation is not yet of highest rank, it is closed by replacing redundant equations with constraint equations, see figure \ref{fig:ColorRMEOldClosed}.
 } \label{fig:ColorRMEOld}
\end{figure}
Each block is a linear system for all levels of one ionization stage, plus the continuum of that ionization stage.
Each system is closed by a constraint equation of the form
\begin{equation} \label{eq:RMEConstraintOld}
 \sum_{i=1}^{c_j} n_{i_j} = \mathscr{N}_j \equiv \sum_{i=1}^{c_j} n_{i,{\rm old}}
\end{equation}
with $1_j$ and $c_j$ being the ground and continuum level of ionization stage $j$, and $\mathscr{N}_j$ the total number density of ion $j$ plus continuum.
In this notation the ground level of the neutral ionization stage $1_1$ is the lowest energy level, denoted with $1$ in equation \eqref{eq:RMEConstraint}.
The system with constraint equations is shown schematically in figure \ref{fig:ColorRMEOldClosed}.
\begin{figure}
\input{colorRateMatrixOldClosed.tex}
\caption{
 The block-diagonal RME of figure \ref{fig:ColorRMEOld} is closed by constraint equations \eqref{eq:RMEConstraintOld}, one for each block.
 \newline
 } \label{fig:ColorRMEOldClosed}
\end{figure}
The constraint equations \eqref{eq:RMEConstraintOld} provide the coupling between the separated ionization stages.

The $\mathscr{N}_j$ do not directly depend on the rates.
This allows to solve the linear system for each ion separately.
The rank of each of the linear systems is much smaller than the rank of the full RME \eqref{eq:RateMatrixEquation}.
Small systems are much faster to solve, which saves computation time.
Furthermore, only relatively small amounts of memory are needed at a time, since the number of matrix positions scales with the square of the systems rank.

Analysis showed that this method is not accurate for the physical conditions of typical atmospheres modeled in this work.
The most obvious problem is that if departures from LTE are large, then also the ionization balance of the species under consideration strongly departs from the LTE balance.
However, the ionization balance is not directly solved for.
Iteratively, the ionization balance can change, via the constraint equations.
This is achieved by shifting the ion population from the ground state to the continuum, or back, which is very ineffective if the initial ionization balance is far off from the true solution.
Thus the convergence properties of this scheme are poor.
The extra iterations required to roughly converge the solution of the RMEs are expensive and cost a multiple of the computation time gained by splitting the RMEs.

The ionization stages are connected by the \emph{key levels} that are the continuum for one stage and the ground state for the next higher stage.
Splitting the rates of the key levels is physically only valid if the net rates to and from the key levels are exactly zero for \emph{each part of the rates separately} (the parts related to the higher \emph{or} the lower ionization stage).
This requirement is satisfied by the solution of the RME, but that solution is just to be found.

In the method artificial levels are added to the atom, the counterparts of the actual key levels.
These additional levels require additional equations in order to close the system.
Those are the equations \eqref{eq:RMEConstraintOld}.
Mapping the results of this extended system back to the actual system is not possible, since the actual key level number density is not defined.
Defining it as the sum or the mean of the twin levels generally violates number density conservation.
Defining the key levels as just one of the pair, neglecting the other one is problematic, since there is no physical preference for one or the other.

Only in the special case where the exact solution of the RME is used in the definition of the constraint equations \eqref{eq:RMEConstraintOld}, the number densities of both twins are exactly equal, and then the twins can be interpreted as one identical level.
So when starting from the right solution, this method yields the right result.
Especially, in the regions where LTE is a good approximation the method gives reasonable results when starting with LTE as initial guess.
On the contrary, where departures from LTE become larger, the initial condition for the RME solver becomes worse, and thus the solution is not exact, and the rate of convergence degrades.

\begin{figure}
 \centerline{ \includegraphics[width=\textwidth]{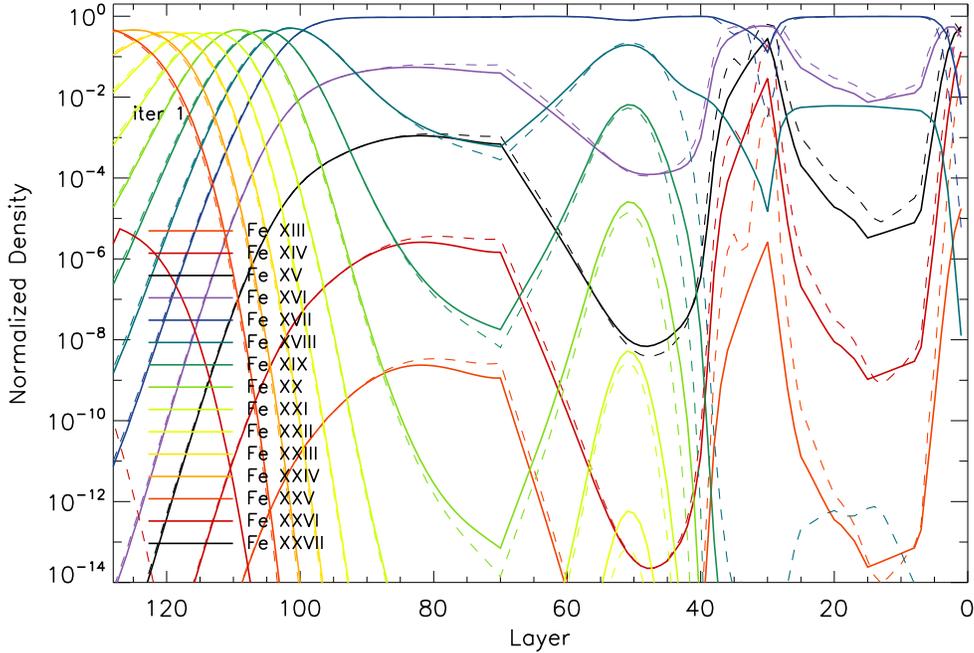}}
 \caption{
  This plot shows the Fe ionization balance of a model of stage 3 (see table \ref{tab:NLTEStages}), in which Fe was just added.
  The $n_i$ of all other atomic species are kept fixed to stabilize the radiation field while the $n_i$ of Fe adapt to it via the RME (solid curves), from their initial LTE values (dashed curves).
  The method to solve the RME that is used in this model is not exact.
  \newline
  Except for Fe\,{\sc xviii} in the outer regions (small layer numbers) all deviations from LTE are relatively small.
  \newline
  In subsequent iterations the ionization balance is refined, see figure \ref{fig:FeRMEOldDelta}.
 } \label{fig:FeRMEOld}
\end{figure}

\begin{figure}
 \centerline{ \includegraphics[width=\textwidth]{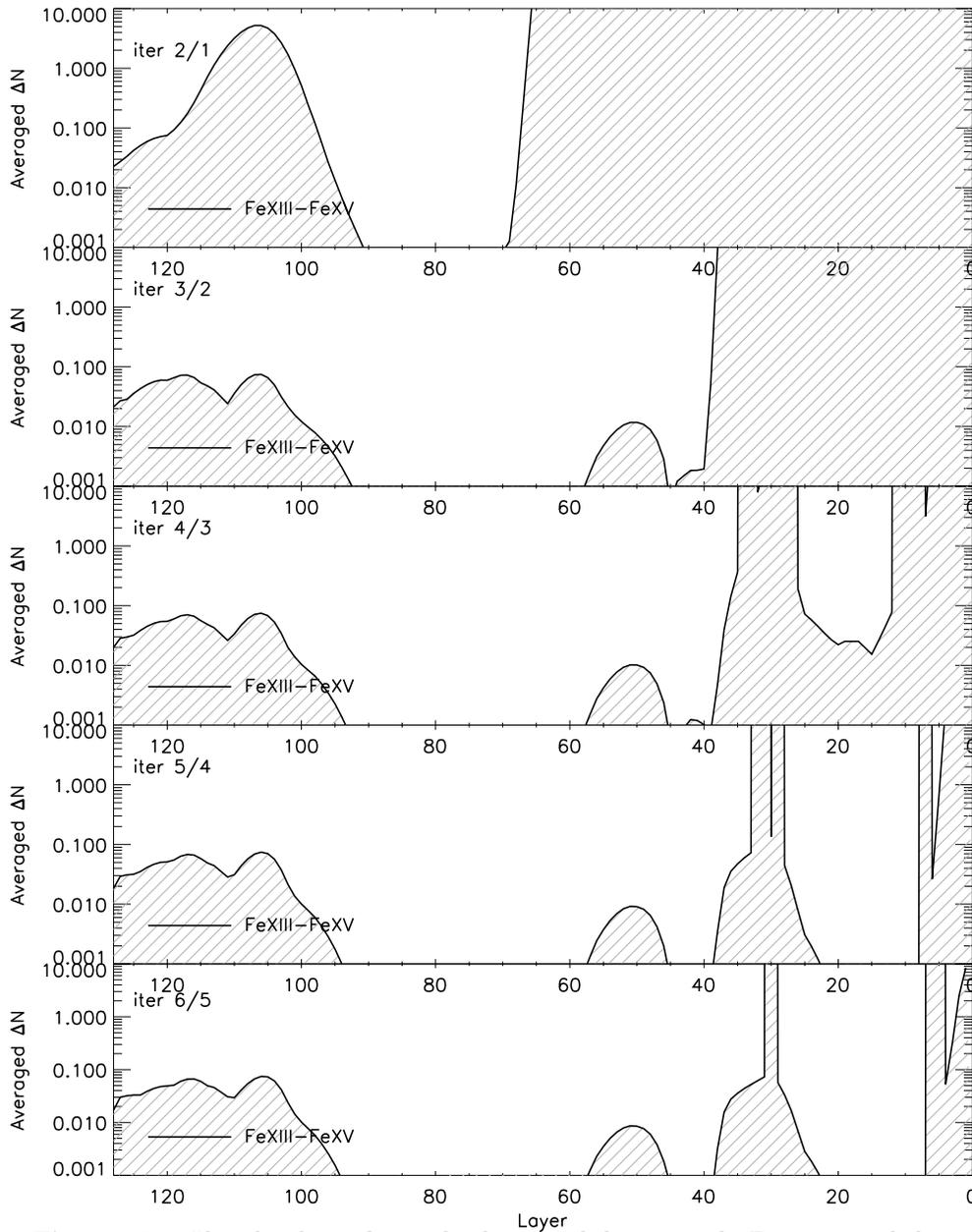}}
 \caption{
  This plot shows the weighted averaged changes to the Fe ionization balance $\overline{\Delta N}_{\rm Fe}$, defined by equation \eqref{eq:MeanDeltaN}, for 6 subsequent iterations.
  In each iteration the $n_i$ of all elements other than Fe are kept fixed, as well as the atmospheric structure.
  Clearly, in the outer regions, where the departure from the initial LTE ionization balance is large, the convergence rate is poor.
  The result from the subsequent corrections to the $n_i$ in the inner regions is visible in figure \ref{fig:FeRMEOldLast}, where the ionization balance after these 6 iterations is shown.
 } \label{fig:FeRMEOldDelta}
\end{figure}

\begin{figure}
 \centerline{ \includegraphics[width=\textwidth]{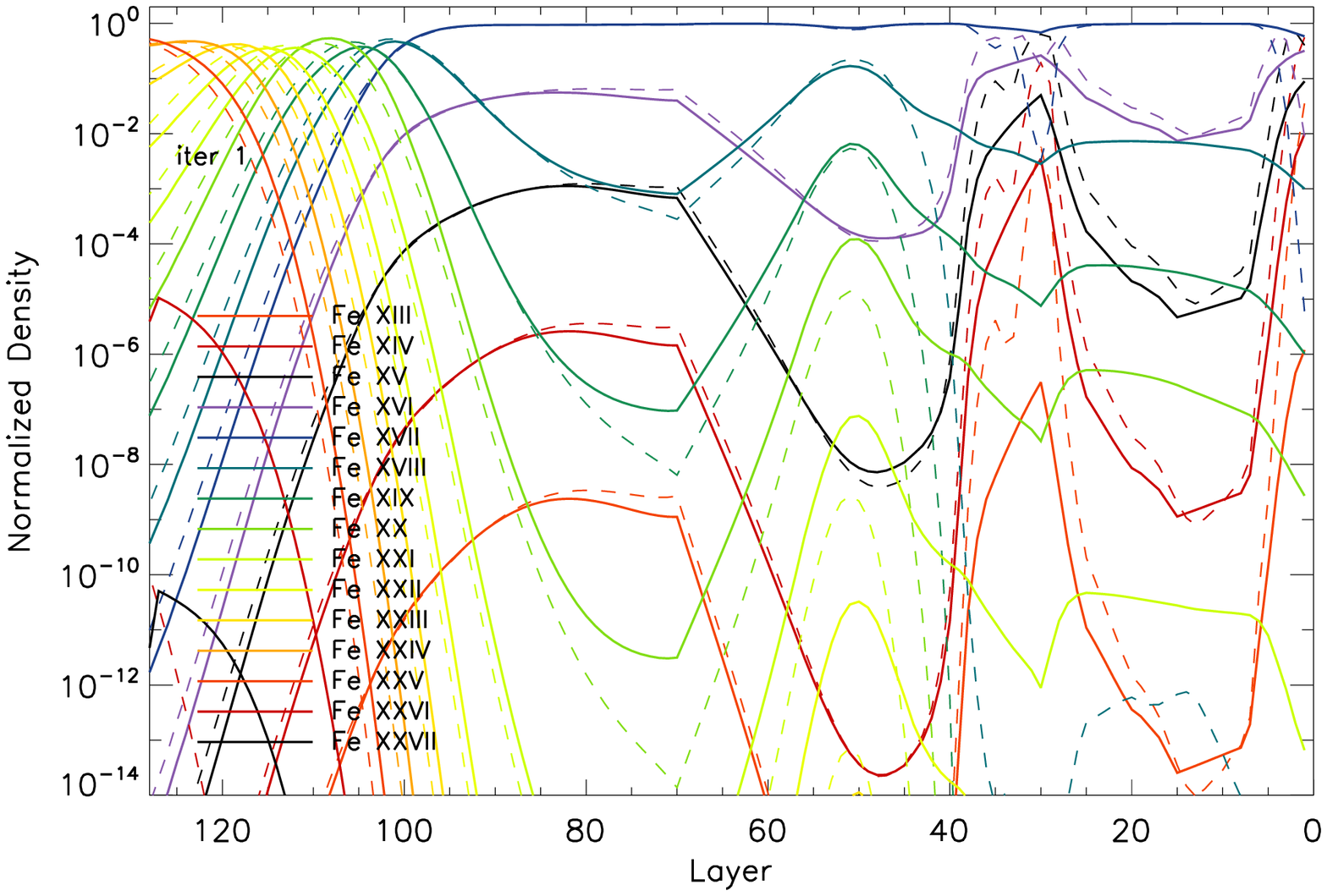}}
 \caption{
  The ionization balance of Fe after 6 iterations is quite different from the balance of the first solution of the RME.
  The LTE ionization balance (initial assumption) is shown by the dashed curves.
  Comparison with the exact solution of the next section, figure \ref{fig:FeRMENew}, shows that there are still significant differences.
  \newline
  Around layer 35 the lower ionization stages Fe\,{\sc xiii}-{\sc svi} are by far not underabundant enough.
  In the inner regions, where the highest ionization stages Fe\,{\sc xxvi} and Fe\,{\sc xxvii} are yet out of LTE, the whole ionization balance is driven out off the correct LTE balance.
  \newline
  Because of the poor convergence properties and the extra iterations needed with this method this method is replaced by a new one, described in section \ref{sec:RMENewSystem}.
 } \label{fig:FeRMEOldLast}
\end{figure}

This is shown by the example of a (typical) model of stage 3 (see table \ref{tab:NLTEStages}) typical for this work in figures \ref{fig:FeRMEOld} and \ref{fig:FeRMEOldDelta}.
In stage 3 Fe is newly added to the model and the population numbers $n_i$ of the other atomic species are kept fixed.
The initial guess for the Fe population numbers is the Saha-Boltzmann distribution, which is unrealistic for the outer regions.
The ionization balance from the first solution of the RME is plotted in figure \ref{fig:FeRMEOld}, in comparison with the initial LTE ionization balance.
Except for Fe\,{\sc xviii} in the outer regions the departures from the LTE ionization balance are small.
The change in the ionization balance in the following iterations, in which the $n_i$ of all other elements and the atmospheric structure are kept fixed, is shown in figure \ref{fig:FeRMEOldDelta}.
The resulting ionization balance in the last of these 6 iterations is plotted in figure \ref{fig:FeRMEOldLast}.
Comparison of this iterated result with the exact solution from the next section, figure \ref{fig:FeRMENew}, shows that the overionization of the higher stages in the outer regions is slow, but the underionization of the lower stages is even slower.

\subsection{The full system}
The most straight forward, generally valid way of solving the linear system of rate equations is described in section \ref{sec:RateMatrixEquation}.
The schematic form of this system is shown in figure \ref{fig:ColorRMEClosed}.
\begin{figure}
\input{colorRateMatrixClosed.tex} 
\caption{
 The system of rate equations is closed by the constraint equation of number density conservation.
 This equation replaces one of the redundant rate equations.
\newline
 The choice of equation to replace has an important influence on the numerical accuracy of the solution.
 A good choice is a key level $k$ with a large expected $n_k$ in the solution of the system.
 } \label{fig:ColorRMEClosed}
\end{figure}
One equation of the raw (unclosed) system is redundant and is replaced by the constraint equation \eqref{eq:RMEConstraint}.
The choice of the equation to replace has a strong influence on the numerical precision of the solution.
If, for example, the constraint equations replaces the equation of matrix row $k$ and the number density $n_k$ is very small $n_k \ll N_E$, then analytically
\begin{equation}
 n_k = N_E - \sum_{i\ne k} n_i
\end{equation}
is the difference of two almost equal numbers.
This choice is very prone to errors from the limited significance of numerical variables.

Experiment shows, that the best choice is the key level with the largest number density.
This is not known prior to solving the system but a guess can be based upon a Saha-Boltzmann distribution, or better, if available, on the solution of the system from previous iterations (the current $n_i$).

\begin{figure}
 \centerline{ \includegraphics[width=\textwidth]{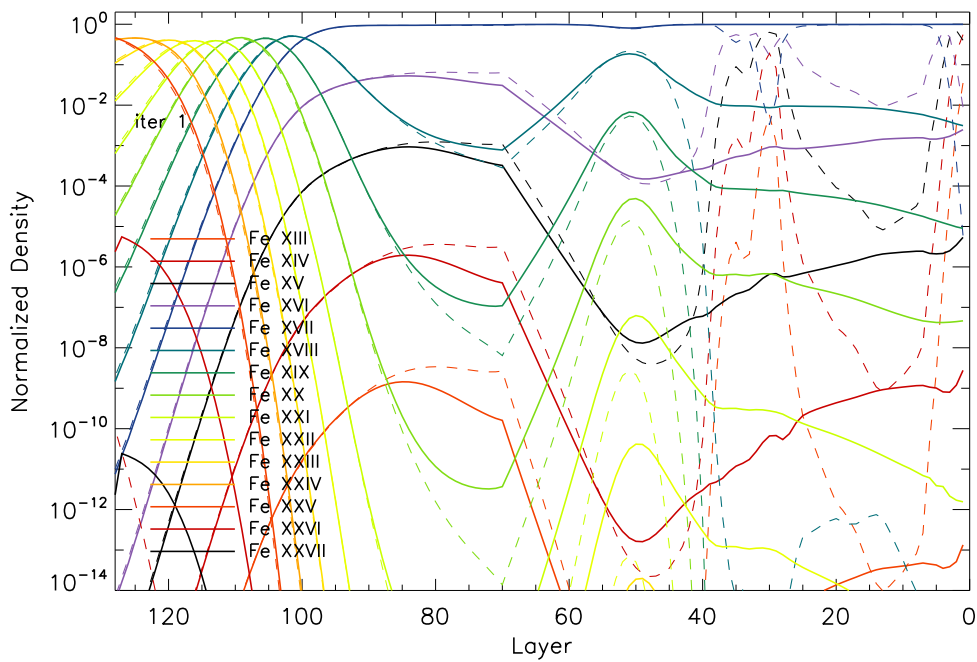}}
 \caption{
  This Fe ionization balance (solid curves) is computed with the full RME \eqref{eq:ColorRMEClosed}, for the same model as in figure \ref{fig:FeRMEOld}.
  For comparison also the LTE balance is shown (dashed curves).
  Solving the full RME is computationally very expensive, but yields the exact solution.
  This solution is used to check the results from other, more resource saving methods.
 } \label{fig:FeRMENew}
\end{figure}

This method yields the exact solution of the RME.
The resulting ionization balance from the first solution of the RME for the same model as in figure \ref{fig:FeRMEOld} is shown in figure \ref{fig:FeRMENew}.
The departures from the original LTE ionization balance are much larger than with the old method of the previous section.
In subsequent iterations, in which the $n_i$ of other elements and the atmospheric structure are kept fixed, the Fe ionization balance hardly changes.
\begin{figure}
 \centerline{ \includegraphics[width=\textwidth]{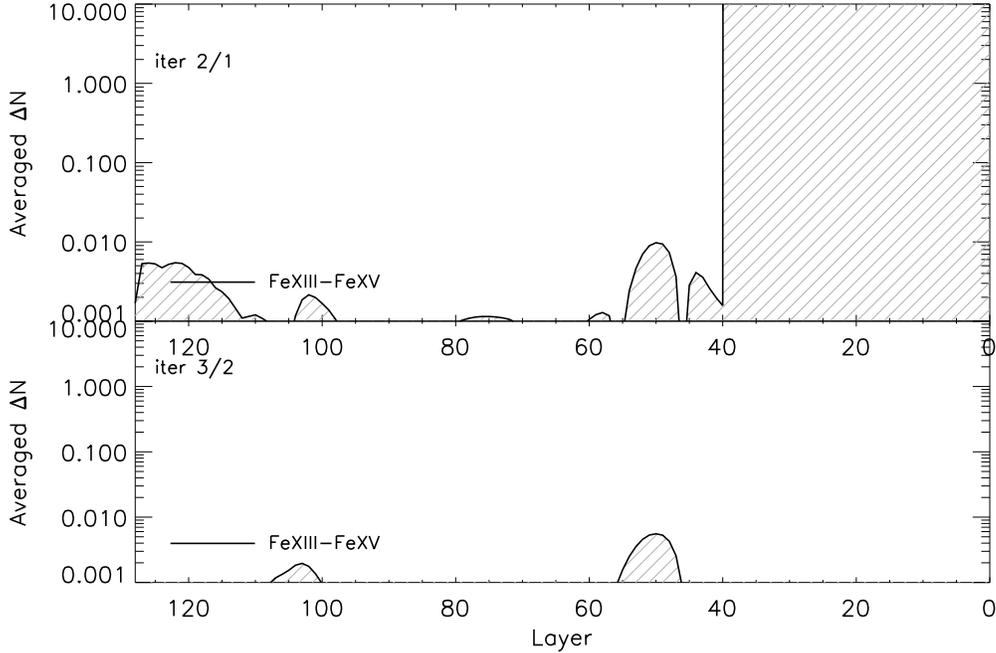}}
 \caption{
  After the first iteration the weighted averaged changes to the Fe ionization balance $\overline{\Delta N}_{\rm Fe}$ are very small in all regions of the atmosphere.
  It shows that despite strong deviations of the $n_i$ for Fe, the radiation field was good enough to yield accurate rates for the largest $n_i$ (that weigh most in the average).
  This results from keeping all other $n_i$ fixed when adding Fe.
 } \label{fig:FeRMENewDelta}
\end{figure}

\subsection{The new system} \label{sec:RMENewSystem}
For the reasons discussed in section \ref{sec:RMESolverWeight} the straight forward method of the previous section is computationally very expensive.
A different approach is required in order to reduce memory consumption and calculation time to a level appropriate for presently available computer power.
The following method provides both an exact solution of the RME \emph{and} a significant reduction in memory consumption and CPU time.

The raw RME \eqref{eq:RateMatrixEquation} without the closing constraint equation \eqref{eq:RMEConstraint} does not have a unique solution.
The solution of the closed system $\vec{n}$ satisfies the open system, but also the scaled solution $a\vec{n}$ does, with $a$ some constant.
So if any such vector $a\vec{n}$ can be found, then it can be scaled to the unique solution $\vec{n}$ using the constraint equation.
Thus a different constraint can be used to solve the matrix.

One possibility is the simple constraint of fixing one of the number densities to some constant $n_k = 1$.
If the level $k$ that is being fixed is chosen to be one of the key levels, then this completely removes the dependency between the ionization stages above and below that key level.
In other words, this choice splits the linear system in two independent parts.
A block diagonal matrix is obtained, with two blocks.
This method can be called the \emph{2-block diagonal grouping method}.
This is shown schematically in figure \ref{fig:ColorRMESplit}.
\begin{figure}
\input{colorRateMatrixSplit.tex}
\caption{
 The unclosed system, figure \ref{fig:ColorRME}, does not have a unique solution.
 However, since its right hand side is the zero vector $\vec{b} = 0$, its linear properties allow to quickly find one arbitrary solution.
 That arbitrary solution of the unclosed system can then be scaled to satisfy the closing constraint equation.
 With this method the unique solution is found in a much faster way than when the closed system is solved.
 \newline
 The modification to the RME that quickly yields one arbitrary solution is shown here.
 One redundant equation is replaced by a fixation of one of the key variables.
 Thus the system of rank 9 becomes block diagonal, and splits into a system of rank 6 and a system of rank 2, which are solved faster than the system of rank 9.
 \newline
 Large atoms with many ionization stages possess many key levels, which brings the ranks of the two blocks closer together.
 So the splitting is more efficient for large atoms than for small atoms, like in this example.
 However, those are already easy to solve.
 } \label{fig:ColorRMESplit}
\end{figure}

The maximum gain from this method is achieved if the system can be split in two parts with equal rank.
In that best case scenario the memory savings are a factor of 4.
In praxis the most centered key level is often not exactly in the middle, especially for small atoms, with only few ionization stages.
But small atoms are easy to solve.
More importantly, for large atoms with many ionization stages, the gain is often not far from the theoretical maximum.

The savings in computation time needed to solve the system are even better.
Comparative tests performed for Fe, by far the largest atom used in this work, showed that the CPU time with the new method is a factor of 7-15 smaller than with the full system method of the previous section.
The exact gain depends on the ionization balance in the converged solution of the RME.
The highest gain is achieved in the case where the ionization balance is centered right at the centermost key level $m_1$ that was kept fixed, i.e. if in the solution $n_{m_1} > n_l$ for any other key level $l \ne m_1$.

This allows to fine tune the choice of key level to fix.
The exact center lies (most often) between two key levels.
From the current population $n_i$ the best populated key level $l$ in the solution of the RME is guessed.
If the second best centered key level $m_2$ is not much further off center than the best centered key level $m_1$ \emph{and} $m_2$ is closer to $l$ than $m_1$, then $m_2$ is kept fixed rather than $m_1$.
This yields a slightly worse gain in memory consumption, but a significant better gain in required CPU time.

Because the CPU time depends on the ionization balance it depends on the region in the atmosphere.
Therefore, if the RME solver is parallelized over layer, then a round robin scheme is better than a clustered (blockwise) spreading.
Thus the layers with a disadvantageous ionization balance are distributed over the workers.

\subsection{Dealing with numerical inaccuracies} \label{sec:RMENumericalAccuracy}
As described in section \ref{sec:RMESolverWeight} the dynamical range in the rate matrix and in the level populations is large.
The RME are first solved using the LU factorization method.
This is an analytical method.
The accuracy is only limited by the numerical precision of the variables with a given condition number of the matrix\footnote
{The condition number is a property of the matrix itself, not of the algorithm or the floating point accuracy.
 It specifies how numerically well-conditioned the equation is.
 In the case of the linear system $Ax=b$, a high condition number (ill-suited for numerical computation) would mean that even a small error in $b$ would yield a large error in the solution $x$.
}.
Iterative linear system solver methods, like the Jacobi method or the Gauss-Seidel method start from an initial guess that is iteratively refined.
These methods are very slow\footnote
{When a Matrix is block diagonal the solution of each block of the matrix equation is independent of the other blocks.
 In the RME considered here, the matrix is nearly block diagonal, so that the solutions of the blocks are not fully independent, but the coupling between the blocks is yet (very) weak.
} for typical large RME if the initial guess is poor.
Therefore, they are not suitable to solve the RME from scratch.
But they are very useful to refine the result from the analytic solver.
The numerical inaccuracies (like negative $n_i$) are removed from the solution of the RME within a few steps.

Thus the combination of the analytical solver and an iterative solver is very powerful.
And another benefit can be drawn from the solver combination.
Since numerical inaccuracies from the analytical solver can be corrected afterwards, this allows to reduce the precision of the variables used by the analytical solver.
Instead of QD variables double-double precision (DD) or even standard double precision (SD) can be used.
Afterwards the precision is checked with the iterative solver, and if the inaccuracies are too large to be corrected within a few iterations, then the system is solved analytically from scratch with a higher precision.
The computational overhead of higher precision variables is so large, that the wasted CPU time of the lower precision solver step is relatively small.
Going from QD to DD saves a factor of 7 in CPU time, and going from DD to SD saves \emph{another} factor of 20-30!
The savings can be attributed to less flops per arithmetic operation, better cache performance, and better performing compiler optimization (especially when going from DD to SD).
These savings combined with the gain from the 2-block diagonalization method of section \ref{sec:RMENewSystem} result in an extreme reduction of CPU time.

With this result, the CPU time consumption for solving the RMEs, instead of being the dominating and limiting part of the atmosphere model calculation, becomes almost negligible against the other parts (e.g. opacity computation and radiation transport).
This result is also important for other massive NLTE models, like supernova models, where the highly abundant iron-peak elements Fe, Co and Ni have very many levels.

\section{The Active Damping method}
One of the challenges of atmosphere modeling is to find a correction procedure that converges for a large variety of atmospheres.
Normally, the price for such universality is that the rate of convergence is not optimal for different types of atmospheres.
But a good convergence rate is important to save calculation time.

\begin{figure}
\centerline{\includegraphics[height=\hsize,angle=90]{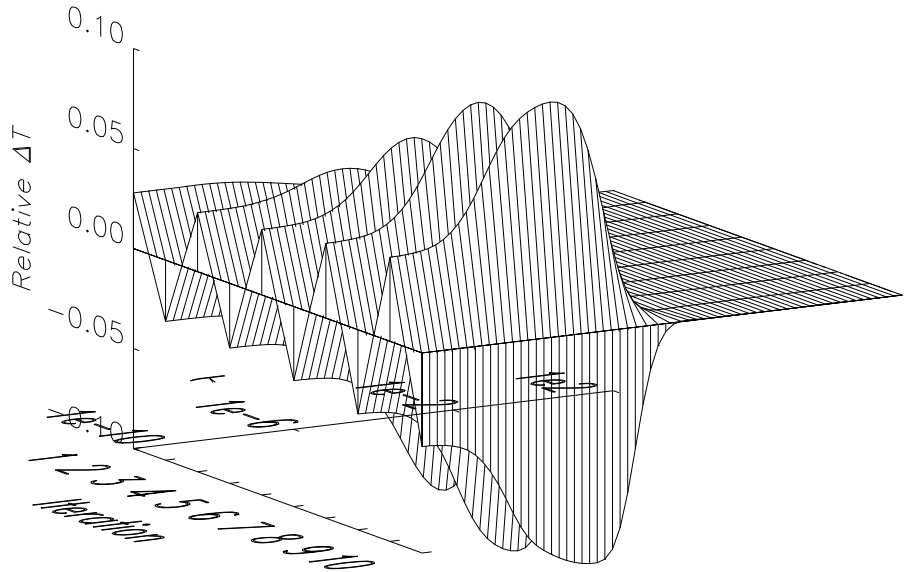}}
\caption{
 An example of oscillations that occur in realistic models when an undamped UL temperature correction is used.
 Here the relative correction to the Temperature is plotted for 64 atmospheric layers (spanning the optical depth range $10^{-10} \le \tau \le 10^2$) over 10 model iterations.
} \label{fig:ULOscillations}
\end{figure}

Analyzing the Uns\"old-Lucy (UL) temperature corrections for realistic models shows that undercorrections and overcorrections (oscillations) occur.
An example of oscillation is given in figure \ref{fig:ULOscillations}.
A number of damping and acceleration methods were proposed in the literature, e.g., \cite{Lucy64}, \cite{Dreizler03}, \cite{Hamann03}, but all of them either need to be individually tuned or are ineffective.
In this section a new ``Active Damping'' (AD) method is described, in which the optimal damping factor for each layer and each iteration is determined on-the-fly from the values of previous corrections.

\begin{figure}
\centerline{\includegraphics[height=.8\textwidth,angle=90]{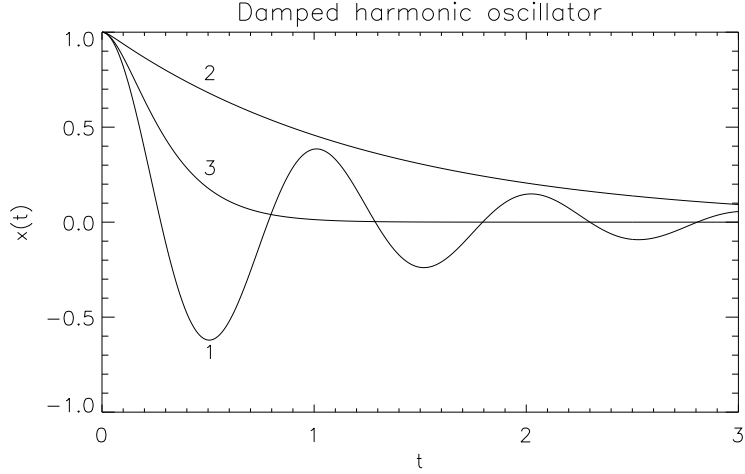}}
\caption{
 Three cases of damped harmonic oscillators:
 1. Underdamped, oscillation occurs;
 2. Overdamped, the equilibrium is approached slowly from one side;
 3. Critical damping, the equilibrium is approached in the fastest possible way.
} \label{fig:CriticalDamping}
\end{figure}

In classical mechanics, a system with a restoring force can be overdamped, underdamped or critically damped, see figure \ref{fig:CriticalDamping}.
The UL correction acts like a restoring force to the temperature driving it towards equilibrium and in analogy there exists an optimal damping factor to this correction.

\subsection{Active damping for linear corrections} \label{sec:LinUL}
As a first approximation the UL correction $\Delta B$, in the following written as $\Delta$, is assumed to be positively proportional to the deviation $d$ from the final temperature (for which energy conservation is satisfied).
\begin{equation} \label{eq:LinUL}
 \Delta_i = f \cdot d_i
\end{equation}
with $f$ the amplification factor and the index $i$ labels the iteration number.
In the case $f=1$ the undamped UL correction would converge within a single iteration.
In praxis, this rarely happens to single layers, let alone to the complete atmospheric structure.

In order to develop the formalism, assume that a damped correction $\Delta_i'$ is applied
\begin{align} \label{eq:ArbitrarilyDampedCorr}
 \Delta_i' &= \beta_i \Delta_i \\
 \beta_i &> 0
\end{align}
with $\beta_i$ being the damping factor for iteration $i$.
The restriction to positive $\beta$ reflects the assumption that $\Delta$ is positively proportional to the deviation $d$.
After the first correction $d$ becomes
\begin{equation}
 d_2 = d_1 + \Delta_1' = (1 - \beta_1 f) d_1
\end{equation}
and using equation \eqref{eq:LinUL} the second correction becomes
\begin{equation}
 \Delta_2 = -f (1-\beta_1 f) d_1
\end{equation}
For $f$ one finds the expression
\begin{equation} \label{eq:Linf}
 f = \frac{\Delta_1 - \Delta_2}{\Delta_1'} = \frac{\Delta_1 - \Delta_2}{\beta_1 \Delta_1}
\end{equation}
The AD-factor $\alpha$ is defined to compensate for deviations from $f=1$, so that the correction damped with $\alpha$ cancels $d$ completely
\begin{equation}
 \alpha_i \Delta_i \equiv -d_i
\end{equation}
In order to calculate the AD-factor, values of two subsequent UL corrections are used.
Using equation \eqref{eq:LinUL} and \eqref{eq:Linf} and generalizing for all iterations one finds
\begin{equation} \label{eq:LinAD}
 \alpha_i = \frac{-d_i}{\Delta_i} = \frac{1}{f_i} = \frac{\beta_{i-1} \, \Delta_{i-1}}{\Delta_{i-1} - \Delta_i}
\end{equation} 
where $f_i$ is the generalization of equation \eqref{eq:Linf}
\begin{equation} \label{eq:LinfGen}
 f_i = \frac{\Delta_{i-1} - \Delta_i}{\beta_{i-1} \Delta_{i-1}}
\end{equation}
The damping factor $\beta_i$ can now be set equal to this AD-factor $\alpha_i$.
In the first iteration, as $\Delta_{i-1}$ is not yet known, some preset value must be used, e.g. $\beta = 1$ for an undamped correction.
Note that in equation \eqref{eq:LinUL} $f$ is assumed to be constant, whereas in equation \eqref{eq:LinAD} $f_i$ changes from iteration to iteration.
$f_i$ is the constant value that fits to iterations $i-1$ and $i$.
With this value of $f$, the UL corrections for these two iterations would be proportional to $d_i$, as in equation \eqref{eq:LinUL}.
This proportionality is never exact, so that after the second iteration the UL correction is found to be not yet completely converged, thus $\Delta_3 \ne 0$.

\subsection{Active Damping for non-linear corrections}
The assumption that the UL correction depends linear, equation \eqref{eq:LinUL}, can be rewritten as
\begin{equation}
 \Delta_i = f \cdot s_{d} \, |d_i|
\end{equation}
with
\begin{equation}
 s_{d} \equiv \mathrm{sign}(d_i)
\end{equation}
Assume that the UL correction satisfies a power law instead of a linear relation
\begin{equation} \label{eq:NLUL}
 \Delta_i = f \cdot s_{d} \, |d_i|^\gamma
\end{equation}
where $0 < \gamma < \infty$.
Note that $d_i$ and its sign are not known a priori.
In analogy to the linear case, equation \eqref{eq:ArbitrarilyDampedCorr}, in every iteration the correction $\Delta_i$ is damped with some factor $\beta_i$
\begin{equation}
 \Delta'_i = \beta_i \cdot f_i \cdot s_{d} \, |d_i|^{\gamma_i}
\end{equation}
$f_i$ and $\gamma_i$ are the values of $f$ and $\gamma$ for the current iteration $i$.
Multiple iterations will be needed since, like in the linear case, the non-linear assumption does not hold exactly.
Some algebra yields the expression for $f_i$
\begin{equation}
 f_i(\gamma) = s_f \left| \frac{s_{i-1} |\Delta_{i-1}|^{1/\gamma} - s_i |\Delta_i|^{1/\gamma}}
     {\beta_{i-1} \Delta_{i-1}} \right|^\gamma
\end{equation}
with
\begin{align}
 s_{i-1} &\equiv \mathrm{sign}(\Delta_{i-1}) \\
 s_{i} &\equiv \mathrm{sign}(\Delta_i) \\
 s_f &\equiv \mathrm{sign} \left( \frac{s_{i-1} |\Delta_{i-1}|^{1/\gamma} - s_i |\Delta_i|^{1/\gamma}} {\beta_{i-1} \Delta_{i-1}} \right)
\end{align}
The parameter $\gamma_i$ is determined by the requirement that (in the assumption of equation \eqref{eq:NLUL}) the amplification factor $f(\gamma)$ is a constant for some value of $\gamma$
\begin{equation} \label{eq:NLconstf}
 f_{i-1}(\gamma) \stackrel{!}{=} f_i(\gamma)
\end{equation}
This equation can be solved numerically using the UL corrections of three subsequent iterations $\Delta_{i-2}$, $\Delta_{i-1}$ and $\Delta_i$, which yields both $f_i$ and $\gamma$, in the following called $\gamma_i$.

Finally, the expression for the AD-factor $\alpha$ in the non-linear case is 
\begin{equation} \label{eq:NLAD}
 \alpha_{i} \equiv \frac{-d_{i}}{\Delta_{i}} = 
                   s_f \frac{|\Delta_i|^{(1-\gamma_i)/\gamma_i}}{|f_i|^{1/\gamma_i}}
\end{equation}
Note that for $\gamma_i = 1$ both $f_i$ and $\alpha_i$ reduce to the linear versions of equations \eqref{eq:LinfGen} and \eqref{eq:LinAD}.

\subsection{Implementation}
When the AD method is applied to stellar atmosphere models a few difficulties arise.
These are caused by the fact that the assumptions of equations \eqref{eq:LinUL} and \eqref{eq:NLUL} do not hold exactly.
In fact, there exists no relation that holds exactly because of the assumptions of the Uns\"old-Lucy scheme.

\subsubsection{Problems}
\begin{enumerate}
\item One problem is related to the singularity in equations \eqref{eq:LinAD} and \eqref{eq:NLAD} for $\Delta_{i} = \Delta_{i-1}$.
The AD-factor behaves asymptotically in the neighborhood of the singularity.
Asymptotic jumps often occur when the corrections $\Delta$ become small, as in this case numerical inaccuracies in $\Delta$ become large.
These jumps in the correction factor have no physical justification.

\item Negative AD-factors occur when the UL correction gets larger than in the preceding iteration and is in the same direction (either positive or negative)
\begin{equation} \label{eq:NegativeAD}
 \frac{\mathrm{sign}(\Delta_i)}{\mathrm{sign}(\Delta_{i-1})} \cdot |\Delta_i| > |\Delta_{i-1}|
\end{equation}
Assuming that the UL corrections are always restoring, i.e. directed towards the equilibrium, these negative corrections are invalid.

\item Large jumps in the AD-factors of adjacent layers cause unphysical jumps in the resulting temperature structure.

\item Large AD-factors (large extrapolations) yield inaccurate results and for very small factors the calculation of the subsequent AD-factor becomes inaccurate.

\item The solutions $\gamma$ of equation \eqref{eq:NLconstf} can be inappropriate, i.e. $\gamma \gg 1$ or $\gamma \ll 1$.
The UL correction is not extremely non-linear.
\end{enumerate}

\subsubsection{Basic solution scheme}
The following scheme proved to handle the problems described above:
\begin{enumerate}
\item Compute $\gamma(l)$ for all layers $l$.
\item Reject extreme values $\gamma(l) < 0.5$ or $\gamma(l) > 1.3$.
\item Treat layers without valid $\gamma$ as linear $\gamma(l) = 1$.
\item Smooth $\gamma$ according to $\gamma(l) = \left(\gamma(l-1) + \gamma(l) + \gamma(l+1)\right)/3$
\item Compute $f(l)$ using the smoothed $\gamma(l)$.
\item Mark 5 layers\footnote
{When less than 100 layers are used, it may be better to mark 3 layers around layer $l$.
} $(l-2, \cdots ,l+2)$ around each layer where $f(l) < 0$.
\item Linearly interpolate $f$ over the marked ranges.\label{item:interp}
\item Calculate the AD-factors $\alpha(l)$.
\item Limit the new damping factors $\beta(l)$ according to $0.25 < \beta(l) < 2.5$ 

\item Finally the new damping factors $\beta_i(l)$ should be smoothed over 5 layers according to
\begin{equation}
 \beta_i(l) = \frac{ 4\beta(l) + 2\beta(l-1) + 2\beta(l+1) + \beta(l-2) + \beta(l+2) } {10}
\end{equation}
\end{enumerate}
All limiting values in this scheme were determined experimentally.

\subsubsection{Limitations}
In some cases the basic scheme fails
\begin{enumerate}
\item step \ref{item:interp} is not possible for ranges that extend to the outermost/innermost layer.
\item if the range to be interpolated over in step \ref{item:interp} is very large, i.e. larger than $\# \rm{layers}/4$
\item if layers with large jumps of the resulting damping factor $\beta$ exist, i.e. when two or more of the following four ranges are exceeded\footnote
{This is appropriate for models with up to 100 layers.
 When more layers are used, values closer to unity (e.g. 4/3 and 3/2 respectively and their inverses) may give better results.
}:
\begin{align}
 2/3 < |\beta(l)/\beta(l \pm 1)| < 3/2 \\
 1/2 < |\beta(l)/\beta(l \pm 2)| < 2/1
\end{align}
\end{enumerate}
Instead of AD the following more primitive method can be used to calculate the damping factor.

\subsection{The alternative: Passive Damping}
In the occasions that active damping fails, one has to revert to another method to determine an appropriate damping factor.
The primitive method described in the following is called ``passive damping'' (PD), indicating that the damping factor is not intended to completely cancel the deviations $d$ in one step in contrast to active damping.
The scheme for passive damping is as follows
\begin{enumerate}
\item For each of those layers $l$ the damping factor of the previous iteration $\beta_{i-1}(l)$
 is taken, and discretized to the values [0.3, 0.42, 0.6, 0.8, 1.0, 1.25, 1.6, 2.0].
\item[-] If strong oscillation occurs ($-1 < \Delta_i/\Delta_{i-1} \ll 0$),
 the value is stepped down by one.
\item[-] If strong overdamping occurs ($ 0 \ll \Delta_i/\Delta_{i-1} < 1$),
 the value is stepped up by one.
\item[-] If oscillation or overdamping is very small ($|\Delta_i/\Delta_{i-1}| \ll 1$),
 the value is retained.
\item[-] If the correction increases ($|\Delta_i/\Delta_{i-1}| > 1$),
 interpolate the damping factor (on the discrete grid) over this layer using the two closest layers for which this is not the case.
For the outermost/innermost layer then the center value of 0.8 can be used.
\item Now the differences in the damping factors between the layers must be reduced to be one step on the discrete grid at most.
 This is done starting from the two enclosing layers for which the AD-factors are still used, going towards the center of the range.
 Only in case there are less layers than steps needed to go from one border to the other, differences larger than one step are to be left over.
\end{enumerate}

\subsection{Convergence properties}
In the calculation of model atmospheres the UL temperature correction is found to converge much faster with AD than with fixed damping constants.
For all types of models tested thus far (white dwarf, brown dwarf, red giant, nova and supernova models) a speed up of a factor of two or more is achieved.

\begin{figure}
 \centerline{\includegraphics[width=.45\textwidth,bb=0 10 255 630]{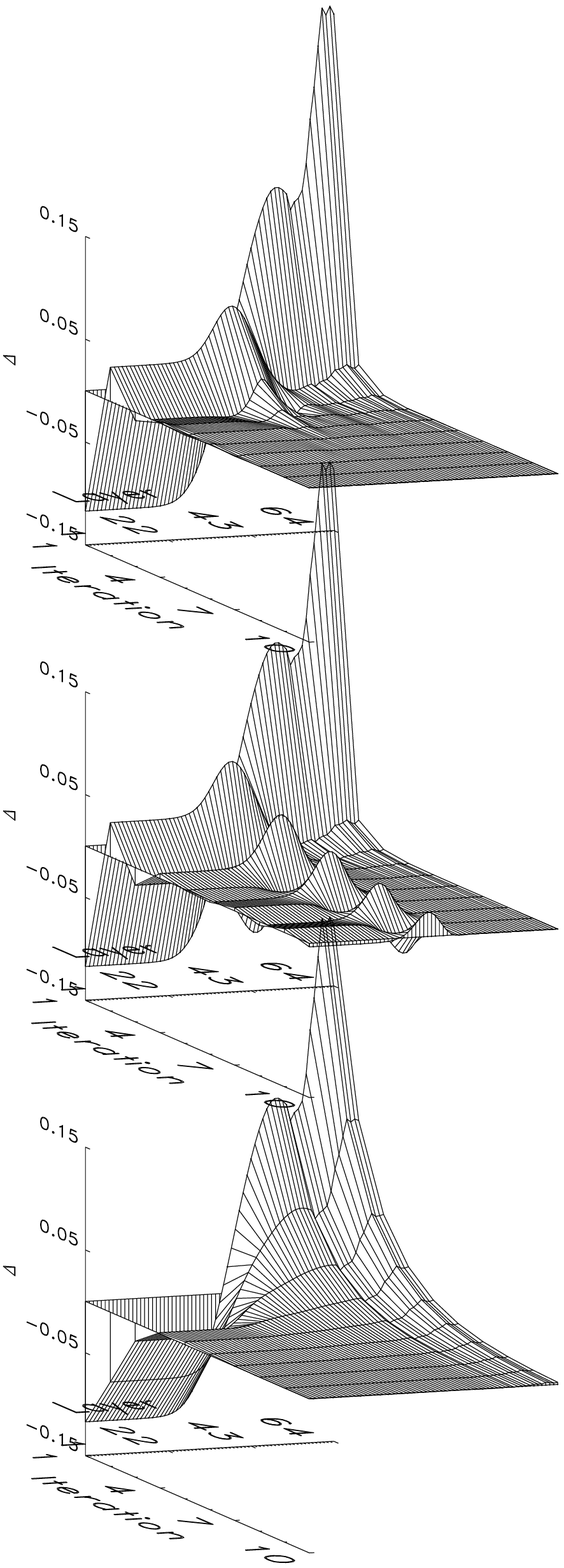}
              \includegraphics[width=.45\textwidth,bb=0 18 255 630]{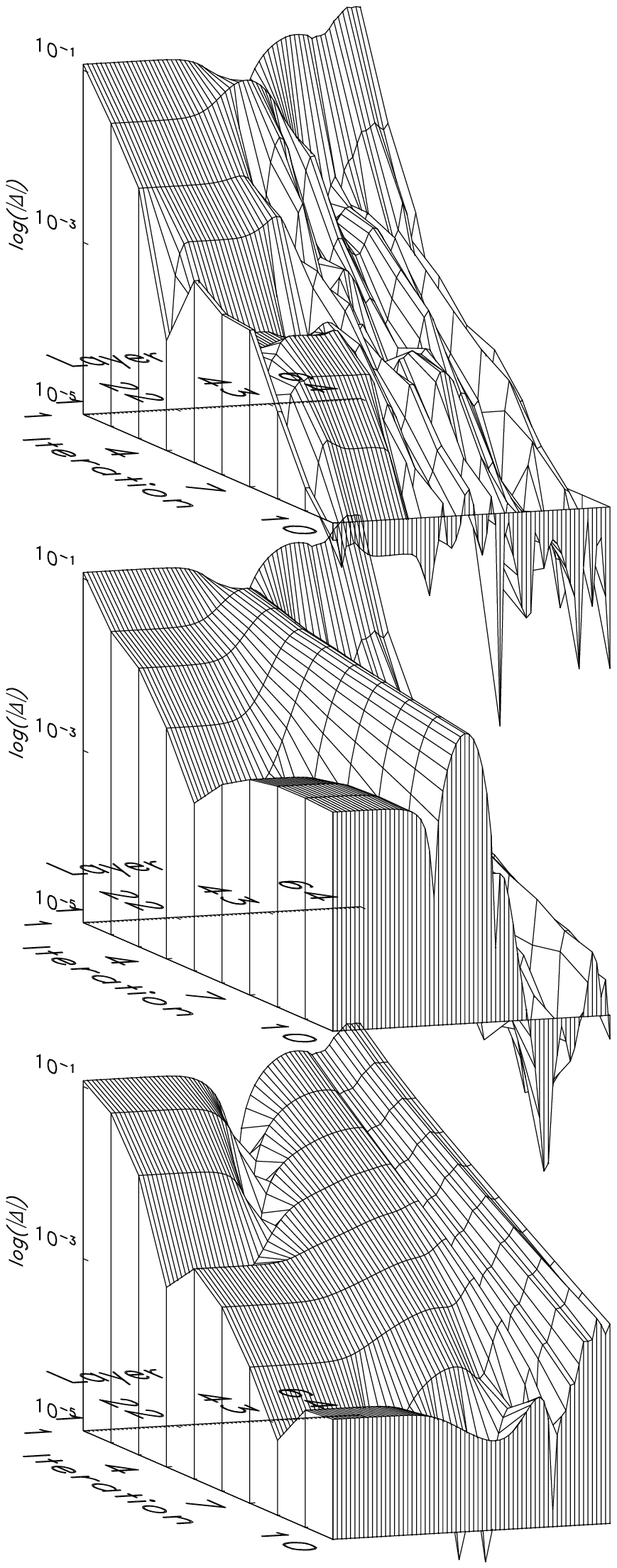}}
 \caption{
  These plots show typical model convergence properties of the UL correction with different damping methods: Active Damping (top), undamped ($\beta(l)=1$) (middle) and constant damping $\beta(l)=0.5$ (bottom).
  On the left the undamped relative correction $\Delta \propto \Delta T/T$ computed with the Uns\"old-Lucy correction procedure is shown for 64 model layers and 10 iterations.
  On the right the same data is plotted logarithmically $\log |\Delta|$ on a scale from $10^{-5}$ to $10^{-1}$.
  \newline
  The rate of convergence with AD is larger than with conventional constant damping: after 3 iterations an accuracy of 0.6\% is reached, for which 9 iterations are needed in case of $\beta=0.5$ and even more in the undamped case.
  The large rate of convergence is maintained even when the corrections become very small.
  After 10 iterations the largest $\Delta$ with AD is a factor of 100 smaller than the largest $\Delta$ without AD.
 } \label{fig:AD_T503}
\end{figure}

Very quick models, i.e. models that converge within 1-3 iterations without the use of AD, do not benefit from the new method, because it needs 1-2 iterations to initialize (for the linear or the non-linear case respectively).

For the comparison of convergence rates, data points in two dimensions must be analyzed, one dimension for spatial layers and one for temperature correction iterations.
A convergence of all model layers is required.
However, an error in a single layer is less significant than an error extending over multiple layers.

In figure \ref{fig:AD_T503} the convergence properties are shown for a hydrostatic LTE model.
This type of model is suitable for the purpose of demonstration.
Firstly, in a hydrostatic atmosphere, being relatively compact, the diversity in physical condition in the atmosphere is smaller than for the radially extended type atmospheres introduced in chapter \ref{sec:NovaStructure}.
Furthermore, the NLTE generalization of the Uns\"old-Lucy temperature correction method, see section \ref{sec:ULNLTE}, requires special care to deal with the outer regions in the atmosphere, where the coupling between the radiation field and the thermal pool is very weak.
These procedures have been implemented in the active damping part of the code, and not in the classical globally constant damping methods.
Therefore, the models typically do not converge at all with the constant damping methods.
But the effect achieved with active damping, as shown by example in figure \ref{fig:AD_T503}, is found in every type of stellar atmosphere model.

The sample model in figure \ref{fig:AD_T503} has been calculated for three cases: using a damping factor of $\beta(l) = 0.5$, undamped $\beta(l) = 1.$ and actively damped $\beta(l) = \alpha(l)$.
$\beta = 0.5$ has proven to be a good value, being small enough to cancel most oscillations.
For simplicity, let us call a model converged when the corrections get below 1\% in every layer (for production models more detailed convergence criteria are used).
Then 8 iterations are needed for $\beta=0.5$, 18 iterations for $\beta=1$ and 3 iterations with AD.
So in this model the convergence speed up is about a factor of 2.5.

Note that $\beta=0.5$ is not fine-tuned, a factor of $\beta=0.7$ yields convergence within 5 iterations for this simple model.

\subsection{Conclusions}
The Uns\"old-Lucy temperature correction can be improved significantly by appropriate damping.
A constant damping factor can be fine-tuned in order to optimize the rate of convergence.
However, as the situation changes from iteration to iteration, a constant damping factor never is optimal.
Furthermore, different regions in atmospheres behave differently, which is again not supported by a constant damping factor.

For the majority of atmospheres a speed-up of about a factor of two or more is achieved with the Active Damping method, without the need for any fine tuning.
However, for simple models that converge within a few iterations using a constant damping factor, the gain is lower than the factor of two, because the method needs 1-2 iterations to initialize.

Generally, the Active Damping method is not limited to the Uns\"old-Lucy temperature correction, but can be applied to any physical system in which a single independent quantity must be brought into some equilibrium.

\section[The NLTE-generalized temp-correction method]{NLTE generalization of the Uns\"old-Lucy temperature correction method}
\subsection{The formalism} \label{sec:ULNLTE}
The derivation of the Uns\"old-Lucy method (section \ref{sec:ULTC}) requires the representation of the source function $S$ as a `weighted' sum of $B$ and $J$ (equation \eqref{eq:ULTC_LTE_Source})
\begin{align}
 S_\l &= (\kappa_\l B_\l + \sigma_\l J_\l)/(\kappa_\l + \sigma_\l) = \varepsilon B_\l + (1-\varepsilon) J_\l \label{eq:ULTCWeigthedSource} \\
 \varepsilon &\equiv \frac{\kappa_\l}{\kappa_\l + \sigma_\l}
\end{align}
This expression is valid for LTE, but not for NLTE.
For example, in NLTE the source function $S$ can exceed the sum of $B$ and $J$ (and thus also the weighted sum of equation \eqref{eq:ULTCWeigthedSource}) or fall below both $B$ and $J$.
Clearly, this is not compatible with the assumed expression of the source function, equation \eqref{eq:ULTC_LTE_Source}.

In general the (macroscopic) source function is built from the (microscopic) rates.
The rates are functions of the radiation field, the level populations, and the electron density, and can not be attributed to thermal or non-thermal (see section \ref{sec:KirchhoffPlanck}) processes.
Hence $S$ cannot be represented in the classical form, equation \eqref{eq:ULTC_LTE_Source}, if the rates are used to calculate $S$.

However, if, in the classical view, at a specific location in the atmosphere thermal processes play a dominant role over non-thermal processes, then the atmosphere is forced into a state close to LTE.
On the other hand, if non-thermal processes are dominant, then the departures from LTE will become large.
Now, reversing these implications, one can estimate the balance between thermal and non-thermal emission by the order of departure from LTE.

Writing the emissivity as the sum of a thermal and a non-thermal fraction
\begin{equation} \label{eq:EmissivityThermal}
 \eta_\l = \eta_\l^{\rm th} + \eta_\l^{\rm Nth}
\end{equation}
allows to use equation \eqref{eq:ULTC2a} as it stands, but now with the averaged opacities $\kappa_B$ and $\kappa_J$ written as
\begin{align}
 \kappa_B &= \frac{1}{B} \int \eta_\l^{\rm th} \,\ud\l \label{eq:ULNLTE1} \\
 \kappa_J &= \frac{1}{J} \int \chi_\l J_\l - \eta_\l^{\rm Nth} \,\ud\l \label{eq:ULNLTE2} 
\end{align}
and $\chi_H$ is defined by equation \eqref{eq:OpacityIntegral3}.

Let each bound-bound transition have a statistical fraction $\theta$ of emission that is thermal
\begin{align}
\begin{split}
 \eta^{\rm bb}_\l &\equiv \eta^{\rm bb,th}_\l + \eta^{\rm bb,Nth}_\l \\
  &= \sum_{i \ne j} \left( \eta^{\rm th}_{\l,ij} + \eta^{\rm Nth}_{\l,ij} \right) \\
  &= \sum_{i \ne j} \big( \theta_{ij} \eta_{\l,ij} + (1-\theta_{ij}) \eta_{\l,ij} \big)
\end{split}
\end{align}
where $i$ and $j$ are the levels of all transitions that contribute to the emissivity at wavelength $\l$.
Then the thermal and non-thermal terms of equation \eqref{eq:EmissivityThermal} can be written as
\begin{align}
 \eta_\l^{\rm th} &= \eta^{\rm ff}_\l + \eta^{\rm fb}_\l + \eta^{\rm bb,th}_\l \label{eq:EtaThermal}\\
 \eta_\l^{\rm Nth} &= \sigma_\l J_\l + \eta^{\rm bb,Nth}_\l \label{eq:EtaNonThermal} 
\end{align}

The emissivity from bound-free and free-free transitions can be assumed to be fully thermal.
In both cases the interaction takes place with a free electron that has a Maxwellian velocity distribution \cite{Mihalas78}.
Note that the NLTE expression for free-bound emissivity \eqref{eq:FBEta} depends on the LTE occupation number of the bound state to which the recombination takes place.

For bound-bound transitions the thermal fraction $\theta_{ij}$ can be estimated as
\begin{equation}
 \theta_{ij} = {\rm min} ( b_i/b_j, b_j/b_i )
  = {\rm min} ( b^*_i/b^*_j, b^*_j/b^*_i )
\end{equation}
The minimum function reflects the fact that departures from LTE can manifest as underpopulation or overpopulation.
If the upper level and the lower level have the same departure $b$ from LTE, then for the transition a pseudo LTE situation exists.
The thermal fraction must then be close to unity.

\subsection{Inferior alternatives} \label{sec:ULNLTEAlt}
The new NLTE generalization of section \ref{sec:ULNLTE} extends the applicability of the Uns\"old-Lucy temperature method to situations where the emissivity is not fully thermal.

Originally, in \phx\ there were two distinct ways to treat the NLTE opacities in the temperature correction procedure.
In the first variant the NLTE opacities are left out from the integrals of the expressions for $\kappa_B$, $\kappa_J$ and $\chi_H$ (equations \eqref{eq:OpacityIntegral1} to \eqref{eq:OpacityIntegral3}).
Since the non-thermal opacities originate from NLTE opacities only, this circumvents the problem.

This is a useful approximation if, as usual, the bulk of the opacities is LTE opacities.
But, as described in section \ref{sec:PureNLTE}, the models performed in this work are pure NLTE models, meaning that there are no LTE opacities at all.
However, the non-thermal opacities originate from the bound-bound transitions only.
Therefore, in analogy to the original idea, the NLTE bound-bound opacities are ignored, and the free-free and bound-free opacities are used as approximation in the opacity means.
From equations \eqref{eq:OpacityIntegral3}, \eqref{eq:ULNLTE1}, \eqref{eq:ULNLTE2}, \eqref{eq:DetailedExtinction}, \eqref{eq:EtaThermal} and \eqref{eq:EtaNonThermal}, omitting the bound-bound terms, it follows
\begin{align}
 \kappa_B &= \frac{1}{B} \int \eta^{\rm ff}_\l + \eta^{\rm fb}_\l \,\ud\l
  = \frac{1}{B} \int \chi^{\rm ff}_\l B_\l + \eta^{\rm fb}_\l \,\ud\l \label{eq:ULNLTEVar1a}\\
 \kappa_J &= \frac{1}{J} \int \left( \chi^{\rm ff}_\l + \chi^{\rm bf}_\l \right) J_\l \,\ud\l \\
 \chi_H &= \frac{1}{J} \int \left( \chi_\l - \chi^{\rm bb}_\l \right) H_\l \,\ud\l \label{eq:ULNLTEVar1c} 
\end{align}

In the second original variant the NLTE emissivities are treated like LTE opacities so that the opacity means $\kappa_B$ and $\kappa_J$ (equations \eqref{eq:OpacityIntegral1} and \eqref{eq:OpacityIntegral2}) become
\begin{align}
 \kappa_B &=  \frac{1}{B} \int (\chi_\l - \sigma_\l) B_\l \,\ud\l \label{eq:ULNLTEVar2a}\\
 \kappa_J &=  \frac{1}{J} \int (\chi_\l - \sigma_\l) J_\l \,\ud\l \label{eq:ULNLTEVar2b}
\end{align}
where $\chi$ is given by equation \eqref{eq:DetailedExtinction}. $\chi_H$ is defined by equation \eqref{eq:OpacityIntegral3}.

\subsection{Comparison}
In order to test the assumptions made in the NLTE generalization of section \ref{sec:ULNLTE} the method must be tested and compared to the alternatives of section \ref{sec:ULNLTEAlt}.
In all variants the temperature corrections are determined by the same equations \eqref{eq:ULTC6a} and \eqref{eq:ULTC6b}, but different definitions for the opacity means are used.

\begin{figure}
 \centerline{\includegraphics[height=\textwidth,angle=90]{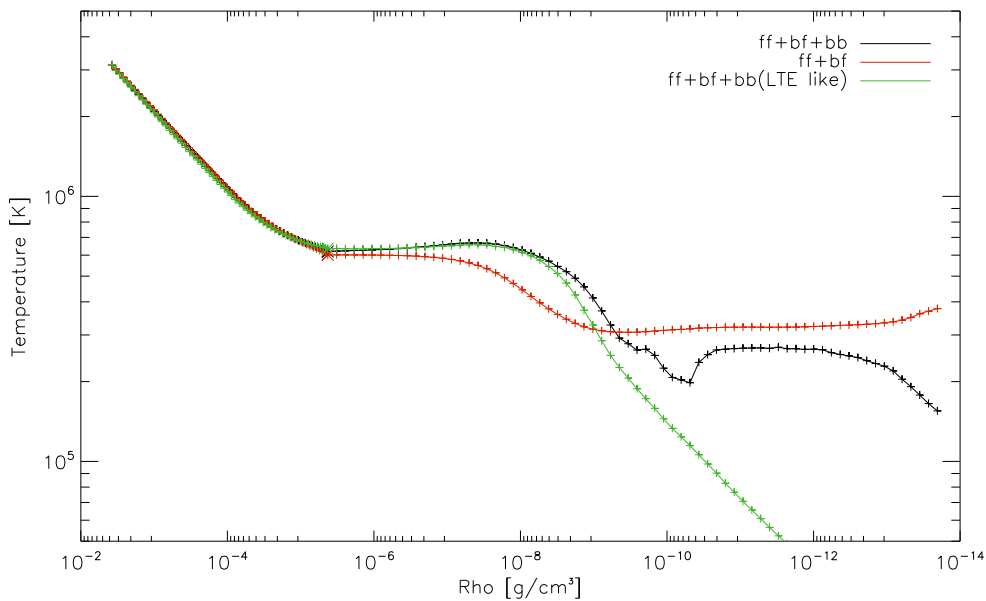}}
 \caption{
 The temperature structures, plotted against the density, for a typical NLTE model resulting from three different variants of temperature correction procedure:
 1. the new NLTE generalization of section \ref{sec:ULNLTE} (black),
 2. an approximation neglecting all bound-bound opacities (red), and
 3. an approximation where the emissivities are treated like in LTE (green).
 \newline
 In the inner regions, where departures from LTE are small, the three methods give very similar results.
 In the outer regions departures are large but with the small densities the opacities are small.
 The most important differences are in the middle regions, with densities between $10^{-4}$ and $10^{-10}$ g/cm$^{-2}$.
 } \label{fig:ULNLTETempStructs}
\end{figure}

The temperature structures obtained from the three variants are plotted in figure \ref{fig:ULNLTETempStructs}.
In the optically thin outer regions the temperature structures differ significantly.
But in the deeper ranges they are almost identical.
There the departures from LTE are small, and thus the non-thermal part of the emission goes to zero.
Note that in LTE, where all $b_i = 1$, the NLTE generalization of section \ref{sec:ULNLTE} reduces to the second original variant.
Furthermore, $J_\l$ becomes similar to $B_\l$, so that the definitions of $\kappa_B$, $\kappa_J$ and $\chi_H$ become identical.

\begin{figure}
 \centerline{\includegraphics[width=\textwidth]{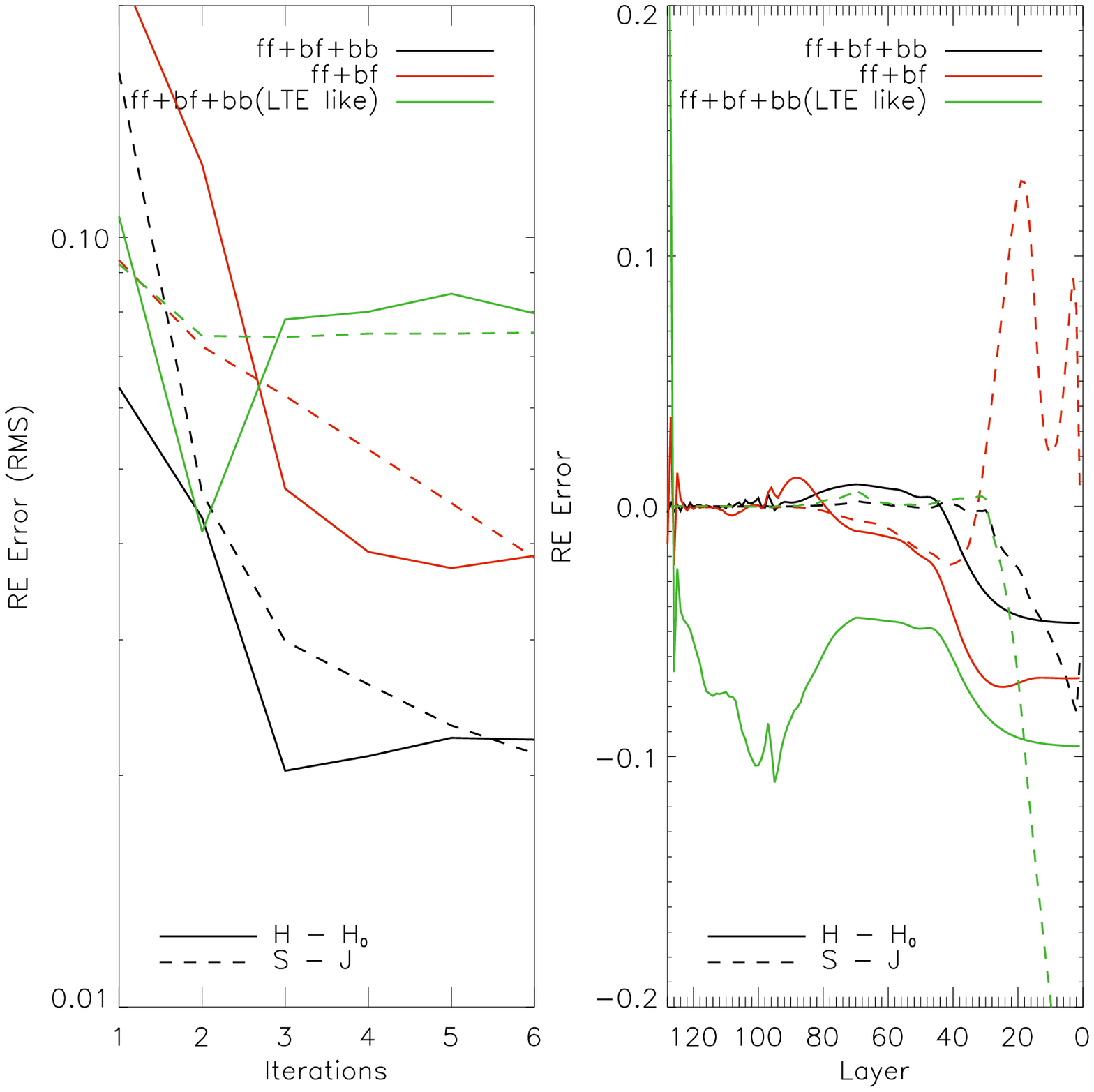}}
 \caption{
  These plots show the radiative equilibrium errors for the same model shown in figure \ref{fig:ULNLTETempStructs}.
  The errors are defined in equations \eqref{eq:REErrorDif} and \eqref{eq:REErrorInt} as
  $\varepsilon_{\rm RE,int}$, corresponds to the integral form ($H - H_0$), and $\varepsilon_{\rm RE,dif}$, corresponding to the differential form ($S-J$) of the RE condition.
  The left plot shows the root-mean-square (RMS) of the errors over all model layers for subsequent temperature correction iterations on a logarithmic scale.
  The errors in the last iteration (number 6) are plotted against the model layer number on the right.
  \newline
  After a number of iterations the errors converge.
  The errors with the two alternative methods are significantly larger than those with the new NLTE generalization method of section \ref{sec:ULNLTE}.
  The layers 40-100 have the most important influence on the spectrum, being between optically thick and thin.
  Right in this range the NLTE generalization method performs better.
 } \label{fig:ULNLTEConv}
\end{figure}

One important test is if the temperature correction method converges to radiative equilibrium, as described in section \ref{sec:RadiativeEquilibrium}.
For the model shown in figure \ref{fig:ULNLTETempStructs} the two measures for relative deviations from RE (defined by equations \eqref{eq:REErrorDif} and \eqref{eq:REErrorInt}) are plotted in figure \ref{fig:ULNLTEConv}.
After a number of temperature correction iterations the errors converge.
The errors with the two alternative methods are significantly larger than those with the new NLTE generalization method of section \ref{sec:ULNLTE}.

\subsection{Limitations of the NLTE generalization} \label{sec:ULNLTELimitations}
The NLTE generalization of the Uns\"old-Lucy temperature correction method treats the opacities in a physically better way than the old alternatives.
It was shown that also the convergence performance is superior.
These two properties make the NLTE generalization the preferred temperature correction method, and throughout this work this method is used for all model computations shown in this work, unless specified otherwise.

There are, however, still a few limitations that are inherent to the UL method.
In the UL method it is assumed that the proper temperature of the gas can be derived from the radiation transport equation (see equations \eqref{eq:ULBasic} and the following).
This can only work if the opacities do significantly dependent on the temperature.
As shown in section \ref{sec:NiScaling} there are atmospherical conditions in which such temperature dependence is not given.
This is especially the case for regions where the departures from LTE are large.
The population numbers are then so far off from their TE values, and so are the opacities, that it becomes very difficult to derive the temperature from it.

In the NLTE generalization scheme this problem is apparent in the definition of $\kappa_J$, equation \eqref{eq:ULNLTE2}, where the integrand can become very small or even negative if the non-thermal part of the emissivity becomes large.
A small difference of two large numbers is very prone to errors.
But the UL method is based on the assumption that the ratio of the opacity \emph{integrals}, $\kappa_J/\kappa_B$, are independent of the temperature, i.e. constant over temperature correction iterations.
In such cases, where the error in $\kappa_J$ becomes large the formalism cannot work accurately.

However, in those cases where the UL method fails, because the opacities (not the integrals) are insensitive to the temperature, the temperature is unimportant for the state of the gas, and so for the radiation transport.
Therefore, where the temperature cannot be derived accurately, there the temperature \emph{does not need to be determined accurately}.

\subsubsection{A `resort' temperature method}
In order to determine the temperature approximately in those regions where the UL method fails, another method must be resorted to.
A very simple assumption can be made using the Stefan-Boltzmann law (equation \eqref{eq:StefanBoltzmannLaw}) for a luminosity that is constantly extrapolated for outer regions.
The following temperature relation follows
\begin{equation}
 T(r) = \sqrt{ \frac{r_{\rm ref}}{r} \left(T(r_{\rm ref})\right)^2 }
\end{equation}
with $r_{\rm ref}$ being the outer reference layer for which the UL scheme is still working reliably.
In temperature structures shown in this work this is apparent by a sudden very smooth, almost linear decline of the temperature against the density in double logarithmic plots.
For example, in figure \ref{fig:ULNLTETempStructs} this occurs from layer 6 outwards, where layer 6 was the reference layer (the outermost layer that was yet determined from the NLTE generalized UL method).

\subsubsection{Validation condition}
One thing left to define is a validation condition for the UL method.
The following condition has proven to work well.
The UL method can be applied if the opacity integrals are approximately constant.
This is tested by the condition that the opacity integrals $\kappa_J$ and $\chi_H$ need to have been reasonably constant for the actual and the previous temperature correction iteration, where for reasonably constant these must satisfy both
\begin{equation}
 1/2 \le \frac{\kappa_J}{\kappa_{J,\rm old}} \le 2
\end{equation}
and
\begin{align}
 2/3 &\le \frac{\kappa_J}{\kappa_{J,\rm old}} \le 3/2 \\
 4/5 &\le \frac{\chi_H}{\chi_{H,\rm old}} \le 5/4
\end{align}
The weaker condition for $\kappa_J$ reflects the fact that the uncertainties in $\kappa_J$ are larger due to the difference in its definition, equation \eqref{eq:ULNLTE2}, as discussed above.

\subsection{Line cooling/heating effects} \label{sec:LineCooling}
The monochromatic integrand of the radiative equilibrium condition, equation \eqref{eq:RadiativeEquilibrium}
\begin{equation}
 x_\l = \eta_\l - \chi_\l J_\l = \chi_\l (S_\l - J\l)
\end{equation}
here denoted with $x$, is called the \emph{net radiative cooling rate}, see also \cite{Rutten95}.
When $S_\l > J_\l$, then locally an overdose of radiation is produced, representing an energy loss for this wavelength.
If $x$ integrated over a line profile is positive then the line has a local cooling effect, called \emph{line cooling}.
The opposite is called line heating.
Also, the continuum can have a local cooling or heating influence.

\begin{figure}
 \centerline{\includegraphics[width=\textwidth]{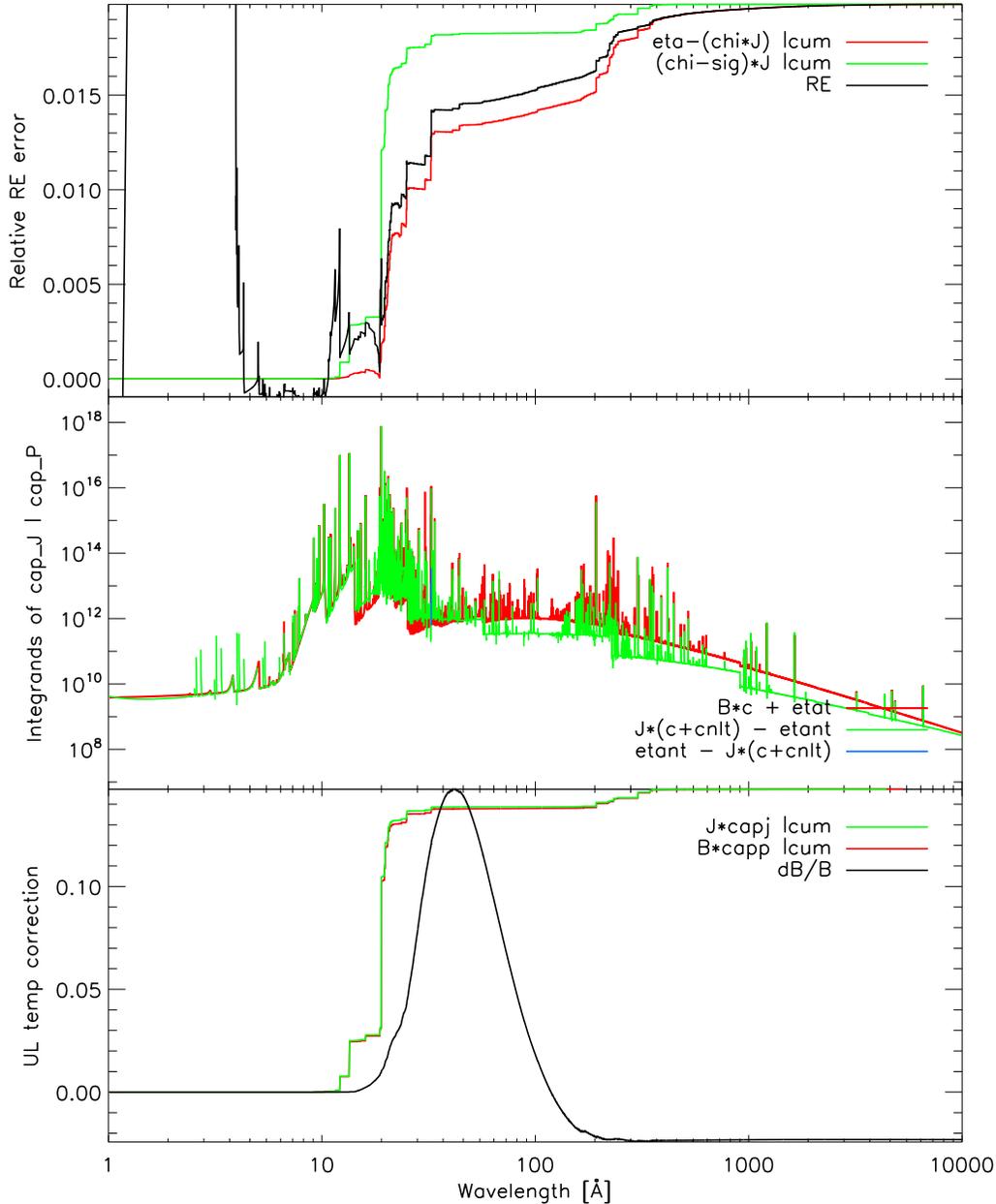}}
 \caption{ \label{fig:LineCooling}
 These graphs show the effect of line cooling (see also the next figures).
 \newline
 In the \emph{upper graph} the gradual development over wavelength of relative radiative equilibrium error $\varepsilon_{\rm RE,dif}$ (equation \eqref{eq:REErrorDif}) is plotted.
 The nominator and denominator are plotted cumulatively ($|$cum) (red and green).
 The ratio of these sums is evaluated per wavelength (black).
 \newline
 The \emph{lower graph} shows something similar to the upper graph.
 Here not the RE is evaluated, but the UL temperature correction, specifically only the $\Delta B_1$ term, equation \eqref{eq:ULTC6a}, that is important in the outer regions where line cooling effects come into play.
 The constituents $B \cdot \kappa_B$ and $J \cdot \kappa_J$ are plotted cumulatively (red and green).
 The ratio of these two minus 1 (equal to $\Delta B_1/B$) is evaluated per wavelength (black).
 \newline
 The \emph{middle graph} shows the integrands of $\kappa_P$ and $\kappa_J$ (not cumulatively) on a logarithmic scale (red and green).
 Negative values of the latter are plotted positively in blue.
 }
\end{figure}

\begin{figure}
 \centerline{\includegraphics[width=\textwidth]{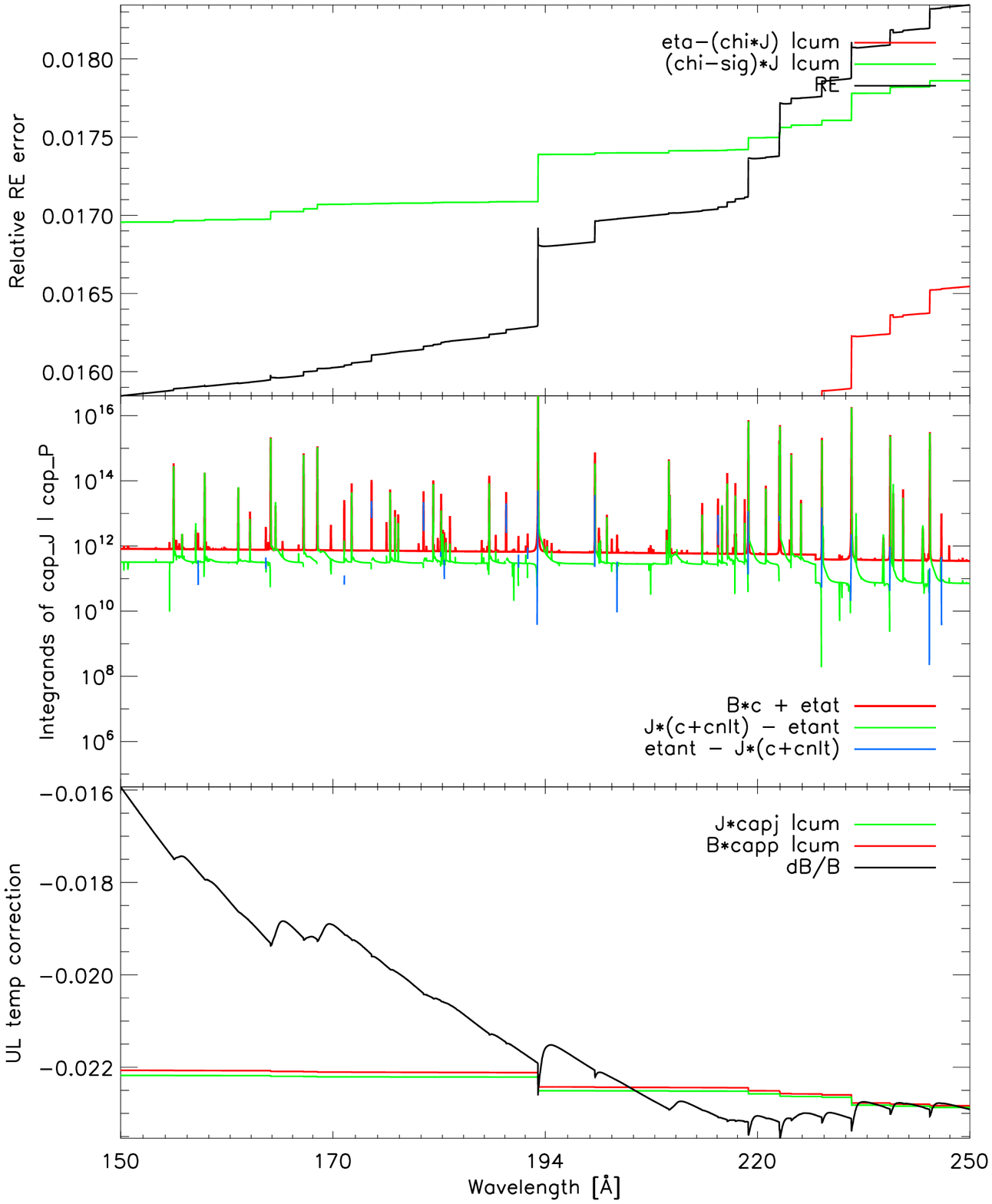}}
 \caption{ \label{fig:LineCoolingZoom}
 This figure shows a zoom in from figure \ref{fig:LineCooling} to a smaller wavelength range.
 The optical depth for this layer is $\tau_{\rm ref} = 2\cdot 10^{-4}$.
 At this depth, the atmosphere starts to become optically thin.
\newline
 Comparing the upper and lower graphs of figure \ref{fig:LineCooling}, it is seen that the relative RE error is more sensitive to line cooling effects than the UL temperature correction.
 For this reason the UL temperature correction is more stable and features better convergence properties than a Newton-Raphson scheme based on RE.
\newline
 The lower graph shows that the strong lines in this wavelength range are heating whereas the continuum is cooling.
 Note that in the upper graph, the continuum is also cooling (the RE err increases) but here the lines are cooling instead of heating.
 }
\end{figure}

\begin{figure}
 \centerline{\includegraphics[width=\textwidth]{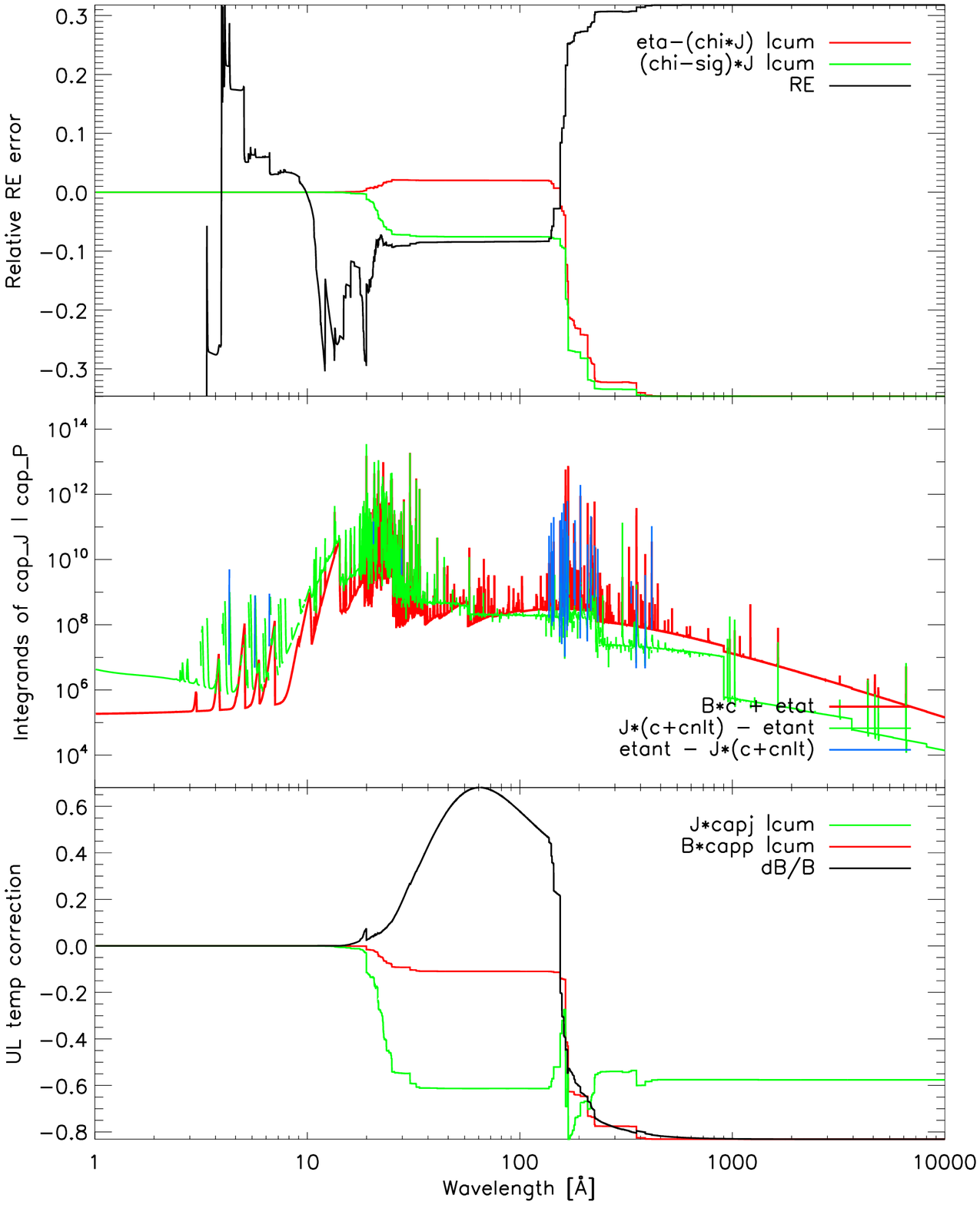}}
 \caption{ \label{fig:LineCoolingDominant}
 Here the same plots are shown as in figure \ref{fig:LineCooling}, but now for a layer further out, with optical depth $\tau_{\rm ref} = 2\cdot 10^{-5}$.
 At this shallow depth the lines start to dominate the heating/cooling influence of the radiation field to the matter.
 In this region also the UL temperature correction starts to become unreliable, because of inaccuracies in the mean intensity in the line centers of the lines that dominate the cooling effect.
 The problem arises at wavelengths where the monochromatic contribution to $\kappa_J$ becomes strongly negative.
 This effect is inherent to a weak coupling of the radiation field to the matter in NLTE models.
 It is discussed in section \ref{sec:ULNLTELimitations}.
 }
\end{figure}

In figures \ref{fig:LineCooling} to \ref{fig:LineCoolingDominant} this line cooling effect is demonstrated.
If line cooling/heating gets dominant over the contribution of continuum cooling/heating then deducing the temperature from the interaction with the radiation field becomes difficult, since the line cooling/heating contributions strongly depend on the run of the mean intensity over the line profile.
In the case of NLTE, the radiation field is iteratively brought into equilibrium with the radiative rates so that especially in the line centers the mean intensity is subject to changes.
This further complicates the derivation of the temperature from line cooling/heating effects in regions that are optically thin apart from the lines that contribute to the cooling/heating.

\clearpage
\chapter{First results} \label{sec:Results}
The primary accomplishment of this work is the framework described in the previous sections.
This framework allows to compute radially extended, massive (atoms levels) NLTE models.
Once this framework is ready and working, it remains to show that it produces useful results.
For this purpose a \emph{modest amount} of models has been computed, as many as available computational power allowed for in the time frame of this work.
These first results are shown in this chapter.

An important test for atmosphere models is if systematic variations of single model parameters produce systematic results.
This test was performed with a few small model series (model grids).
The size of these grids is \emph{yet small}, limited by computation time, and further expansion is planned for near future.
Apart from the purpose of testing the new framework, these grids give the opportunity to explore the parameter space of the models.
Furthermore, careful analyzation of the impact of single model parameters on the results is an important preparation for a detailed \emph{spectral analysis} of the available observational data, for which the models are fine-tuned to each observation individually.

In section \ref{sec:Grids} the results are shown for a few small model grids, for both the hydrostatic and the hybrid type (nova) atmosphere models.
Subsequently, in section \ref{sec:LTEvsNLTE} LTE and NLTE models are compared in order to confirm that a NLTE treatment is necessary for this type of models.
Finally, in section \ref{sec:Fits} the models are compared to the currently available `high-res' X-ray observations of novae in the super-soft X-ray state (V4743 Sgr, RS Oph, and V2491 Cyg).
It is stressed at this point, that so far \emph{only a small amount of models has been computed}.
Therefore, the models are \emph{far from} fine-tuned to fit the observations.
Nevertheless, these first results fit the observations remarkably well, and a vast improvement is achieved with respect to previous work with \phx\ \cite{PetzPhd}, where models \emph{were} fine-tuned.
These results can be seen as a very promising starting point for a thorough analysis of the observations through a detailed modeling process.

\section{Model grids} \label{sec:Grids}
So far, parameter grids have been computed around the most basic model atmosphere parameters.

Grids are computed for \emph{pure hydrostatic} models and for the \emph{hybrid} models as they are described in chapter \ref{sec:NovaStructure}.
The former type of models has two advantages.
Firstly, there are only two basic model parameters ($T_{\rm eff}$ and $\log(g)$) in contrast to five for the hybrid models ($T_{\rm eff}$, $\log(g)$, $\dot{M}$, $v_\infty$ and $\beta$).
Therefore, the dimension of the grid is much smaller and thus the number of models to be computed.
Secondly, the computational demands are significantly lower than for the nova-type hybrid models.
The two main reasons for this are:
\begin{enumerate}
\item Hydrostatic atmospheres are relatively (radially) compact, so there is less diversity in physical conditions in the atmosphere. Consequently, a smaller number of layers is needed to sample the atmosphere.
\item Since the velocity field is zero there is no coupling between the wavelength points so that numerically each point can be treated independently. This is an ideal situation for parallel computing. It features almost perfect scalability (to multiple parallel tasks).
\end{enumerate}

For the hydrostatic models the basic parameters are the classical effective temperature $T_{\rm eff}$ and gravity $\log(g)$.
The effective temperature is run from 450kK (= 450,000K) to 1000kK, in 50kK steps for the hydrostatic models.
In the grid of nova-type hybrid models the effective temperatures are 550kK, 600kK and 650kK (extension to lower and higher values is work in progress).
This range is based on comparison with observations from novae in a supersoft X-ray state (see section \ref{sec:Fits}).

The log($g$) is varied along with $T_{\rm eff}$, accounting for the fact that the radiation pressure increases with the luminosity and thus with $T_{\rm eff}$.
In a hydrostatic atmosphere the effective gravitation determines the radial extension (see chapter \ref{sec:NovaStructure}).
Therefore, in order to compare atmospheres with a comparable effective gravitation, the following relation \cite{Mihalas78} was used to define a default $\log(g)$
\begin{equation} \label{eq:MihalasGmin}
 g \gtrsim 65 (T_{\rm eff}/10^4)^4
\end{equation}
from which then small systematic deviations were made in both directions.

In a later stage, it was found from the original derivation of relation \eqref{eq:MihalasGmin} \cite{Underhill49} that the right hand side is a factor of 10 too high.
The \emph{right version} is
\begin{equation} \label{eq:Gmin}
 g \gtrsim 6.5 (T_{\rm eff}/10^4)^4
\end{equation}
By that time a major part of the grid computations was already completed.
From then on, only the lower $\log(g)$ parts of the grid have been continued, so that in the grid most models are computed around $g$ being half of the (wrong) lower boundary given by equation \eqref{eq:MihalasGmin}.
Consequently, all models shown in this section feature a rather high $\log(g)$.

\begin{figure}
 \centerline{\includegraphics[height=\textwidth,angle=90]{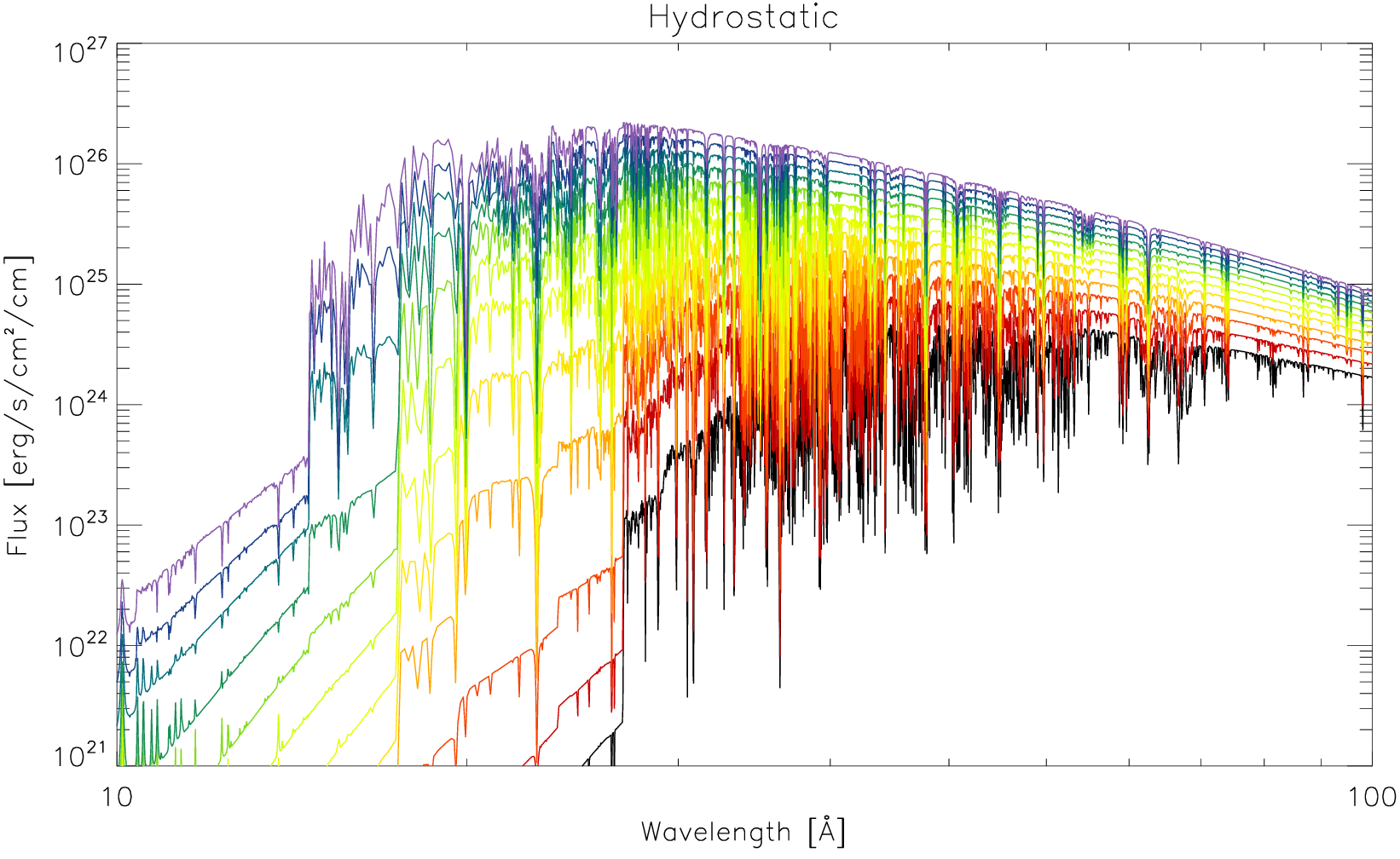}}
 \centerline{\includegraphics[height=\textwidth,angle=90]{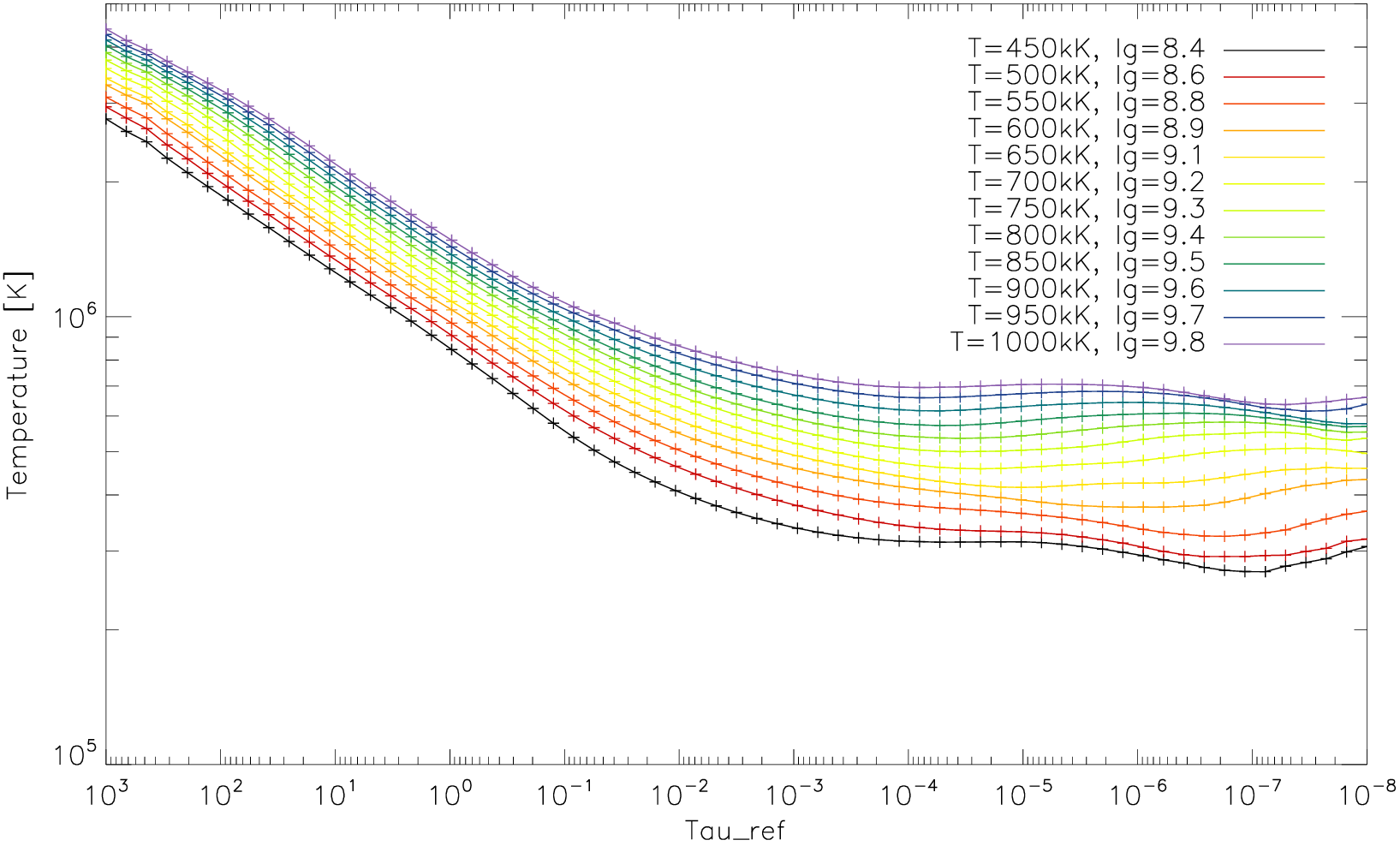}}
 \caption{ \label{fig:TeffGridPP}
 The effective temperature $T_{\rm eff}$ is varied and a corresponding approximate effective gravitation is kept fixed using equation \eqref{eq:MihalasGmin}.
 The upper graph shows the spectra, the lower graph the temperature structures on the optical depth reference scale.
 \newline
 A systematic change in the spectra and the temperature structures can be observed from changing the parameter $T_{\rm eff}$.
 }
\end{figure}

\begin{figure}
 \centerline{\includegraphics[height=\textwidth,angle=90]{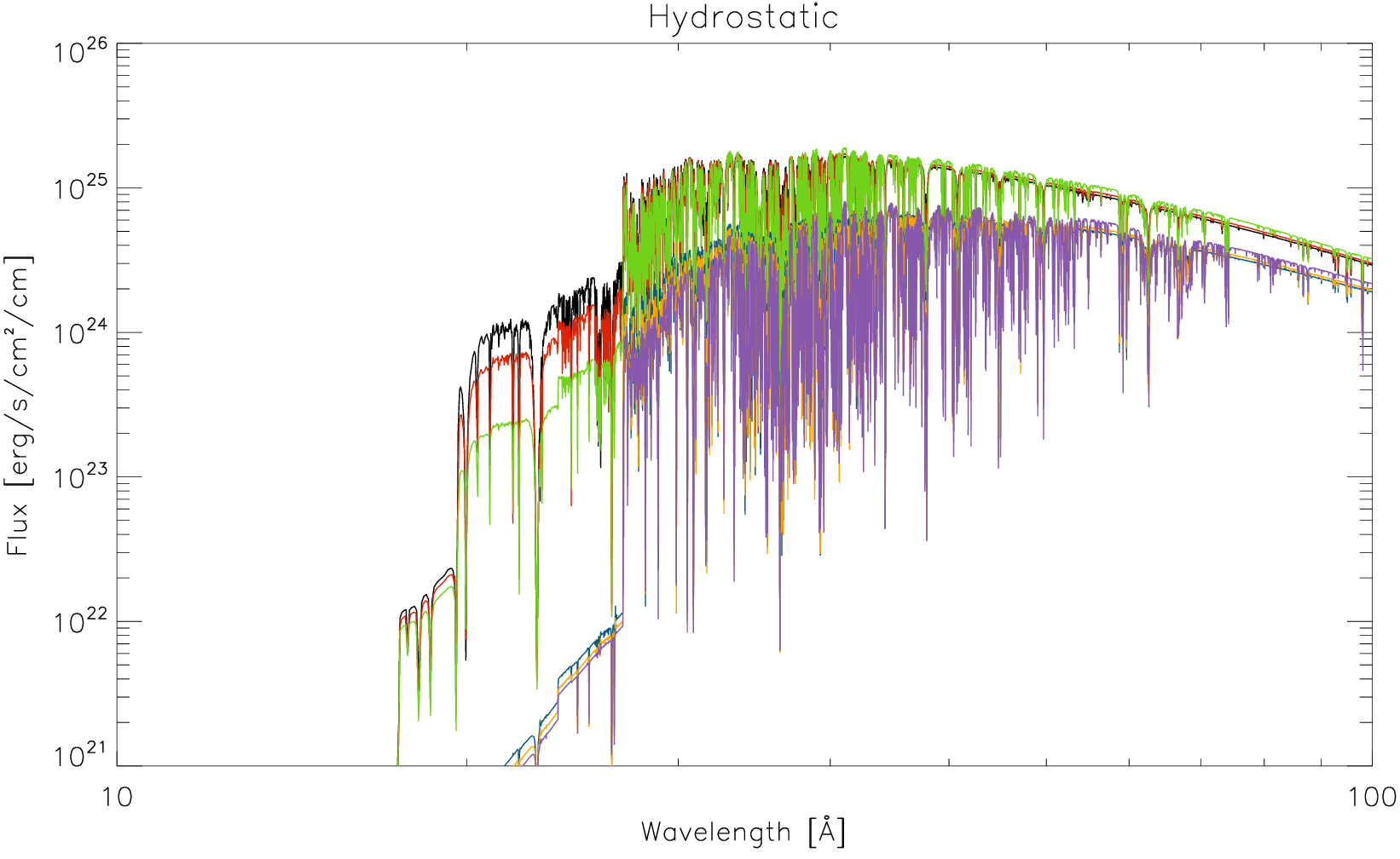}}
 \centerline{\includegraphics[height=\textwidth,angle=90]{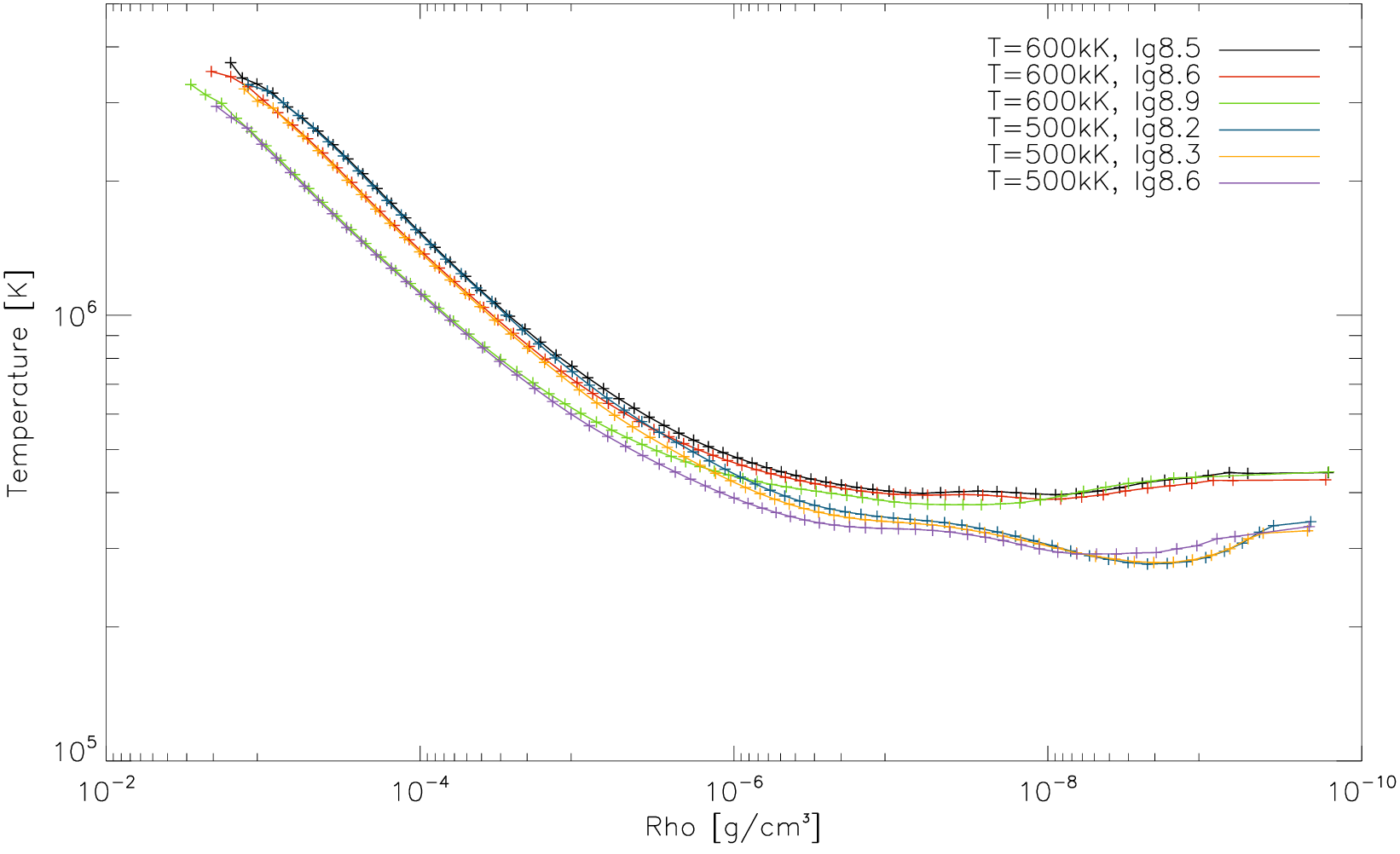}}
 \caption{ \label{fig:LoggGridPP}
 The $\log(g)$ is varied for two fixed effective temperatures.
 The temperature structures are plotted logarithmically against the density.
 \newline
 There are two branches in the temperature structures in the tenuous part of the atmosphere, corresponding to the two effective temperatures, and three branches can be discerned in the high density region, which correspond to the three values of $\log(g)$ that were used.
 Apparently, in these models the impact of these two parameters varies with depth in the atmosphere.
 In the 600kK spectra the $\log(g)$ affects the strength of the C\,{\sc vi} ionization edge, which differentiates the spectra.
 In the 500kK spectra the influence of $\log(g)$ is very small.
 }
\end{figure}

\begin{figure}
 \centerline{\includegraphics[height=\textwidth,angle=90]{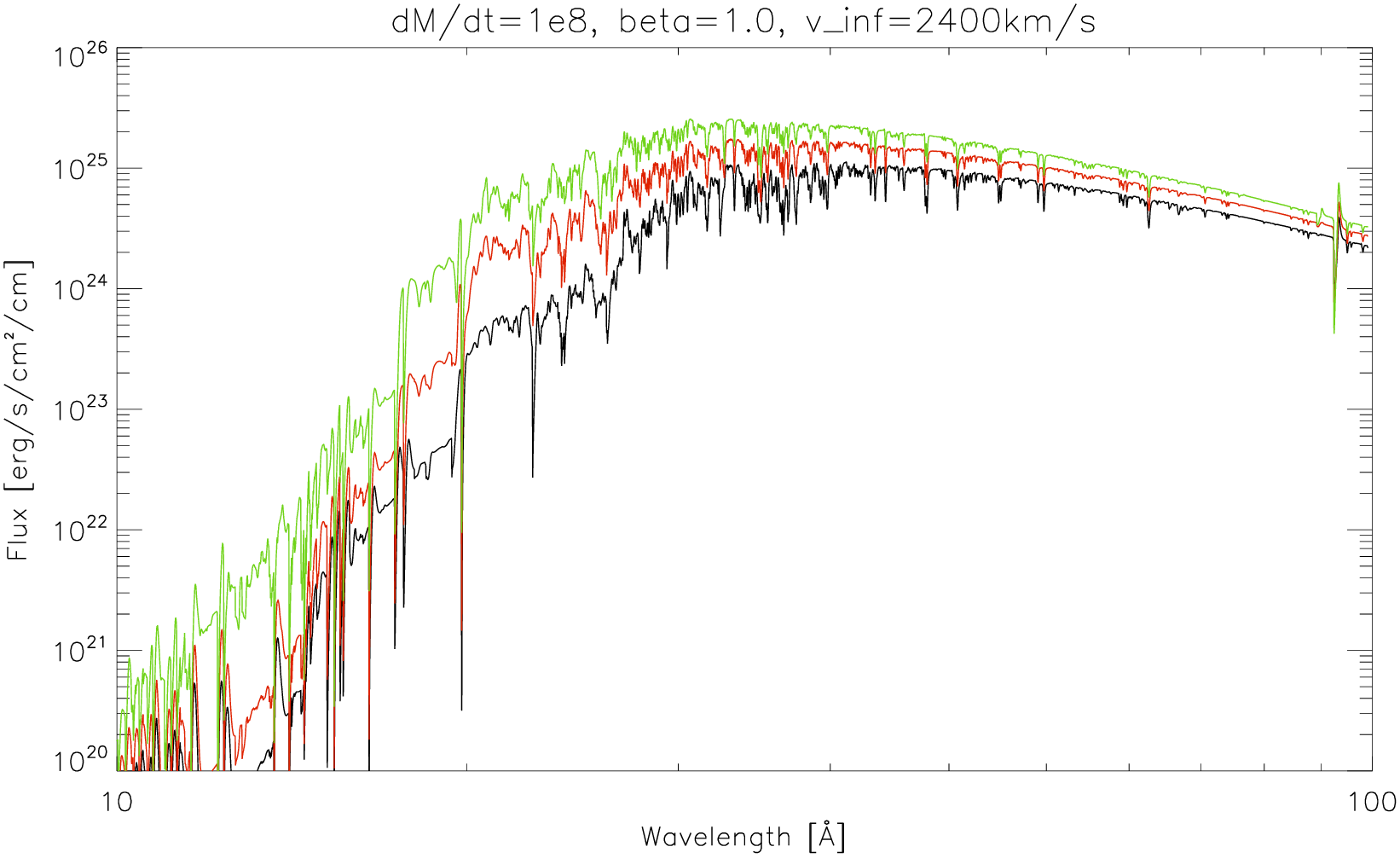}}
 \centerline{\includegraphics[height=\textwidth,angle=90]{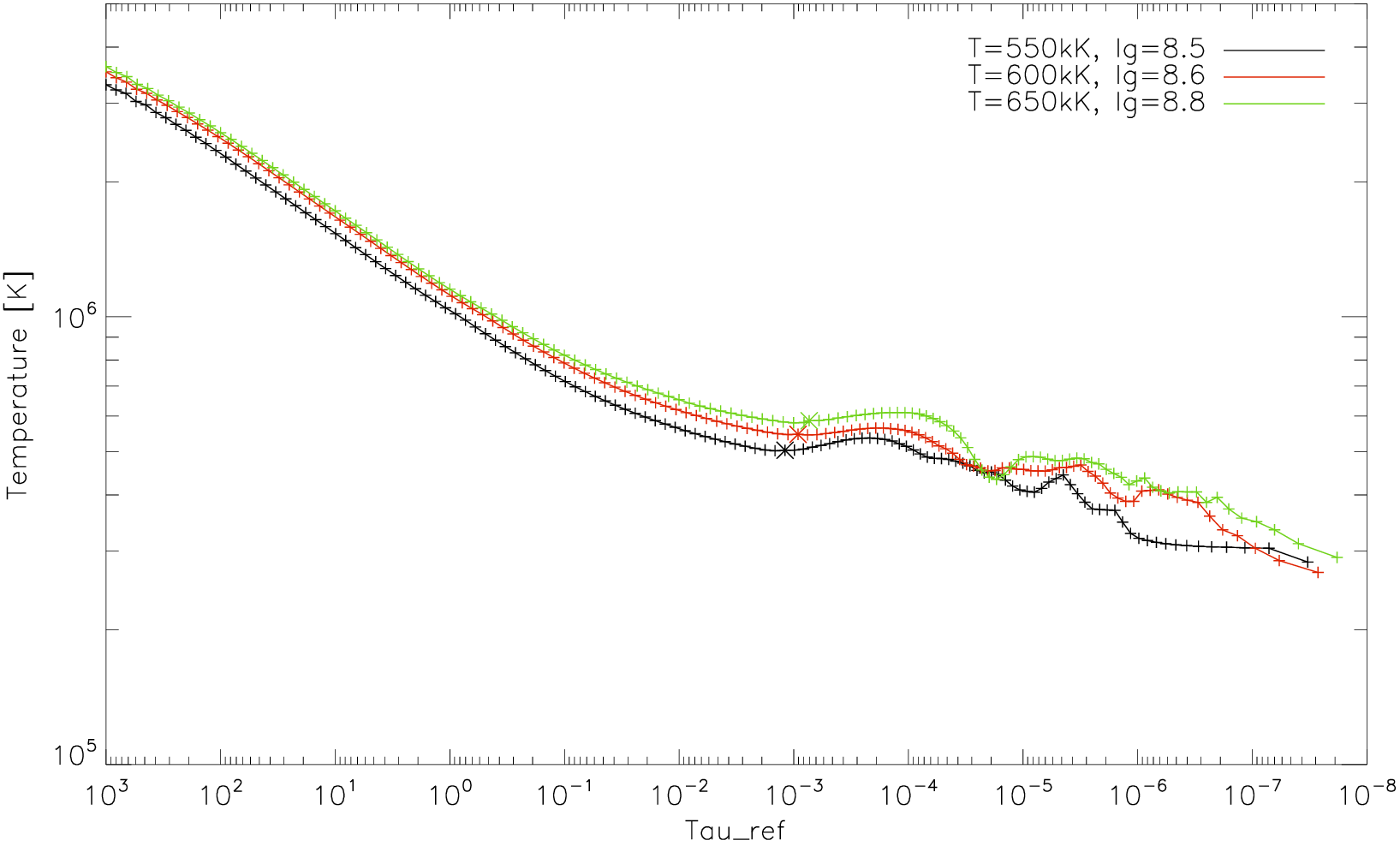}}
 \caption{ \label{fig:TeffGrid}
 In this figure the effective temperature $T_{\rm eff}$ is varied and a corresponding approximate effective gravitation is kept fixed using equation \eqref{eq:MihalasGmin}.
 Also, all velocity field parameters are identical.
 The temperature structures are a bit more variable than in the hydrostatic models.
 Around a density of $\rho = 10^{-9} g/cm^3$ a dip develops with rising temperature.
 But there is a gradual development in the spectra with $T_{\rm eff}$, comparable to that for the hydrostatic spectra of figure \ref{fig:TeffGridPP}.
 }
\end{figure}

\begin{figure}
 \centerline{\includegraphics[height=\textwidth,angle=90]{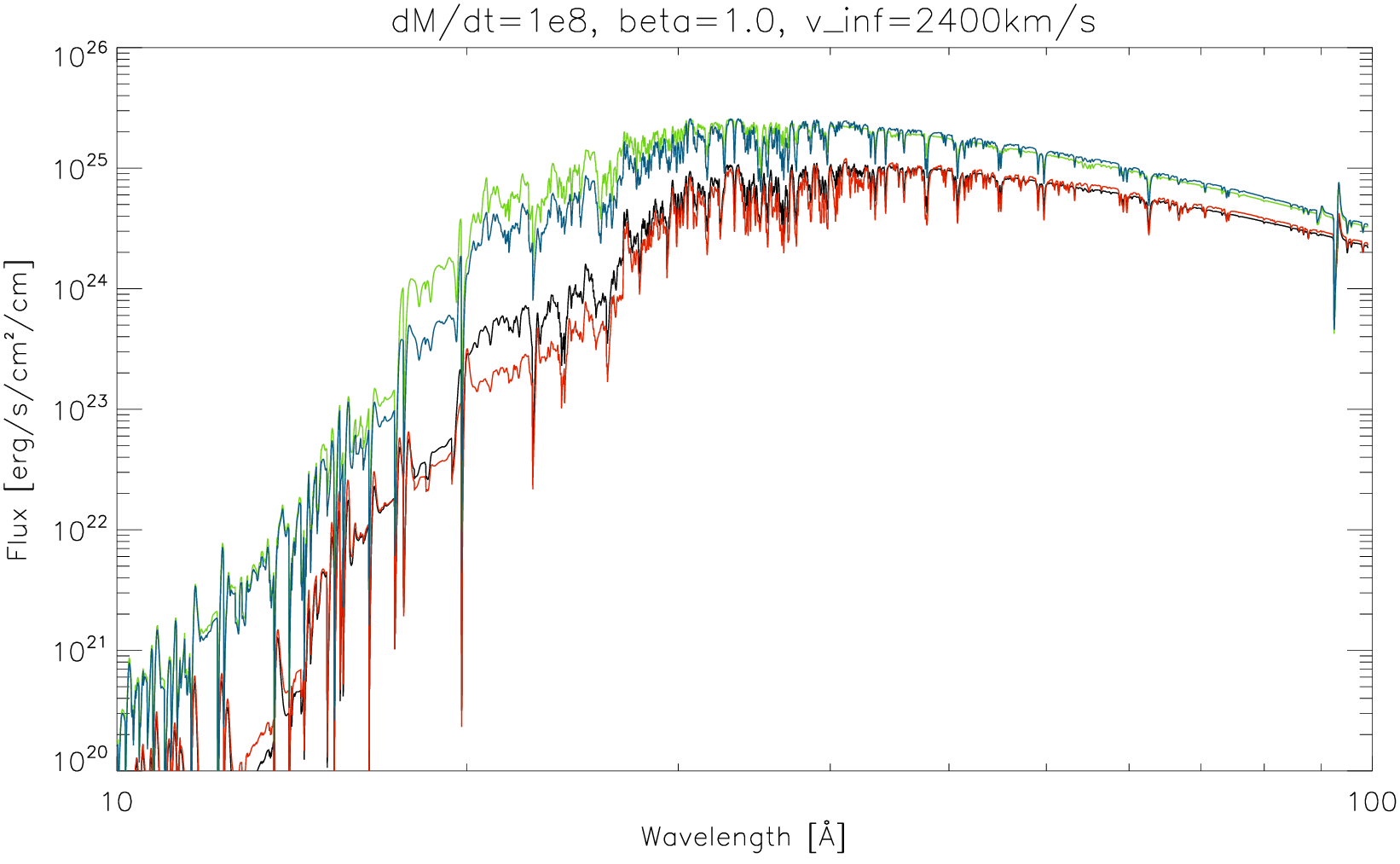}}
 \centerline{\includegraphics[height=\textwidth,angle=90]{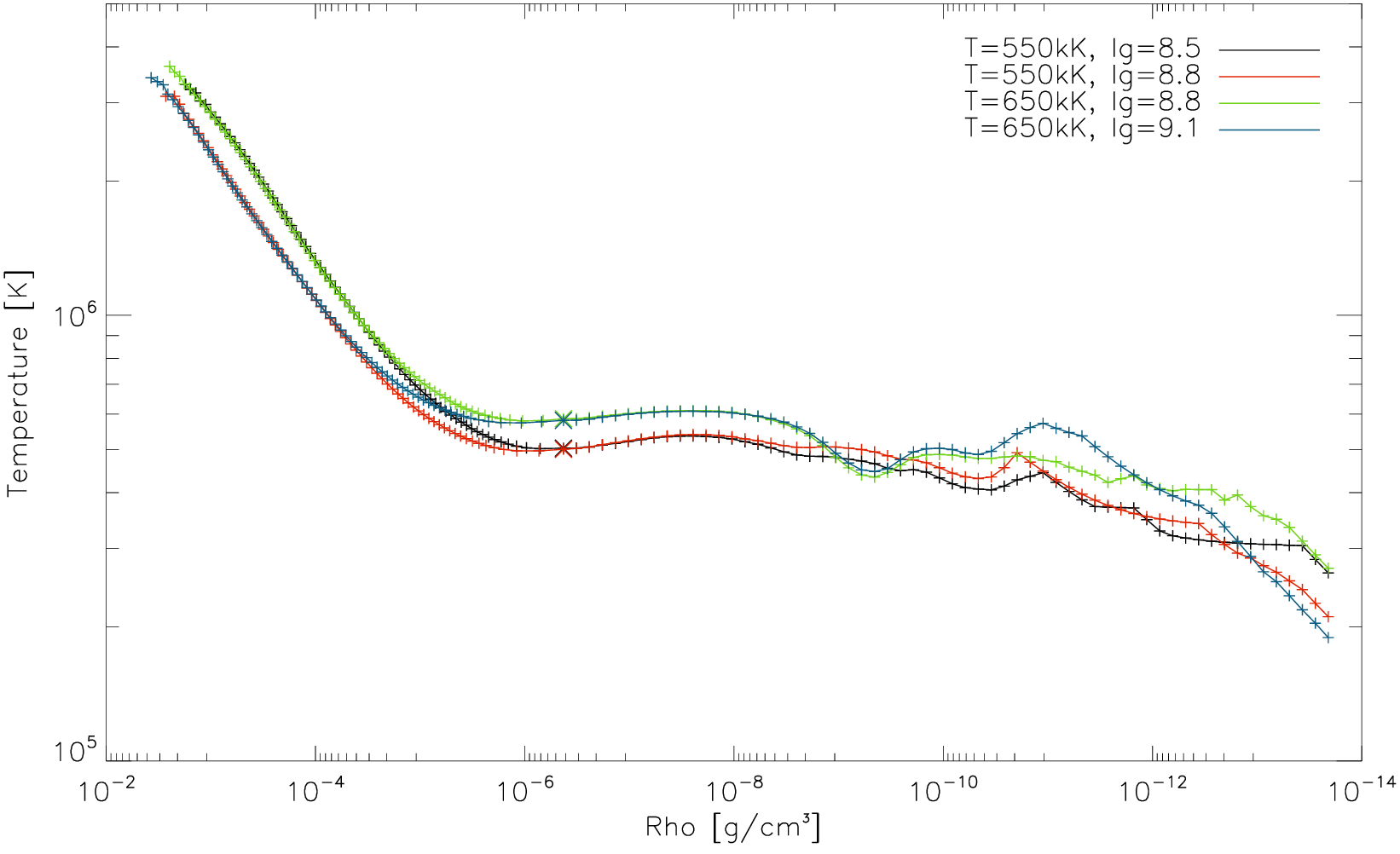}}
 \caption{ \label{fig:LoggGrid}
 The $\log(g)$ is varied for two fixed effective temperatures and velocity field.
 The temperature structures bundle in two in the tenuous part of the atmosphere, corresponding to the two effective temperatures, and in two other branches in the high density region, which correspond to the values of $\log(g)$.
 The same effect was observed for the hydrostatic models (figure \ref{fig:LoggGridPP}).
 The differences in spectra caused by $\log(g)$ are the largest between 14.2 and 25.3\AA{} (due to the O\,{\sc viii} and C\,{\sc vi} edges respectively), which also corresponds to the hydrostatic case.
 }
\end{figure}

\begin{figure}
 \centerline{\includegraphics[height=\textwidth,angle=90]{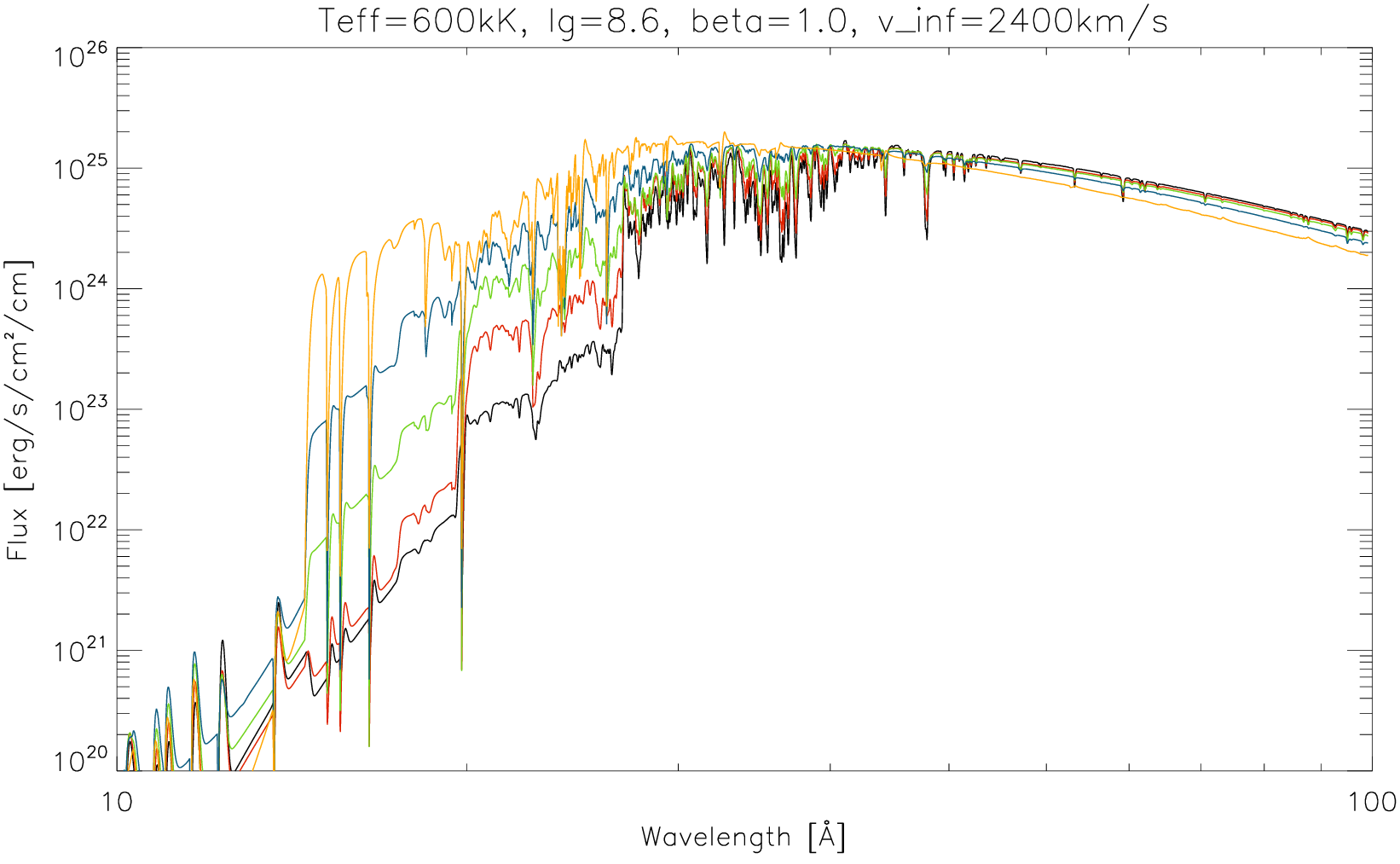}}
 \centerline{\includegraphics[height=\textwidth,angle=90]{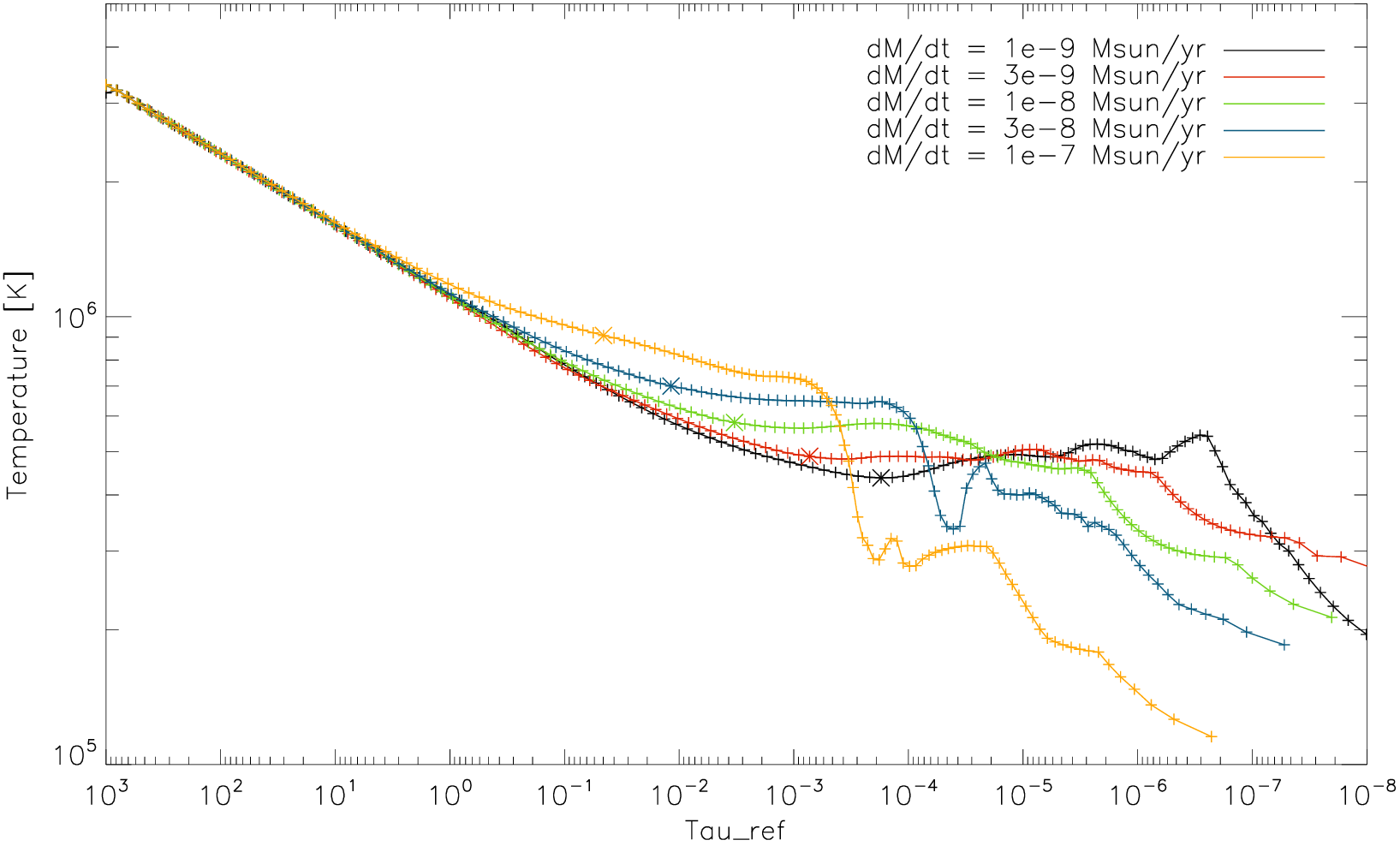}}
 \caption{ \label{fig:dMGrid}
 The mass loss rate $\dot{M}$ is varied for a fixed $v_\infty$ and $\beta$.
 This parameter obviously has a large influence on the spectrum, increasing the short wavelength flux.
 \newline
 A gradual development can be observed longwards of the 14.2\AA{} O\,{\sc viii} ionization edge, in the continuum and in the lines.
 Also, the temperature structures show a systematic behavior.
 With increasing mass loss rates the density of the wind and thus the optical depth at the bottom wind layer (marked with a large cross) increases.
 Outwards of  $\tau_{\rm ref}=10^{-5}$ no systematic temperature development can be expected (see section \ref{sec:ULNLTE}).
 The models are stable for the whole range of mass loss rates tested.
 }
\end{figure}

\begin{figure}
 \centerline{\includegraphics[height=\textwidth,angle=90]{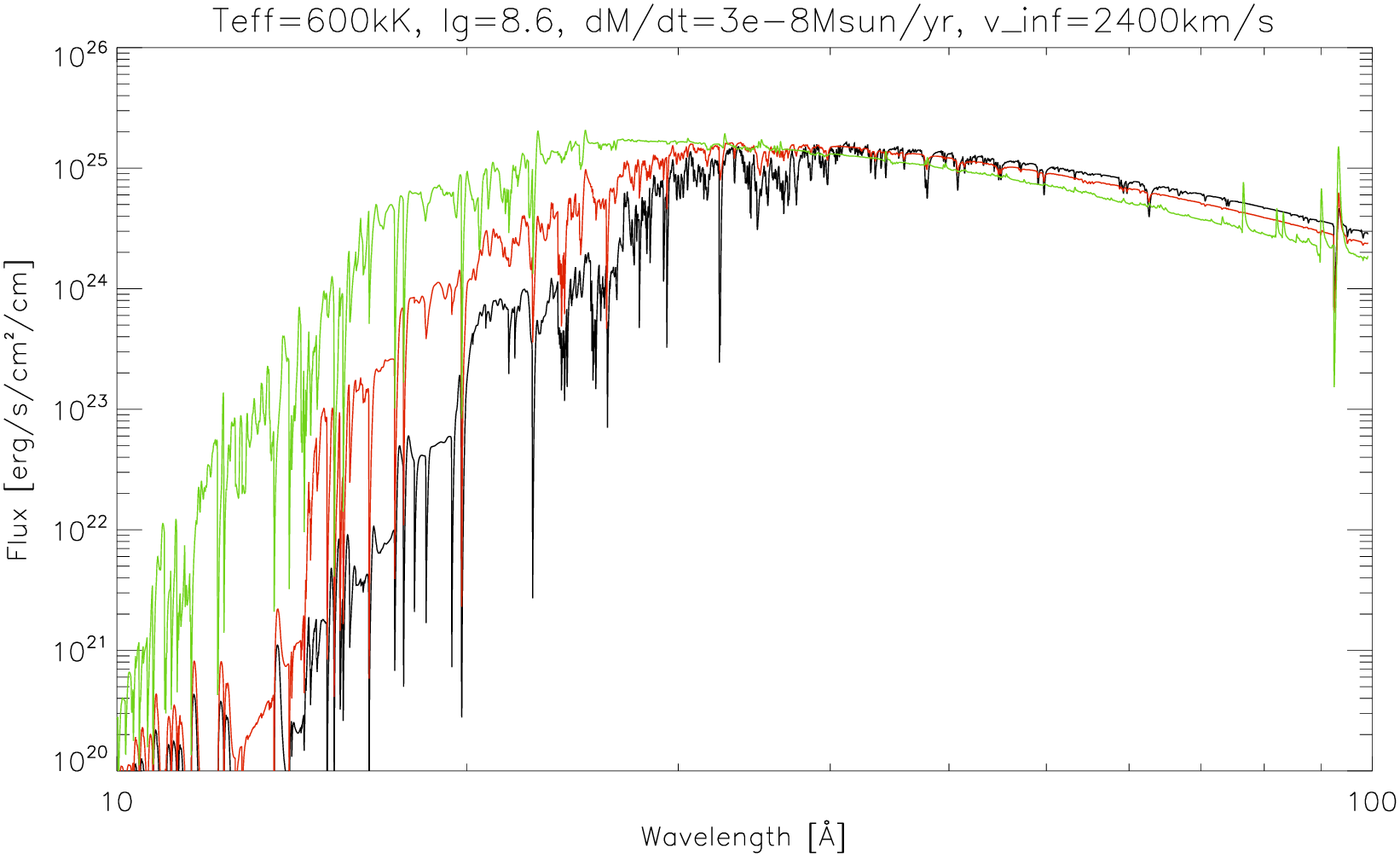}}
 \centerline{\includegraphics[height=\textwidth,angle=90]{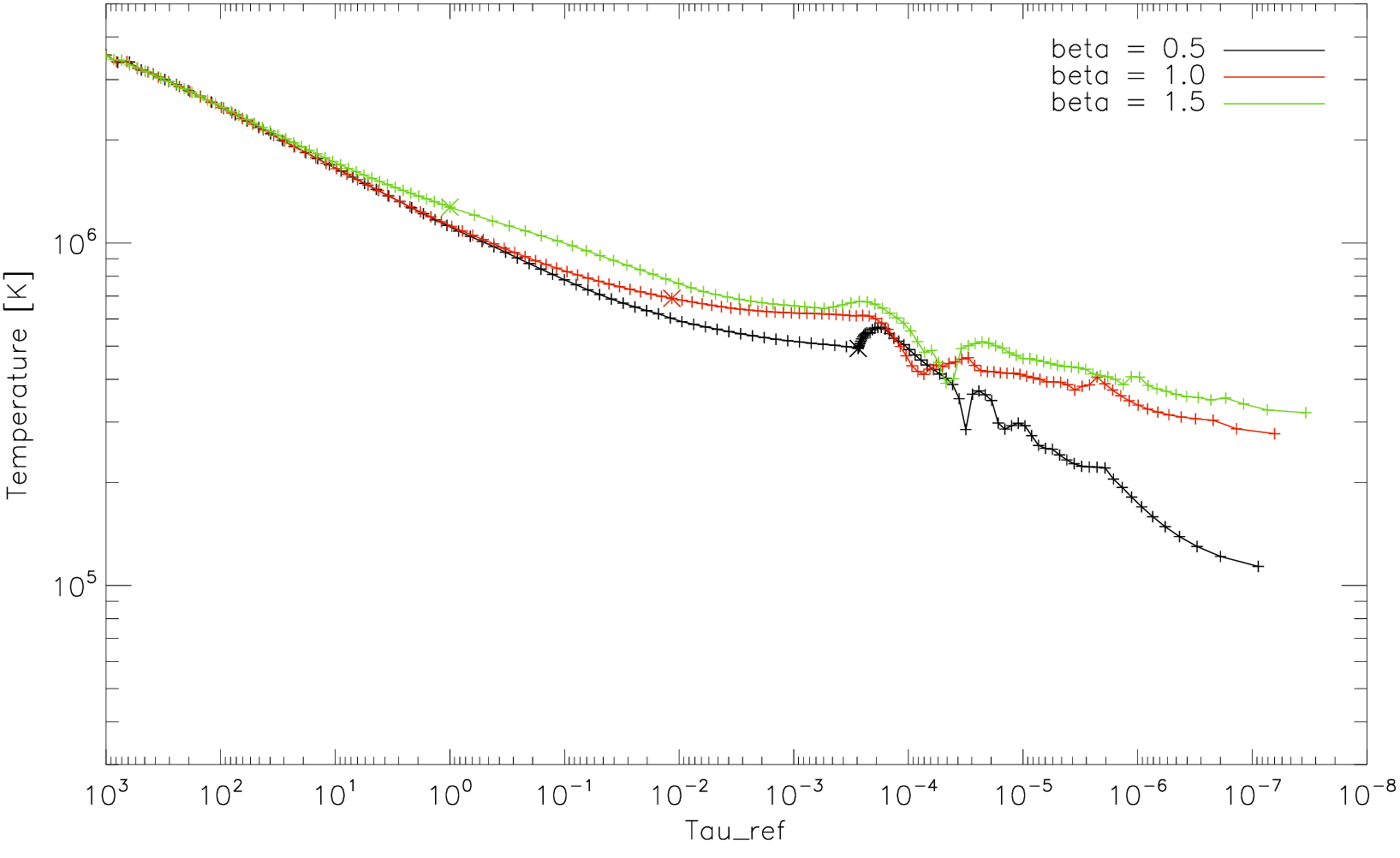}}
 \caption{ \label{fig:BetaGrid}
 The $\beta$ velocity field parameter is varied for a fixed mass loss rate and terminal velocity.
 This parameter has a large influence on the spectra.
 \newline
 With a larger $\beta$ the velocity increases more slowly outwards, and from the continuity equation \eqref{eq:Continuity} it follows that then also the density falls more slowly.
 Therefore, the optical thickness of the expanding shell is larger.
 This can be seen in the lower graph, where the optical depth of the bottom layer of the wind (marked with a large cross) is larger for the higher values of $\beta$.
 With the larger optical depth also the temperature at the transition layer increases through a stronger backwarming effect, and thus the spectrum appears hotter.
 }
\end{figure}

\begin{figure}
 \centerline{\includegraphics[height=\textwidth,angle=90]{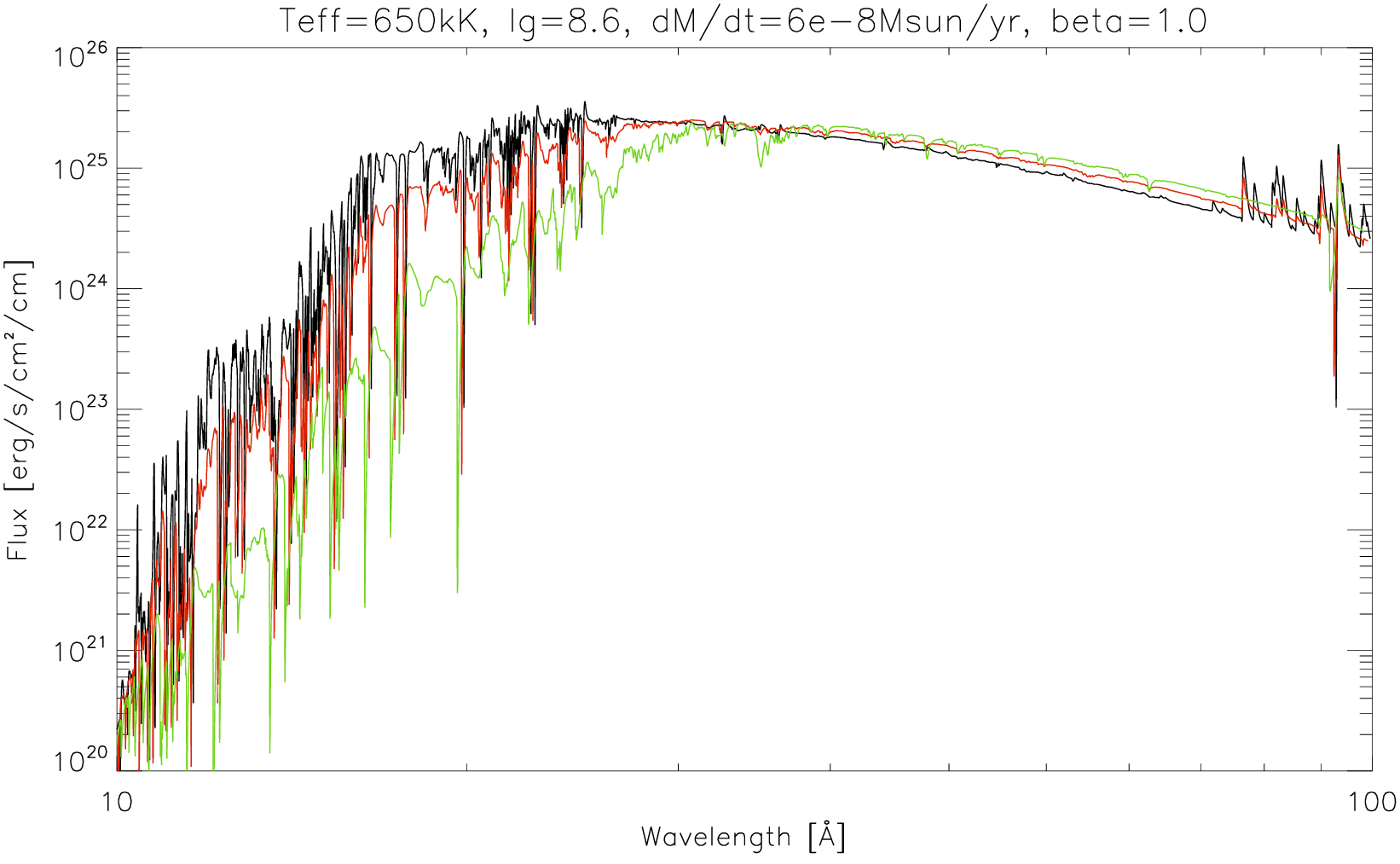}}
 \centerline{\includegraphics[height=\textwidth,angle=90]{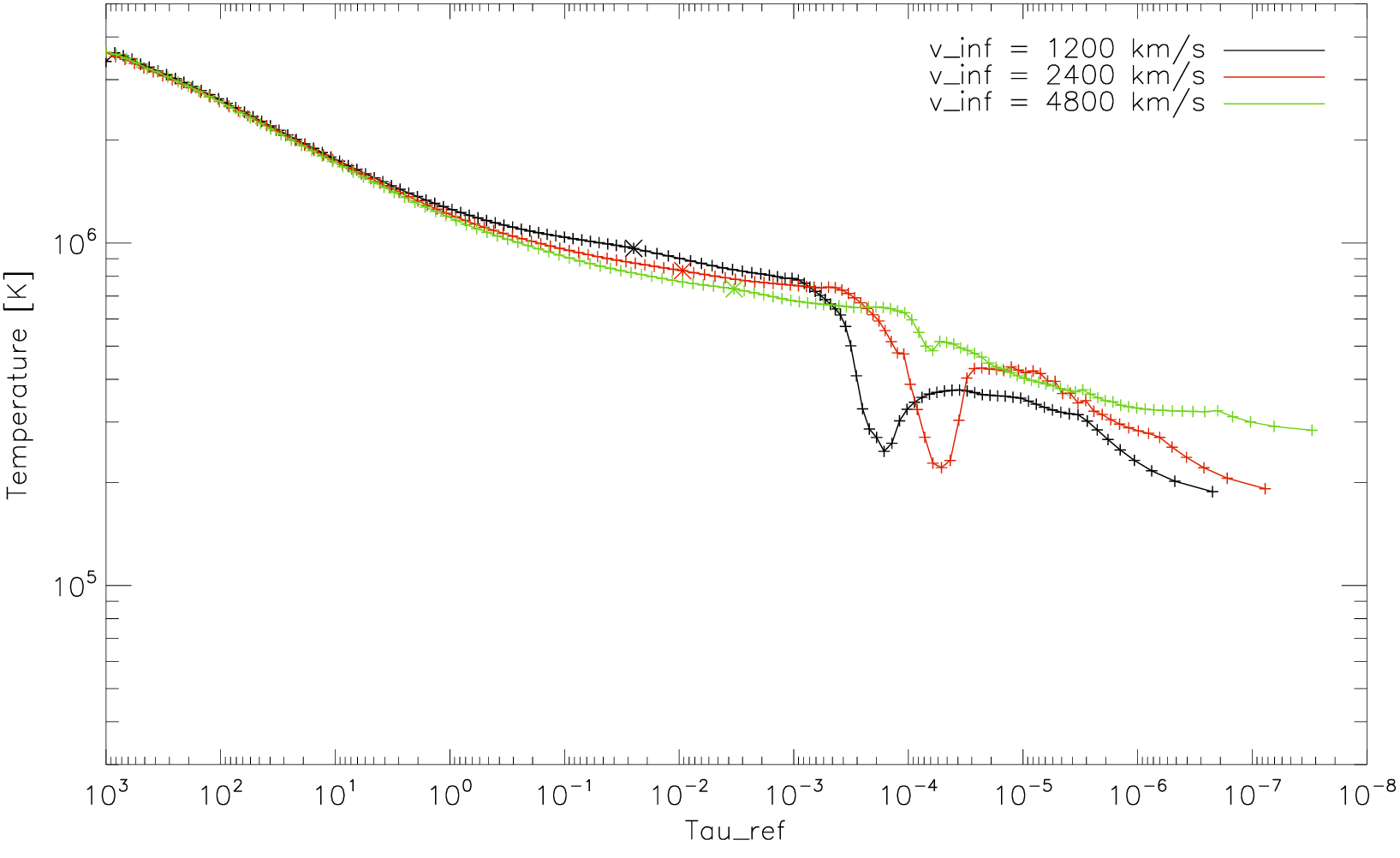}}
 \caption{ \label{fig:V0Grid}
 In this figure the asymptotic velocity $v_\infty$ is varied and the mass loss rate $\dot{M}$ and $\beta$ are kept fixed.
 This seems to have major impact on the spectra: higher velocities result in decreased flux in the short wavelength range, so the spectrum appears cooler.
 In the lower graph it can be seen, that indeed the temperature in the optically important region $10^{-3} < \tau_{\rm ref} < 10^{0}$ is lower for the higher wind velocity models.
 \newline
 This behavior can be understood in the same way as the plots in figure \ref{fig:BetaGrid}.
 An increase in the velocity for a fixed mass loss rate leads to an increase in the density according to the continuity equation \eqref{eq:Continuity}.
 The opacity scales (roughly) with the density.
 As a consequence of the increased opacity then the temperature rises (backwarming effect).
 \newline
 Compare also figure \ref{fig:dMv0Grid}, where $\dot{M}$ is scaled linearly with $v_\infty$, so that the density structures are constant.
 }
\end{figure}

\begin{figure}
 \centerline{\includegraphics[height=\textwidth,angle=90]{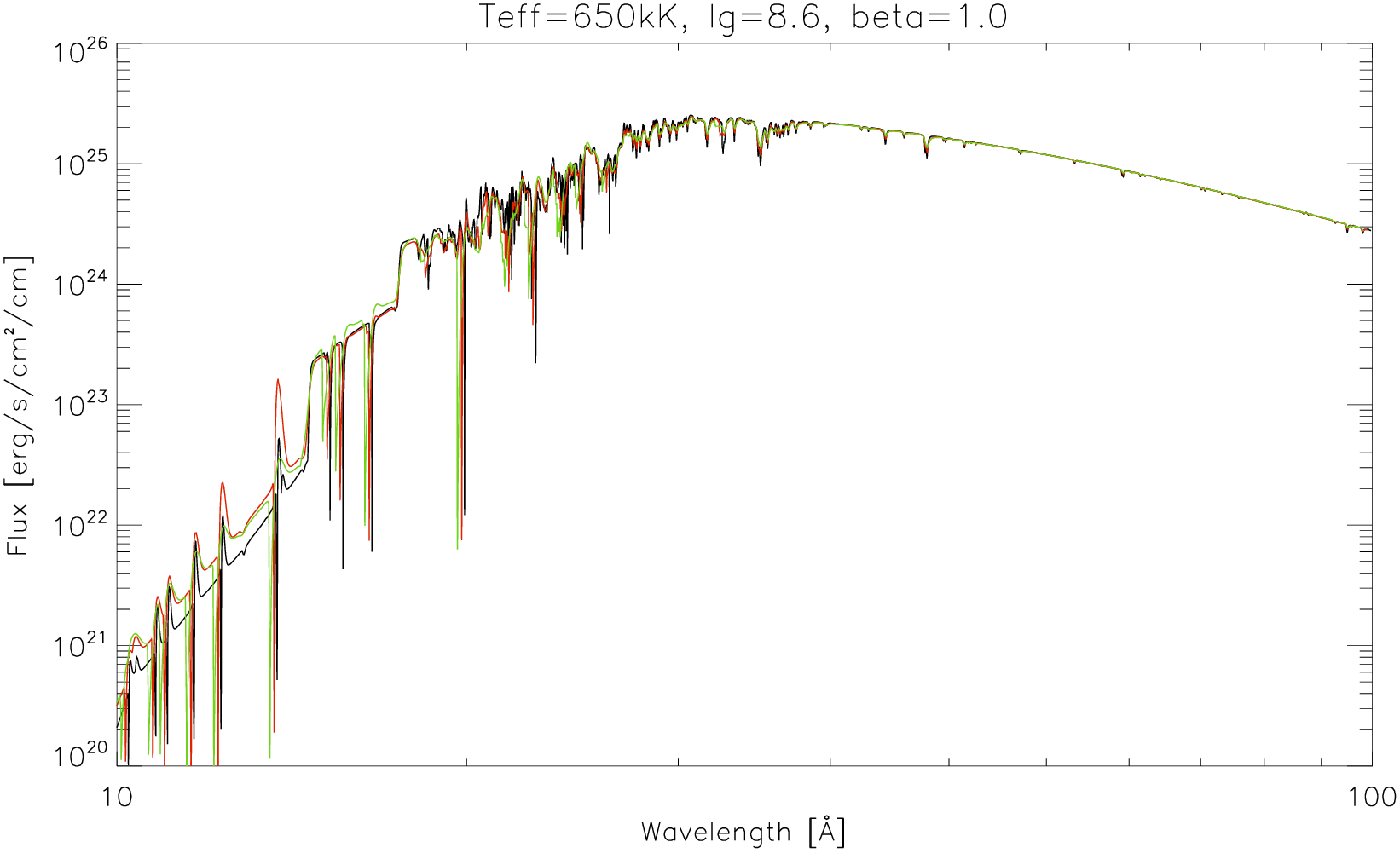}}
 \centerline{\includegraphics[height=\textwidth,angle=90]{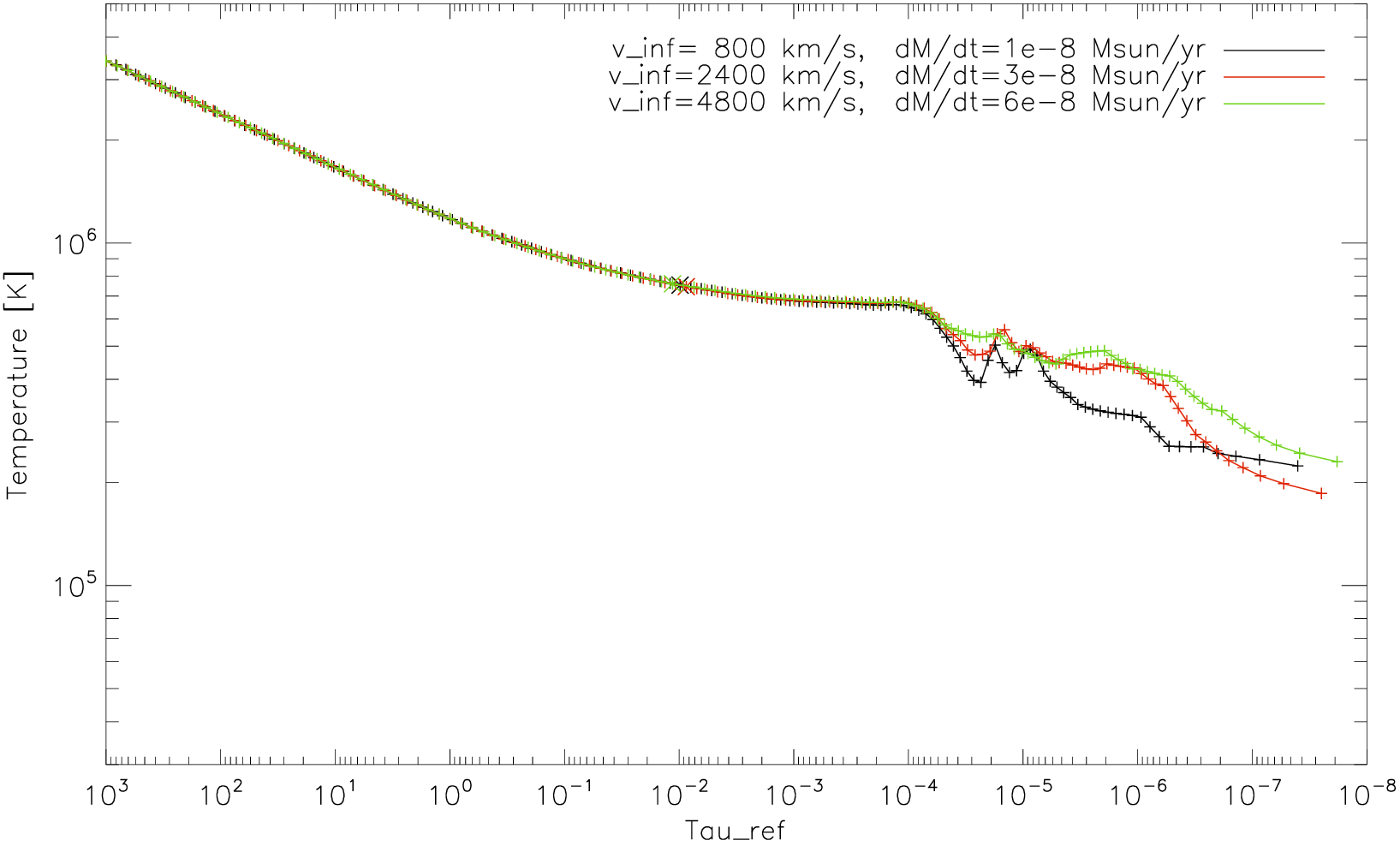}}
 \caption{ \label{fig:dMv0Grid}
 In this figure the terminal velocity $v_\infty$ is changed, but at the same time the mass loss rate $\dot{M}$ is scaled by the same amount, so that the resulting density structures for the expanding envelopes are identical (see equation \eqref{eq:Continuity}).
 \newline
 Compared with the differences due to only $v_\infty$ variation (see figure \ref{fig:V0Grid}) the differences between the models in this figure are very small.
 The spectra are very similar, see figure \ref{fig:dMv0GridZoom} for a zoom in on a smaller part of the spectrum.
 Also, the temperature structures are very similar, except for the outer region that is optically very thin with $\tau_{\rm ref} < 3 \cdot 10^{-5}$.
 In those regions the wavelength averaged opacities become dominated by the contributions of single strong atomic lines, and therefore also the radiative equilibrium and the temperature are.
 This effect is called \emph{line cooling/heating}, see section \ref{sec:LineCooling}.
 The velocity field does influence the line cooling/heating effect, and therefore also the temperature structure.
 }
\end{figure}

The results of these grids are shown in the figures \ref{fig:TeffGridPP} and \ref{fig:LoggGridPP} for the hydrostatic models and \ref{fig:TeffGrid} to \ref{fig:dMv0Grid} for the hybrid models.
The figures show the effects of the following parameters to the model spectra and temperature structures.
\begin{enumerate}
\item figure \ref{fig:TeffGridPP}: hydrostatic -- effective temperature $T_{\rm eff}$
\item figure \ref{fig:LoggGridPP}: hydrostatic -- gravitation constant $\log(g)$
\item figure \ref{fig:TeffGrid}: hybrid -- effective temperature $T_{\rm eff}$
\item figure \ref{fig:LoggGrid}: hybrid -- gravitation constant $\log(g)$
\item figure \ref{fig:dMGrid}: hybrid -- mass loss rate $\dot{M}$
\item figure \ref{fig:BetaGrid}: hybrid -- beta velocity parameter $\beta$
\item figure \ref{fig:V0Grid}: hybrid -- terminal velocity $v_\infty$
\item figure \ref{fig:dMv0Grid}: hybrid -- $\dot{M}$ together with $v_\infty$
\end{enumerate}
\clearpage

\clearpage

\begin{figure}
 \centerline{\includegraphics[height=\textwidth,angle=90]{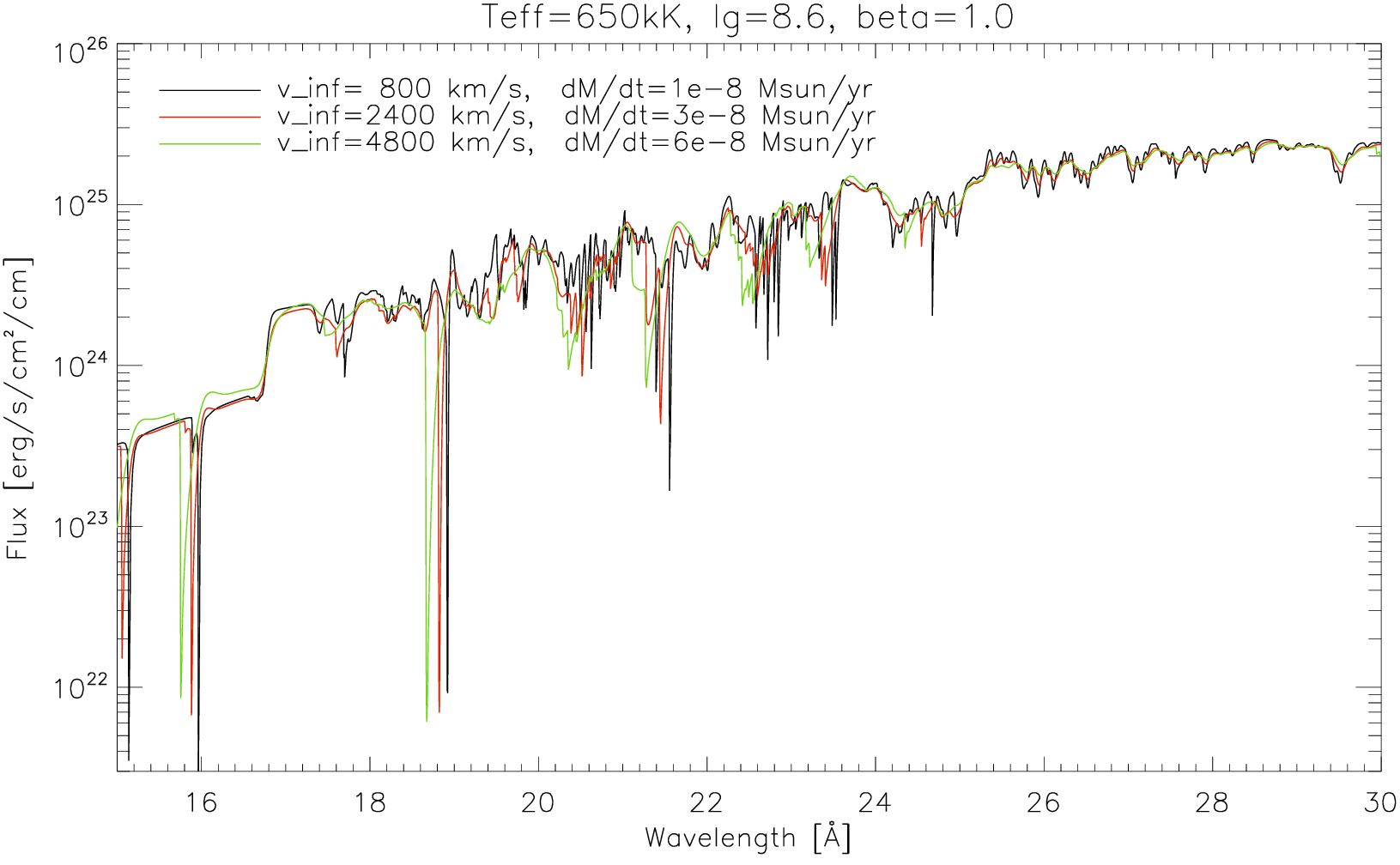}}
 \caption{ \label{fig:dMv0GridZoom}
 This figures shows the same spectra as in figure \ref{fig:dMv0Grid}, but now zoomed in to a smaller wavelength range and on a linear wavelength scale.
 \newline
 In the spectra the continuum fluxes are almost the same.
 There is a stronger blue shift in the absorption lines for the models with higher velocities.
 Also, the details (narrow lines) seen in the low velocity spectrum (black curve) are smeared out with higher velocities.
 Therefore, the spectrum for the highest velocity in this example (green curve) is very smooth.
 }
\end{figure}

\section{LTE vs. NLTE} \label{sec:LTEvsNLTE}
The major part of the theory and methods described in sections \ref{sec:RadiationTransport} and \ref{sec:GoodNLTE} dealt with the problem of computing NLTE models.
The extra complications in atmosphere modeling (in both methodology \emph{and} computation time), introduced by relaxing the assumption of LTE, are yet to be justified.

Whether LTE is a useful approximation for a specific atmosphere or not can \emph{only} be ascertained from the direct comparison with a NLTE model.
If the deviations from NLTE are small, then LTE is a good approximation, otherwise it is not.
This direct comparison is done here for the expanding nova type and hydrostatic type models.
Since the nova type models are based upon a hydrostatic core, a pure hydrostatic atmosphere is recovered in the limit for tenuous winds, i.e. very small mass loss rates.

\subsection{Expanding models}
\begin{figure}
 \centerline{\includegraphics[height=\textwidth,angle=90]{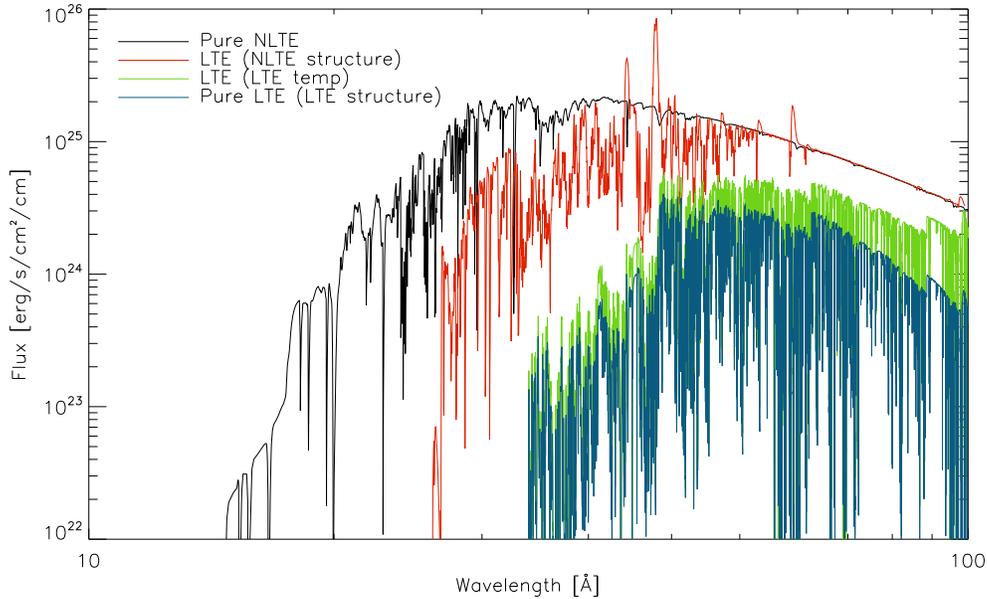}}
 \caption{ \label{fig:LTEvsNLTE}
  Here the spectra are shown for a typical nova-type model for four different situations.
  The models were computed in descending order (as shown in the legend), starting from the `Pure NLTE' model (black), each of them using the previous model result as initial condition.
  The first two spectra (black and red) are computed from the NLTE atmospheric structure.
  For these two all atmospheric conditions were identical, except for the population numbers, being in statistical equilibrium (black) or in LTE (red).
  In the green model the temperature was released to adapt to the LTE opacities, but the density structure was kept fixed.
  The blue model is the pure LTE case, where the whole structure is released to adapt to the LTE opacities, i.e. the temperature, density and all derived quantities (like partial pressures).
  \newline
  For the radially extended nova atmospheres treated in this work obviously LTE is a poor approximation.
 }
\end{figure}

In figure \ref{fig:LTEvsNLTE} a LTE spectrum is shown in comparison with a NLTE spectrum for a typical radially extended nova atmosphere of the type described in chapter \ref{sec:NovaStructure}.
If a LTE spectrum is computed from the NLTE atmospheric structure, so that the atmospheric conditions for the models are identical except for the population numbers, then the result is already very different from the pure NLTE spectrum.
If only the temperature structure is released to adapt to the LTE opacities, then the resulting spectrum becomes already more different.
%
%
%
\begin{figure}
 \centerline{\includegraphics[height=\textwidth,angle=90]{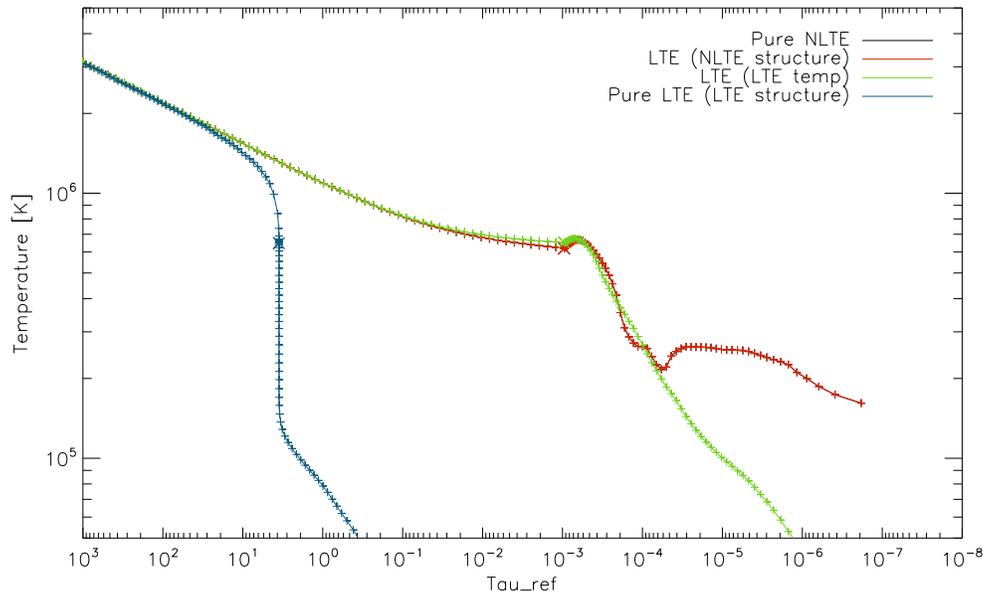}}
 \caption{ \label{fig:LTEvsNLTEStruct}
 These are the temperature structures for the different NLTE and the LTE spectra compared in figure \ref{fig:LTEvsNLTE}.
 The NLTE temperature structure for the black and red spectrum \emph{coincide}.
 With the LTE structure (blue curve) the outer parts of the atmosphere are already optically thick at very low densities, because the low ionization stages these temperatures and LTE yield very effective X-ray absorbers.
 }
\end{figure}
If the atmospheric structure is released to adapt to the LTE opacities instead of using the NLTE structure, then the resulting spectra become even more different, see figure \ref{fig:LTEvsNLTE}.
The temperature structures for these two cases are plotted in figure \ref{fig:LTEvsNLTEStruct}.
Obviously, the NLTE treatment is essential for this type of models.

\clearpage
\subsection{Hydrostatic models}
\begin{figure}
 \centerline{\includegraphics[height=\textwidth,angle=90]{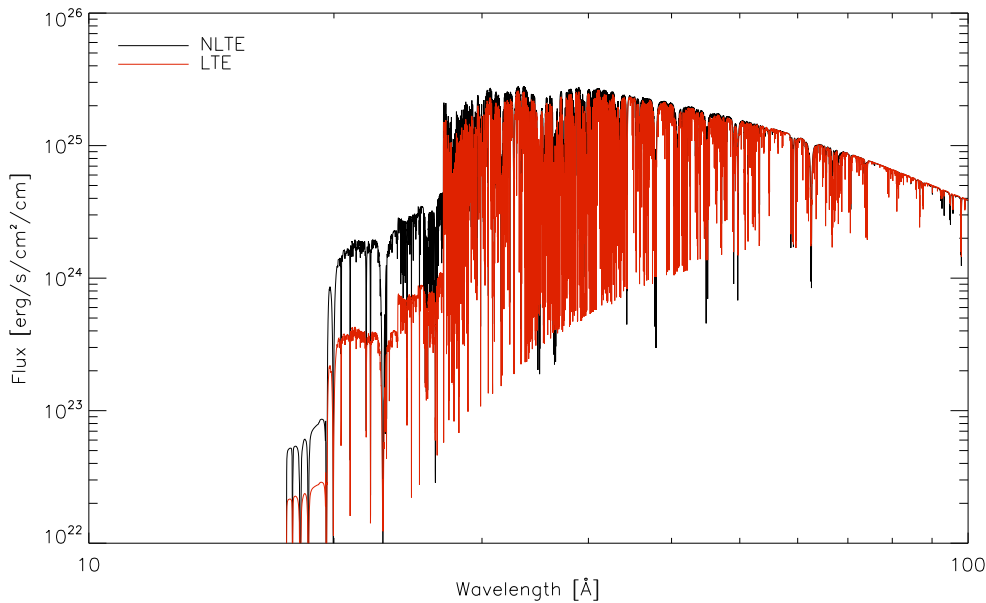}}
 \caption{ \label{fig:LTEvsNLTE_PP}
 These spectra are for a hydrostatic atmosphere.
 The structures for these spectra are identical except for the population numbers.
 The most obvious difference is the strength of the ionization edges, a result that was also found for the nova-type atmospheres in figure \ref{fig:LTEvsNLTE}.
 This is mainly caused by the ionization balance shown in figure \ref{fig:LTEvsNLTEIonBal_PP}.
 }
\end{figure}

\begin{figure}
 \centerline{\includegraphics[height=\textwidth,angle=90]{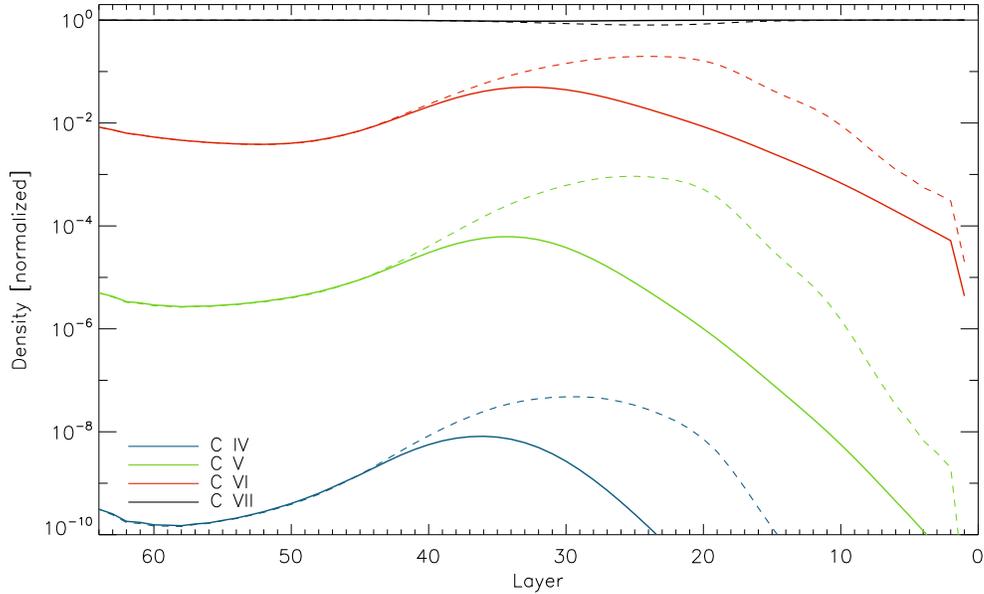}}
 \caption{ \label{fig:LTEvsNLTEIonBal_PP}
 This plot shows the ionization balances for the model of figure \ref{fig:LTEvsNLTE_PP} for LTE (dashed curves) and NLTE (solid curves).
 NLTE effects underpopulate the C\,{\sc vi} ionization stage.
 }
\end{figure}

\begin{figure}
 \centerline{\includegraphics[height=\textwidth,angle=90]{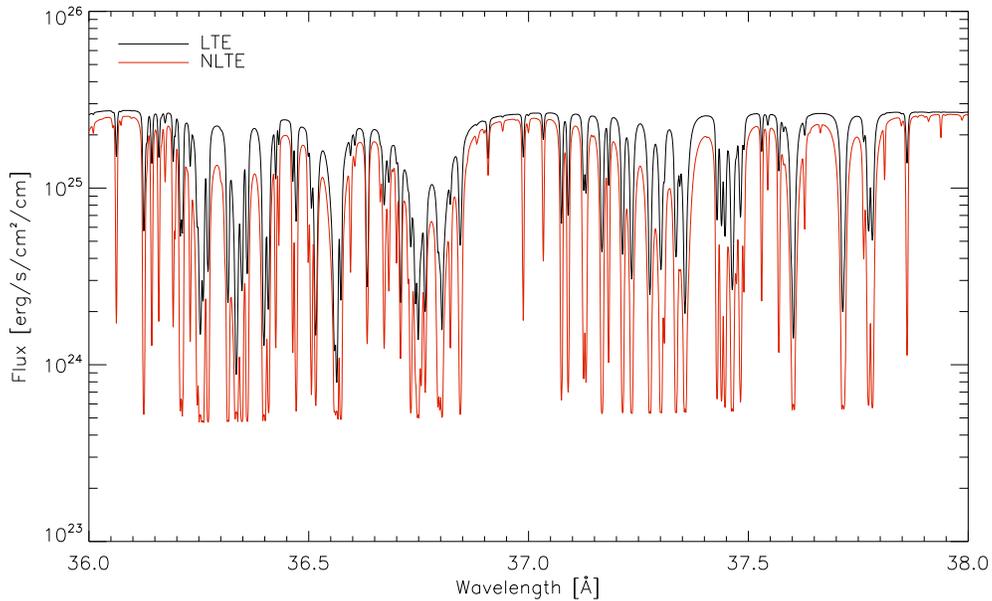}}
 \caption{ \label{fig:LTEvsNLTEzoom_PP}
 A zoom into a smaller wavelength range is plotted for the spectra of figure \ref{fig:LTEvsNLTE}.
 This displays that differences between LTE and NLTE for compact hydrostatic atmospheres are not limited to strong ionization edges but are also apparent in the line features.
 Note that the LTE spectrum is computed from the NLTE structure.
 From pure LTE much larger differences can be expected (see also figure \ref{fig:LTEvsNLTE}, where this was shown for nova-type models).
 }
\end{figure}

Whereas the departures from LTE are large for the hybrid nova-type models, this is not necessarily the case for other types of models, like the hydrostatic ones examined here.
These hydrostatic atmospheres are very compact in contrast to the radial extension of nova-type models.
In figure \ref{fig:LTEvsNLTE_PP} a LTE spectrum is shown in comparison with a NLTE spectrum.
The differences between LTE and NLTE are much smaller than for the nova-type atmosphere in figure \ref{fig:LTEvsNLTE}.
However, the discrepancies are still large enough to disqualify LTE as a good approximation, the more since the underlying structure was computed in full NLTE.
So even if a good structure is available, then LTE is still a poor assumption.
The most striking difference in the spectra is the strength of the C\,{\sc vi} ionization edge at 25.3\AA{}.
The ionization balances for LTE and NLTE for this hydrostatic model are shown in figure \ref{fig:LTEvsNLTEIonBal_PP}.
Clearly, C\,{\sc vi} is suppressed by NLTE effects in the outer regions of the atmosphere.
But differences are found not only in the ionization edges but also the detailed line features.
This becomes visible when zooming in to a smaller wavelength range, which is shown in figure \ref{fig:LTEvsNLTEzoom_PP}.

\clearpage
\section{Transformation of counts to flux spectra}
Where in the previous sections it was shown that systematic results are obtained from the new framework, the question if these models give a realistic description of nova atmospheres is yet open.
In order to compare theoretical spectra with X-ray observations the interstellar extinction and instrumental characteristics must be considered.

\subsection{Interstellar extinction} \label{sec:ISModel}
Accurate estimates of interstellar (IS) extinction are necessary for the analysis and interpretation of almost all astronomical soft X-ray observations.
The IS extinction in the X-ray wavelength range is far from gray.
The soft X-rays are more strongly absorbed than the hard X-rays.
Therefore, the IS extinction does not only influence the estimated bolometric luminosity (from a given radius and distance) of the object, but also the hardness ratio, i.e. the ratio of hard and soft photon counts.

In this work, an IS model is used that is based on the absorption cross sections compiled by \cite{Balucinska92} for 17 astrophysically important species.
Specifically, the effective extinction curve is calculated with Balucinska-Church \& McCammon's {\tt FORTRAN} code\footnote
{Available at ftp://adc.astro.umd.edu/pub/adc/archives/catalogs/6/6062A/
}, assuming solar abundances.
The total extinction is then proportional to the hydrogen column density.
The model includes (continuum) absorption from bound-free transitions only.
Although this code has not been updated since 1999, it is still the most widely used in literature.

\begin{figure}
 \centerline{\includegraphics[width=\textwidth,bb=0 0 498 372]{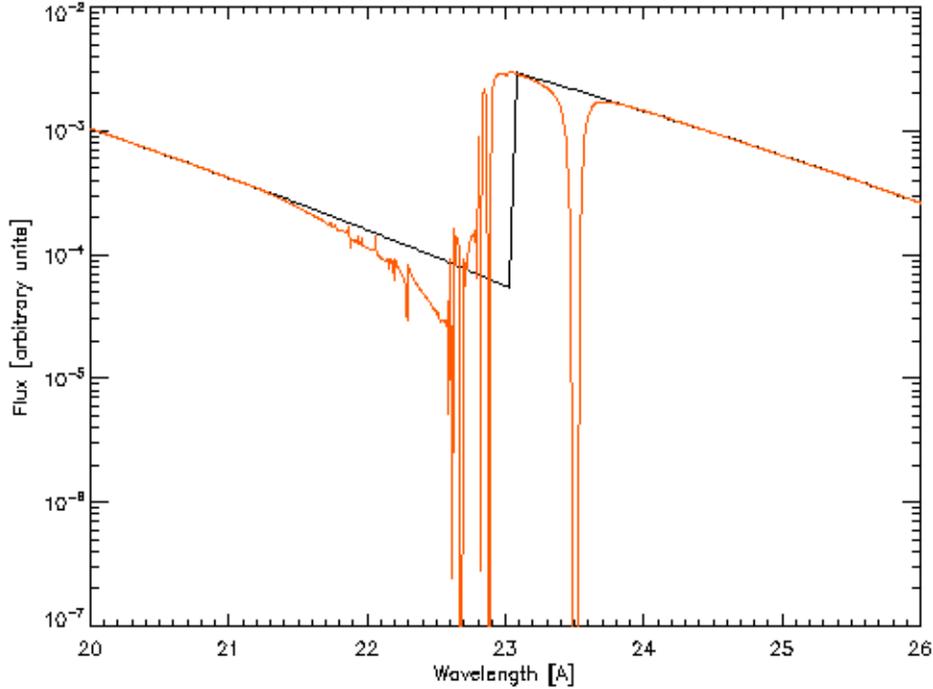}}
 \caption{ \label{fig:Tbabs}
 This plot compares the O\,{\sc i} ionization edge from a new version (red) of the interstellar absorption package {\tt tbabs} with an oder version (black).
 This code is newer than commonly used code of \cite{Balucinska92}, that also used in this work.
 Detailed comparison and possibly inclusion of this newer package into the code used for this work is planned for future.
 From: http://astro.uni-tuebingen.de/$\sim$wilms/research/tbabs/
 }
\end{figure}

Also, a newer code exists, called {\tt tbabs} \cite{Wilms00}, that includes more fancy absorption features, like absorption lines from neutral elements.
An example given by the authors that compares the absorption coefficients from an old and a new version of {\tt tbabs} is shown in figure \ref{fig:Tbabs}.
Inclusion of these features is important when a chi-square method is used to determine the ``goodness'' of a fit.
But the chi-square method poses some fundamental problems when used for IS absorbed spectra (see section \ref{sec:FitProcedure}), so that it is not used in this work and the fits are done ``by eye''.
Apart from improving the fit goodness value $\chi^2$, the line absorption features provide a better basis for tuning the assumed interstellar chemical composition set from the comparison of observations with synthetic spectra.
{\tt tbabs} has not yet been used in this work.
It is designed for XSPEC, a large spectral analysis package, and is not out-of-the-box suitable include in this work's code.
Private communication with the authors and a detailed comparison of the IS absorption cross sections and its influences on the model fits is planned for the future.

Whereas the slope of the `Wien' tail of the spectrum 
The IS extinction inclines so strongly to longer wavelengths, that the slope

\subsection{Instrumental characteristics}
The `high-res' X-ray observations used in this work come from the X-ray grating satellites Chandra and XMM-Newton, with the LETGS and RGS spectographs respectively.
After the data reduction process, that deals with most of the instrumental characteristics, one obtains the number of photons that were detected per energy bin in the total exposure time.
This reduction process was done by \cite{JanUwePrivat} for all observations.
The reduced data still contain higher order contributions \emph{or} detector gaps that must be dealt with (see below).

Given the effective area $\sigma$ of each bin $i$ with bin width $w$ and the total exposure time $\tau$, the photon count $n$ can be converted to the observed (absorbed) flux $F$ in the center wavelength $\l$ of the bin according to
\begin{equation} \label{eq:CountsToFlux}
 F^{\rm abs}_{\l_i} = \frac{ hc}{\l_i} \, \frac{ n_i - n^{\rm bg}_i }{ \sigma_{\l_i} \tau w_{\l_i}}
\end{equation} 
with $n^{\rm bg}$ being the background counts.

With the transmission coefficient $T$, being a function of the hydrogen column mass and the chemical composition $x_j$, the unabsorbed flux becomes
\begin{equation}
 F_\l = \frac{F^{\rm abs}_\l }{ T_\l(N_{\rm H},x_j) }
\end{equation}
The inverse transformation, from unabsorbed flux to counts, thus becomes
\begin{equation} \label{eq:FluxToCounts}
 n_i = F_{\l_i} T_{\l_i} \sigma_{\l_i} \tau w_{\l_i} \frac{\l}{hc} + n^{\rm bg}_i
\end{equation}
The resolution of the synthetic spectra is significantly higher than that of the observed spectra.
Therefore, the conversion process is done on the observed wavelength grid, and the synthetic flux points must be spread over that grid.
For the comparison with observations, synthetic spectra must be transformed from the Lagrangian to the Eulerian frame, and for the Lorentz transformation no wavelength parallelization can be done.
In order to save unnecessary computation time, the synthetic spectra are computed with not many more wavelength points than necessary, i.e. a spectral resolution\footnote
{Note that the synthetic resolution can vary with wavelength.
} of roughly 3 times the observed resolution.
This means, that in the spreading of the synthetic flux over the observed wavelength grid, not only the flux points that fall within the detector bin must be treated, but also the proportional parts of the next closest flux points on either side of the bin, according to their distance to the bin boundary.

\subsubsection{Chandra specific}
\begin{figure}
 \centerline{\includegraphics[height=\textwidth,angle=-90]{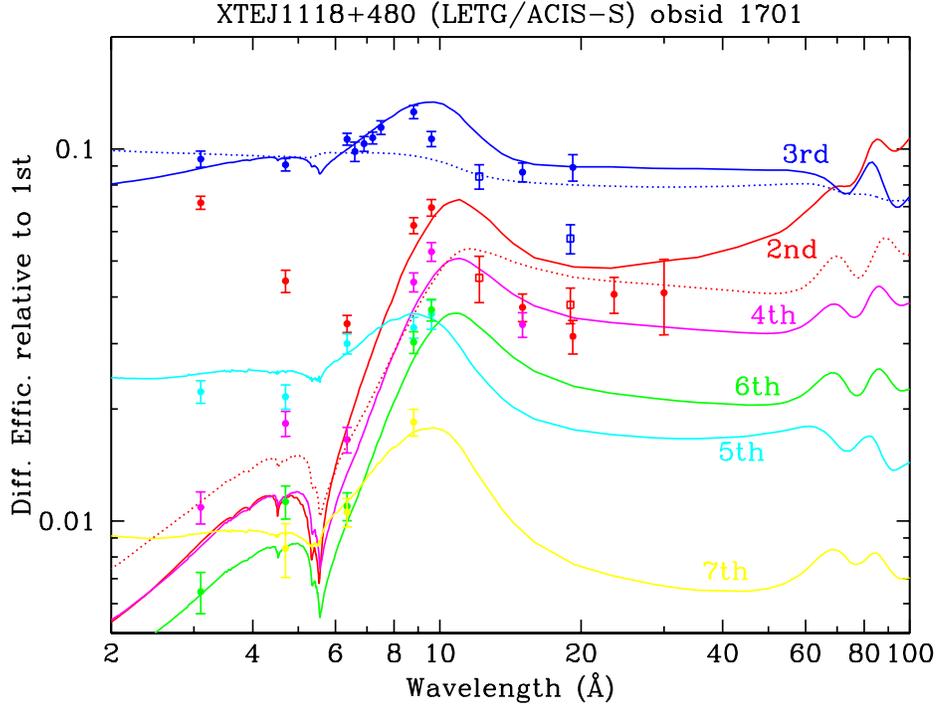}}
 \caption{ \label{fig:ChandraEff}
 This plot shows Chandra's higher order diffraction efficiencies relative to the 1st order efficiency.
 The dotted curves show the pre-launch model, the solid curves the current model predictions.
 From the Chandra X-ray Center homepage.
 }
\end{figure}

However, the Chandra LETGS spectra contain the overlapping contributions of the first \emph{and} higher diffraction orders.
The relative strengths of the different orders are plotted in figure \ref{fig:ChandraEff}.
The photons in the higher orders are measured in the spectrograph at the double, triple, etc. of the actual wavelength.
Writing the photon counts in bin $i$ as the sum of the contributions of the first three orders
\begin{equation}
 n_i - n^{\rm bg}_i = n^1_i + n^2_i + n^3_i
\end{equation}
and using equation \eqref{eq:FluxToCounts} one obtains
\begin{multline}
 F_{\l_i} = \frac{hc}{\l_i} \frac{1}{T_{\l_i} \sigma_{\l_i} w_{\l_i}}
 \Bigg( n_i
  - F_{\l_i/2} T_{\l_i/2} \sigma^{2{\rm nd}}_{\l_i} w_{\l_i/2} \frac{\l/2}{hc} \\
 - F_{\l_i/3} T_{\l_i/3} \sigma^{3{\rm rd}}_{\l_i} w_{\l_i/3} \frac{\l/3}{hc}
 \Bigg)
\end{multline}
This equation involves simultaneous values of $F$ for different wavelengths.
But due to the systematic wavelength shift in the orders, the equation can be solved recursively, starting at the small wavelength boundary.

Disentangling the higher orders is very helpful in determining the proper IS extinction, parameterized by $N_{\rm H}$, as the influence of $N_{\rm H}$ is the largest for the long wavelengths in which also the contributions of higher orders are most significant.
It is stressed at this point, that the fits are very sensitive to $N_{\rm H}$, so that a small change in $N_{\rm H}$ results in a large change in other fit parameters, like $T_{\rm eff}$.

\begin{landscape}
\begin{figure}
 \centerline{\includegraphics[width=\textwidth,angle=90]{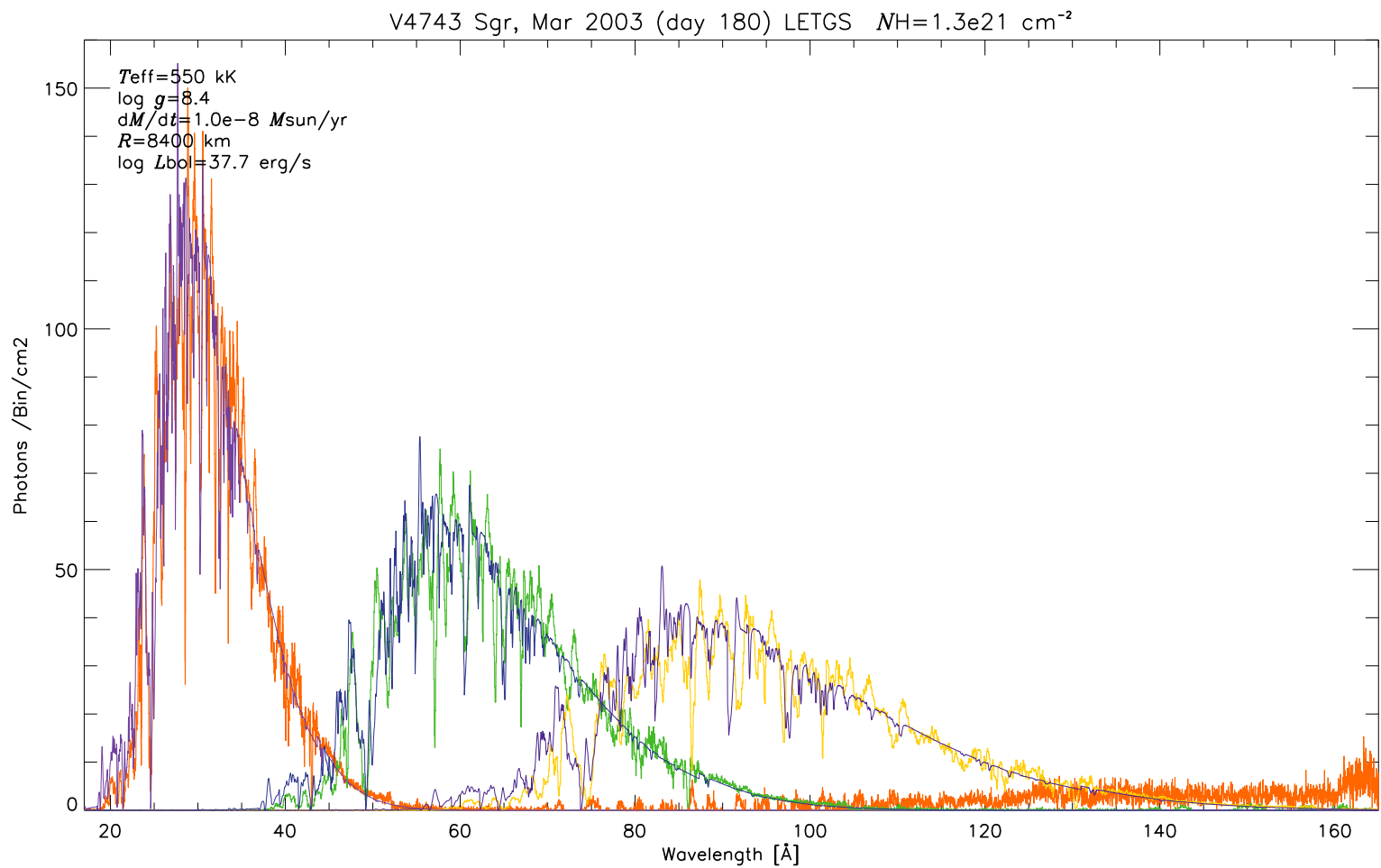}}
 \caption{ \label{fig:HigherOrder}
 The first 3 orders extracted from the spectrum of figure \ref{fig:V4743Mar}. $\sigma_\l$, $\sigma^{\rm 2nd}_\l$ and $\sigma^{\rm 3rd}_\l$ are normalized, thus the relative strengths are unrealistic.
 }
\end{figure}
\end{landscape}

An example of the result from extracting the first, second and third order is shown in figure \ref{fig:HigherOrder}.
The missing treatment of higher becomes apparent in the rise of first order flux for long wavelengths $\l > 100$\AA{}.

\subsubsection{XMM-Newton specific}
The RGS spectra do not feature higher orders, but they have detector gaps (due to hot pixels and defective CCD chips), i.e. wavelength ranges in which the effective area is 0.
These appear as strong spikes in the observed data.
With $\sigma=0$ the conversions between the representations cannot be applied.
When the observed counts from the two independent detector channels, RGS1 and RGS2, are combined, then the number of gaps is strongly reduced and the quality of the spectra improves significantly.
In order to cope with these gaps, both the effective area's and the observed counts are interpolated linearly.
Apart from the gaps, still jumps in the effective area as large as a factor of 2 occur.

In order to ease the interpretation the linear counts/bin spectra, these are converted to a constant effective area of $\sigma=1$, thus removing the pseudo absorption features in the observational data.
That means that the counts are plotted as they would have been seen with an instrument that is equally sensitive over the whole wavelength range.

\subsection{The fitting procedure} \label{sec:FitProcedure}
In order to compare the theoretical and observational data, the data must be transformed into one of both representations: unabsorbed flux or IS absorbed counts/bin.
In literature, the counts/bin representation is the most common.
The advantage of this representation is, that the transformations are easiest to do in forward direction, like simulating (instead of disentangling) the overlapping higher orders of diffraction.
But it is stressed here, that it is important to do the comparison between synthetic and observed spectrum in \emph{both representation simultaneously}.

The counts/bin spectra plotted on a linear scale reveal the statistical nature of low counts detections, that are important in X-rays.
It gives an immediate impression in which wavelength ranges the detection was strong or weak with respect to the background signal.
Also, it allows to show the number of counts that are interpreted as background in the same plot.

On the other hand, the unabsorbed flux, plotted logarithmically, is very sensitive to the IS extinction.
As in the red tail of the (absorbed counts/bin) observation the actual flux from the object is attenuated typically by a factor 1000, the signal that is left in the observer representation is very small, so that deviations of a factor of 2 or more do not catch the eye.
In the flux representation, where the observed signal is corrected for IS extinction by the same large factor, even small deviations in the absorbed model become apparent.

\subsubsection{`By eye' instead of chi-square}
Often it is argued, defending the applicability of the chi-square method as fitting procedure, that the reproduction by the synthetic spectra in the fit is the most important in \emph{those} regions where the signal is the strongest.
This is true for pure statistical significance tests, but not for comparing two physical models:
while the reproduction of the observation \emph{is} important in those regions where the \emph{original, unabsorbed} flux would have been the strongest (the IS extinction has no physical relation to the actual flux of the object we want to model), also the reproduction of the overall shape of the spectrum, including the weak fluxes, is required for a physically acceptable model.

Using the chi-square method has the very welcome effect, that $N_{\rm H}$ can be treated as an additional free parameter in the modeling process.
This yields supposedly better fits in the absorbed counts/bin representation.
A bad $N_{\rm H}$ in such a fit manifests in the red (absorbed) tail of the spectrum, where the slope of the model does not meet the slope of the observation.
In this tail the model counts can be off by a relatively large factor, a factor that would certainly catch the eye in the region of maximum counts.
On the other hand, constraining the $N_{\rm H}$ accurately from both the counts/bin \emph{and} logarithmic flux plots, poses a strong restriction on the determination of a good fitting model spectrum, and leads to the disqualification of fits that `look better' (at first sight) in a linear counts/bin plot only with a wrong value of $N_{\rm H}$.

\clearpage
\section{First `fits' to X-ray nova observations} \label{sec:Fits}
In sections \ref{sec:Grids} and \ref{sec:LTEvsNLTE} it was shown that with the improved theory and the new methods developed in this work (chapters \ref{sec:RadiationTransport} and \ref{sec:GoodNLTE}) systematic results are achieved from the NLTE nova models.
This is the prerequisite for useful modeling of observational data.

Once this is made sure, the model spectra can be compared to observational data, as the final goal of this work is to improve our understanding of the nature of the observed objects: the classical novae.
The procedure to attain this goal is to construct a model, physically as realistic as possible, and vary the assumptions in the model and compare the results with the observations.
The assumptions in the model are being constrained by the generally accepted theoretical and observational results.
The basic assumptions about the structure for nova models in this work are described in chapter \ref{sec:NovaStructure}.

Another important assumption concerns the abundances of the chemical elements in the model atmosphere.
The abundances do not only have a major impact on the opacities, and thus on all model results, but also play an important role in the evolution of the nova.
Sophisticated nova evolution models exist that can simulate nova outbursts, the TNR scenario (see section \ref{sec:NovaTheory}).
These depend sensitively on the abundances of the pre-outburst configuration.
With such models the nucleosynthetic yields can be traced throughout the outburst, and thus detailed chemical composition sets can be derived for the ejected material, see \cite{Starrfield74}, \cite{Prialnik78}, \cite{Prialnik79} and \cite{Starrfield09}.

However, in the SSS phase the ejecta become optically thin and deeper layers become visible.
These have a history that can be very different than the history of the matter ejected during the initial outburst.
In general, there are three different sources of material with different chemical compositions.
\begin{enumerate}
\item Material that is accreted from the companion, which is commonly assumed to have solar abundances.
\item Material that is dredged up from the white dwarf.
This material can be either CO enhanced, or ONeMg enhanced.
\item Previously thermo-nuclearly processed material from the nova that is reaccreted after the outburst. This material is strongly NO enhanced. The C abundance of the ejecta depends on the white dwarf material, but even novae with a ONeMg white dwarf are expected to have a supersolar C abundance.
\end{enumerate}
Being a mixture of these components, the chemical composition of the  deep layers that become visible in the SSS phase is not unlikely to be different from the composition sets available in literature, which are derived from either theory or observations for \emph{earlier stages} after the outburst.

The SSS phase allows a `deep' insight in the inner regions of the nova atmosphere, `close' to the region where nuclear burning takes place.
Therefore, it will be very interesting to accurately determine the chemical compositions that are `seen' in this phase.

However, variation of the chemical abundances introduces a large number of additional free parameters to the model.
And given the computational requirements per model, the spectral analysis (which involves fine-tuning the abundances) needs to be done in a systematic step-by-step way.
The `bulk grid computation' method used in section \ref{sec:Grids} to explore the parameter space of the basic atmosphere structure parameters, is practically not suitable because the dimensions of such grids become too large.
Detailed spectral analysis has not yet been done in the scope of this work but \emph{is} planned for the near future.

The models that are compared to the observations in this section all use solar abundances \cite{Asplund05}.
It is a standard set of abundances, that is suitable to test the models, but is possibly not very realistic.

\subsection{Expanding models} \label{sec:ExpandingFits}
In figures \ref{fig:V4743Mar} to \ref{fig:V2491Apr2} the synthetic spectra for the expanding nova type models computed so far are compared with all well exposed `high-resolution' observations of novae in a supersoft X-ray state available to date:
\begin{itemize}
 \item figure \ref{fig:V4743Mar}: V4743 Sgr, March 2003, day 180, LETGS
 \item figure \ref{fig:V4743Apr}: V4743 Sgr, April 2003, day 196, RGS
 \item figure \ref{fig:V4743Jul}: V4743 Sgr, July 2003, day 302, LETGS
 \item figure \ref{fig:V4743Sep}: V4743 Sgr, September 2003, day 371, LETGS
 \item figure \ref{fig:V4743Sep}: V4743 Sgr, February 2004, day 526, LETGS
 \item figure \ref{fig:RSOphMar}: RS Oph, March 2006, day 39.7, LETGS
 \item figure \ref{fig:RSOphApr1}: RS Oph, April 2006, day 54.3, RGS
 \item figure \ref{fig:RSOphApr2}: RS Oph, April 2006, day 66.9, LETGS
 \item figure \ref{fig:V2491Apr1}: V2491 Cyg, April 2008, day 39.9, RGS
 \item figure \ref{fig:V2491Apr2}: V2491 Cyg, April 2008, day 49.7, RGS
\end{itemize}

The V4743 Sgr models have a terminal velocity of $v_\infty = 2400$ km/s, the RS Oph models have $v_\infty = 1200$ km/s, and the models for V2491 Cyg have $v_\infty = 4800$ km/s.
These velocities have not been tuned.
They originated in the grids made to explore the parameter space, shown in section \ref{sec:Grids}, as the half and the double of $2400$ km/s, which was originally chosen as default value with the observed line shifts for V4743 Sgr \cite{Ness03} in mind.
All models have $\beta=1.5$.

The distances to the objects are assumed to be 3.9, 1.5 and 10.2 kpc for V4743 Sgr, RS Oph and V2491 respectively.
These distances only affect the fit parameter $R_{\rm WD}$ that scales the synthetic flux to the observed flux.

\clearpage
\begin{landscape}
\begin{figure}
 \centerline{\includegraphics[width=\textwidth,angle=90]{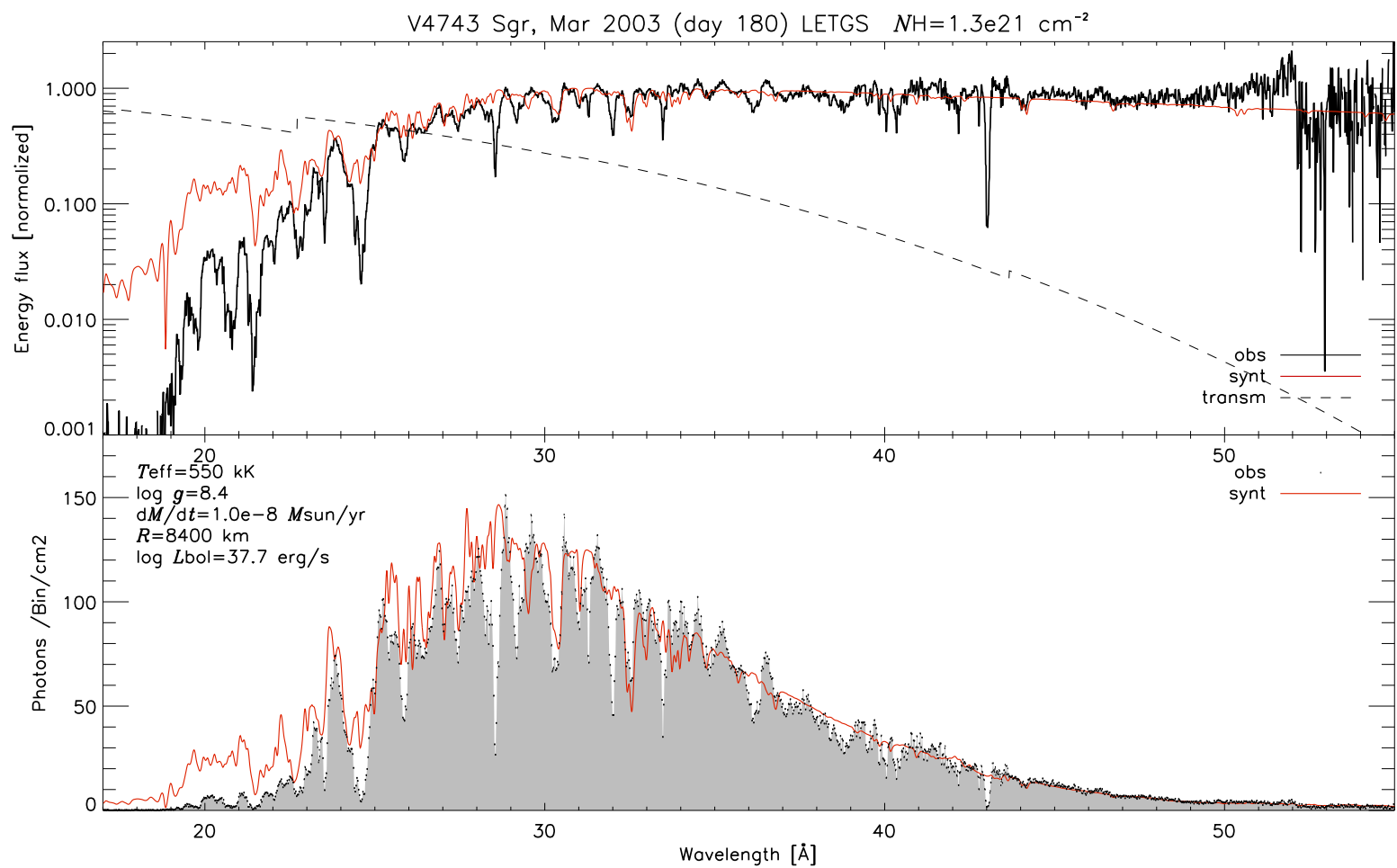}}
 \caption{ \label{fig:V4743Mar} Expanding model}
\end{figure}
\begin{figure}
 \centerline{\includegraphics[width=\textwidth,angle=90]{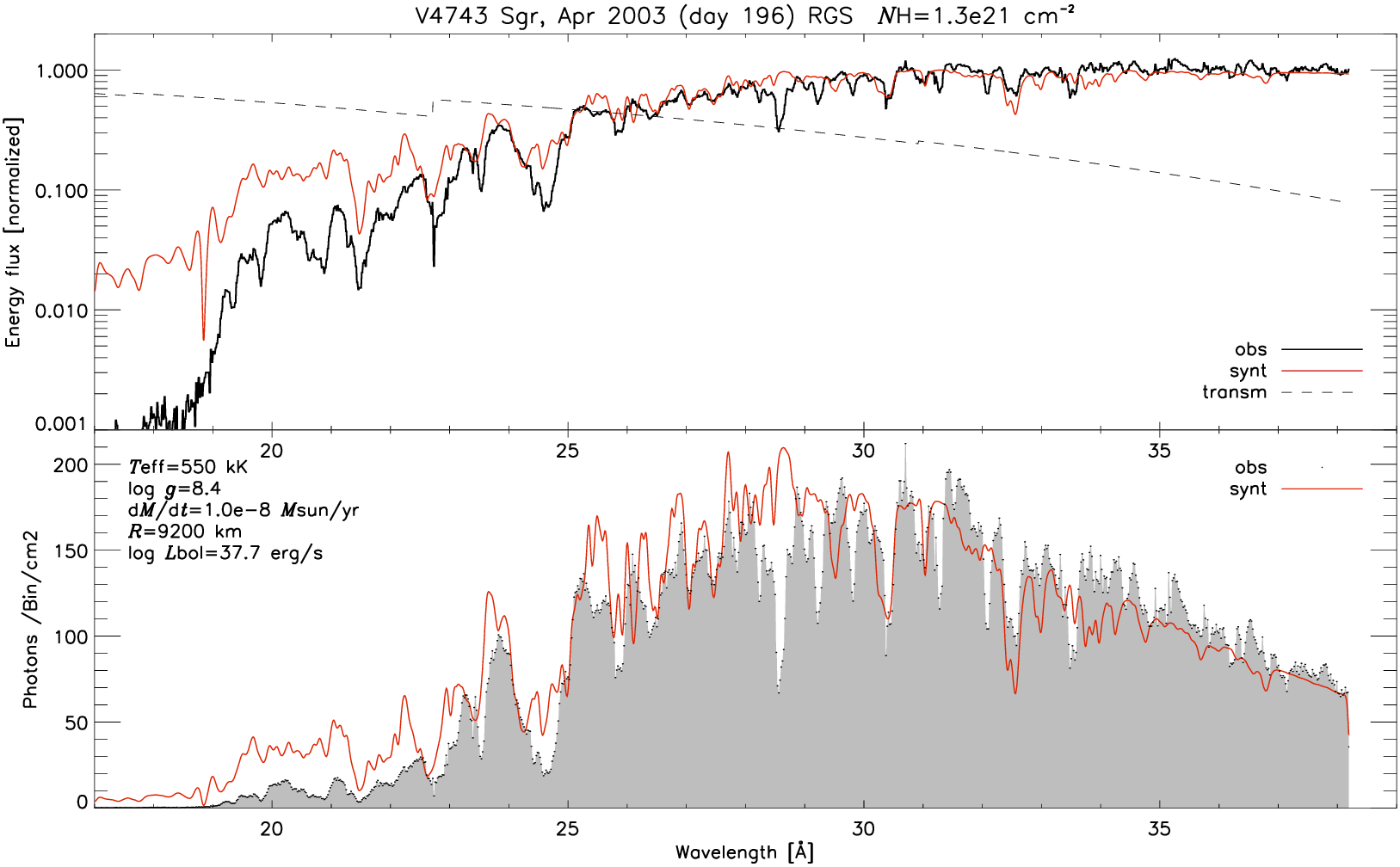}}
 \caption{ \label{fig:V4743Apr} Expanding model}
\end{figure}
\begin{figure}
 \centerline{\includegraphics[width=\textwidth,angle=90]{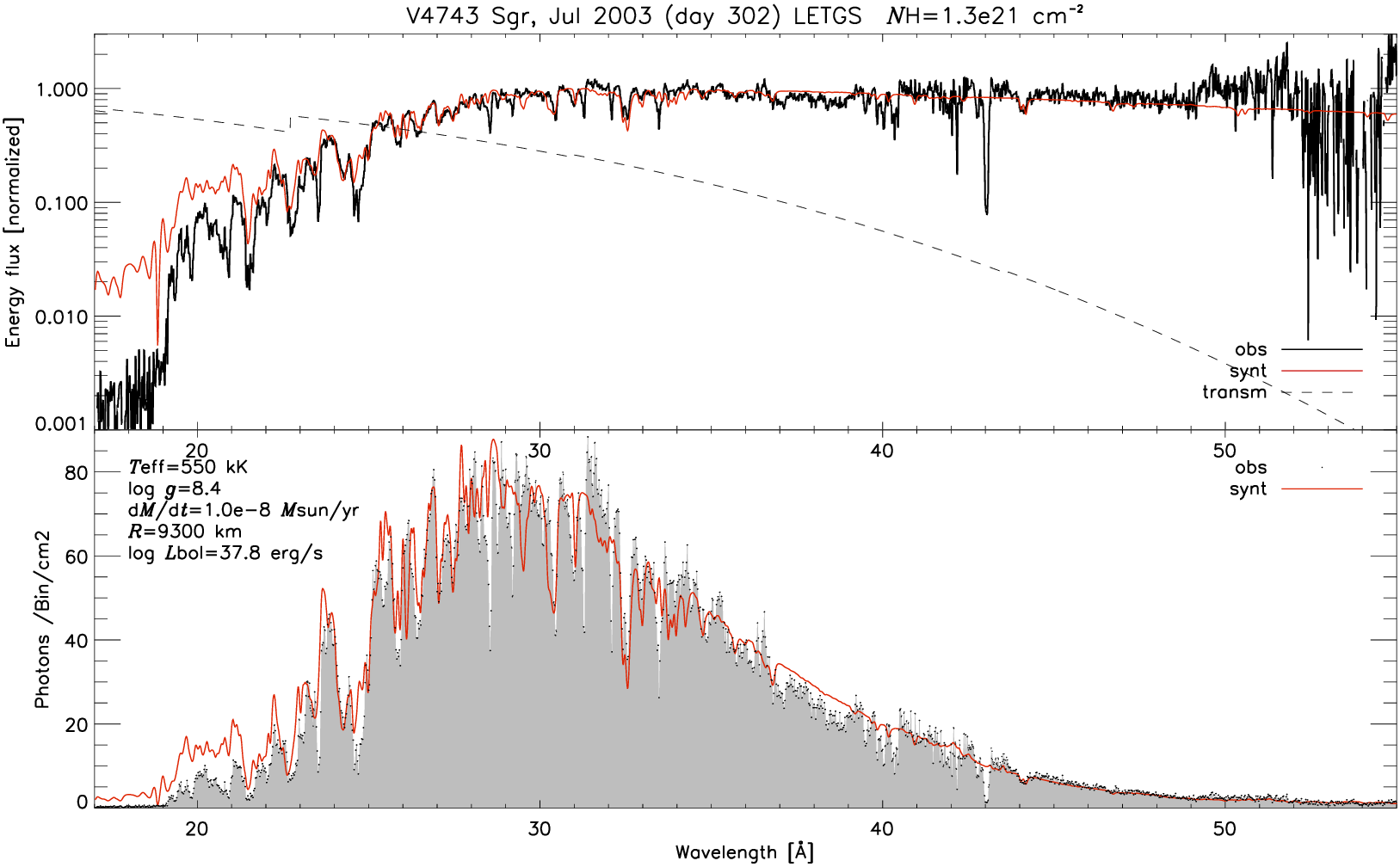}}
 \caption{ \label{fig:V4743Jul} Expanding model}
\end{figure}
\begin{figure}
 \centerline{\includegraphics[width=\textwidth,angle=90]{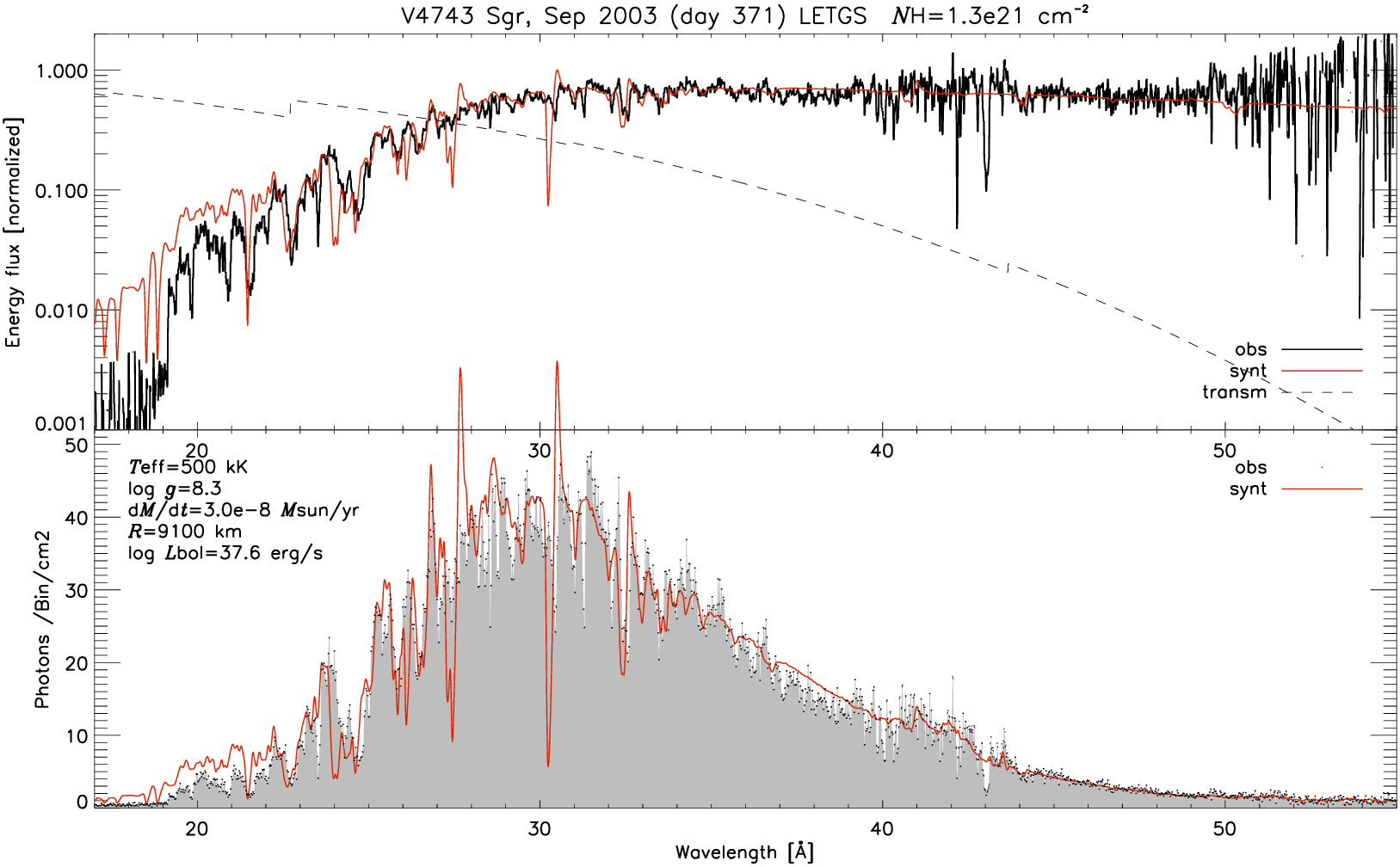}}
 \caption{ \label{fig:V4743Sep} Expanding model}
\end{figure}
\begin{figure}
 \centerline{\includegraphics[width=\textwidth,angle=90]{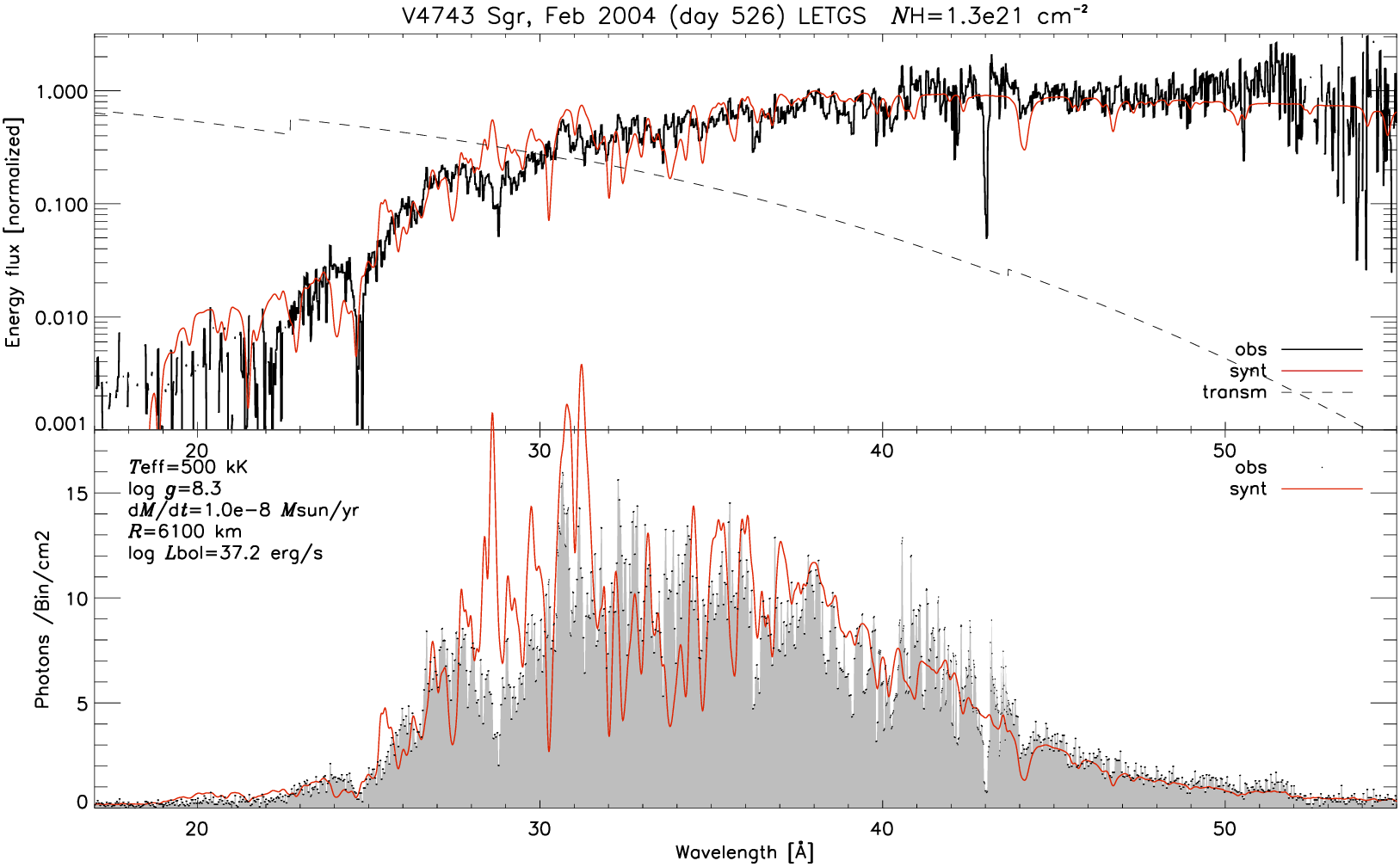}}
 \caption{ \label{fig:V4743Feb} Expanding model}
\end{figure}
\begin{figure}
 \centerline{\includegraphics[width=\textwidth,angle=90]{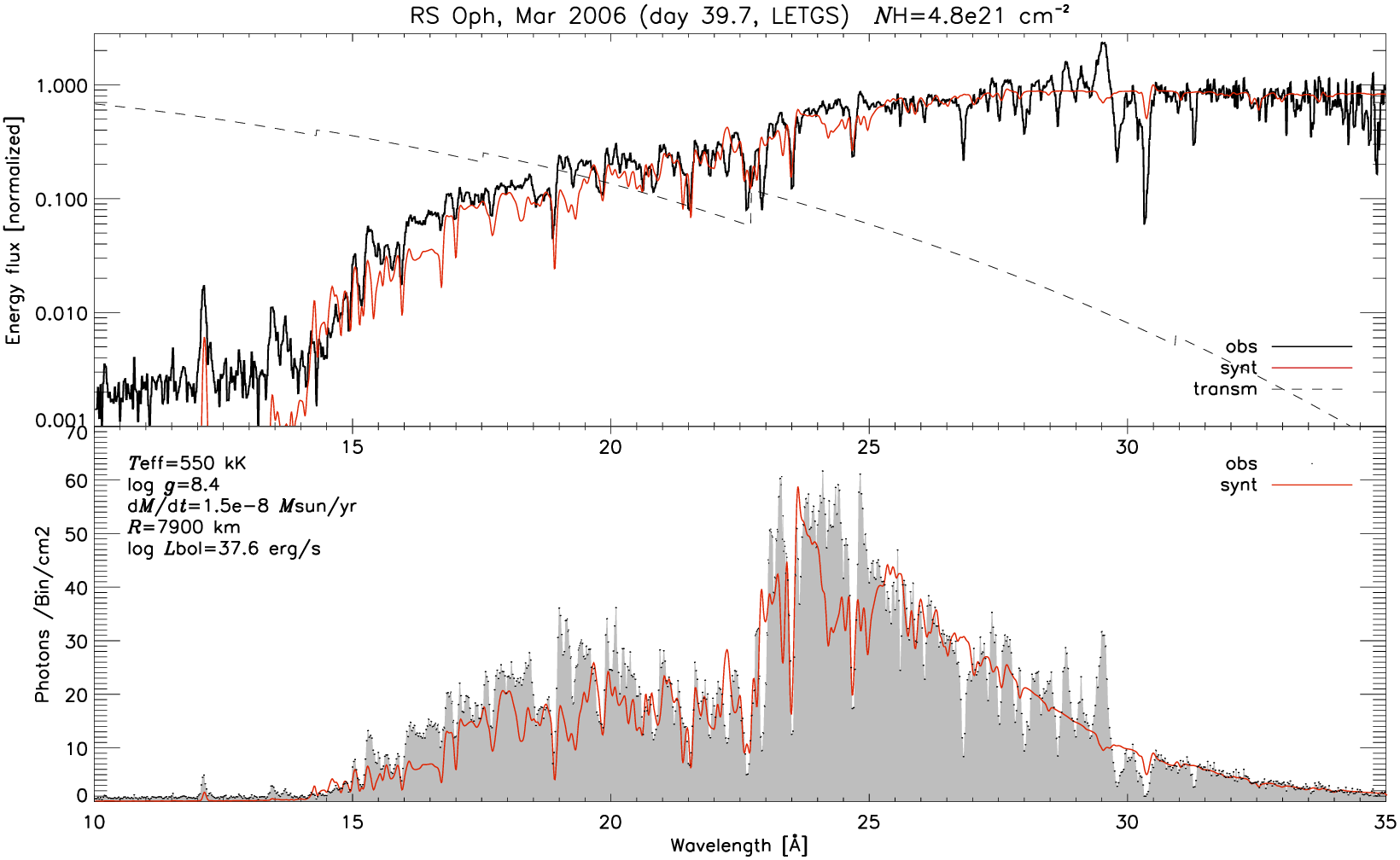}}
 \caption{ \label{fig:RSOphMar} Expanding model}
\end{figure}
\begin{figure}
 \centerline{\includegraphics[width=\textwidth,angle=90]{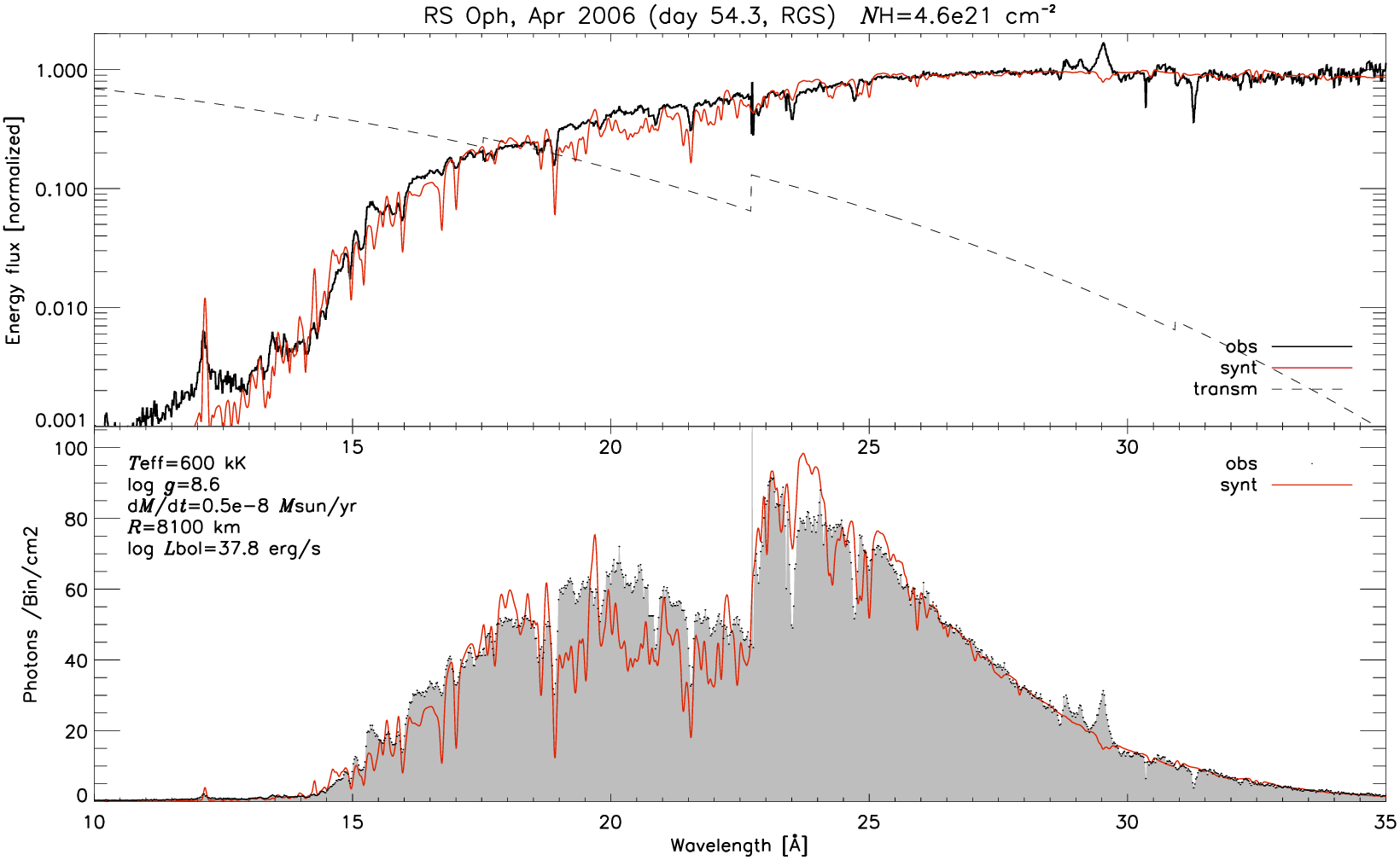}}
 \caption{ \label{fig:RSOphApr1} Expanding model}
\end{figure}
\begin{figure}
 \centerline{\includegraphics[width=\textwidth,angle=90]{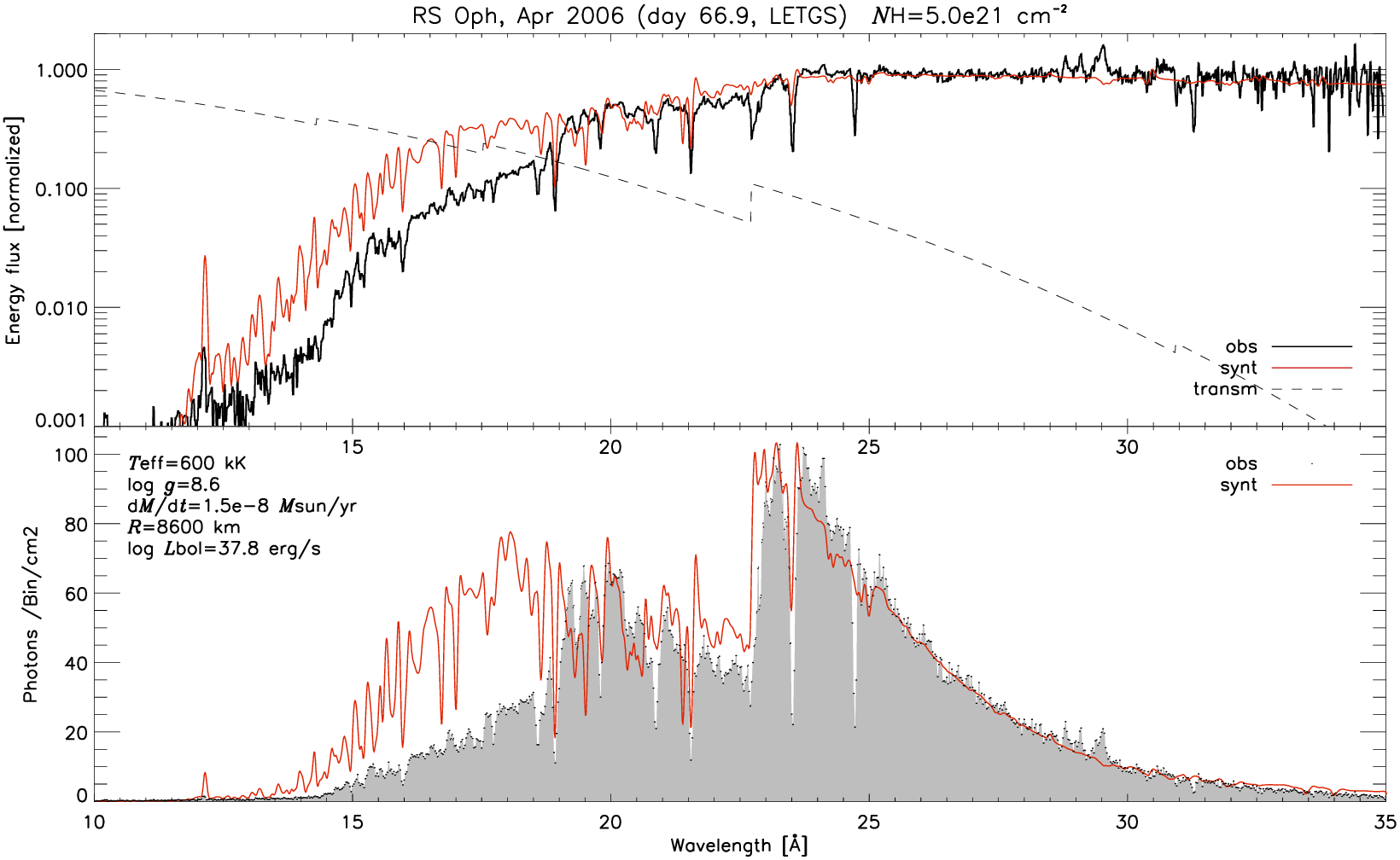}}
 \caption{ \label{fig:RSOphApr2} Expanding model}
\end{figure}
\begin{figure}
 \centerline{\includegraphics[width=\textwidth,angle=90]{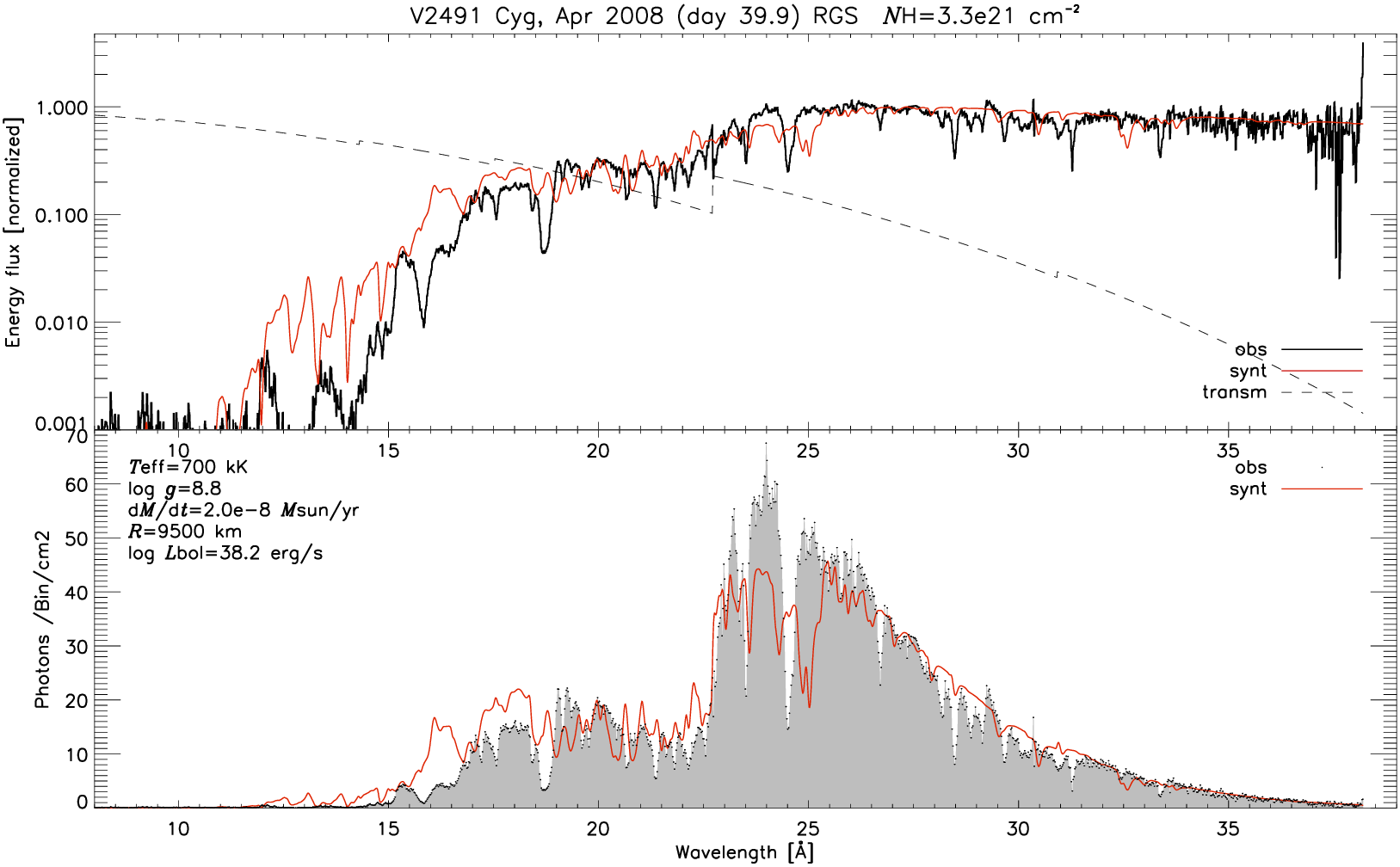}}
 \caption{ \label{fig:V2491Apr1} Expanding model}
\end{figure}
\begin{figure}
 \centerline{\includegraphics[width=\textwidth,angle=90]{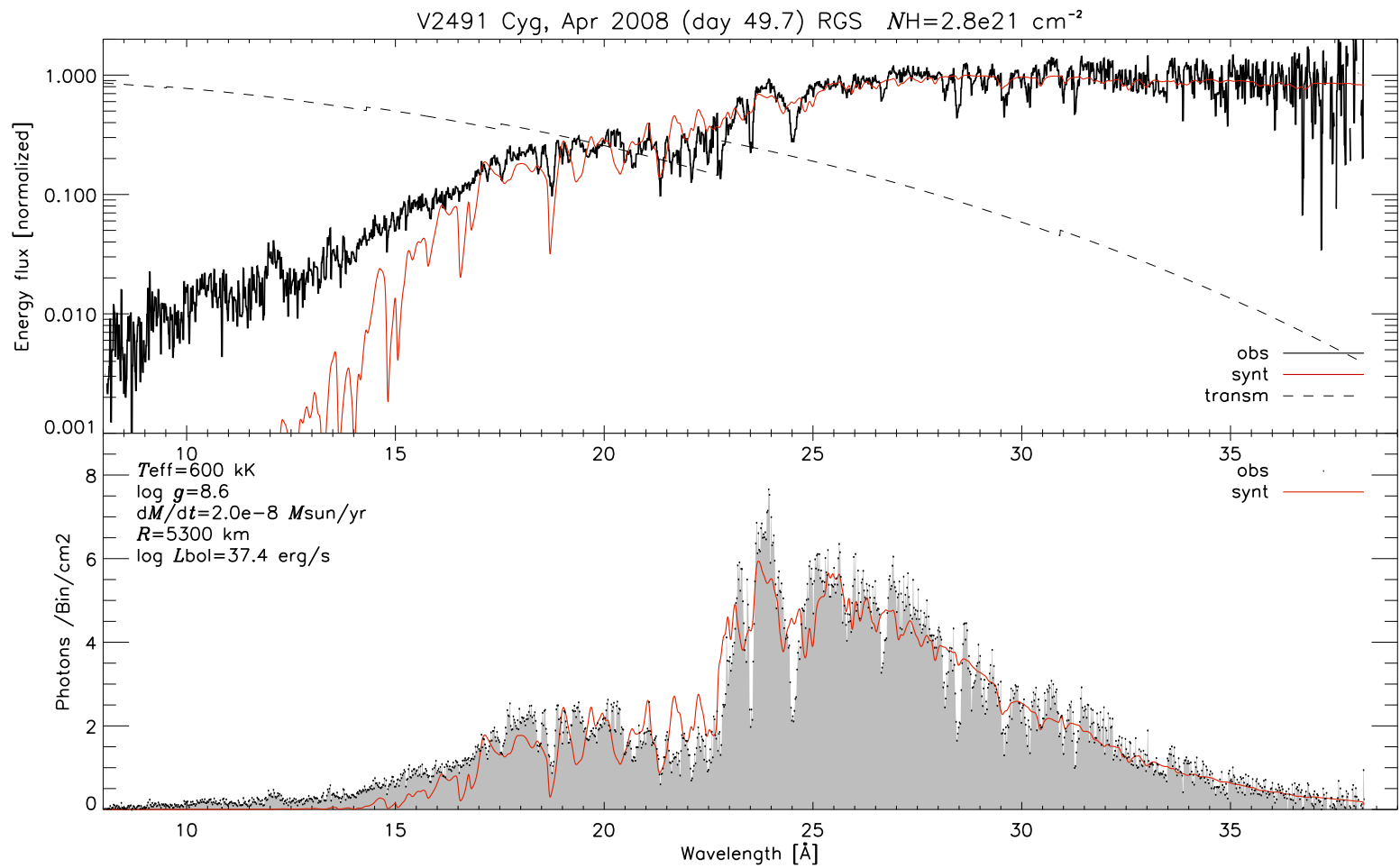}}
 \caption{ \label{fig:V2491Apr2} Expanding model}
\end{figure}
\end{landscape}

\clearpage
\subsection{Discussion: Expanding} \label{sec:DiscussionExpanding}
In the comparison between the models and the observations all X-ray grating (Chandra and XMM-Newton) spectra were used that are available to date.
For none of the observations the models have been fine tuned, but rather it has been shown, that for all available observations a model can be found that at least reproduces the general shape and some detail of the observation.

For the interpretation of the fit quality, it must be kept in mind, that the steps in the values of the basic model parameters in the grid were large.
In the process of finding the `best fit' model, this often results in one model being off in one direction, and the next model in the grid being off in the other, so that an increase in the parameter resolution in the grid would certainly improve the quality of the `best fit'.

\subsubsection{V4743 Sgr}
The V4743 Sgr models have a terminal velocity of $v_\infty = 2400$km/s.
In all of them the flux in the blue tail of the model spectra is too strong.
This is most apparent shortwards of 22.5\AA{}, which exactly matches the N\,{\sc vi} ionization edge, so that increasing the N abundance to super solar could fix this problem.
The models with $T_{\rm eff} = 500$kK for the September and April observations show a somewhat too strong emission component in some of the P Cycni profiles (around observed wavelengths of 27.4\AA{} and 30.2\AA{}).
On the other hand, for other lines these emission components fit the observations very well (around observed wavelengths of 26.8\AA{} and 32.6\AA{})

\subsubsection{RS Oph}
In all RS Oph models, that have a terminal velocity of $v_\infty = 1200$km/s, but especially in comparison with the last two (April) observations, some absorption lines are too narrow and too deep.
This is probably due to the neglected instrumental response matrix.
The instrumental accuracy is usually not exactly as high as the resolution, as photons of a certain wavelength are smeared out over a few bins in the detector.
Such a smearing effect would make the lines in the model a bit broader and less deep.

The middle observation, of day 54.3, is much smoother than the first and the third observation.
The spectral features in the model are also found in the observations, but there they are much smoother.
For this observation a periodic variability was discovered \cite{Osborne06}.
It would be interesting to divide this observations in multiple time frames, in order to see if there is a temporal development in the observation that causes details to be smeared out in the time integrated data.

For the day 54.3 and day 66.9 observations, when going from lower to higher wavelengths there is an obvious step up (increase) in the observed flux at 18.6\AA{}, being the ionization edge of N\,{\sc vii}.
This is not followed by the models.
Again, increasing the N abundance would probably fix this problem.
In the first model, for day 39.7, this problem doesn't seem to exist.
It is hard to believe, that the abundances would have changed very strongly since then.
It will be interesting to find out if an abundance set that suits the later observations will also supply models that fit the March observation.

\subsubsection{V2491 Cyg}
V2491 Cyg is the fast expanding nova. The models were computed with $v_\infty = 4800$km/s, twice the speed as used from V4743 Sgr.
For this outburst only two grating observations exist.
And these observations show very large differences.
In contrast to the other two sources, where the bolometric luminosity (displayed in the legend of each plot) is almost constant for the observations, the luminosity decreases by a factor of three.
The first observation, on day 39.9, was done at almost maximum X-ray brightness (see the lightcurve in figure \ref{fig:V2491Lightcurve}).
After maximum brightness, the count rate quickly declined.

In the model for the first observation, with $T_{\rm eff}=700$kK, the O\,{\sc vii} ionization edge at 16.8\AA{} is too weak.
Also, the characteristic O absorption lines at 19.0\AA{} and 21.6\AA{} are much too weak in the model.
However, in the cooler model for the second observation the edge and the lines are strong enough.
The influence of changes to the abundance can be quite different for such different models, but the difference needed here is very large.
Probably not only the abundances need to be tuned but also the model structure.

Furthermore, for both observations the N\,{\sc vi} edges at 22.5\AA{} are too weak and also the N\,{\sc vii} line at 24.8\AA{} and the N\,{\sc vi} line at 28.8\AA{}.

The C\,{\sc vi} absorption edge at 25.3\AA{} is too strong in the model for day 39.9 and a bit too strong in the model for day 49.7.

\subsection{Hydrostatic models} \label{sec:HydrostaticFits}
In figures \ref{fig:V4743Mar_PP} to \ref{fig:V2491Apr2_PP} the spectra from hydrostatic models computed so far are compared with all well exposed `high-resolution' observations of novae in a supersoft X-ray state available to date:
\begin{itemize}
 \item figure \ref{fig:V4743Mar_PP}: V4743 Sgr, March 2003, day 180, LETGS
 \item figure \ref{fig:V4743Apr_PP}: V4743 Sgr, April 2003, day 196, RGS
 \item figure \ref{fig:V4743Jul_PP}: V4743 Sgr, July 2003, day 302, LETGS
 \item figure \ref{fig:V4743Sep_PP}: V4743 Sgr, September 2003, day 371, LETGS
 \item figure \ref{fig:V4743Sep_PP}: V4743 Sgr, February 2004, day 526, LETGS
 \item figure \ref{fig:RSOphMar_PP}: RS Oph, March 2006, day 39.7, LETGS
 \item figure \ref{fig:RSOphApr1_PP}: RS Oph, April 2006, day 54.3, RGS
 \item figure \ref{fig:RSOphApr2_PP}: RS Oph, April 2006, day 66.9, LETGS
 \item figure \ref{fig:V2491Apr1_PP}: V2491 Cyg, April 2008, day 39.9, RGS
 \item figure \ref{fig:V2491Apr2_PP}: V2491 Cyg, April 2008, day 49.7, RGS
\end{itemize}
Since the models do not treat the expansion, the spectra are blue shifted with 1000, 2400 and 4000km/s for the V4743 Sgr, RS Oph, and V2491 Cyg observations respectively.

\clearpage
\begin{landscape}
\begin{figure}
 \centerline{\includegraphics[width=\textwidth,angle=90]{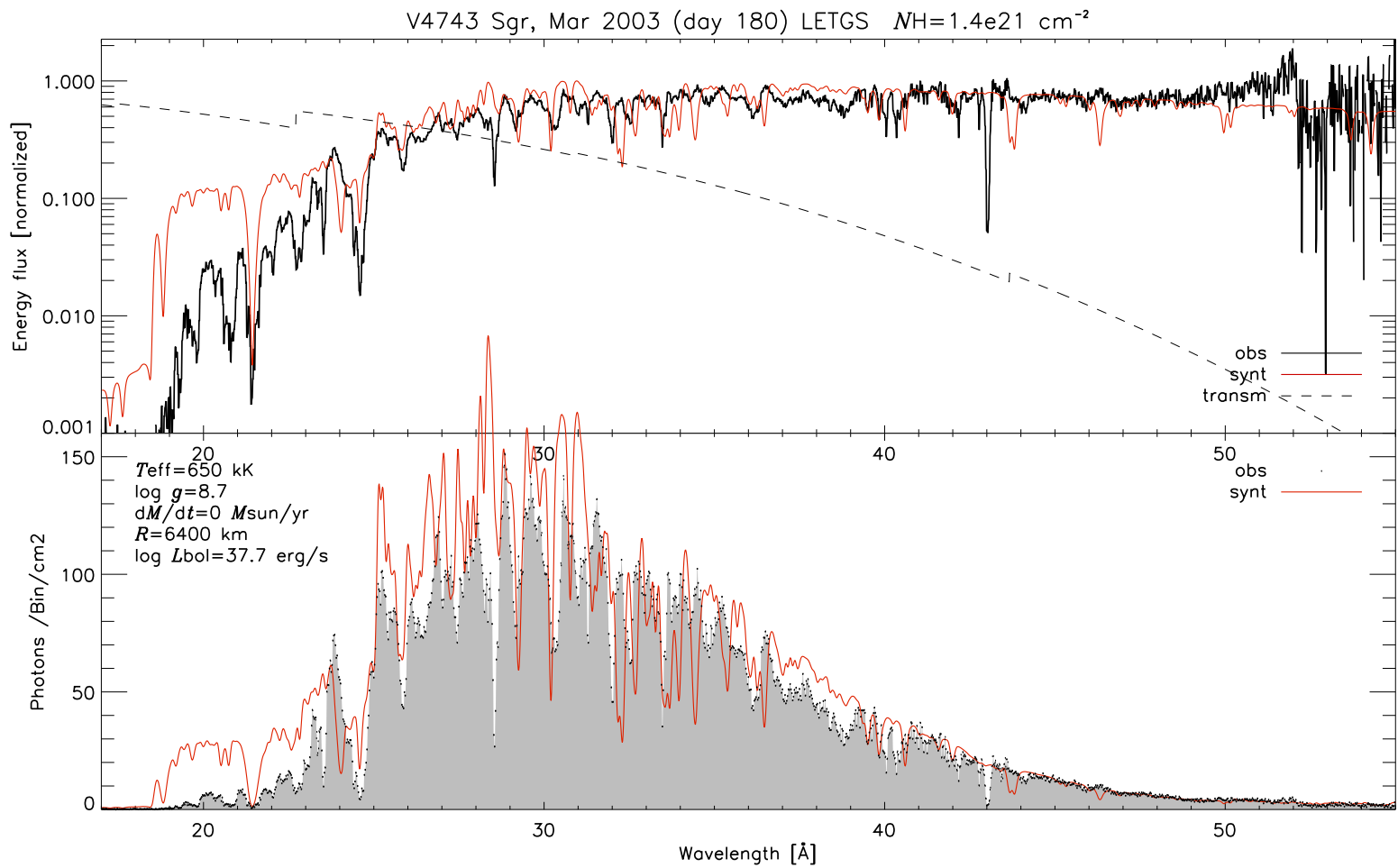}}
 \caption{ \label{fig:V4743Mar_PP} Hydrostatic model}
\end{figure}
\begin{figure}
 \centerline{\includegraphics[width=\textwidth,angle=90]{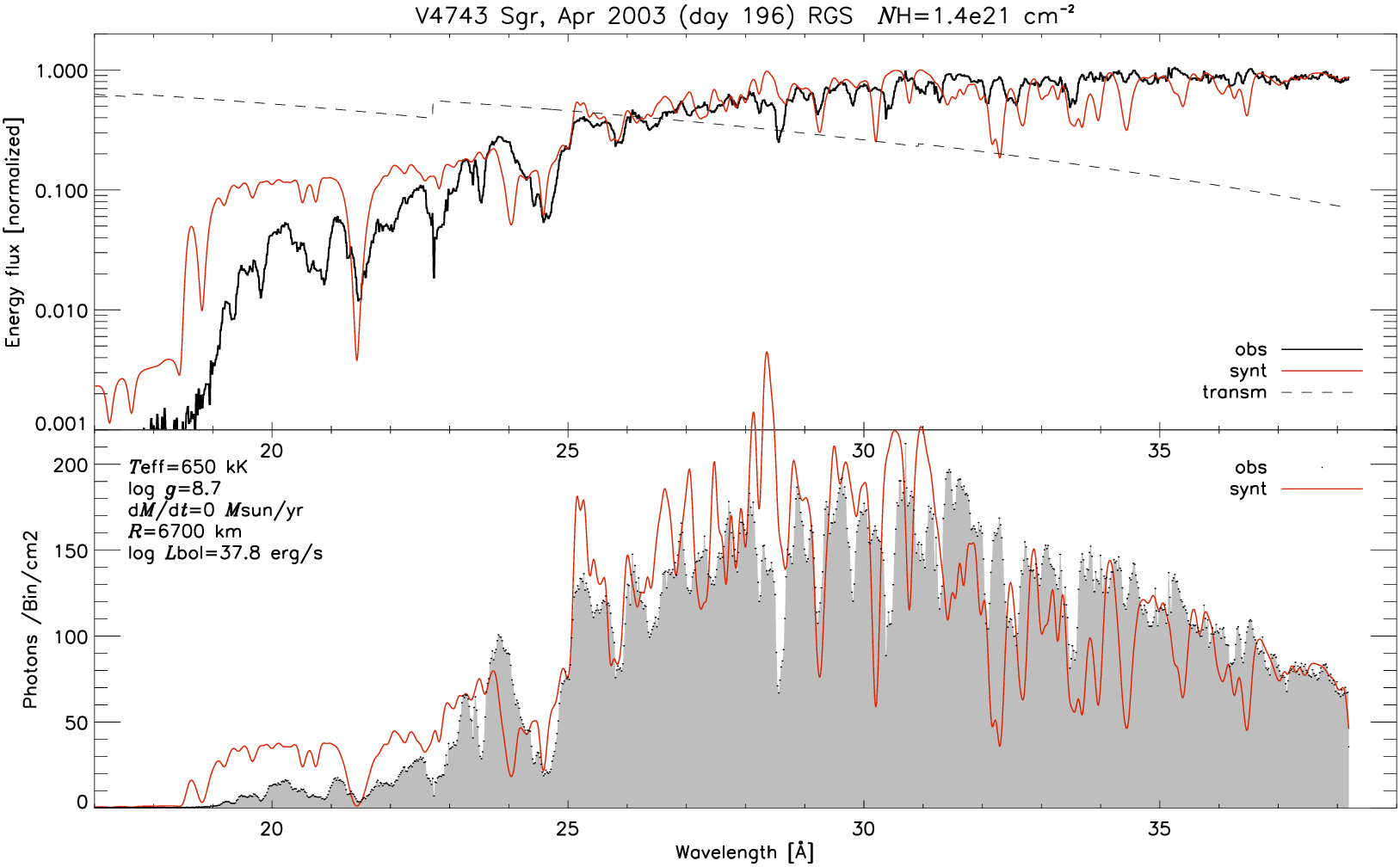}}
 \caption{ \label{fig:V4743Apr_PP} Hydrostatic model}
\end{figure}
\begin{figure}
 \centerline{\includegraphics[width=\textwidth,angle=90]{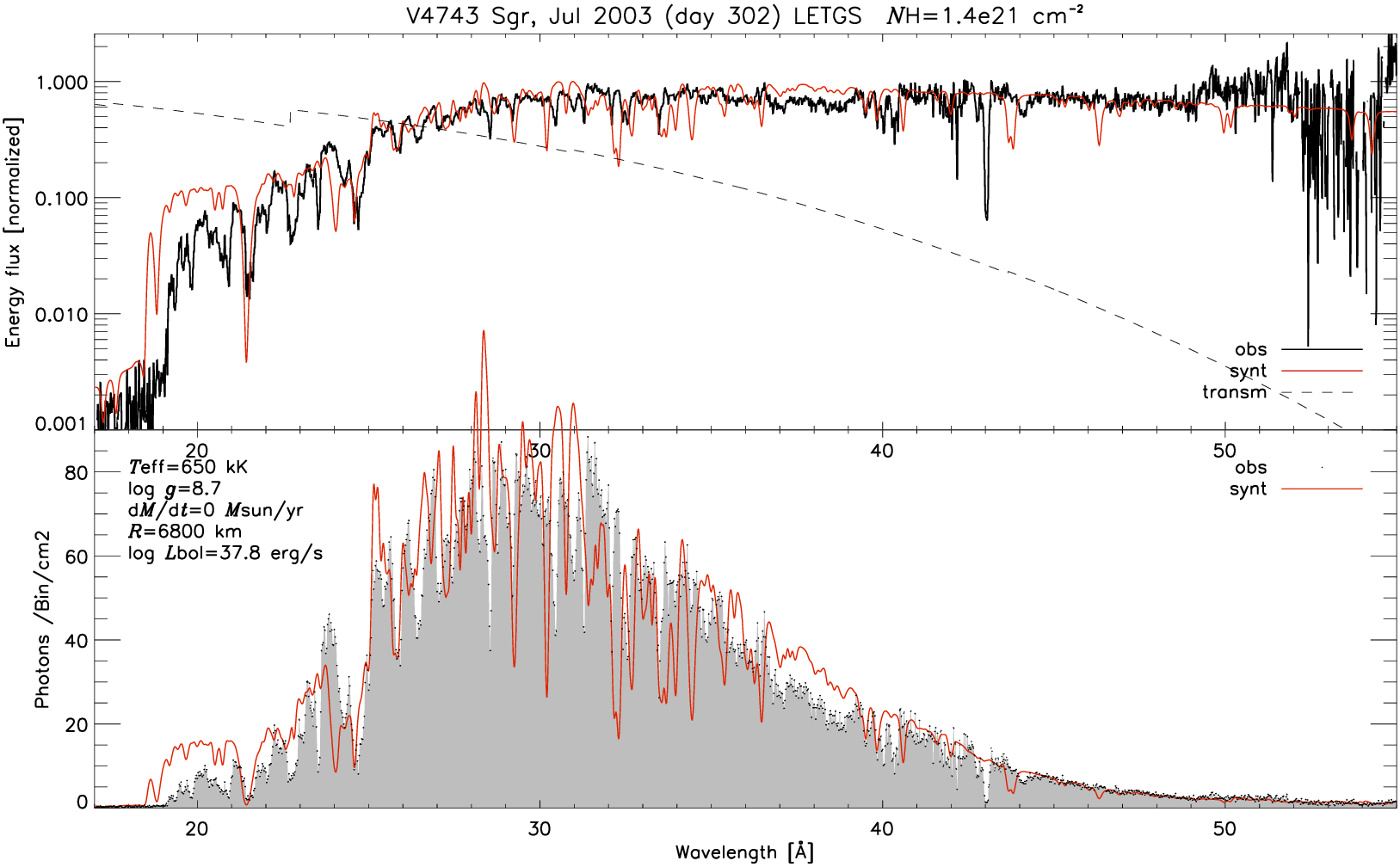}}
 \caption{ \label{fig:V4743Jul_PP} Hydrostatic model}
\end{figure}
\begin{figure}
 \centerline{\includegraphics[width=\textwidth,angle=90]{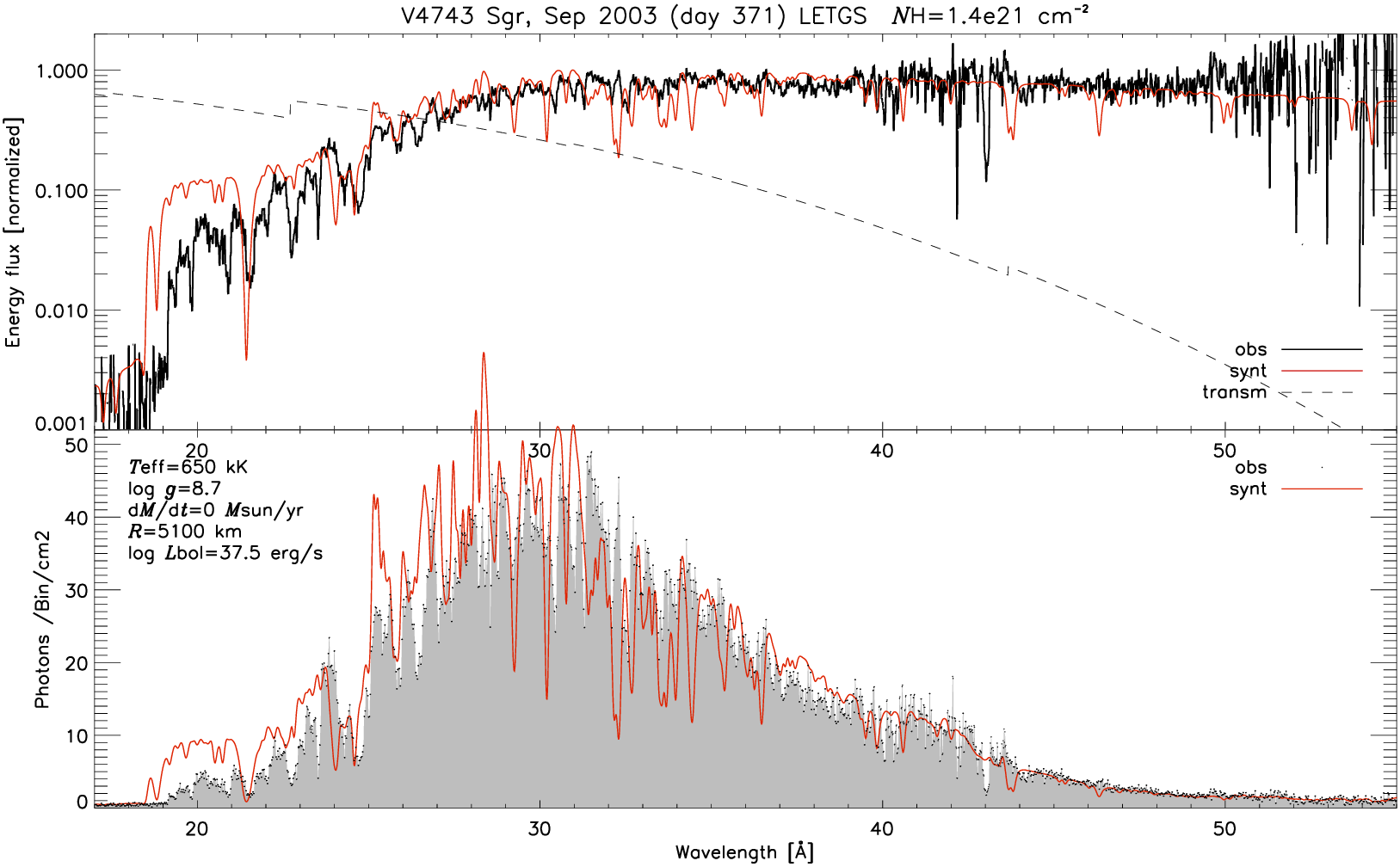}}
 \caption{ \label{fig:V4743Sep_PP} Hydrostatic model}
\end{figure}
\begin{figure}
 \centerline{\includegraphics[width=\textwidth,angle=90]{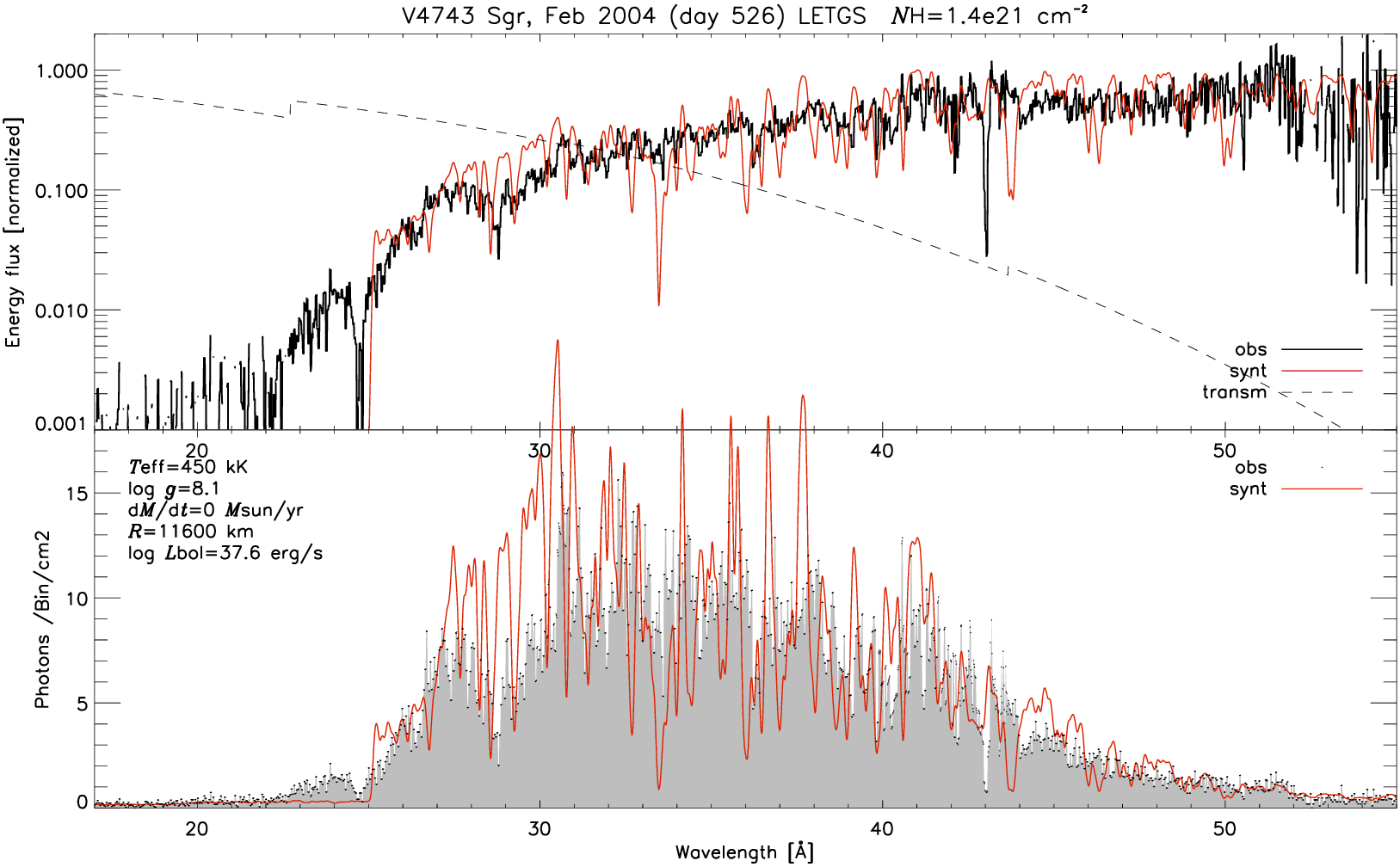}}
 \caption{ \label{fig:V4743Feb_PP} Hydrostatic model}
\end{figure}
\begin{figure}
 \centerline{\includegraphics[width=\textwidth,angle=90]{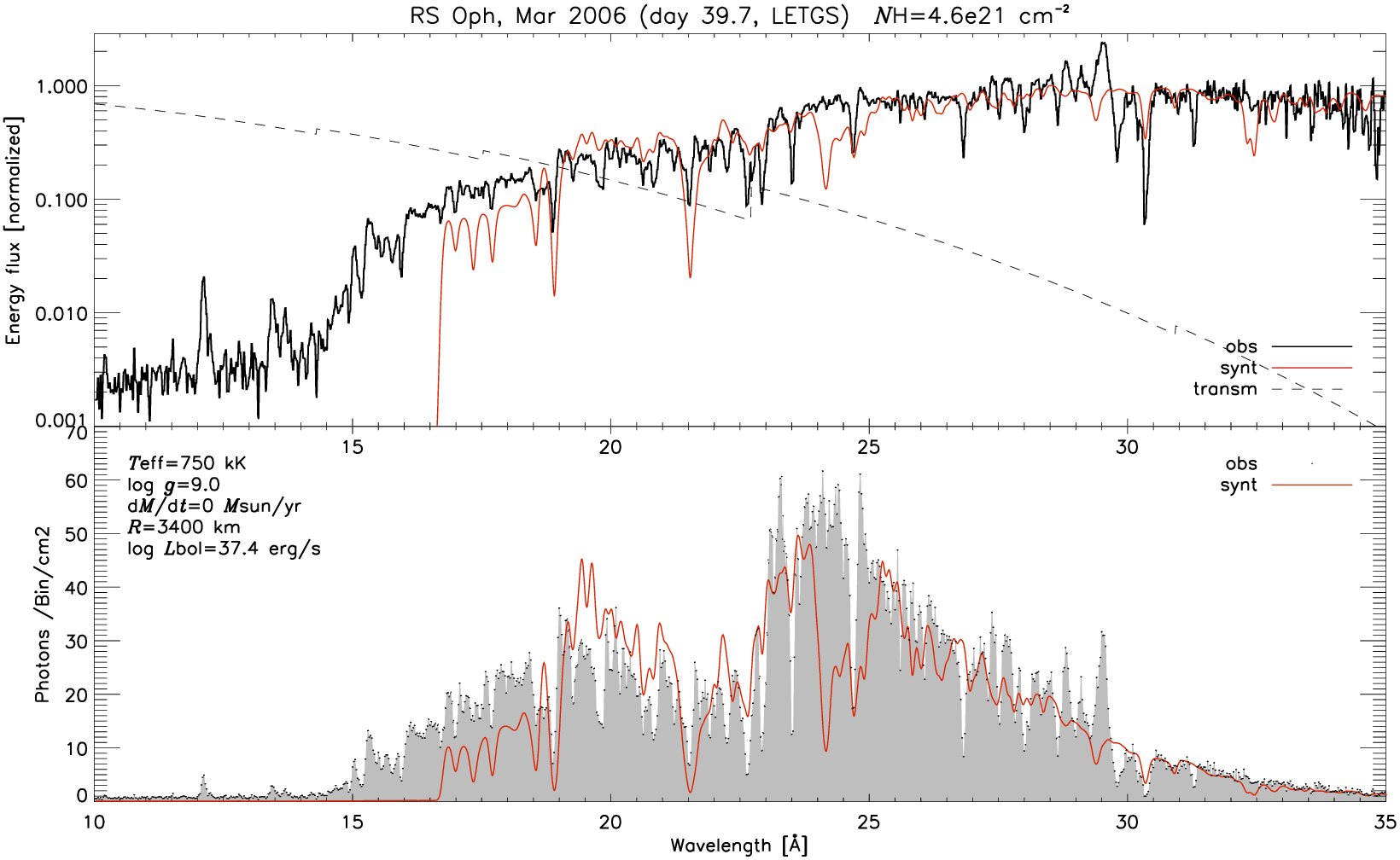}}
 \caption{ \label{fig:RSOphMar_PP} Hydrostatic model}
\end{figure}
\begin{figure}
 \centerline{\includegraphics[width=\textwidth,angle=90]{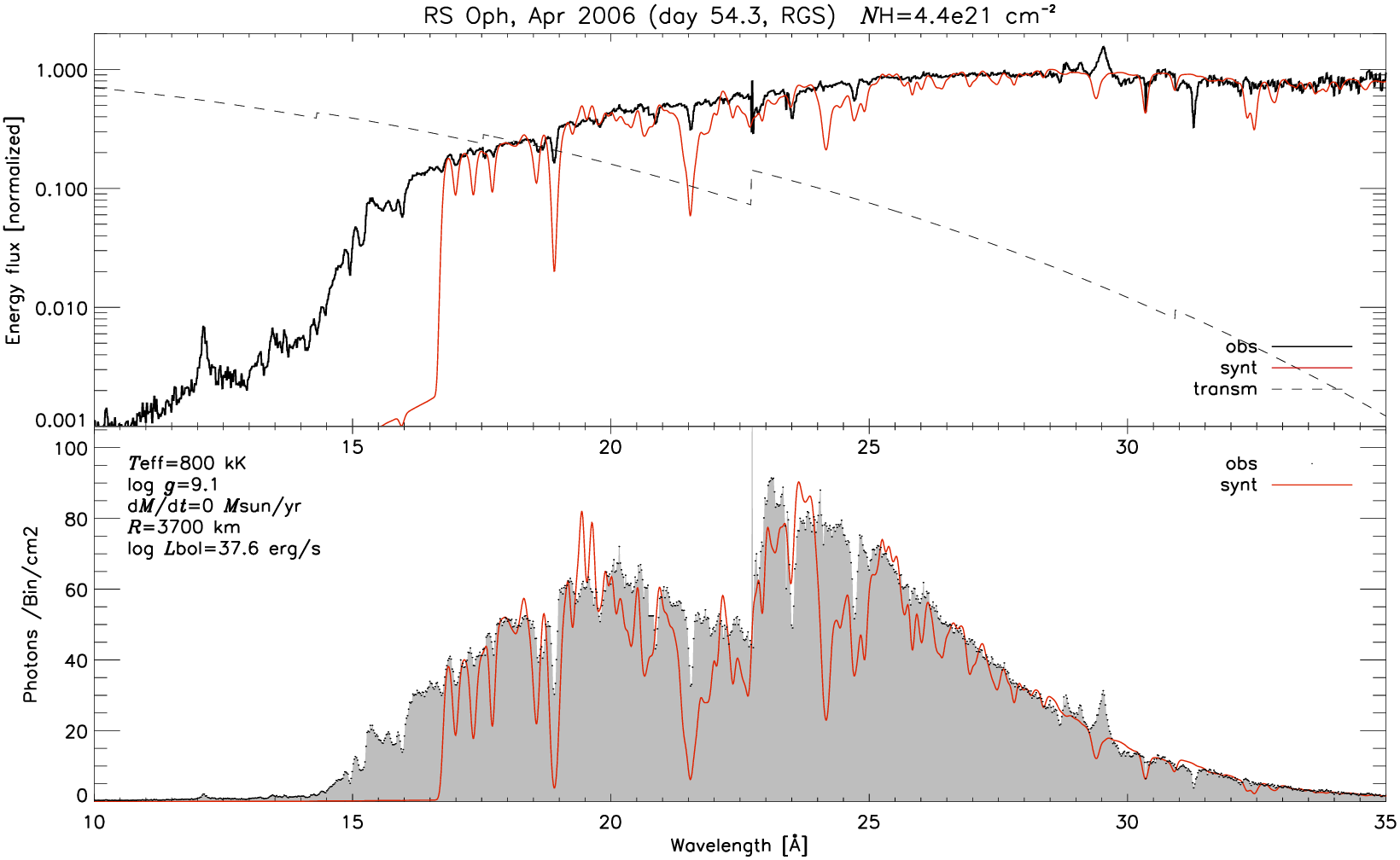}}
 \caption{ \label{fig:RSOphApr1_PP} Hydrostatic model}
\end{figure}
\begin{figure}
 \centerline{\includegraphics[width=\textwidth,angle=90]{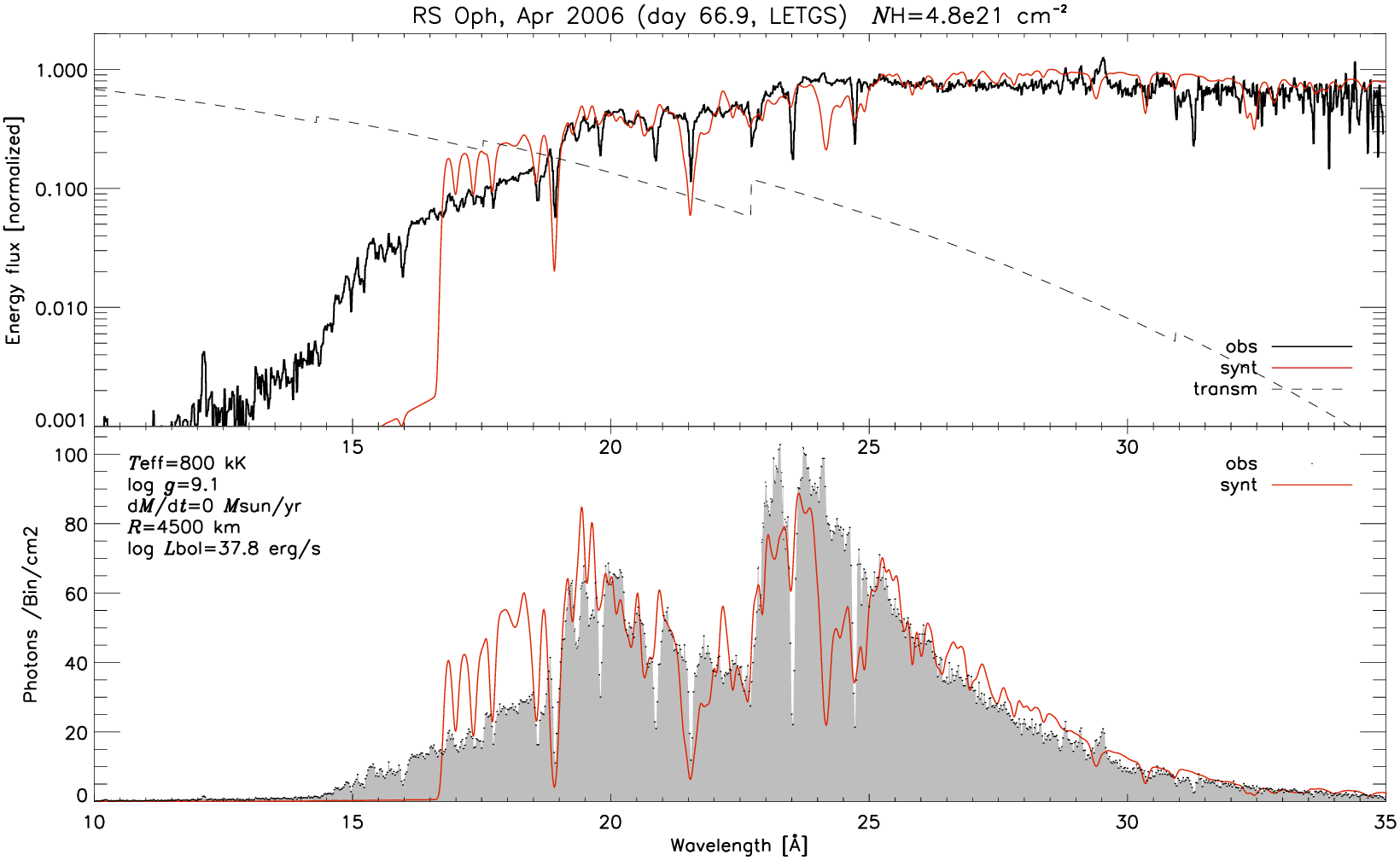}}
 \caption{ \label{fig:RSOphApr2_PP} Hydrostatic model}
\end{figure}
\begin{figure}
 \centerline{\includegraphics[width=\textwidth,angle=90]{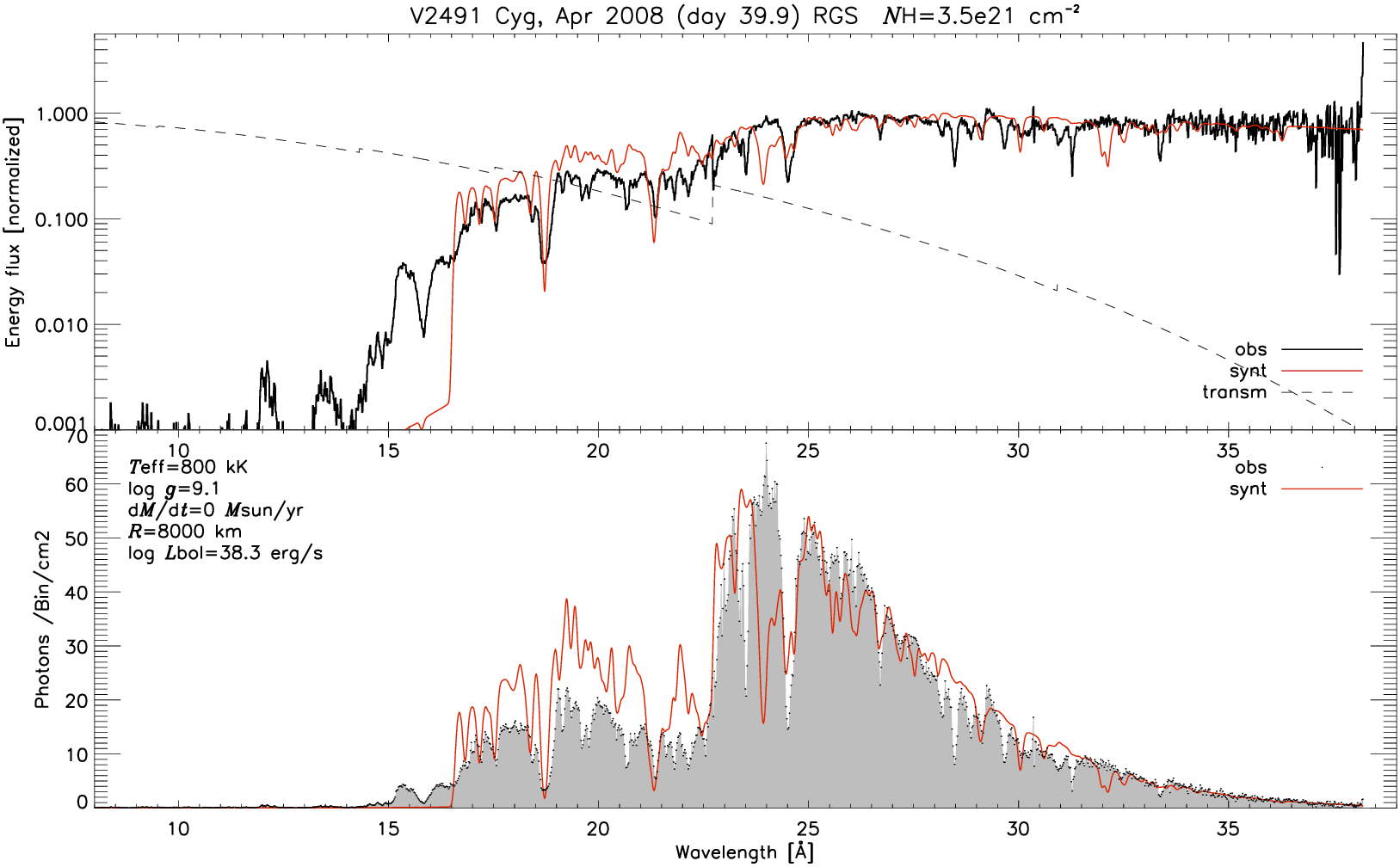}}
 \caption{ \label{fig:V2491Apr1_PP} Hydrostatic model}
\end{figure}
\begin{figure}
 \centerline{\includegraphics[width=\textwidth,angle=90]{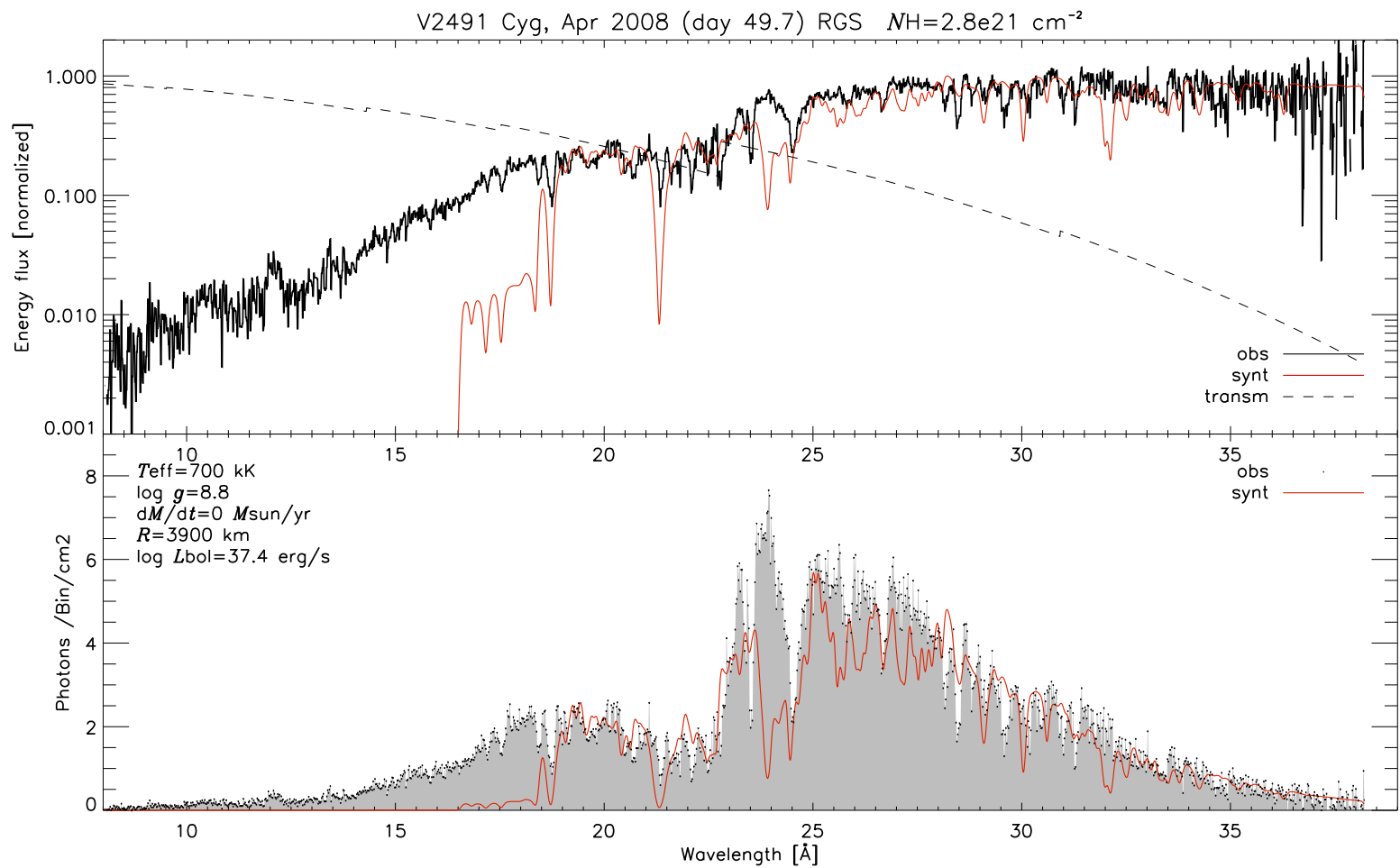}}
 \caption{ \label{fig:V2491Apr2_PP} Hydrostatic model}
\end{figure}
\end{landscape}
\clearpage

\subsection{Discussion: Hydrostatic} \label{sec:DiscussionHydrostatic}
Just like for the expanding models shown in section \ref{sec:ExpandingFits}, the hydrostatic models are compared to all X-ray grating (Chandra and XMM-Newton) data that are available to date.
The models have not been fine tuned, they all use the same solar abundances as the expanding models.
The parameter resolution in the grid (e.g. the step size in $T_{\rm eff}$ is 50kK) is identical with the expanding models grid.

\subsubsection{V4743 Sgr}
The hydrostatic models reproduce the observations of V4743 Sgr quite well, in this respect the quality is comparable to the expanding models, except for the last observation, day 526.
The model with $T_{\rm eff} = 450$kK features a strong C\,{\sc vi} ionization edge at 25.3\AA{} that is not present in the observations and very weak in the expanding models.

However, the reproduction of the detailed spectral features is much worse than the expanding models.
There is hardly any absorption feature that matches the observation in line shape, line strength and line center.
The ad-hoc blue shift does move the lines closer to the observed line centers but since lines are formed in different regions in the wind with different local velocities, the global constant shift is not accurate.

\subsubsection{RS Oph}
The hydrostatic models compared with RS Oph all show a much so strong O\,{\sc vii} ionization edge at 16.8\AA{}, so that shortwards of this edge the flux much too weak.
Surely, this could be compensated with a reduced O abundance, but it is interesting that a sub-solar O abundance is not required for the expanding models for RS Oph.
Also, for all observations the O\,{\sc vii} absorption line at 21.6\AA{} is much too strong in the model spectra, which supports the idea that the O abundance should be decreased in order to improve the fit.

\subsubsection{V2491 Cyg}
Just like for RS Oph, the hydrostatic models show a much too strong O\,{\sc vii} ionization edge and too strong O absorption lines, like the O\,{\sc vii} line at 21.6\AA{}.
Decreasing the O abundance by a factor of 10-100 would certainly improve the fit.
But again, this is not found for the expanding models.
On the contrary, the O abundance was concluded to be too low in for the expanding models.

Furthermore, for both observations the N\,{\sc vi} edges at 22.5\AA{} are a bit too weak and also the N\,{\sc vii} line at 24.8\AA{} and the N\,{\sc vi} line at 28.8\AA{}.
This was also found for the expanding models.

The C\,{\sc vi} absorption edge at 25.3\AA{} is too strong in the model for day 49.7 and a bit too strong in the model for day 39.9.
Also, in the expanding models a too strong C\,{\sc vi} edge was found, however there the effect was clearly stronger for day 39.9 instead of for day 49.7.

\clearpage
\chapter{Conclusion and Outlook} \label{Conclusion}
In this work nova atmospheres are treated as a hybrid type atmosphere consisting of a hydrostatic core with an expanding envelope on top of it, see chapter \ref{sec:NovaStructure}.
In section \ref{sec:PureNLTE} it was shown that these nova models need to be treated as \emph{pure NLTE} models.
This requirement introduces a number of complications that have lead to new methods and improvements to the code.
In section \ref{sec:Grids} it was shown that systematic results are obtained from the new framework.
This is a prerequisite for doing any useful modeling of observed spectra.
However, this is not yet a guarantee for successful modeling of observations.
Therefore, the next step has been to compare the model results with observations in section \ref{sec:Fits}.

It is found, that for all well exposed grating observations available to date models can be found in the test model grid that match the observations quite well, section \ref{sec:ExpandingFits}.
Also, comparison of the observations was done with hydrostatic models that were computed with the new version of code, section \ref{sec:HydrostaticFits}.
Generally, the general shape of the observations can be reproduced reasonably well, but they are by far inferior to the expanding models in matching the detailed spectral features.
A very interesting finding from the two sets of comparisons is that often contradicting conclusions are drawn from the expanding and hydrostatic models about the corrections that need to be made to the abundances in order to improve the fit, sections \ref{sec:DiscussionExpanding} and \ref{sec:DiscussionHydrostatic}.

The new framework appears to be ready and working.
The first results from the new models show remarkable agreement with the observations.
With the final goal to understand the physical nature of novae, this is a promising start in the important task of determination of the physical conditions of nova atmospheres in the late SSS stage.
But a lot of work is yet to be done and numerous improvements and refinements to the models are outstanding.

\subsection{Observational objectives}
\paragraph{Tuning the models}
\begin{itemize}
\item The parameter resolution (stepsize) used in the model grids is yet poor, so when the observation falls right between two adjacent models then the fit is far from optimal.
Therefore, the grid resolution must be increased.

\item All models computed thus far used solar abundances.
Although the precise chemical composition that is `seen' in this late SSS phase of the nova outburst is not yet well determined, it can be expected from measurements of earlier stages in the outburst that solar abundances are not very realistic.
In the process of fine tuning the abundances presumably also the values of the basic atmosphere structure parameters will change, so that this is an iterative process.

\item In this work no attention has been payed to the details of the atomic data.
As shown in previous work on X-ray nova with \phx\ \cite{PetzPhd} the atomic data can have a major influence on the spectrum.
However, the new models of this work significantly differ from the models in the previous work.
Therefore, the conclusions drawn must be tested and revised where needed.

\item The interstellar extinction model used in this work is rather basic.
Newer, more fancy IS models exist in literature \cite{Wilms00}, see also section \ref{sec:ISModel}.
The influence of the newer model must be analyzed.
Given the very large impact of the IS extinction, improvements in this field could have an effect on the fit quality of synthetic spectra to observations.
And, more importantly, in the process of tuning a model to an observation errors from a wrong IS Model could lead to wrong fit parameters.

\item Another important thing to do is to determine the uniqueness of a good fit, or in other words, to determine the `error bars' on the fit parameters.

\end{itemize}

\paragraph{Interrelation with nova evolution models}
The results of good model fits could provide useful constraints for the development of hydrodynamical nova evolution models, like \cite{Starrfield09}.
And those, in turn, could provide constraints for the radiative transport models. Collaboration in this field of work looks very prospective.

\paragraph{Fragmentation of long exposure times}
Many of the grating spectra are exposed well enough to allow for segmentation of the data into smaller time frames.
Each time frame can then be analyzed separately, as done for example in \cite{Ness07}.
Several observations show puzzling short timescale variations in the lightcurve.
Investigating these variations would be interesting to begin with.

\paragraph{New observations}
Although the grating satellites are already coming to age, it is not easy to get observation time for X-ray novae.
The new models presented in this work will provide the means to analyze and interpret the observations.
This is a very important argument in the race against competing research field, and due to the lack of this ability it has become very hard to get allocation time for nova observation.

\paragraph{Other observational data}
Apart from grating spectra, there is a huge amount of low-resolution X-ray spectra from the Swift satellite.
Although these data do not provide much the spectral detail, they could still be valuable, especially when grating spectra are available for the same object.
The grating spectra could be used to determine a good fine tuned model and the Swift spectra could provide the information to determine changes in the atmosphere from that point.

\paragraph{Multi-wavelength band studies}
Since X-ray observations involve some typical difficulties, e.g. the uncertainties introduced by unconstrained IS extinction, extension to multi-wavelength band studies could provide very interesting new insights.
Swift does do simultaneous multi-wavelength band observations already.

\subsection{Theoretical objectives}
Apart from the numerous objectives that involve comparison with observations, there is a number of outstanding theoretical questions that need to be elaborated further.

\begin{itemize}
\item The line profiles used in the models must be sophisticated.
As described in section \ref{sec:BBFrequency} the stark broadening treatment as currently implemented is not accurate for the atmospheric conditions close to the white dwarf boundary or below.

\item In section \ref{sec:SourceInterpolation} a method has been proposed that combines the linear and parabolic interpolation of the source function over the characteristic rays, between the intersection points with the concentric model layers.
As shown, that method will improve the accuracy of the solution of the radiation field.
This has two advantages.
In the first place, this could save idle iterations because the consistency between radiation field and radiative rates will be reached more quickly.
Secondly, this method should reduce the number of ALI iterations required to reach the prescribed accuracy.
On itself the radiation transport part of the code is computationally very quick, but this part of the code must be computed sequentally for all wavelength points.
Therefore, when scaling the code to many processors, the radiative transport is in the end the bottleneck.
Speeding up the small radiative transport piece of code with the new interpolation method would yield a large overall performance gain.

\item The target flux for static atmospheres is constant with radius.
For moving atmospheres the target flux can be transformed to the Lagrangian frame with the equations given in section \ref{sec:ULNonStatic}.
However, the gas also cools when it expands.
This effect has silently been ignored in the derivation of the approximate non-static generalization of the UL temperature correction method.
Also, the $\partial \mathscr{H}_0 / \partial r$ term in \eqref{eq:ULTC6} that vanishes in the static case has not been discussed.
These effects could have a significant impact on the temperature structure, and thus on all model results.
The impact depends on the velocity and density fields of the wind.

\item It was shown in section \ref{sec:ULNLTE} that in NLTE the applicability of the UL temperature correction method is limited on physical grounds.
The method is based on the assumption that the temperature of the gas can be derived from the radiation transport equation.
This can only work if the opacities, that is where the properties of the gas enter, depend significantly on the temperature.
In case of LTE that is given.
In the other extreme of large NLTE effects it is not, and the UL method fails.

In literature a completely different method is known that is based on the thermal balance of electrons \cite{Kubat99}.
This method has already been implemented in the code, but could not yet be tested or used due to time constraints.
For the nova models the temperature in the outer regions is does not have a large influence on the models (see section \ref{sec:ULNLTELimitations} for a discussion), but (also for other types of models) this might yet be interesting to test and compare with the current temperature methods.

\end{itemize}

\subsection{3D models}
The 1D spherical symmetry that has been assumed in this work clearly is a poor approximation given the binarity of a nova system.
However, at present there are no 3D hydrodynamical models for novae that could provide a realistic atmospheric structure for 3D radiation transport.
But the transition to 3D radiative transport models will be important as soon as such 3D structures become available.

\subsection{Application of the new methods to supernova models}
Finally, the framework described in this work has been developed for X-ray novae.
But none of the methods is limited in application to this specific class of objects.
On the contrary, most of the improvements will help as much in other types of models.
Especially supernova models will be very interesting to model given their similarities with nova models: high expansion velocities, large radial extension and strong NLTE effects.
The more, since it has never been possible before to treat all species in pure NLTE, including all relevant ionization stages of Fe, Co and Ni (see sections \ref{sec:RMENewSystem} and \ref{sec:RMENumericalAccuracy}).

\newpage
\vspace*{\fill}
\cleardoublepage

\begin{appendix}
\chapter{Problems with broad line theory} \label{sec:BBWavelength}
Radiation can equivalently be described on the wavelength scale or on the frequency scale.
The radiation field has a certain energy distribution that is interpreted in $\ud\l$ units or in $\ud\nu$ units.
However, one has to be careful with assumptions that are made in one representation or the other.
Especially when dealing with broad atomic line profiles, a number of problems arise.
Before discussing the bigger problem with the Einstein coefficients, line profile symmetries are discussed, as an example of the difference between assumptions in different representations.

\section{Line profile asymmetry}
If a line profile is symmetric around the line center in the wavelength representation, it is not symmetric in the frequency representation, and vice versa.
This is caused by the non-linear transformation between the two representations.
For very narrow line profiles, like Gaussian Doppler profiles, this asymmetry is a negligible effect.
However, for broad profiles, like Lorentz or Voigt profiles (equation \eqref{eq:LorentzProfile} or \eqref{eq:FrequencyVoigt}), this asymmetry becomes apparent.

The frequency-Lorentz profile, equation \eqref{eq:LorentzProfile}, converted to the wavelength representation is given by
\begin{equation}
\begin{split}
 \psi_\l &= \psi_\nu \frac{c}{\l^2}
  = \frac{\Delta \nu_d/\pi}{\left(\frac{c}{\l} - \frac{c}{\l_0}\right)^2 + (\Delta \nu_d)^2} \frac{c}{\l^2} \\
  &= \frac{1}{\pi} \frac{[\Delta \nu_d \l \l_0/c] \frac{\l_0}{\l} }{(\l_0 - \l)^2 + [\Delta \nu_d \l \l_0/c]^2}
  = \frac{1}{\pi} \frac{\Delta \l_d}{(\l_0 - \l)^2 + (\Delta \l_d \l/\l_0)^2} \label{eq:PsiFreqToWavel}
\end{split}
\end{equation}
where in the first step equation \eqref{eq:PsiConversion} was used and in the last step the Lorentz width parameter converts as
\begin{equation} \label{eq:LorentzWidthConversion}
 \Delta \nu_d = \frac{c}{\l_0^2} \Delta \l_d
\end{equation}
This profile is still area normalized but not symmetric around $\l_0$ in the wavelength scale.

The wavelength-Lorentz profile is given by
\begin{equation}
 \psi_\l = \frac{1}{\pi} \frac{\Delta \l_d}{(\l_0 - \l)^2 + (\Delta \l_d)^2}
\end{equation}
and comparison with the converted frequency-Lorentz profile, equation \eqref{eq:PsiFreqToWavel}, shows that they differ by a factor of $\l/\l_0$ in one of the terms in the denominator.
This factor is approximately 1 when close to the line center, but for the line wings these factors significantly influence the profile shape.

\section{The Einstein coefficients $B$}
A very important difference between the wavelength and the frequency representation lies in the definition of the Einstein coefficient for absorption and induced emission $B$.
In short, for broad lines, if $B$ is defined constant in the frequency representation, then it is wavelength dependent in the wavelength representation and vice versa.
In section \ref{sec:BBFrequency}, where the bound-bound rates and opacities were developed, $B$ was defined per intensity in the frequency representation $I_\nu$, equation \eqref{eq:FrequencyBBRate}.
In order to see the difference to $B$ defined per intensity in the wavelength representation $I_\l$ here the expression for the upwards rates $R_{ij}$ are derived in the wavelength representation.

The absorption rate per particle, in analogy to equation \eqref{eq:FrequencyBBRate}, is
\begin{equation} \label{eq:WavelengthBBRate}
 R_{ij} = B_{ij}^\l \int_0^\infty \phi_\l J_\l \,\ud \l
\end{equation}
where $B_{ij}^\l$ is the Einstein coefficient defined as the rate per units of intensity in the wavelength representation.
Using equation \eqref{eq:BBUpRates} the following conversion rule for $B_{ij}$ follows
\begin{equation} \label{eq:EinsteinBConversion}
 B_{ij}^\l = B_{ij} \frac{\l^2}{c}
\end{equation}
This means that if $B_{ij}$ is defined constant with $\l$ then $B_{ij}^\l$ is not, or if $B_{ij}^\l$ is defined constant then $B_{ij}$ is not.

However, the Einstein relations derived in the wavelength representation, in analogy to equation \eqref{eq:EinsteinRelationDeriv}, become
\begin{align}
 A_{ji} &= \frac{2hc^2}{\l_0^5} B_{ji}^\l  \label{eq:WavelengthEinsteinRel1}\\
 B_{ji}^\l &= g_i/g_j B_{ij}^\l \label{eq:WavelengthEinsteinRel2}
\end{align}
Note that $A_{ji}$ is independent of the representation in which the radiation field is described.
Using equation \eqref{eq:EinsteinRel1} a different conversion rule for $B_{ij}$ follows
\begin{equation} \label{eq:EinsteinBConversionAlternative}
 B_{ij}^\l = B_{ij} \frac{\l_0^2}{c}
\end{equation}

The contradiction between equations \eqref{eq:EinsteinBConversion} and \eqref{eq:EinsteinBConversionAlternative} results from the assumption of narrow lines made in the derivation of the Einstein relations.
If lines are infinitely narrow, then the contradiction is resolved, as the rates and opacities are only evaluated at the point $\l = \l_0$.
But for broad lines, obviously, the assumption of narrow lines made in the derivation of the Einstein relations is not consistent with the conversion of the frequency representation result to the wavelength representation.
Since different representations of the radiation field are physically equivalent, and in the conversion between them no additional assumptions are made, the conclusion drawn here is that for broad lines the Einstein relations are not valid.

In some literature, e.g. \cite{Mihalas78}, the Einstein relations are derived as
\begin{align}
 A_{ji} &= \frac{2h\nu^3}{c^2} B_{ji} \label{eq:EinsteinRel1Wrong}\\
 B_{ji} &= g_i/g_j B_{ij} \label{eq:EinsteinRel2Wrong}
\end{align}
However, this is inconsistent with using the Boltzmann level distribution to write the ratio of the TE level populations $n_i$ and $n_j$ in equation \eqref{eq:EinsteinRelationDeriv} as
\begin{equation}
 \frac{n_i^\TE}{n_j^\TE} = \frac{g_i}{g_j} e^{h\nu_{ij}/kT} \neq \frac{g_i}{g_j} e^{h\nu/kT}
\end{equation}
The reason why the first Einstein relation is written in the form of equation \eqref{eq:EinsteinRel1Wrong} is to impose a uniformity in the expressions of opacities and rates between the bound-bound and bound-free cases.
For the latter indeed possesses the factor $\frac{2h\nu^3}{c^2}$.

\section{Wavelength vs. frequency representation} \label{sec:WavelengthVsFrequencyRepresentation}
The physically right way to deal with this problem would be to separately derive expressions for the $A$ and $B$ coefficients, instead of relating them using equilibrium arguments.
However, these would depend on the precise physical conditions of the system.
Generally, they should have to be derived from a fully quantum mechanical description.
But such treatment is already quite complicated for Hydrogen, see \cite{Mihalas78}, let alone for much larger atoms as oxygen or even Iron.
This goes far beyond the scope of this work.
And data of this quality are not available at the present time.

But even if the Einstein relations are not exactly right, they might still give useful approximations.
In this work, the two variants of equations \eqref{eq:EinsteinBConversion} and \eqref{eq:EinsteinBConversionAlternative} were examined.

\subsection{Frequency variant}
The first variant was derived in the frequency representation in section \ref{sec:BBFrequency}.
It uses frequency-Voigt profile functions, equation \eqref{eq:FrequencyVoigt}.
Those are symmetric around the line center on the frequency scale.
The rates and opacities expressed in the $A$ coefficients, that are independent of the choice of representation, for the frequency variant are
\begin{align}
 R_{ij} &= A_{ji} \frac{\l_0^3}{2hc} \frac{g_j}{g_i} \int_0^\infty \phi_\l \frac{\l^2}{c} J_\l \,\ud\l \label{eq:UpFrequencyRates}\\
 R_{ji} &= A_{ji} \left(1 + \frac{\l_0^3}{2hc} \int_0^\infty \psi_\l \frac{\l^2}{c} J_\l \,\ud\l \right) \label{eq:DownFrequencyRates}\\
 \eta_{ij,\l} &= \frac{hc}{4\pi\l_0} A_{ji} n_j \psi_\l \label{eq:EtaFrequencyVariant}\\
 \chi_{ij,\l} &= \frac{hc}{4\pi\l_0} A_{ji} \frac{\l_0^3 \l^2}{2hc^2} \frac{g_j}{g_i} \left( n_i\phi_\l - n_j \frac{g_i}{g_j} \psi_\l \right) \label{eq:ChiFrequencyVariant}
\end{align}

\subsection{Wavelength variant}
The second variant is based on the wavelength representation of the radiation field.
The Einstein $B$ coefficients are defined as the rates per units of intensity in the wavelength representation.
In this variant wavelength-Voigt profile functions are assumed, which are symmetric around the line center on the wavelength scale.
The expression for the wavelength-Voigt profile are similar to the frequency-Voigt.
The dispersion width parameter converts as equation \eqref{eq:LorentzWidthConversion}, and similarly the Doppler width
\begin{equation}
 \Delta \nu_D = \frac{c}{\l_0^2} \Delta \l_D
\end{equation}
The $u$ parameter becomes the dimensionless wavelength offset instead of the dimensionless frequency offset, but the Voigt damping parameter $a$, equation \eqref{eq:VoigtDampingParameter}, is equivalent in both representations.
\begin{align}
 \psi(a,u) = \psi(a,u) &= \frac{H(a,u)}{\Delta \l_D \sqrt{\pi}} \label{eq:WavelengthVoigt1} \\
 u &= \frac{\l - \l_0}{\Delta \l_D} \\
 a &= \frac{\Delta \l_d}{\Delta \l_D} = \frac{\Delta \nu_d}{\Delta \nu_D} \label{eq:WavelengthVoigt3}
\end{align}
The rates and opacities expressed in the $A$ coefficients, that are independent of the choice of representation, for the wavelength variant are
\begin{align}
 R_{ij}^\l &= A_{ji} \frac{\l_0^5}{2hc^2} \frac{g_j}{g_i} \int_0^\infty \phi_\l J_\l \,\ud\l \label{eq:UpWavelengthRates}\\
 R_{ji}^\l &= A_{ji} \left(1 + \frac{\l_0^5}{2hc^2} \int_0^\infty \psi_\l J_\l \,\ud\l \right) \label{eq:DownWavelengthRates}\\
 \eta_{ij,\l}^\l &= \frac{hc}{4\pi\l_0} A_{ji} n_j \psi_\l \label{eq:EtaWavelengthVariant}\\
 \chi_{ij,\l}^\l &= \frac{hc}{4\pi\l_0} A_{ji} \frac{\l_0^5}{2hc^2} \frac{g_j}{g_i} \left( n_i\phi_\l - n_j \frac{g_i}{g_j} \psi_\l \right) \label{eq:ChiWavelengthVariant}
\end{align}

\subsection{Flux comparison}
\begin{figure}
 \centerline{\includegraphics[width=\textwidth]{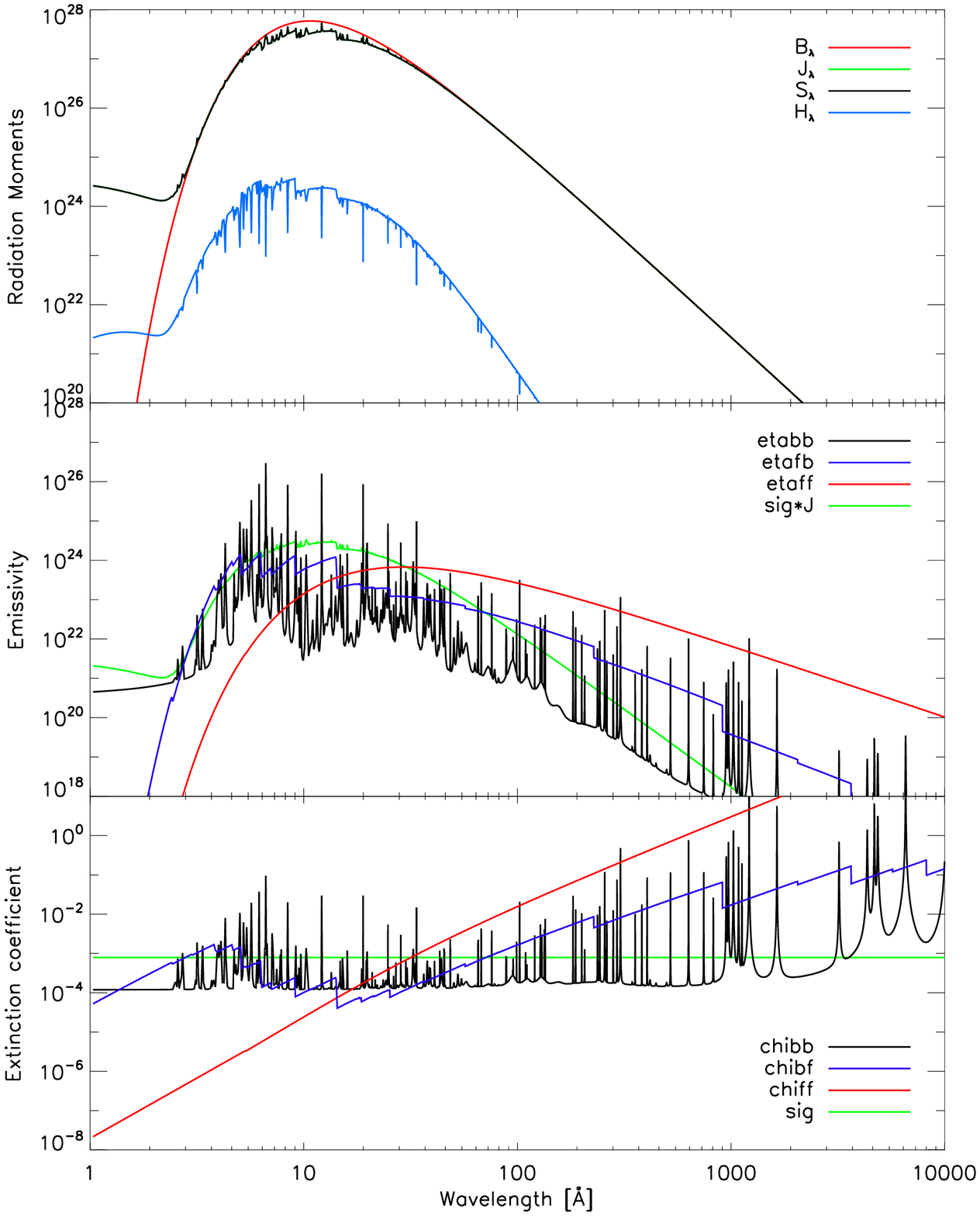}}
 \caption{
  The top graph shows the source function $S_\l$, the related radiation moments $J_\l$ and $H_\l$ and the Planck function $B_\l$ for a deep layer in the atmosphere, layer 120 of 128.
  In the two lower graphs the opacities are plotted, which generate the source function, equations \eqref{eq:DetailedEmissivity} and \eqref{eq:DetailedExtinction}.
  Here the wavelength variant of the bound-bound opacities, equations \eqref{eq:ChiWavelengthVariant} and \eqref{eq:EtaWavelengthVariant} are used.
  \newline
  This variant causes trouble in the inner parts of the atmosphere.
  There the expected approximate Planckian shape of $S_\l$ is not reproduced.
  This is caused by the fact that the blue tails of the line wings are so strong that the bound-free opacities no longer dominate the continuum extinction.
  Furthermore, the bulk of the flux in deep layers is at short wavelengths and is therefore efficiently absorbed.
  Radiation transport shows, that this lowers the integrated Eddington flux $H$ by a factor of 4 in comparison with the frequency variant.
  \newline
  In figure \ref{fig:InFreqRadiation} the same plots are shown for the frequency variant.
 } \label{fig:InWavelRadiation}
\end{figure}

If the wavelength variant is used then a number of problems arise in the deeper atmosphere layers.
The extinction of far line wings in the wavelength variant is almost independent of the wavelength, equation \eqref{eq:ChiWavelengthVariant}, since there the profile function is very flat, see figure \ref{fig:InWavelRadiation}.
\begin{figure}
 \centerline{\includegraphics[width=\textwidth]{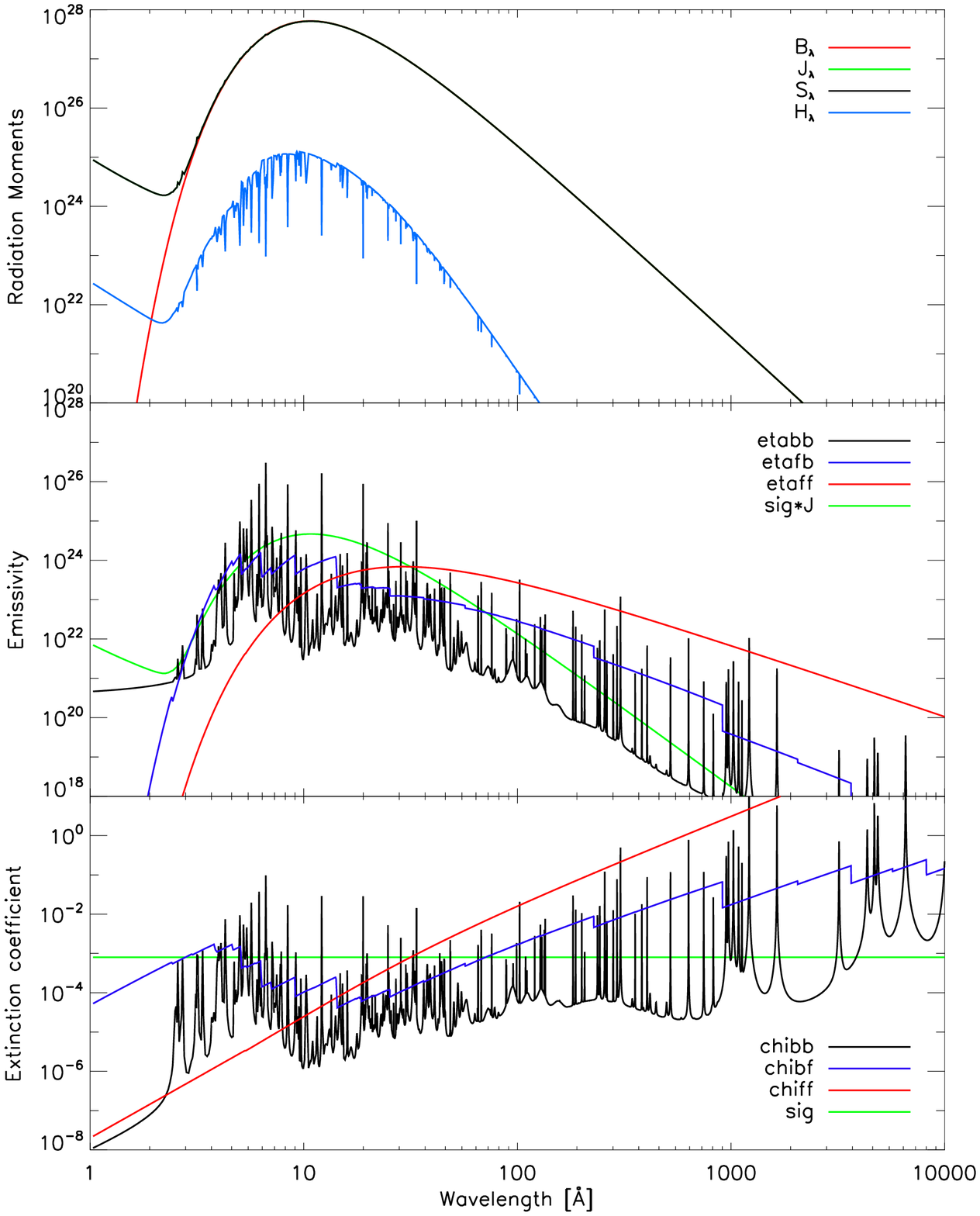}}
 \caption{
  This figure shows the same quantities as figure \ref{fig:InWavelRadiation} with the only difference that the bound-bound opacities are computed in the frequency variant, equations \eqref{eq:ChiFrequencyVariant} and \eqref{eq:EtaFrequencyVariant}.
  \newline
  This variant is superior to the wavelength variant.
  It produces an accurate Planckian source function above 3\AA, where the bulk of the radiative energy lies.
  Below 3\AA{} the wings of the bound-bound emission profiles are stronger than the bound-free emissivity, leading to a departure from the pure LTE source function.
  The only wavelength dependence in the emissivity comes from the profile functions.
  So if this departure of the source function is wrong, then the profile functions $\psi_\l$ are.
  The same effect occurs in the wavelength variant, of figure \ref{fig:InWavelRadiation}.
  But since the radiation field is weak below 3\AA{} this does not have a significant influence on the models.
 } \label{fig:InFreqRadiation}
\end{figure}
\label{pag:StrongBlueLineWing}
The line profiles have wavelength symmetry, so they easily extend to small wavelengths\footnote
{For a line with wavelength symmetry and the center at $\l_0=100$\AA{} extending the line from 10\AA{} to 1\AA{} makes the wing just a factor of 0.1 broader, whereas for a line with frequency symmetry and the center at $\nu_0=c/(100{\textrm \AA})$ extending the line from $c/10$\AA{} to $c/1$\AA{} makes the wing a factor of 10 broader.
}.
At the same time, the energy of the radiation field per $\ud\l$ increases linearly with decreasing $\l$, so that the extincted energy from the far blue line wings increases linearly towards short wavelengths.
In the deep layers of the atmosphere, the radiation field is approximately Planckian, and as the temperatures are high, the peak values are at short wavelengths.
Because of the efficient far wing extinction in this wavelength range the bulk of the flux is strongly absorbed.
Comparing the same layer in both variants shows that this effect decreases the integrated Eddington flux $H$ in the wavelength variant by a factor of 4.

\subsection{Mean radiation field comparison}
Since the bound-bound contribution to the source function is only planckian in the line center, see section \ref{sec:BBFrequency}, the bound-free opacities are responsible for generating a planckian continuum source function.
But in the wavelength variant the bound-free opacities do not dominate the bound-bound opacities of the line wings at short wavelengths, because the blue line wings are so strong.
Thus the source function in the continuum does not accurately approximate the Planck function, and so the mean intensity does not.
In LTE the radiation field $I_\l$ is not exact planckian since then the Eddington flux $H_\l$ would be zero, like described in section \ref{sec:LTE}.
So in approximate LTE indeed deviations may exist.
However, in the mean intensity $J_\l$ the deviations should be small, much smaller than those produced in the wavelength variant.
In the frequency variant this problem does not occur, shown in figure \ref{fig:InFreqRadiation}.
This is a strong indication that the wavelength variant of the expressions for the opacities, equations \eqref{eq:ChiFrequencyVariant} and \eqref{eq:EtaFrequencyVariant}, are unusable.

\subsection{Departure coefficient comparison}
\begin{figure}
 \centerline{ \includegraphics[height=\textwidth,angle=90]{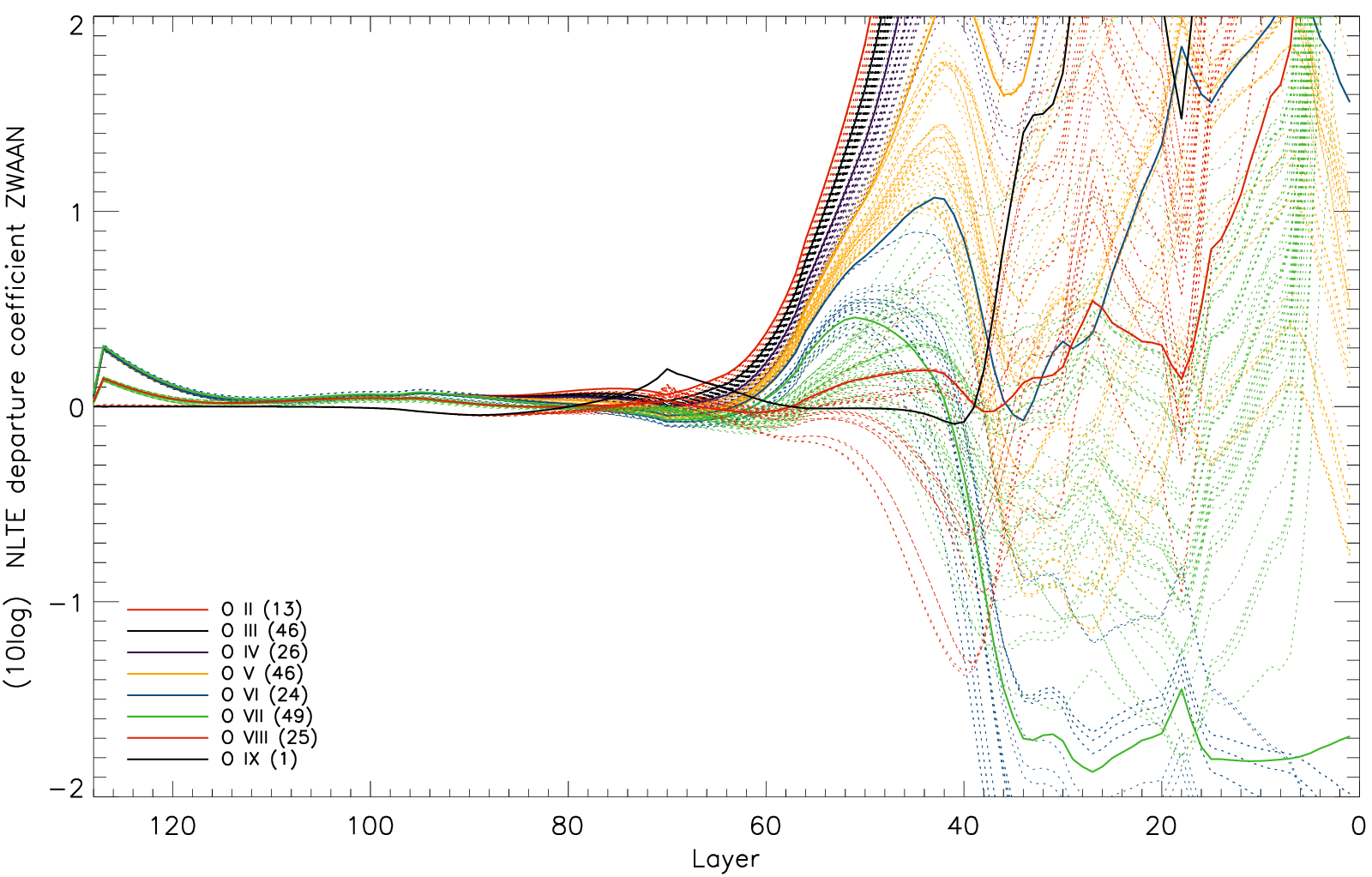}}
 \centerline{ \includegraphics[height=\textwidth,angle=90]{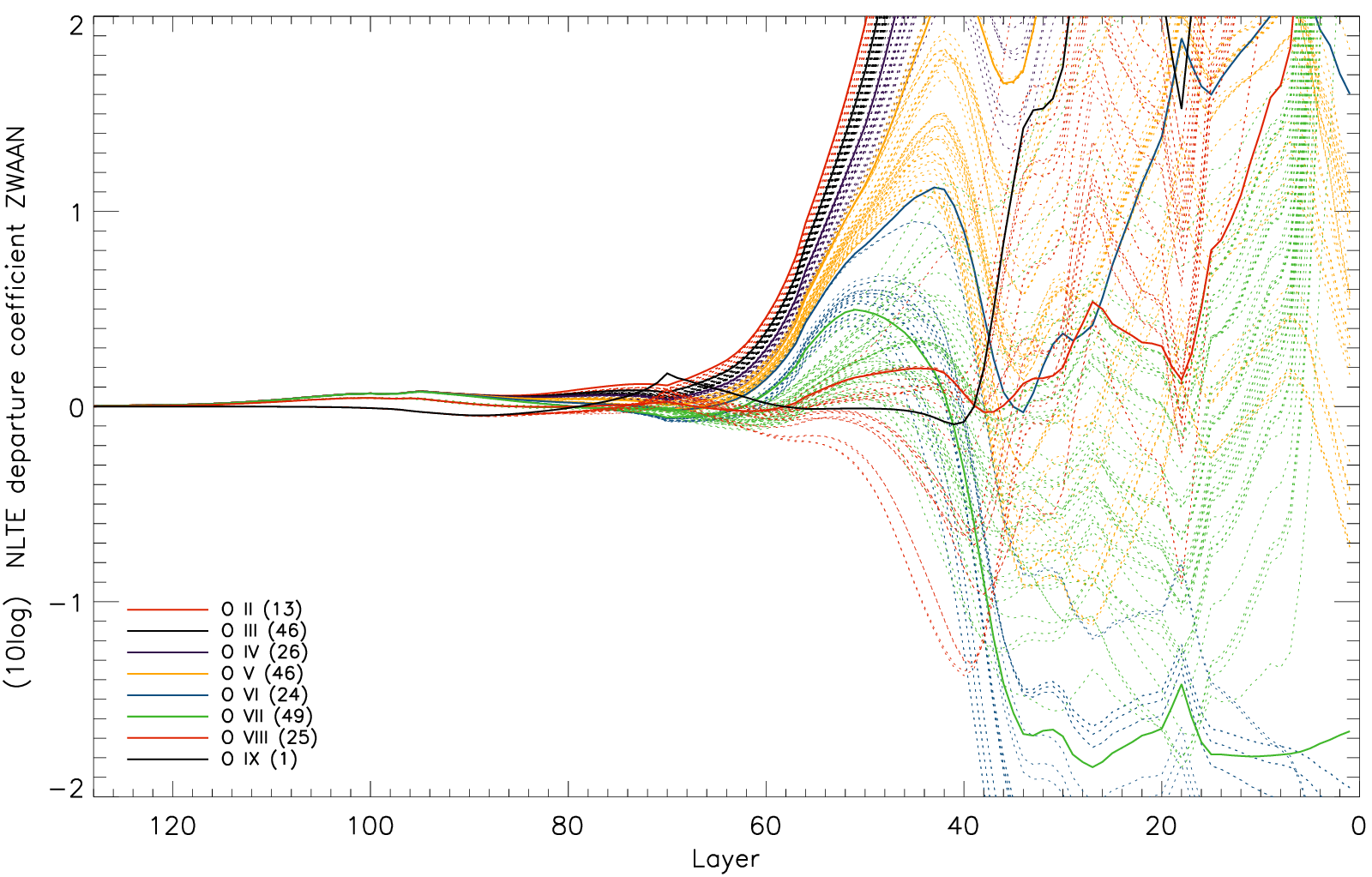}}
 \caption{
  These graphs show the NLTE departure coefficients $b_i$ on a logarithmic scale for all levels of oxygen against the layer number in the atmosphere model.
  The layer number increases going inwards.
  The number of levels in each ionization stage are written in parentheses.
  \newline
  The wavelength variant (top) of the bound-bound rates and opacities, equations \eqref{eq:ChiWavelengthVariant} and \eqref{eq:EtaWavelengthVariant}, are problematic in the inner regions of the atmosphere.
  There, the departure coefficients do not converge to $b_i = 1$.
  That is the LTE approximation that holds better the deeper the layer in the atmosphere is.
  \newline
  The frequency variant, equations \eqref{eq:ChiFrequencyVariant} and \eqref{eq:EtaFrequencyVariant}, does not show this problematic behavior and for this and other reasons, see the text, is superior to the other variant.
 } \label{fig:InWavelRates}
\end{figure}

The inaccuracy in the mean intensity, together with the wavelength variant of the rate equations \eqref{eq:UpFrequencyRates} and \eqref{eq:DownFrequencyRates}, leads to radiative rates that bring the population numbers out of their LTE balance in the innermost regions of the atmosphere.
This is shown by the NLTE departure coefficients $b_i$ for all levels of oxygen in figure \ref{fig:InWavelRates}.
With the wavelength variant, when going from the outer layer 1 inwards the departure coefficients converge towards $b_i = 1$ up till around layer 100.
From there on inwards they diverge again.
With the frequency variant, the $b_i$ converge smoothly in the deepest regions of the atmosphere.

\subsection{Conclusion}
From the three problems that go along with the wavelength variant it is concluded that the description derived from the frequency representation and transferred to the wavelength scale is superior.
This means that the Einstein coefficients $B$ favorably are interpreted as constant in the frequency domain, so that they become wavelength dependent after conversion to the wavelength domain.

\section{Frequency vs. original variant}
Originally in \phx\ neither of the two variants described in the previous section were used.
In the original variant used in \phx\ the $B_{ij}^\l$ are independent of wavelength and defined as the rates per unit of Intensity in the wavelength representation.
Thus far, the approach is analog to the Wavelength variant of the previous section.

But in the expressions for the opacities and rates the $A_{ji}^o$ and $B_{ji}^o$ (where the $o$ stands for 'original') are derived from the $B_{ij}^\l$ using another form of Einstein relations
\begin{align}
 A_{ji}^o &= \frac{2hc^2}{\l^5} B_{ji}^o  \label{eq:PhxOriginalEinsteinRel1}\\
 \begin{split}
  B_{ji}^o &= \frac{g_i}{g_j} e^{\frac{hc}{\l_0 kT}} e^{-\frac{hc}{\l kT}} B_{ij}^\l
            = \frac{n_i^\LTE}{n_j^\LTE} e^{-\frac{hc}{\l kT}} B_{ij}^\l
            = \frac{n_i^*}{n_j^*} e^{-\frac{hc}{\l kT}} B_{ij}^\l \label{eq:PhxOriginalEinsteinRel2}
 \end{split}
\end{align}
This form follows from equation \eqref{eq:EinsteinRelationDeriv1} when the summations over all atomic energy states $i$ and $j$ of all atomic species are neglected.
This means that, instead of assuming that all interactions of the radiation with the matter together reproduce the TE radiation field, each transition on itself reproduces the TE field.
This requirement is stronger than the requirement of detailed radiative balance.

The exponential terms in \eqref{eq:PhxOriginalEinsteinRel2} originate from using the Boltzmann law, which is exactly the difference to equation \eqref{eq:EinsteinRel2Wrong}.
Note that both $A_{ji}^o$ and $B_{ji}^o$ are strongly wavelength dependent.
In the line center this form of the Einstein relations yields the wavelength variant of equations equation \eqref{eq:WavelengthEinsteinRel1} and \eqref{eq:WavelengthEinsteinRel2}.

As constant $A_{ji}$ coefficients are the line strength input data to \phx\ these must be converted to $B_{ij}^\l$.
In order to do this the $A_{ji}$ are interpreted as the line center $A_{ji}^o$ values and converted to $B_{ij}^\l$ using the Einstein relations for the line centers.

In this variant only the $B_{ij}^\l$ are constant and therefore these are the appropriate coefficients to express the rates and opacities in
\begin{align}
 R_{ij}^o &= B_{ij}^\l \int_0^\infty \phi_\l J_\l \,\ud\l \label{eq:UpPhxOriginalRates}\\
 R_{ji}^o &= B_{ij}^\l \frac{g_i}{g_j} e^{hc/\l_0 kT} \int_0^\infty \left( \frac{2hc^2}{\l^5} + J_\l \right) \psi_\l e^{-hc/\l kT} \,\ud\l \label{eq:DownPhxOriginalRates}\\
 \eta_{ij,\l}^o &= \frac{hc}{4\pi\l_0} B_{ij}^\l n_j \psi_\l \frac{2hc^2}{\l^5} \frac{n_i^*}{n_j^*} e^{-hc/\l kT} \label{eq:EtaPhxOriginalVariant}\\
 \begin{split}
 \chi_{ij,\l}^o &= \frac{hc}{4\pi\l_0} B_{ij}^\l \left( \phi_\l n_i - \psi_\l n_j \frac{n_i^*}{n_j^*} e^{-hc/\l kT} \right) \\
    &= \frac{hc}{4\pi\l_0} B_{ij}^\l n_i^* \left( \phi_\l b_i^* - \psi_\l b_j^* e^{-hc/\l kT} \right) \label{eq:ChiPhxOriginalVariant}
 \end{split}
\end{align}
In this variant the wavelength-Voigt profiles are used, equations \eqref{eq:WavelengthVoigt1} to \eqref{eq:WavelengthVoigt3}, as originally in \phx.

The single line source function for this original variant becomes
\begin{equation} \label{eq:PhxOriginalSLine}
 S_{ij,\l}^{\rm bb,o} = \frac{2hc^2}{\l^5} \left( \frac{b_i^*}{b_j^*} \frac{\phi_\l}{\psi_\l} e^{hc/\l kT} -1 \right)^{-1}
\end{equation}
When the Boltzmann distribution holds and complete redistribution is assumed then this reduces to the Planck function.

\begin{figure}
 \centerline{ \includegraphics[width=\textwidth]{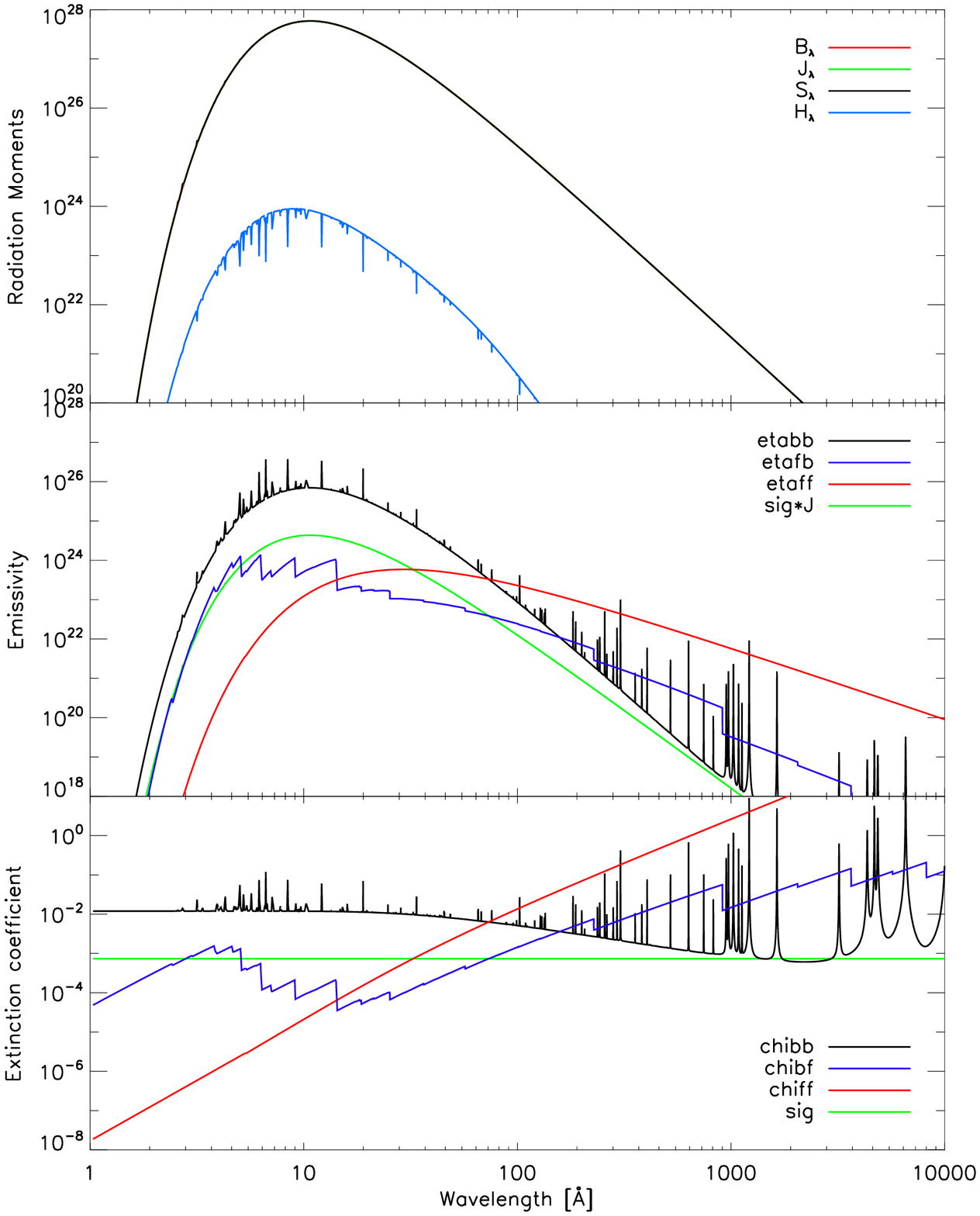}}
 \caption{
  This figure shows the same quantities as figures \ref{fig:InWavelRadiation} and \ref{fig:InFreqRadiation} with the only difference that the bound-bound opacities are computed in the original \phx\ variant, equations \eqref{eq:ChiPhxOriginalVariant} and \eqref{eq:EtaPhxOriginalVariant}.
  \newline
  It produces an almost perfect Planckian source function for the whole wavelength range, because even the single line source functions, equation \eqref{eq:PhxOriginalSLine}, are Planckian as $b_i^* \approx 1$.
  The integrated Eddington flux $H$ is lower by a factor of 12 compared to the frequency variant.
  The reason is that the almost perfect Planckian intensities cancel efficiently in the second moment of the radiation field.
  \newline
  In this variant wavelength-Voigt profiles are used, which feature strong blue line wings.
  These blue wings dominate the short wavelength range of the extinction coefficient.
  Due to the strong wavelength dependence of $A_{ji}^o$ the bound-bound emissivity looks Planckian.
 } \label{fig:InWeirdRadiation}
\end{figure}

The results with this variant for the same layer (deep in the atmosphere) as in figures \ref{fig:InWavelRadiation} and \ref{fig:InFreqRadiation} are shown in figure \ref{fig:InWeirdRadiation}.
The opacities (emission and extinction) are dominated by the bound-bound opacities shortwards of 100\AA.
This is partly due to the wavelength-Voigt profiles.
Like discussed on page \pageref{pag:StrongBlueLineWing} these profiles exhibit a strong blue line wing.
The mean intensity $J_\l$ accurately reproduces the Planck function.
However, the accuracy is so high that in the second moment of the radiation field the accurate intensities efficiently cancel.
For this original \phx\ variant the integrated Eddington flux $H$ for this layer is a factor of 12 smaller than for the frequency variant.
On itself this is not yet a big problem.

\begin{figure}
 \centerline{ \includegraphics[width=\textwidth]{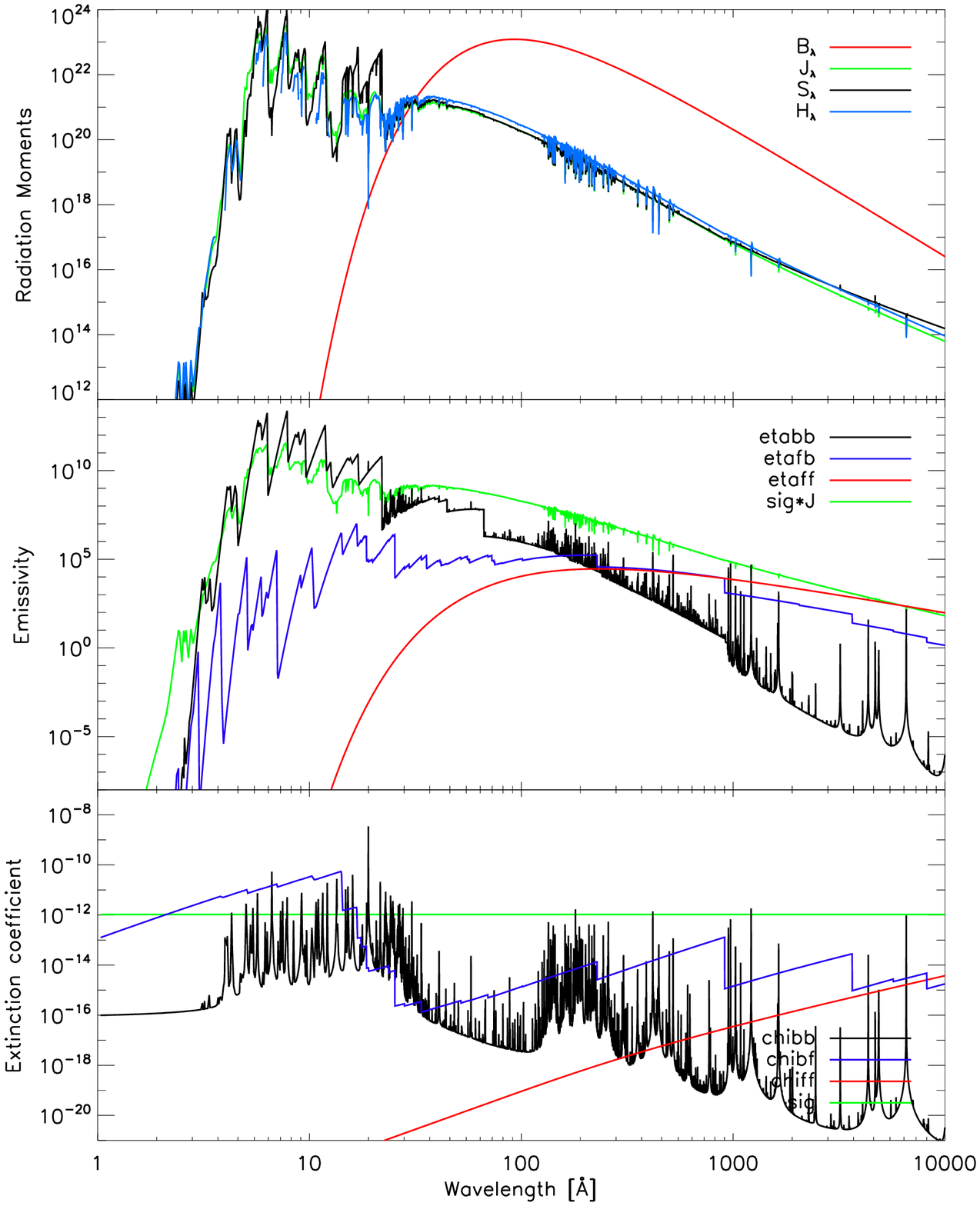}}
 \caption{
  This figure shows the same model as figure \ref{fig:InWeirdRadiation} for an outer layer in the atmosphere (layer 20) on a logarithmic wavelength scale.
  The bound-bound opacities are computed in the original \phx\ variant, equations \eqref{eq:ChiPhxOriginalVariant} and \eqref{eq:EtaPhxOriginalVariant}.
  \newline
  Jumps in the emissivity occur due to omitting transitions that would exhibit lasering.
  Presently, lasering cannot yet be treated in the radiative transport.
  At the jump points the line extinction coefficient smoothly goes to zero, so the run of the extinction is smooth.
  The artifacts in the emissivity also leave their signatures in the flux.
  Comparison of the flux in this layer with the frequency variant, figure \ref{fig:OutFreqRadiation}, shows that the frequency variant is preferable.
 } \label{fig:OutWeirdRadiation}
 
\end{figure}

Big problems with this variant arise in the outer parts of the atmosphere.
Figure \ref{fig:OutWeirdRadiation} shows the same model as in figure \ref{fig:InWeirdRadiation} for a outer layer of the atmosphere.
\begin{figure}
 \centerline{ \includegraphics[width=\textwidth]{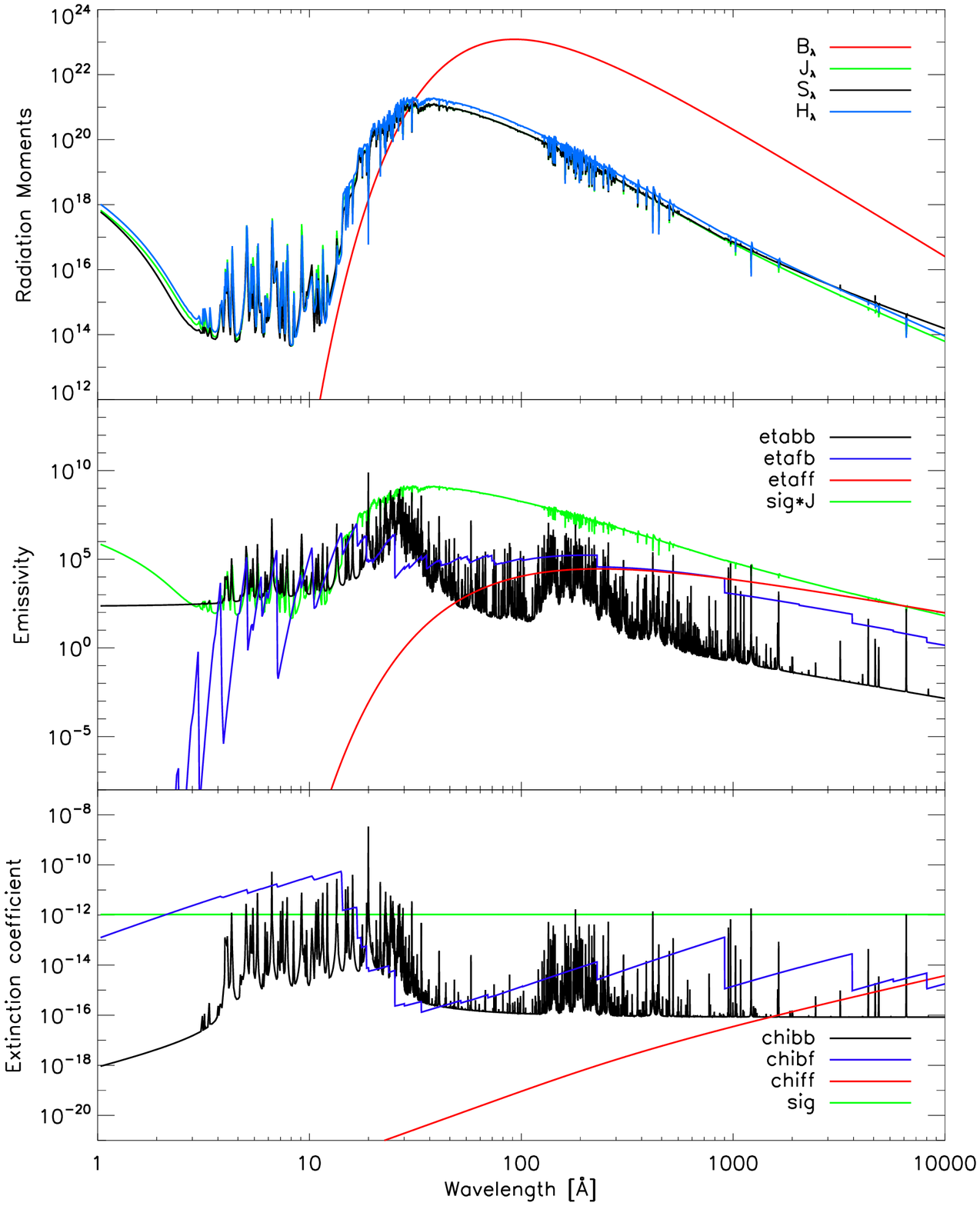}}
 \caption{
  This figure shows the same quantities as figure \ref{fig:OutWeirdRadiation} for the same atmospheric layer on a logarithmic wavelength scale.
  The models are identical except that here the bound-bound opacities are computed in the frequency variant, equations \eqref{eq:EtaFrequencyVariant} and \eqref{eq:ChiFrequencyVariant}.
  \newline
  The main differences lie in the shape of the emissivity and the flux.
  One disadvantage of the frequency variant is that due to unrealistic emission line profiles the emissivity does not fall of steep enough in the short wavelength range.
  This causes the mean intensity and the flux to rise again towards shorter wavelengths, which is clearly unphysical.
  However, the influence of this problem on the model is negligible (see the text).
 } \label{fig:OutFreqRadiation}
\end{figure}
When for a radiative transition the departure coefficient of the upper level is larger than of the lower level $b_j^* > b_i^*$ and the wavelength increases, then at some wavelength the extinction, equation \eqref{eq:ChiPhxOriginalVariant}, becomes negative.
Negative extinction on itself is not unphysical, this phenomenon is called lasering, see \cite{Rutten95}.
However, lasering effects can cause the (total) source function to become negative.
A negative source function in the radiative transport formalism can lead to negative intensities along a characteristic ray.
This is clearly unphysical.
And numerically the operator splitting method does not converge.
Therefore, if for a transition lasering would occur, the opacities of the transition are omitted.
The extinction goes to zero smoothly with wavelength, but the emissivity does not at the same time.
Omitting the transition longwards of the wavelength from which lasering occurs leads to jumps in the bound-bound emissivity.
These jumps are the side effect of not treating lasering in the radiation transport.
The physically right way to deal with this problem would be to find a way to treat the lasering effect in the radiative transport.
This has not yet been accomplished in the scope of this work, but is suspended for future work.

Alternatively, one could omit all opacity of lines with $b_j^* > b_i^*$.
However, this condition is much stronger than the condition $n_j/g_j > n_i/g_i$ that is used in the frequency variant.
In LTE higher levels are energetically unfavored, described by the exponential damping term in the Boltzmann law, equation \eqref{eq:Boltzmann}.
Experiment shows that omitting such lines throws out a lot of lines that are treated in the Frequency variant, for example the vast majority of lines below 20\AA.

The frequency variant in the outer layers, figure \ref{fig:OutFreqRadiation}, shows the same artifact at small wavelengths as in inner layers, figure \ref{fig:InFreqRadiation}. 
The emissivity does not fall off steeply in the Wien tail.
In the frequency variant the only wavelength dependence in the emissivity comes from the profile function.
However, the overall effect of this problem on the model is small.
The rates are not affected significantly, because of the $\l^2/c$ term being small for short wavelengths (equations \eqref{eq:BBUpRates} and \eqref{eq:BBDownRates}).
And also the influence on the opacity integrals of the temperature correction (see section \ref{sec:ULTC}) is negligible.

\chapter{First non-solar fits to V4743 Sgr}
The models presented here are first-go, non-solar abundance fits to the five V4743 Sgr grating spectra that were also shown in section \ref{sec:Fits}.
Their computation just finished the very same day on which this work was finalized.
Today, September 7th 2009, the new supercomputer in the Konrad-Zuse-Zentrum in Berlin became available and was yet completely empty, so that 7x480 processors could be reserved for a few hours even before the opening was officially announced.
With this total of 3360 processors, the last stage (stage 3, see section \ref{sec:ModelingSteps}) of a whole model grid with $v_\infty = 2400$ km/s could be computed at once.

The chosen abundances are solar, except for N and O, which are overabundant by a factor of 100 and 10 respectively.
These abundances are a first endeavor for non-solar abundances, based on the compositions obtained from nova evolution models for the ejecta \cite{Starrfield09}.

As yet, these abundances N 100x solar and O 10x solar have not yet been tuned in any way.
It is the first and only non-solar composition that was computed.
Nevertheless, the models show a very interesting improvement of the fits to the observations with respect to the solar abundance models shown in section \ref{sec:Fits}.

Figures \ref{fig:V4743MarN20O10} to \ref{fig:V4743FebN20O10} show the following observations.
\begin{itemize}
 \item figure \ref{fig:V4743MarN20O10}: V4743 Sgr, March 2003, day 180, LETGS
 \item figure \ref{fig:V4743AprN20O10}: V4743 Sgr, April 2003, day 196, RGS
 \item figure \ref{fig:V4743JulN20O10}: V4743 Sgr, July 2003, day 302, LETGS
 \item figure \ref{fig:V4743SepN20O10}: V4743 Sgr, September 2003, day 371, LETGS
 \item figure \ref{fig:V4743FebN20O10}: V4743 Sgr, February 2004, day 526, LETGS
\end{itemize}
Just like for the solar abundance models the terminal velocity is $v_\infty = 2400$ km/s, and $\beta = 1.5$.  
The the non-solar models are plotted in green along with the solar models, for comparison, in red.

The most significant improvements from the non-solar abundances with 100x N and 10x O relative to solar, is the stronger N VII ionization edge at 22.5AA{} and absorption lines at 24.8AA{} (N\,{\sc vii}) and 21.6AA{} (O\,{\sc vii}).

\clearpage
\begin{landscape}
\begin{figure}
 \centerline{\includegraphics[width=\textwidth,angle=90]{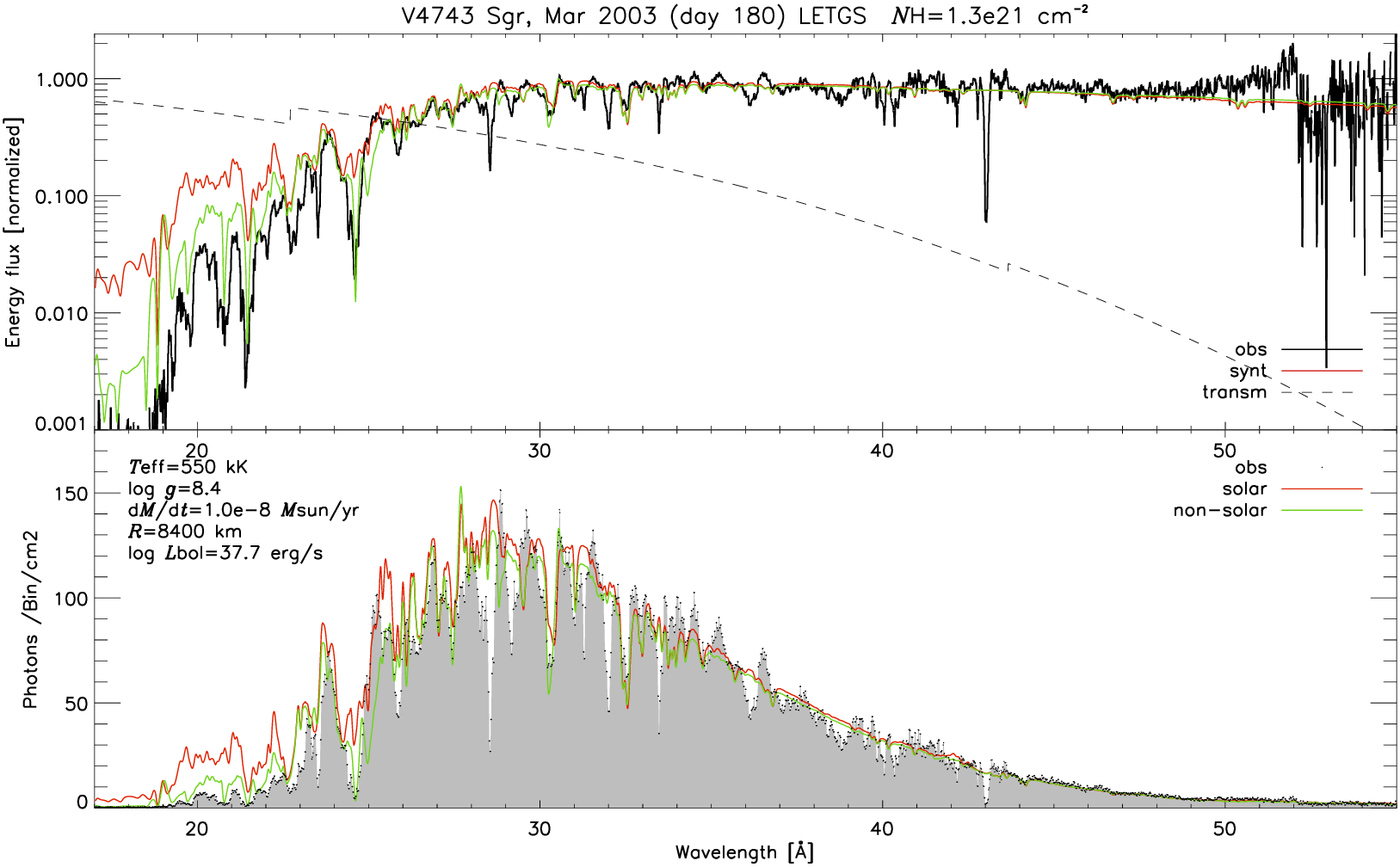}}
 \caption{ \label{fig:V4743MarN20O10} Expanding model}
\end{figure}
\begin{figure}
 \centerline{\includegraphics[width=\textwidth,angle=90]{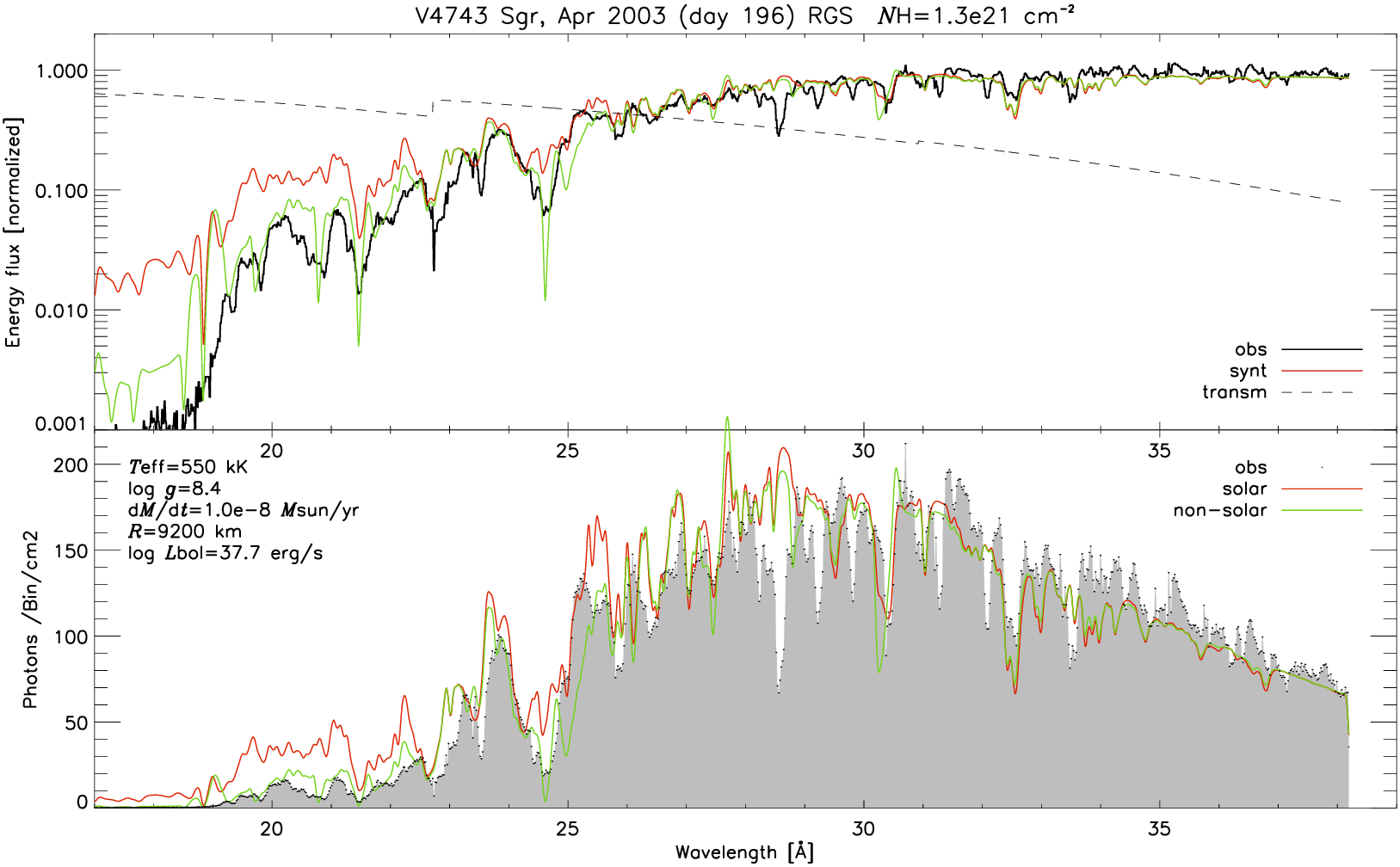}}
 \caption{ \label{fig:V4743AprN20O10} Expanding model}
\end{figure}
\begin{figure}
 \centerline{\includegraphics[width=\textwidth,angle=90]{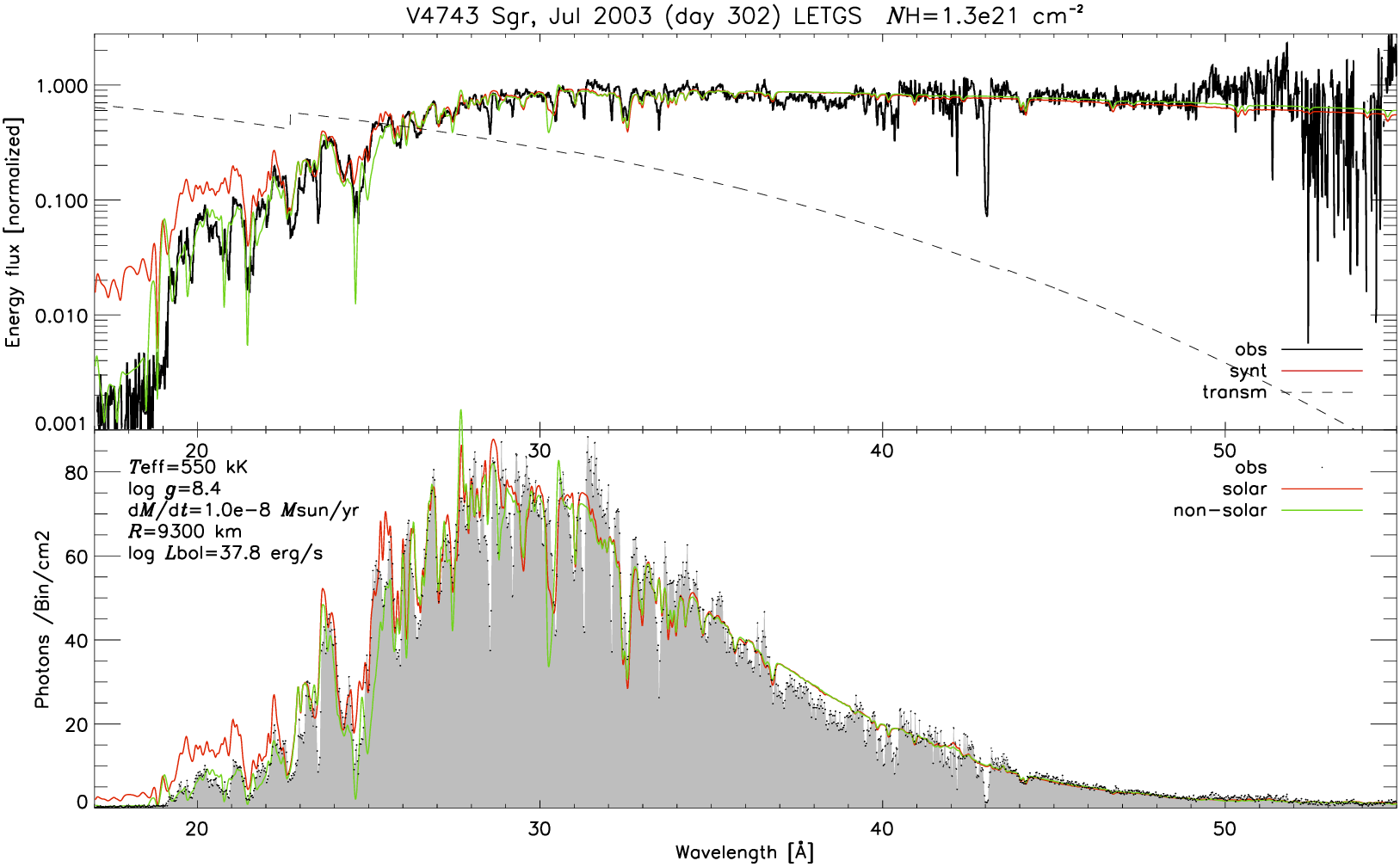}}
 \caption{ \label{fig:V4743JulN20O10} Expanding model}
\end{figure}
\begin{figure}
 \centerline{\includegraphics[width=\textwidth,angle=90]{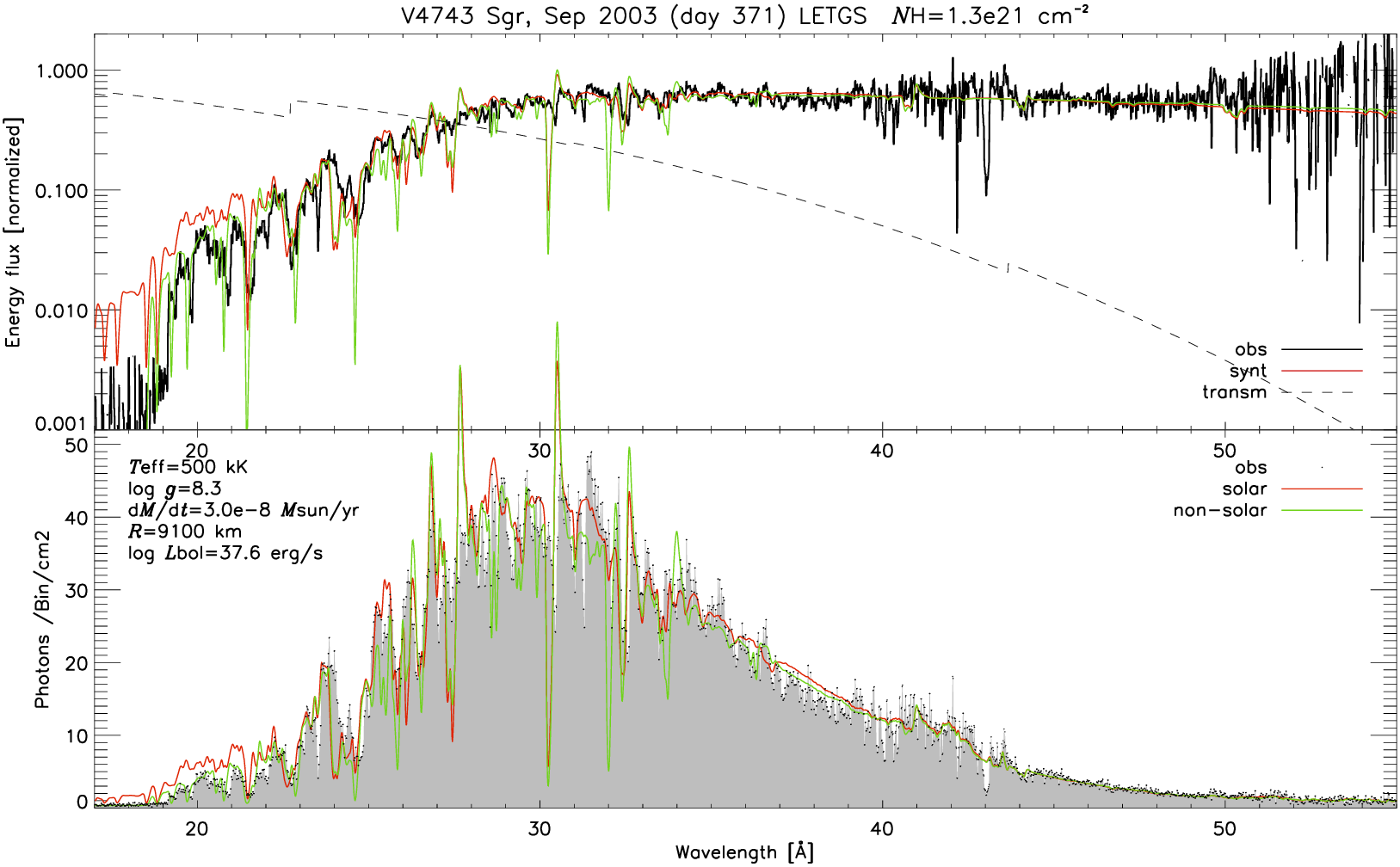}}
 \caption{ \label{fig:V4743SepN20O10} Expanding model}
\end{figure}
\begin{figure}
 \centerline{\includegraphics[width=\textwidth,angle=90]{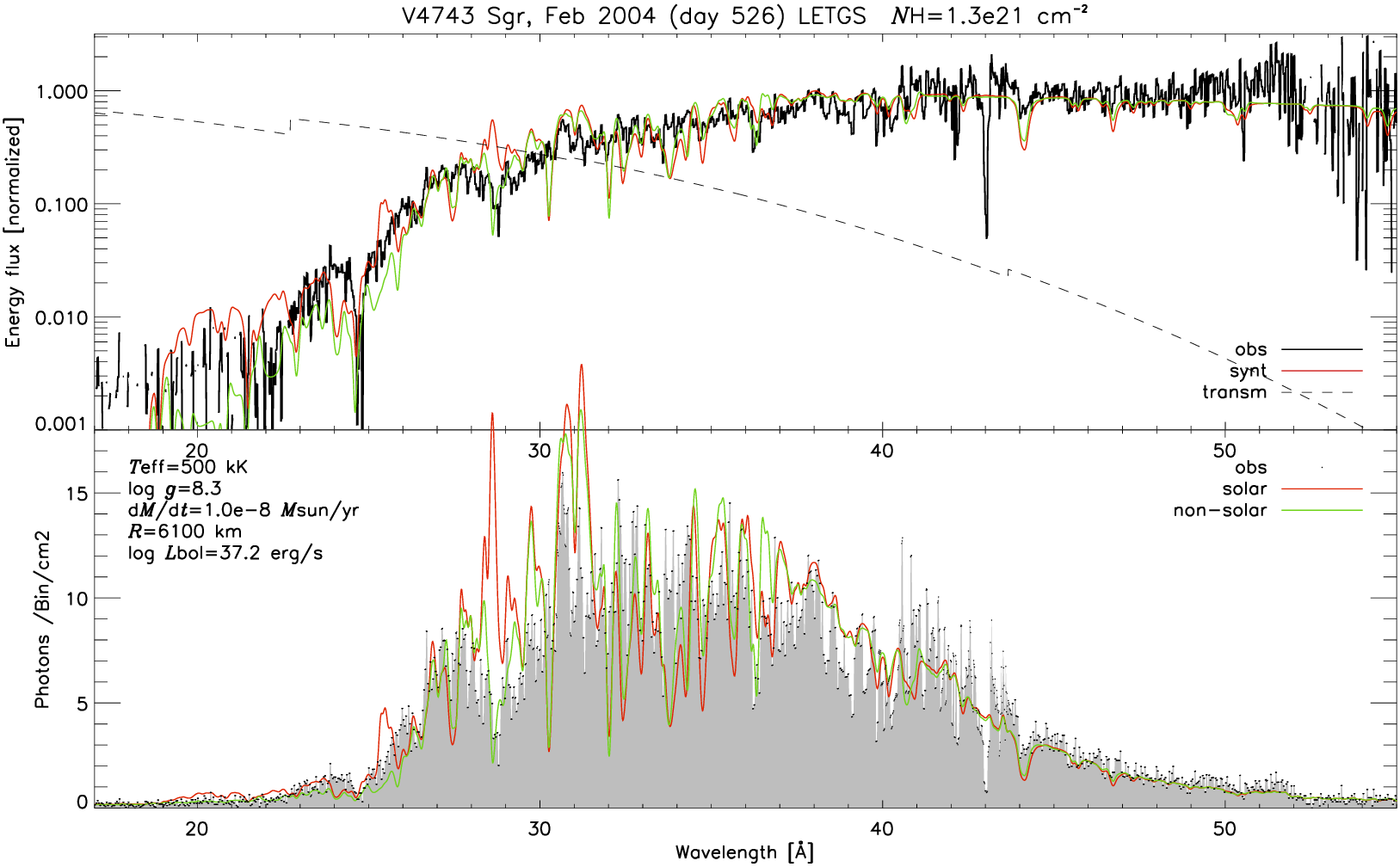}}
 \caption{ \label{fig:V4743FebN20O10} Expanding model}
\end{figure}
\end{landscape}

\end{appendix}

\backmatter
\clearpage
\addcontentsline{toc}{chapter}{References}
\bibliography{RT,Phoenix,TCorr,DATA,Nova}

\clearpage
\chapter*{Acknowledgements}
First, I am delighted to thank Prof. Peter Hauschildt for providing me with the very interesting topic, for the lots of good advice, especially in the last months, and for helping me with the successfull execution of this work.

I thank Prof. Eddie Baron for the very successful collaborations that we had. I look forward to meet you in Chicago.

I thank Prof. Sumner Starrfield for the short and inspiring talk we had in Tempe. There will be a lot of good nova work to do together.

A warm thanks to Knop, my very nice and very helpful roommate, I thank you for all the help you gave me, for taking the time to discuss problems with me.

Thanks to Jan-Uwe Ness, for lots of good collaboration, for providing me with observational data, for your hospitality, for the interesting insights into the scientific world and the nova community, for introducing me to people, motivating me to attend workshops, write papers etc.

Also I am very thankful to Prof. Rob Rutten for the incredibly inspiring graduate course and exercises in radiative transport that he gave in Utrecht 2001. While this was the most interesting course I ever had on university, he gave me the fundamental understanding that I'm still profited by.

I thank Dr. John Dombeck for his great idl color tables.

Then, I want to thank Oom Piet and Tante Cora for the many encouraging talks, for all their support for us, the good and useful advice out of long good experience in a situation like ours.

I thank my parents for all their very important advice for important decisions that were to be made lately, and in the past.

Finally, the person that I want to thank most is my lovely wife Silke, for supporting me with endless patience in the lots of work that was to be done in the last couple of years, months, weeks and days!

\clearpage
\section*{Erklaerung}
\thispagestyle{empty}
\vspace{2cm}
Hiermit versichere ich, Dani\"el R. van Rossum, die vorliegende Arbeit selbstaendig verfasst und nur unter Zuhilfenahme der angegebenen Quellen und Hilfsmittel angefertigt zu haben. Desweiteren erklaere ich mich mit dem Verleih und der Veroeffentlichung dieser Arbeit einverstanden.

\vspace{2cm}
\noindent\underline{\hspace{5cm}} \hspace{1cm} Hamburg, den 31.05.2006\\
Unterschrift

\end{document}